%% file: HardwarePaper.tex
\documentclass[a4paper,11pt]{article}
\pdfoutput=1

\usepackage{jinstpub}
\usepackage{graphicx}
\usepackage{hyperref}
\usepackage[binary-units = true]{siunitx}
\newcommand{\SIadj}[2]{\SI[number-unit-product={\text{-}}]{#1}{#2}} 

\usepackage{upgreek}
\usepackage{textgreek}
\usepackage[titletoc]{appendix}
\usepackage{multirow}
\usepackage{multicol}
\usepackage{longtable}
\usepackage{subfig}
\usepackage{wasysym}
\usepackage[super]{nth}
\usepackage{booktabs}
\usepackage{sidecap}
\usepackage[yyyymmdd]{datetime}

\usepackage{pifont}
\usepackage{color}

\usepackage{comment}

\usepackage[acronym,xindy,toc,nonumberlist]{glossaries}
\setacronymstyle{long-short}
\makenoidxglossaries

\usepackage[intoc,refpage,noprefix]{nomencl}
\makenomenclature

\newcommand{\figref}[1]{Figure~\ref{#1}}
\newcommand{\tabref}[1]{Table~\ref{#1}}
\newcommand{\eqnref}[1]{Equation~\ref{#1}}
\newcommand{\secref}[1]{Section~\ref{#1}}

\def\betael{$\upbeta$-electron}
\def\betaels{$\upbeta$-electrons}
\def\betadec{$\upbeta$-decay}
\def\betaspec{$\upbeta$-spectrum}

\def\pin{\textit{p-i-n}~}

\def\t2{$\text{T}_2$}
\def\d2{$\text{D}_2$}
\def\h2o{$\text{H}_2\text{O}$}
\def\rb{\textsuperscript{83}Rb}
\def\kr{\textsuperscript{83m}Kr}
\def \kthirtytwo {K-32}
\def \mthirtytwos {$\text{M}_2\text{M}_3-32$}
\def \lthreethirtytwo {$\text{L}_3-32$}
\def \lthreeninefour {$\text{L}_3-9.4$}
\def\am241{\textsuperscript{241}Am}

\DeclareSIUnit\year{yr}
\DeclareSIUnit\ppm{ppm} 
\DeclareSIUnit\cps{cps} 

\input{glossary-collection.tex}

\title{The Design, Construction, and Commissioning of the KATRIN Experiment}

\collaboration{
\includegraphics[height=17mm]{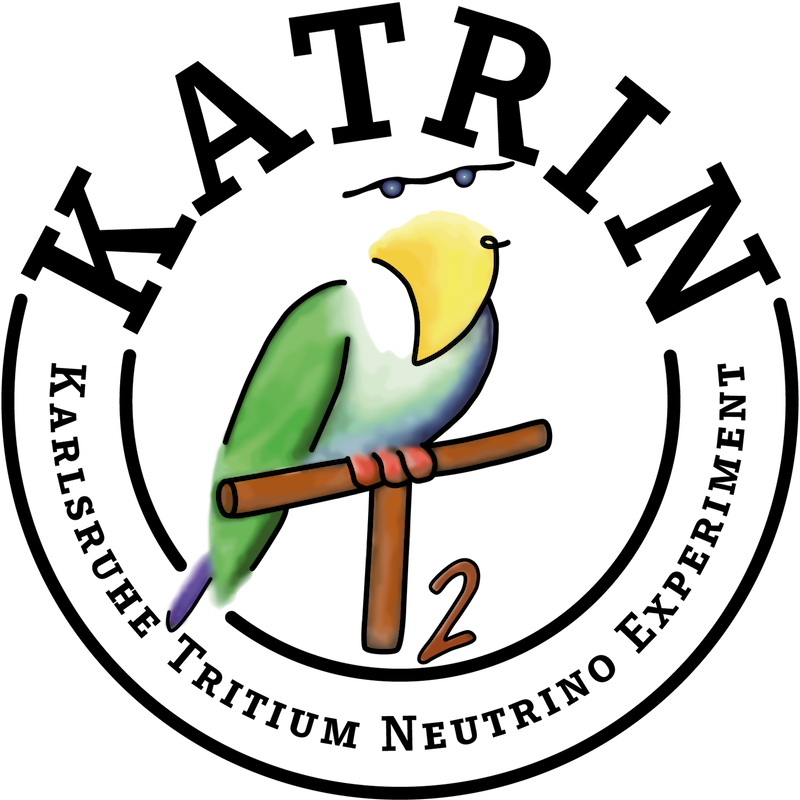}\\[6pt]
KATRIN collaboration}

\input{KATRINauthors2021hw.tex}
\emailAdd{thomas.thuemmler@kit.edu}

\abstract{
\glsresetall

The \acrfull{katrin} experiment, which aims to make a direct and model-independent determination of the absolute neutrino mass scale, is a complex experiment with many components.
More than 15 years ago, we published a \gls{tdr} \cite{KATRIN2005} to describe the hardware design and requirements to achieve our sensitivity goal of \SI{0.2}{\electronvolt} at 90\% C.L. on the neutrino mass.
Since then there has been considerable progress, culminating in the publication of first neutrino mass results with the entire beamline operating \cite{Aker2019-PRL}.
In this paper, we document the current state of all completed beamline components (as of the first neutrino mass measurement campaign), demonstrate our ability to reliably and stably control them over long times, and present details on their respective commissioning campaigns.
}


\begin{document}
\maketitle

\flushbottom

\input{Introduction}

\input{SystemOverview}
\input{WGTS}
\input{TransportSystem}
\input{SpectrometerSystem}
\input{DetectorSystem}
\input{CalibrationAndMonitoringSystem}
\input{DataManagement}
\input{Summary}

\acknowledgments

We are grateful to \acrfull{kit} for hosting and continuously supporting the \gls{katrin} experiment in many different ways.
We are deeply indebted to the staff of all the workshops at \gls{kit} and at all our other institutes for the great developments and constructions, without which this unique experiment would never have been possible.
We greatly appreciate the many services at \gls{kit}, such as project management, quality control, supply of liquid helium, etc.
We would like to thank \gls{kit}, all our institutions, the scientific community and the funding agencies, who have always believed in the success of this experiment, despite the delays of this difficult experiment, which was pushed to the limit of what is technically feasible.
\par
We would also like to add our heartfelt thanks for the invaluable contributions and guidance of collaborators who have passed away, including H.~Bichsel, J.~Bonn, B.~Freudiger, V.\,M.~Lobashev,  E.\,W.~Otten.
\par
We acknowledge the support of Helmholtz Association,
Ministry for Education and Research BMBF (5A17PDA, 05A17PM3, 05A17PX3, 05A17VK2, and 05A17WO3), Helmholtz Alliance for Astroparticle Physics (HAP),  Helmholtz Young Investigator Group (VH-NG-1055),
and Deutsche Forschungsgemeinschaft DFG (Research Training Groups GRK 1694 and GRK 2149, and Graduate School GSC 1085 - KSETA) in Germany;
Ministry of Education, Youth and Sport (CANAM-LM2015056, LTT19005) in the Czech Republic; Ministry of Science and Higher Education of the Russian Federation under contract 075-15-2020-778;
and the United States Department of Energy through grants  DE-FG02-97ER41020, DE-FG02-94ER40818, DE-SC0004036, DE-FG02-97ER41033, DE-FG02-97ER41041, DE-AC02-05CH11231, DE-SC0011091, and DE-SC0019304, and the National Energy Research Scientific Computing Center.

\printnoidxglossaries

\printglossary[type=\acronymtype,title=Abbreviations]

\bibliographystyle{JHEP}
\bibliography{HardwarePaper,RefArXivPublications,RefGeneralPublications,RefKATRINDiplomaMasterTheses,RefKATRINPhDTheses,RefKATRINPublications}

\end{document}

%% file: glossary-collection.tex
\newglossaryentry{katrin-g} {
  name={KATRIN},
  description={The KArlsruhe TRItium Neutrino \acrshort{katrin} experiment, which aims to make a precision measurement of the effective neutrino mass, to a design sensitivity of \SI{0.2}{\electronvolt} at 90\% C.L. after 5 years of taking data},
}
\newacronym[see={[Glossary:]{katrin-g}}]{katrin}{KATRIN}{KArlsruhe TRItium Neutrino \glsadd{katrin-g}}

\newglossaryentry{firsttritium}{
	name={FirstTritium},
	description={The first tritium measurement campaign of \gls{katrin} was performed over a few weeks during spring 2018. Uses only small amounts of tritium, amounting to 0.5\% tritium atoms in a \d2{} carrier gas. The goal was to verify that all safety measures were in place, and perform some commissioning.\cite{Aker2020_firstOperationTritium}}
	}
\newglossaryentry{knm1}{
	name={KNM1},
	description={The first neutrino mass measurement campaign of \gls{katrin} was performed over a 4-week period during spring 2019.\cite{KNM1-PRL-2019}}
	}

\newacronym{rs}{RS}{Rear System}
\newacronym{cms}{CMS}{Calibration and Monitoring System}
\newacronym{wgts}{WGTS}{Windowless Gaseous Tritium Source}
\newacronym{cps}{CPS}{Cryogenic Pumping Section}
\newacronym{dps}{DPS}{Differential Pumping Section}
\newacronym{ps}{PS}{Pre-Spectrometer}
\newacronym{ms}{MS}{Main Spectrometer}
\newacronym{ie}{IE}{inner wire electrode system}
\newacronym{fpd}{FPD}{Focal Plane Detector}
\newacronym{mos}{MoS}{Monitor Spectrometer}
\newacronym{rw}{RW}{Rear Wall}

\newacronym{lara}{LARA}{Laser Raman Spectroscopy}
\newacronym{bixs}{BIXS}{$\upbeta$-Induced X-Ray Spectroscopy}
\newacronym{fbm}{FBM}{Forward Beam Monitor}
\newacronym{uhv}{UHV}{ultra-high vacuum}
\newacronym{tmp}{TMP}{turbo-molecular pump}
\newacronym{neg}{NEG}{non-evaporable getter}
\newacronym{pp2}{PP2}{Pump Port 2}
\newacronym{pp5}{PP5}{Pump Port 5}
\newacronym{bt1}{BT1}{Beam Tube 1}
\newacronym{bt4}{BT4}{Beam Tube 4}
\newacronym{bt5}{BT5}{Beam Tube 5}
\newacronym{v4}{V4}{Valve 4} 
\newacronym{rscm}{RSCM}{Re-condensing Superconducting Magnet}
\newacronym{sdd}{SDD}{Silicon Drift Detector}
\newacronym{fticr}{FT-ICR}{Fourier Transform Ion Cyclotron Resonance}
\newacronym{pulcinella}{PULCINELLA}{Precision Ultra-Low Current Integrating Normalization Electrometer for Low-Level Analysis}
\newacronym{hopg}{HOPG}{Highly-Oriented Pyrolytic Graphite}
\newacronym{egun}{e-gun}{electron gun}
\newacronym{ckrs}{CKrS}{Condensed \kr{} Source}
\newacronym{gkrs}{GKrS}{Gaseous \kr{} Source}
\newacronym{ap}{AP}{analyzing plane}
\newacronym{pae}{PAE}{Post Acceleration Electrode}
\newacronym{orca}{ORCA}{Object oriented Real-time Control and Acquisition}
\newacronym{sipm}{SiPM}{Silicon Photomultipliers}
\newacronym{wls}{WLS}{Wave-Length Shifting}
\newacronym{adei}{ADEI}{Advanced Data Extraction Infrastructure}
\newacronym{pcs7}{PCS7}{SIMATIC PCS 7 distributed control system}
\newacronym{bora}{BORA}{personalized collaBORAtive data display}
\newacronym{beans}{BEANS}{Building Elements for ANalysis Sequence}
\newacronym{kali}{KaLi}{KATRIN Library}
\newacronym{kaas}{KaaS}{KATRIN as a Service}
\newacronym{json}{JSON}{JavaScript Object Notation}
\newacronym{xml}{XML}{eXtensible Markup Language}
\newacronym{kaffee}{KAFFEE}{Katrin Automation Framework for Fpd Examination and Evaluation}
\newacronym{idle}{IDLE}{Intermediate Data Layer for Everyone}
\newacronym{roast}{ROAST}{Realtime Orca Analysis on STreaming}
\newacronym{lfcs}{LFCS}{Low-Field Correction System}
\newacronym{ps1}{PS1}{Pre-Spectrometer magnet 1}
\newacronym{ps2}{PS2}{Pre-Spectrometer magnet 2}
\newacronym{pch}{PCH}{"Pinch magnet"}
\newacronym{det}{DET}{"Detector magnet"}
\newacronym{daq}{DAQ}{Data acquisition}
\newacronym{opc}{OPC}{Open Platform Communication}
\newacronym{crio}{cRIO}{compact Reconfigurable Input Output}
\newacronym{soap}{SOAP}{Simple Object Access Protocol}
\newacronym{rest}{REST}{Representational State Transfer}
\newacronym{sql}{SQL}{Structured Query Language}
\newacronym{emcs}{EMCS}{Earth Magnetic field Compensation System}
\newacronym{ldls}{LDLS}{Laser-Driven Light Source}
\newacronym{psu}{PSU}{Power Supply Unit}
\newacronym{il}{IL}{Inner Loop}
\newacronym{ol}{OL}{Outer Loop}
\newacronym{iss}{ISS}{Isotope Separation System}
\newacronym{bte}{BTE}{beam tube element}
\newacronym{ppe}{PPE}{pump port element}
\newacronym{zeus}{ZEUS}{Central DAQ and Control System}
\newacronym{hmi}{HMI}{Human-Machine Interface}
\newacronym{api}{API}{Application Programming Interface}
\newacronym{kdb}{KDB}{KATRIN Database}
\newacronym{flt}{FLT}{First Level Trigger}
\newacronym{usb}{USB}{Universal Serial Bus}
\newacronym{sbc}{SBC}{Single Board Computer}
\newacronym{http}{HTTP}{Hypertext Transfer Protocol}
\newacronym{tilo}{TILO}{Test of Inner Loop}
\newacronym{rmms}{RMMS}{Radial Magnetic Monitoring System}
\newacronym{vmms}{VMMS}{Vertical Magnetic Monitoring System}
\newacronym{pp2f}{PP2F}{Pump Port 2F}

\newacronym{ln2}{LN2}{liquid nitrogen}
\newacronym{lhe}{LHe}{liquid helium}

\newacronym{mtd}{MTD}{Measurement Time Distribution}
\newacronym{mace}{MAC-E}{Magnetic Adiabatic Collimation with Electrostatic}
\newacronym{fsd}{FSD}{Final State Distribution}
\newacronym{tof}{ToF}{time-of-flight}

\newacronym{tdr}{TDR}{technical design report}
\newacronym{dvm}{DVM}{precision digital voltmeter}
\newacronym{scs}{SCS}{slow control system}
\newacronym{tlk}{TLK}{Tritium Laboratory Karlsruhe}
\newacronym{kit}{KIT}{Karlsruhe Institute of Technology}
\newacronym{npi}{NPI}{Nuclear Physics Institute}

%% file: KATRINauthors2021hw.tex
\affiliation[a]{Institute for Astroparticle Physics~(IAP), Karlsruhe Institute of Technology~(KIT), Hermann-von-Helmholtz-Platz 1, 76344 Eggenstein-Leopoldshafen, Germany}
\affiliation[b]{Technische Universit\"{a}t M\"{u}nchen, James-Franck-Str. 1, 85748 Garching, Germany}
\affiliation[c]{IRFU (DPhP \& APC), CEA, Universit\'{e} Paris-Saclay, 91191 Gif-sur-Yvette, France }
\affiliation[d]{Center for Experimental Nuclear Physics and Astrophysics, and Dept.~of Physics, University of Washington, Seattle, WA 98195, USA}
\affiliation[e]{Helmholtz-Institut f\"{u}r Strahlen- und Kernphysik, Rheinische Friedrich-Wilhelms-Universit\"{a}t Bonn, Nussallee 14-16, 53115 Bonn, Germany}
\affiliation[f]{Institute of Experimental Particle Physics~(ETP), Karlsruhe Institute of Technology~(KIT), Wolfgang-Gaede-Str. 1, 76131 Karlsruhe, Germany}
\affiliation[g]{Institut f\"{u}r Kernphysik, Westf\"{a}lische Wilhelms-Universit\"{a}t M\"{u}nster, Wilhelm-Klemm-Str. 9, 48149 M\"{u}nster, Germany}
\affiliation[h]{Institute for Data Processing and Electronics~(IPE), Karlsruhe Institute of Technology~(KIT), Hermann-von-Helmholtz-Platz 1, 76344 Eggenstein-Leopoldshafen, Germany}
\affiliation[i]{Institute for Nuclear Research of Russian Academy of Sciences, 60th October Anniversary Prospect 7a, 117312 Moscow, Russia}
\affiliation[j]{Max-Planck-Institut f\"{u}r Kernphysik, Saupfercheckweg 1, 69117 Heidelberg, Germany}
\affiliation[k]{Institute for Technical Physics~(ITEP), Karlsruhe Institute of Technology~(KIT), Hermann-von-Helmholtz-Platz 1, 76344 Eggenstein-Leopoldshafen, Germany}
\affiliation[l]{Max-Planck-Institut f\"{u}r Physik, F\"{o}hringer Ring 6, 80805 M\"{u}nchen, Germany}
\affiliation[m]{Department of Physics and Astronomy, University of North Carolina, Chapel Hill, NC 27599, USA}
\affiliation[n]{Triangle Universities Nuclear Laboratory, Durham, NC 27708, USA}
\affiliation[o]{Permanent address: Laboratory of Optoelectronics and Devices, Physics Department, Faculty of Sciences, University of Setif-1 (19000), Algeria}
\affiliation[p]{Department of Physics, Faculty of Mathematics and Natural Sciences, University of Wuppertal, Gau\ss{}str. 20, 42119 Wuppertal, Germany}
\affiliation[q]{Departamento de Qu\'{i}mica F\'{i}sica Aplicada, Universidad Autonoma de Madrid, Campus de Cantoblanco, 28049 Madrid, Spain}
\affiliation[r]{Nuclear Physics Institute of the CAS, v. v. i., CZ-250 68 \v{R}e\v{z}, Czech Republic}
\affiliation[s]{Laboratory for Nuclear Science, Massachusetts Institute of Technology, 77 Massachusetts Ave, Cambridge, MA 02139, USA}
\affiliation[t]{Department of Physics, Carnegie Mellon University, Pittsburgh, PA 15213, USA}
\affiliation[u]{Department of Physics, Swansea University, Swansea SA2 8PP, UK}
\affiliation[v]{Institute for Nuclear and Particle Astrophysics and Nuclear Science Division, Lawrence Berkeley National Laboratory, Berkeley, CA 94720, USA}
\affiliation[w]{University of Applied Sciences~(HFD)~Fulda, Leipziger Str.~123, 36037 Fulda, Germany}
\affiliation[x]{School of Physics and Astronomy, Cardiff University, Cardiff, Wales, CF24 3AA, UK}
\affiliation[y]{Department of Physics, Case Western Reserve University, Cleveland, OH 44106, USA}
\affiliation[z]{Department of Physics, University of California at Santa Barbara, Santa Barbara, CA 93106, USA}
\affiliation[a|]{Institut f\"{u}r Physik, Humboldt-Universit\"{a}t zu Berlin, Newtonstr. 15, 12489 Berlin, Germany}
\affiliation[b|]{Project, Process, and Quality Management~(PPQ), Karlsruhe Institute of Technology~(KIT), Hermann-von-Helmholtz-Platz 1, 76344 Eggenstein-Leopoldshafen, Germany    }

\author[a]{M.~Aker,}
\author[b,c]{K.~Altenm\"{u}ller,}
\author[d]{J.~F.~Amsbaugh,}
\author[e]{M.~Arenz,}
\author[a,f]{M.~Babutzka,}
\author[a]{J.~Bast,}
\author[g]{S.~Bauer,}
\author[a]{H.~Bechtler,}
\author[g]{M.~Beck,}
\author[h]{A.~Beglarian,}
% ...

%% file: Introduction.tex
\section{Introduction}
\label{sec:introduction}


The determination of the neutrino mass plays an important role in cosmology, particle physics, and astroparticle physics.
The investigation of the energy spectrum of tritium \betadec{} currently is the most model-independent method for addressing this topic and gives the most sensitive result \cite{Aker2019-PRL}.

Owing to neutrino flavor mixing, the effective neutrino mass $m_\mathrm{\nu}$ appears as an incoherent sum of the neutrino mass eigenstates contributing to the electron neutrino.
Furthermore, the signature of the neutrino mass is maximal at a few eV below the kinematic endpoint of the tritium \betaspec{} at $E_0 \approx \SI{18.6}{\kilo\electronvolt}$ \cite{Otten2008, Drexlin2013, formaggio2021direct}.
Based on the experience of its predecessor experiments in Mainz \cite{Kraus2005} and Troitsk \cite{Aseev2011}, which determined an upper bound of $m_{\upnu} < \SI{2}{\electronvolt}$ \cite{Olive2014}, the \gls{katrin} experiment aims to improve the sensitivity on  $m_{\upnu}$ by one order of magnitude to \SI{0.2}{\electronvolt} \cite{KATRIN2005}.

\gls{katrin} uses an ultra-luminous, \acrfull{wgts} and an integrating spectrometer of \acrfull{mace} filter type \cite{Lobashev1985, Picard1992} for the energy analysis of the \betadec{} electrons.


This paper, which builds on the original \acrfull{tdr} \cite{KATRIN2005}, will focus on hardware implementation and commissioning results, and various systems for calibration and monitoring. It describes the status of all systems as they were during the first neutrino mass measurement campaign.

The hardware implementation is described in the corresponding sections below, which include the source, transport, spectrometer, and detector sections.
Each of these has undergone its respective commissioning phase.
The level of detail in the individual sections varies depending on the availability of previously published commissioning reports, which will be cited accordingly. 

For a complex system like the \gls{katrin} experiment, the ability to precisely calibrate and monitor its many subsystems is imperative.
There have been several dedicated measurement campaigns over the years to hone different calibration techniques using electron sources, as well as more recent campaigns which focused on building a better understanding of ion safety and plasma effects. 
Additional systems for monitoring every operating parameter --- from the source gas composition and activity to the adjustment of the spectrometer's high voltage --- have successfully demonstrated the \gls{katrin} experiment's ability to meet its design requirements for stability and control.

%% file: SystemOverview.tex
\section{System Overview}
\label{Ch:SystemOverview}

\input{KATRINSetup}

\input{KATRINKeyParameters}

\input{KATRINMeasurementScenarios}

\clearpage

%% file: KATRINSetup.tex
\subsection{Setup of the KATRIN experiment}
\label{Subsection:KATRINSetup}

\begin{figure}[t]
	\centering
	\includegraphics[width=1\textwidth]{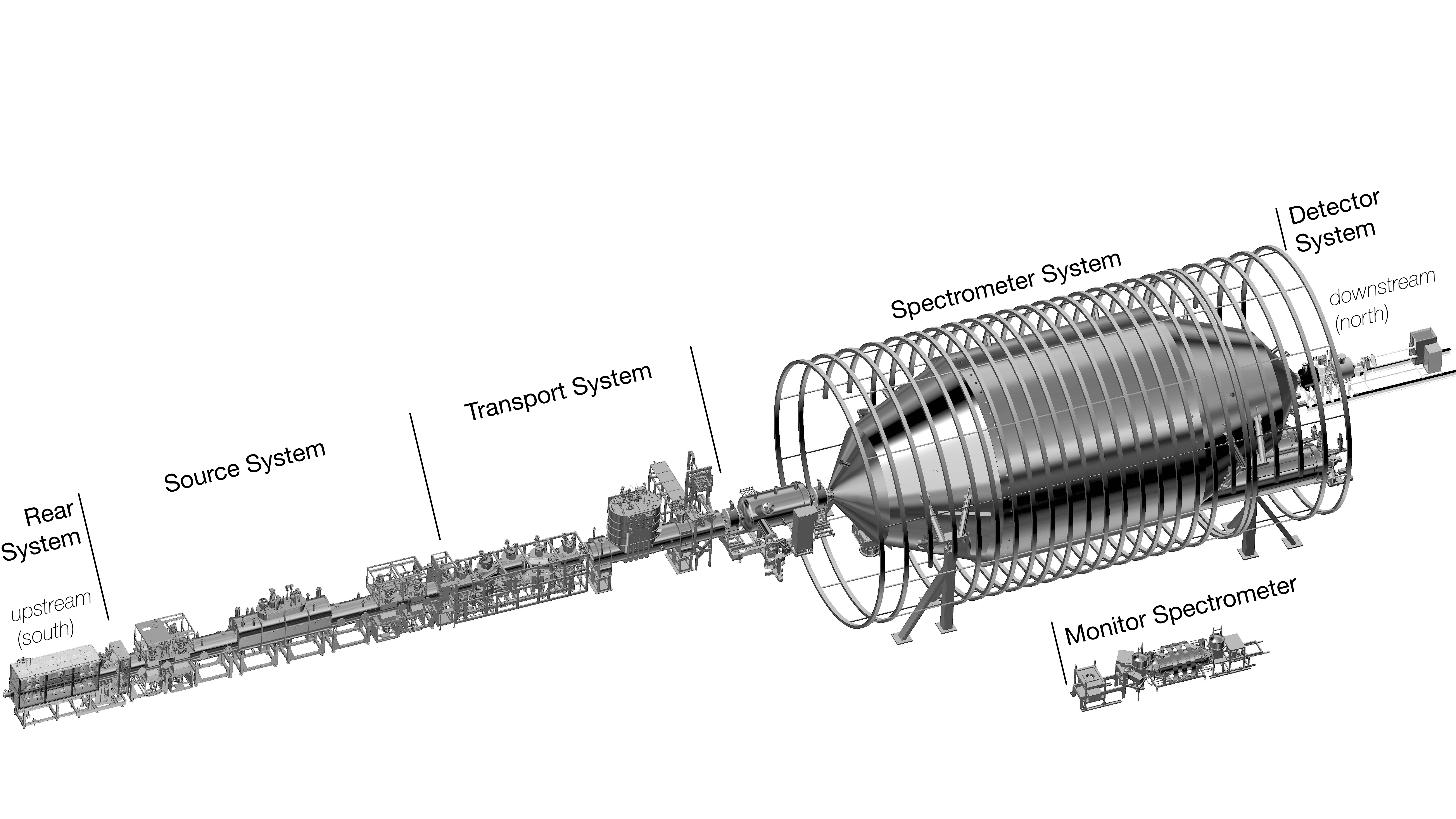}
	\caption{Schematic view of the \gls{katrin} experimental setup. The overall length of the setup is about \SI{70}{m}. The \betaels{} from the tritium source are guided to the spectrometer by the electromagnetic field along the beamline. Electrons which have passed the spectrometer are then counted by the detector. Refer to \secref{SubSection:MonitorSpectrometer} for details about the \gls{mos} in its parallel beamline.
	}
	\label{Figure:KATRIN-Setup}
\end{figure}

The \gls{katrin} tritium \betadec{} experiment consists of five main systems: the source system, the transport system, the spectrometer system, the detector system, and the rear system (\figref{Figure:KATRIN-Setup}) . This section gives a short overview of the tasks the different systems have to perform. Detailed information, including references, can be found in subsequent sections. 

\begin{description}

	\item[Source system]
		The source system (\secref{sec:source_system}) delivers up to \SI{e11}{} tritium \betaels{} per second with high stability.
Its main purpose is to provide the \gls{wgts}.		
		It consists of the source magnet-cryostat system with a 10-m-long source tube, the \acrfull{rw} \footnote{For technical reasons is the \gls{rw} located in the glovebox of the \gls{rs} (\figref{Figure:RearSystem})}, the \gls{bixs} system for monitoring the tritium source strength, and the tritium inner loop system. The inner loop system processes up to \SI{40}{g} of pure tritium per day (\secref{sec:Loop-overview}). 
		
	\item[Transport system] 	
		The transport system (\secref{sec:transport_system}) transports \betaels{} adiabatically from the source to the spectrometer system. The tritium flow rate between the source and the spectrometer system is reduced by more than 12 orders of magnitude to keep the detector background rate from tritium \betadec{} in the \gls{ms} below \SI{1e-3}{cps}. The reduction is achieved by a combination of differential and cryogenic pumping, the \gls{dps} and the \gls{cps}. Whereas the former uses a combination of \glspl{tmp} to reduce tritium flow by 5 orders of magnitude, the latter is a cold argon frost system held at \SIrange{3}{4}{K}, to further reduce tritium flow by more than 7 orders of magnitude.

	\item[Spectrometer system]
		The spectrometer system (\secref{sec:spectrometer_system}) consists of two large-volume \acrshort{mace} filter spectrometers: the \acrfull{ps} and the \acrfull{ms}. In addition, there is the \acrfull{mos} in a parallel beamline (\secref{SubSection:MonitorSpectrometer}). The \acrshort{ps} can be configured to work as a pre-filter.
		Setting its retarding voltage down to \SI{-18.3}{\kilo\volt} it can reject all \betaels{} from the tritium source except those in the region of interest close to the \betaspec{} endpoint $E_0$.
Due to its minimum magnetic field of \SI{20}{\milli\tesla}, the \acrshort{ps} transports electrons adiabatically at all retarding potential settings, even grounded.
The \acrshort{ms} is designed to analyze the energy of the tritium \betaels{} with a resolution of \SI{\sim 1}{eV}.
However, in the standard neutrino mass measurement mode, the energy resolution is set to \SI{2.77}{eV} for \SIadj{18.6}{\kilo\electronvolt} electrons, see \secref{Subsection:KATRINKeyParameters}.

	\item[Detector system]
		The detector system (\secref{sec:detector_system}) counts all \betaels{} and background electrons which pass through the \gls{ms}, and enables systematic investigations of the whole \gls{katrin} experiment. 
Its main component is the \gls{fpd}.		
		It can handle high count rates (on the order \SI{1}{\mega cps} over the entire detector) necessary for test and calibration measurements with sources such as \kr, as well as a timing resolution of $<$ \SI{100}{ns} for time-of-flight mode measurements. 

\begin{figure}[b!]
	\centering
	\includegraphics[width=0.9\textwidth]{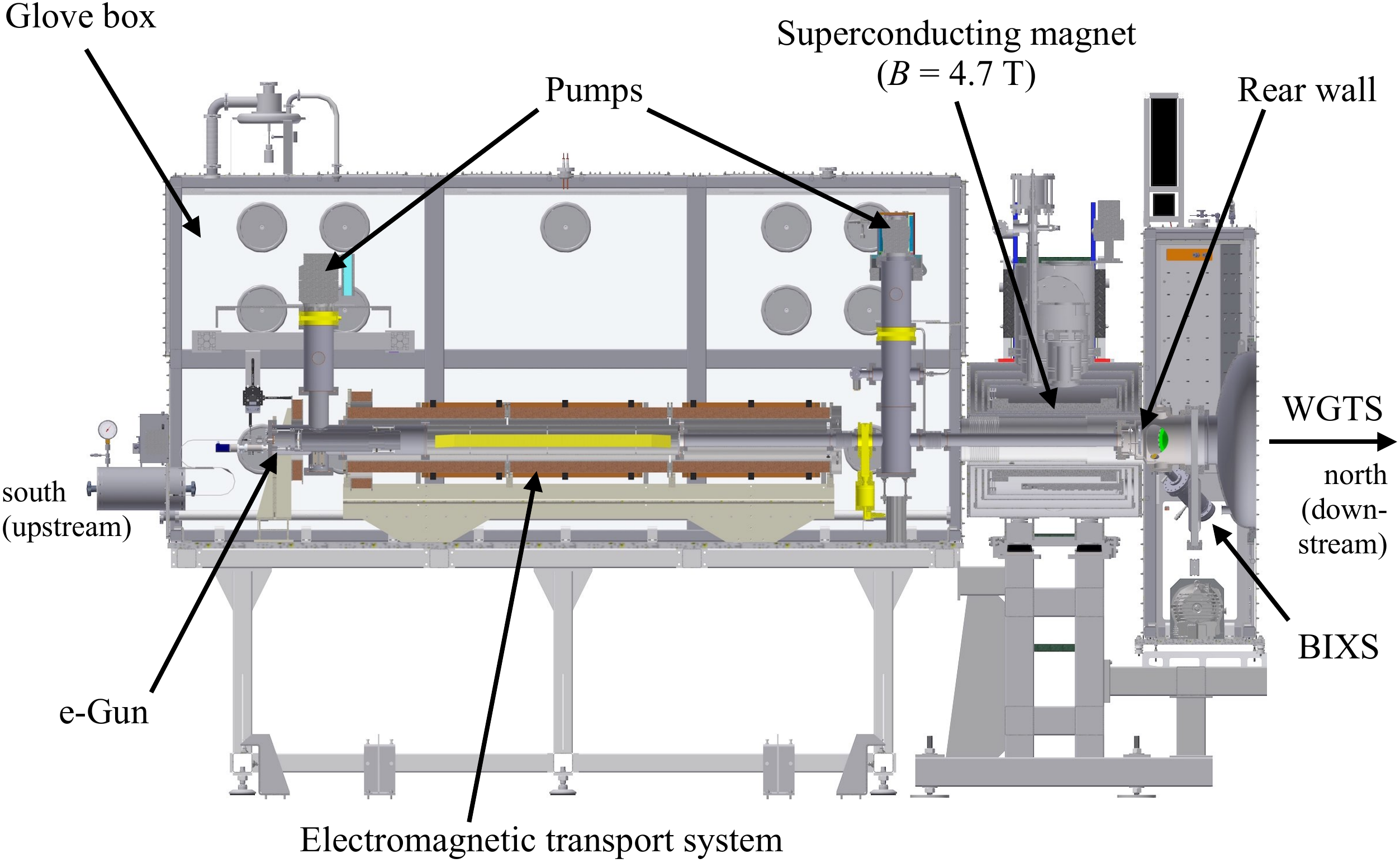}
	\caption{CAD drawing of the \gls{rs} seen from the eastern side. This is the upstream end of the experiment, facing downstream (north) towards the \gls{wgts}.  The two main components are shown here: \gls{egun} system and the superconducting magnet. In addition, the \acrfull{rw} which separates the \gls{rs} from the \gls{wgts}, and the \gls{bixs} system which monitors the tritium source strength by measuring the X-rays from the \gls{rs} are displayed.  The parts which have contact to tritium are enclosed in a glovebox.   
}
	 	\label{Figure:RearSystem}
\end{figure}

	\item[Rear system]
		The rear system is located upstream of source system (\figref{Figure:RearSystem}). It consists of an \gls{egun} system for calibration and monitoring purposes (\secref{sec:calibration_and_monitoring_systems}), and a superconducting magnet to guide the \gls{egun} electrons into the \gls{wgts}. The rear system enables the monitoring of crucial operation parameters, such as source gas composition, source gas activity, and \gls{ms} high voltage stability. 
\end{description}

Numerous interplaying parameters and processes contribute to the overall performance of the \gls{katrin} experiment, which needs to be operated in a reliable, stable and reproducible way over a measurement period of several years. In order to achieve this, both a distributed calibration and monitoring system (\secref{sec:calibration_and_monitoring_systems}) and a complex data and control infrastructure (\secref{Subsection:SCAndRunControl}) have been developed, installed, and commissioned.

%% file: KATRINKeyParameters.tex
\subsection{Key Parameters of the KATRIN Experiment}
\label{Subsection:KATRINKeyParameters}

The key parameters of the experiment were specified such that there are comparable contributions of statistical ($\sigma_\mathrm{stat}$ =  \SI{0.018}{\electronvolt\squared}) and systematic ($\sigma_\mathrm{syst,tot}$ =  \SI{0.017}{\electronvolt\squared}) uncertainties to the neutrino mass squared $m_{\overline{\nu}}^2$ after 3 full years of beam time.

The key operation parameters of the \gls{katrin} experiment are summarized in \tabref{table:KATRINKeyParameters}. The following list provides a short description of each parameter:

\begin{itemize}
	\item Column density $\rho d$: The integrated number of tritium molecules along the beam tube axis per unit of cross-sectional area. 
	\item Tritium purity $\varepsilon_{\text{T}}$: The fraction of tritium atoms in the gas before injection into the \gls{wgts},  
	\begin{equation}
		\label{eq:epsT}
		\varepsilon_{\text{T}} = \frac{N_{\text{T}_2} + \frac{1}{2}(N_{\text{DT}} + N_{\text{HT}})} {  N_{\text{T}_2} + N_{\text{DT}} + N_{\text{D}_2} + N_{\text{HT}} + N_{\text{HD}} + N_{\text{H}_2 }},
		\end{equation}

	where $N_{\text{\{XY\}}}$ is the number of molecules of a given isotopologe, with \{X,Y\} being \{tritium (T), deuterium (D) or hydrogen (H)\}.

	\item HT/DT ratio $\kappa$:
	\begin{equation}
		\label{eq:kappa}
		\kappa=\frac{N_{\textrm{HT}}}{N_{\textrm{DT}}}.
	\end{equation}
	
	\item Source temperature $T_{\text{S}}$: The temperature of the source beam tube. 

	\item Source tube temperature stability $\sigma_{\text{T,S}}$: The stability of the source tube temperature over time.

	\item Source magnetic field $B_{\text{S}}$: The magnetic field inside the source tube. This is the starting magnetic field of the \betaels{}.

	\item Transported magnetic flux $\Phi$: The magnetic flux which has to be guided from the \gls{wgts} through the whole system to the detector free of any obstacles.
	\begin{equation}
		\label{eq:fluxtube}
		\Phi = \int \vec{B} \cdot \,\mathrm{d} \vec{A}
	\end{equation}
	
	where $\vec{B}$ is the magnetic field and $\vec{A}$ the surface element. The transmission of electrons from the \gls{egun} and krypton sources, and hence the correct alignment of the entire beam line, has been successfully achieved in the First Light commissioning campaign 2016 \cite{Arenz2018}.

	\item Maximum magnetic field $B_{\text{max}}$: The maximum value of the magnetic field located at the pinch magnet at the downstream end of the \gls{ms}. The ratio of $B_{\text{S}}$ to $B_{\text{max}}$ defines the maximum accepted pitch angle $\theta_{\text{max}}$ (see \eqnref{eq:thetaMax} and \secref{subsec:specFilterPrinciple}).

	\item Retarding voltage monitoring precision $\sigma_\mathrm{HV}$: The precision to which the spectrometer's retarding voltage can be measured.

	\item Spectrometer energy resolution $\Delta \text{E}$: The energy resolution for a \gls{mace} filter configuration is the amount of energy, which remains in the cyclotron motion even after adiabatic collimation of a transmitted \betael{} from an isotropic source. See \secref{subsec:specFilterPrinciple} for details. 

	\item Background $\text{B}$: The detector signals that are produced by processes other than \betaels{}.
\end{itemize}

\begin{table}[!ht]
\caption{Key operation parameters of the \gls{katrin} experiment. The first column represents the values given in the \gls{tdr} \cite{KATRIN2005}. The second column shows the current standard \gls{katrin} operational settings, where those values marked with a \textbf{*} were determined during commissioning measurements.
 }
\begin{center}
\begin{tabular}{lcc}
\hline
\textbf{Parameter} & \textbf{Design Report} & \textbf{Standard Setting}  \\
\hline
Column density $\rho d$ & \SI{5e17}{\per\centi\meter\squared}  & \SI{5e17}{\per\centi\meter\squared} \\
Tritium purity $\varepsilon_{\text{T}}$ & \SI{> 95}{\percent} & \SI{> 95}{\percent}\textbf{*}\\
Source temperature $T_{\text{S}}$ & \SI{27}{\kelvin}  & \SI{30}{\kelvin}\textbf{*} \\
Source tube temperature stability $\sigma_{\text{T,S}}$ & \SI{< 0.1}{\percent\per\hour} &  \SI{< 0.1}{\percent\per\hour}\textbf{*} \\
Source magnetic field $B_{\text{S}}$ & \SI{3.6}{\tesla} & \SI{2.52}{\tesla} \\
Transported magnetic flux $\Phi$& \SI{191}{\tesla\centi\meter\squared} &  \SI{134}{\tesla\centi\meter\squared}   \\
Maximum magnetic field $B_{\text{max}}$ & \SI{6}{\tesla} & \SI{4.2}{\tesla} \\
\hline
Retarding voltage monitoring precision $\sigma_{HV}$ & \SI{<3e-6}{} & \SI{<3e-6}{}\\ 
Spectrometer energy resolution $\Delta \text{E}$ & \SI{0.93}{\electronvolt} & \SI{2.77}{\electronvolt} \textbf{*} \\
Background $\text{B}$ & \SI{10e-3}{cps} & \SI{300e-3}{cps}\textbf{*} \\
\hline
\end{tabular}
\end{center}
\label{table:KATRINKeyParameters}
\end{table}

The most notable differences compared to the specifications from the design report \cite{KATRIN2005} are a global \SI{70}{\percent} scaling of the magnetic fields and an increased background. The magnetic field reduction was necessary due to the technical limitations of the \gls{cps} magnet system (\secref{Subsection:CryogenicPumpingSystem}). The increased background is due to a new background source which was not observed in previous, less sensitive experiments. The reduction of this background was achieved at the cost of a worsened \gls{ms} energy resolution. The optimal trade-off between background and energy resolution with respect to the neutrino mass sensitivity was found for a background level around \SI{300e-3}{cps} at an energy resolution
of \SI{2.77}{\electronvolt} for \SIadj{18.6}{\kilo\electronvolt} electrons.

%% file: KATRINMeasurementScenarios.tex
\subsection{Measurement Modes}
\label{Subsection:KATRINMeasurementScenarios}
\label{subsec:measurement_scenarios}

The \gls{katrin} experimental setup can be operated with different control parameters that allow various measurement modes, including standard neutrino-mass runs, regular calibration measurements, and dedicated studies on systematic effects. Between measurement phases, the experiment is put into maintenance mode typically at least once per year.

\begin{figure}[t!]
	\begin{center}
	\includegraphics[width=0.9\textwidth]{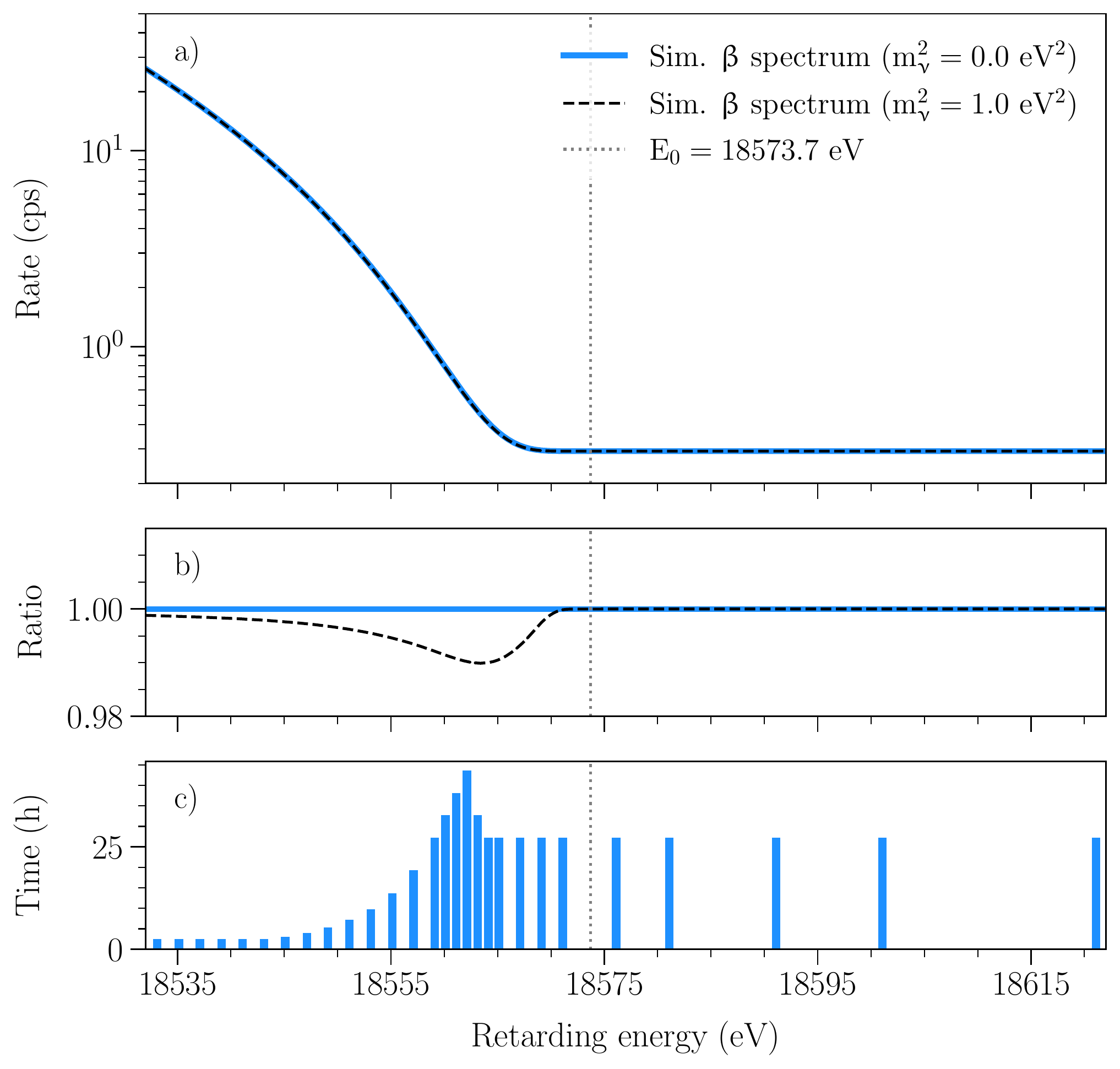}
	\end{center}
	\caption{Scanning scheme and \gls{mtd} for measuring a continuous spectrum with a \gls{mace} filter.
Plot (a) shows two simulated integrated electron energy spectra in a \SIadj{90}{\electronvolt} wide energy window around the endpoint $\mathrm{E_0}$ of tritium \betadec{}.
Simulation input parameters and endpoint energy value are taken from \cite{Aker2019-PRL}.
The two different spectra are compared near the \betadec{} endpoint: one with a squared neutrino mass of \SI{1.0}{\electronvolt^2}, and one with \SI{0.0}{\electronvolt^2}.  
Their ratio is shown in plot (b).
The \gls{mtd} in plot (c) shows how the amount of time spent at each retarding potential is defined, which is needed to discriminate both spectra with the highest sensitivity.
It is optimized for a neutrino mass sensitivity of \SI{1}{\electronvolt}.
The region with the largest amount of time allotted per retarding potential corresponds to the energy range where the neutrino mass causes the largest deviation from the zero-mass spectrum.
}
	\label{fig:sample_mtd}
\end{figure}

\paragraph{Neutrino-mass mode}
This is the standard mode of operation to continually adjust the retarding voltage of the \gls{ms} in the range of $[E_0 - \SI{40}{\electronvolt}; E_0 + \SI{50}{\electronvolt}]$ while tritium is in the system.
This scanning range can be adjusted if required.
The voltage and the time spent at each setting are defined by the \gls{mtd} (\figref{fig:sample_mtd}).
A typical run at a given voltage lasts between \SI{20}{\second} and \SI{600}{\second}; a full scan of the energy range given above takes about \SI{2}{\hour}.

Of these standard neutrino-mass runs, a small portion will be dedicated to sterile neutrino searches. These searches involve scanning much farther (order of \SI{}{\kilo\electronvolt}) below the endpoint $E_0$.

\paragraph{Calibration mode}
To check the long-term system stability, calibration measurements are done regularly.
The neutrino-mass mode is suspended for the duration of these measurement:
\begin{itemize}
    \item An energy calibration of the \gls{fpd} (\secref{sec:detector_system}) is performed weekly, which requires closing off the detector system from the main beamline for about \SI{4}{\hour}.
    \item The offset and the gain correction factor of the low-voltage readout in the high-voltage measurement chain needs to be calibrated based on standard reference sources (\secref{subsec:HV-Distribution-Monitoring}). This requires stopping the precision monitoring of the \gls{ms} retarding potential twice per week for about \SI{0.5}{\hour} each.  
    \item The column density of the \gls{wgts} (\secref{sec:source_system}) is determined by a dedicated calibration electron source. The \gls{ms} is set to a fixed retarding potential while the energy of the dedicated electron source is varied in a \SIadj{200}{\electronvolt} range.
This measurement takes about \SI{2}{\hour}, typically once per week.
\end{itemize}

\paragraph{Systematic-effects modes}
In addition to the regular neutrino-mass and calibration modes, the system is also operated under special conditions to study various systematic effects.
These measurements were performed during system commissioning and are not part of the typical operation of the experiment:
\begin{itemize}
    \item The investigations of the \gls{ms} background require closing off the spectrometer and detector sections from the source and transport sections. The measurement is then performed at a fixed spectrometer voltage, typically for several hours. Multiple runs are often combined to accumulate sufficient statistics.
   Further details on these backgrounds, as well as their mitigation, are discussed later in the paper (\secref{subsection:MSbackgroundMitigation}).
    \item The energy loss of signal electrons due to scattering in the source section \cite{Kleesiek2019} is studied by a mono-energetic photoelectron source at the \gls{rs}.
    These electrons are used as probes to determine the loss at different energies, and are used to form an \emph{in situ} measurement of the energy loss function (\secref{sec:egun}).
    A typical measurement set takes about \SI{12}{\hour} to collect; data from several sets are combined in the analysis. 
    \item Plasma effects, which influence the potential seen by \betaels{} in the \gls{wgts}, were explored using the  \gls{gkrs} in the \gls{wgts} (\secref{Subsubsection:GKrS}). These investigations demonstrated how the \gls{rw} can be used to control and stabilize the plasma effects (\secref{sec:wgts_rear_wall}).
    \item The detection and blocking of ions in the beamline is of vital importance. These ions are a potential source of background electrons in the spectrometer section, and could damage the \gls{fpd} if they are not blocked. Studies on ion detection and blocking techniques were completed as part of an ion safety check (\secref{SubSection:IonBlocking}).
\end{itemize}

\paragraph{Maintenance mode} 
This mode is scheduled at least once a year for dedicated maintenance tasks, such as servicing cryogenic pumps, conditioning the spectrometers' vacuum via bake-out, \gls{cps} regeneration, and other infrastructure upkeep. With one or more subsystems taken offline for maintenance, no data is taken in this mode.

%% file: WGTS.tex
\section{Windowless Gaseous Tritium Source}
\label{sec:source_system}

An ultra-luminous windowless gaseous tritium source (\gls{wgts}) with an activity of up to \SI{e11}{\becquerel} is used in the \gls{katrin} experiment. Such a source was operated by the LANL experiment \cite{Wil87} and developed further by the Troitsk experiment \cite{Belesev1995}.
Section \ref{subsec:WGTS-Principle} describes the working principle of a \gls{wgts} and summarizes the main requirements the \gls{wgts} has to fulfill in the \gls{katrin} experiment. To implement a \gls{wgts}, key systems are needed which are described in the  sections \ref{subsec:WGTS-magnet-cryostat-system}, \ref{sec:TritiumInnerLoopSystem} and \ref{sec:wgts_rear_wall}.

\input{WGTSPrinciple}

\input{WGTSMagnet}

\input{WGTSInnerLoop}

\input{WGTSRearWall}

\clearpage

%% file: WGTSPrinciple.tex
\subsection{WGTS Principle and Basic Requirements}
\label{subsec:WGTS-Principle}

\begin{figure}[b]
    \centering
    \includegraphics[width=0.9\textwidth]{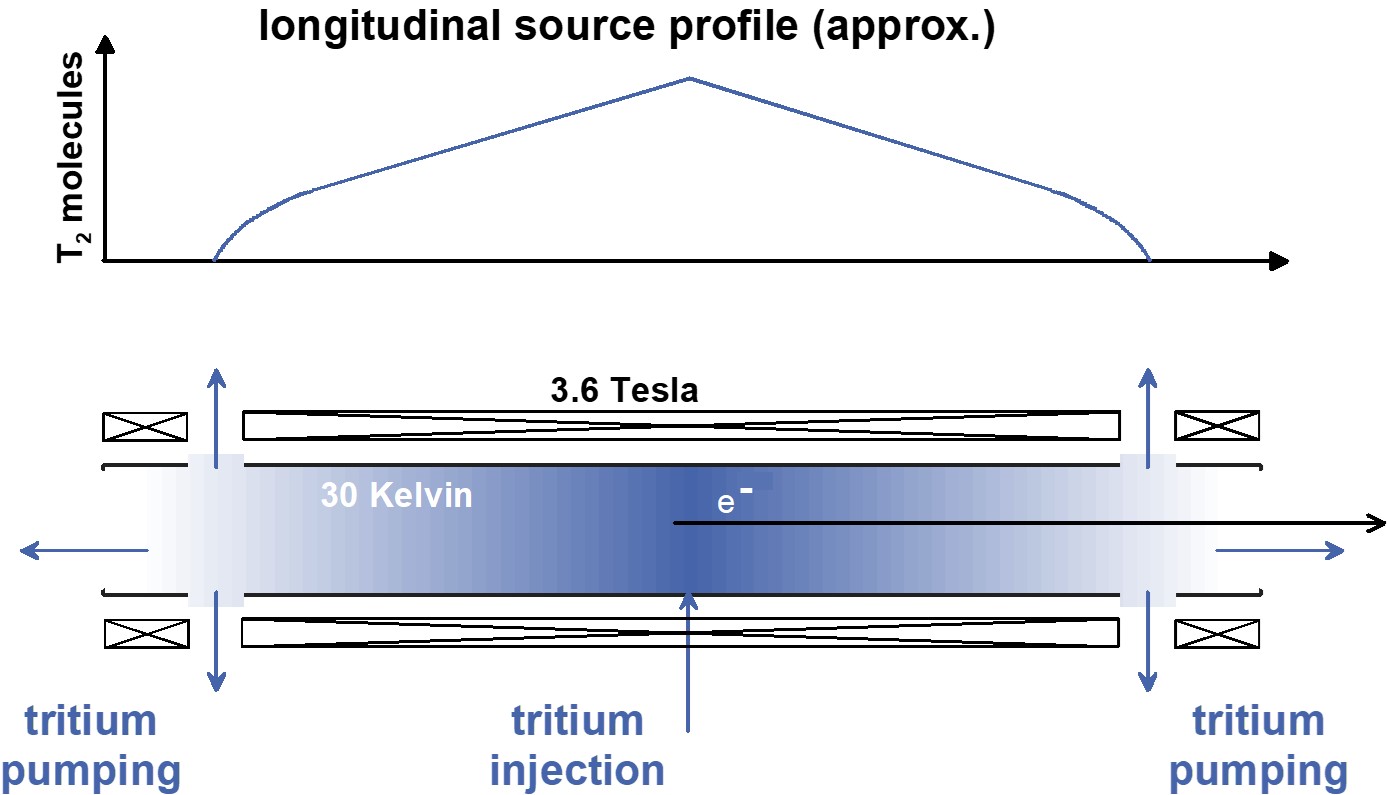}
    \caption{The working principle of the \acrlong{wgts}. The density profile inside the beam tube is kept constant by continuous injection of tritium gas in the middle and pumping it out at both ends. The \gls{wgts} beam tube of \gls{katrin} is \SI{10}{m} long and has a diameter of \SI{9}{cm}.}
    \label{Figure:WGTSPrinciple}
\end{figure}

The working principle of the \gls{wgts} is shown in \figref{Figure:WGTSPrinciple}. A \gls{wgts} consists of a tube filled with high-purity molecular tritium gas. The tritium decays inside the tube and the \betaels{} are adiabatically guided (without disturbing the electron's energy) to the spectrometer by a magnetic field. The tube has no windows at its ends to prevent energy loss of the electrons in the window material\footnote{Energy loss by inelastic scattering of electrons with the tritium gas exists and needs to be taken into account. }. The tritium molecules therefore can freely diffuse from the middle of the tube to both ends, where they are pumped away to prevent the spectrometer from being contaminated. A continuous injection of fresh tritium results in a stable longitudinal density profile of tritium molecules, as shown schematically in \figref{Figure:WGTSPrinciple}. The \gls{wgts} needs to be operated at low temperatures (below \SI{100}{\kelvin}) to reduce the conductance of the tube, and therefore the tritium throughput necessary for maintaining the tritium molecule density. 

The \gls{wgts} tube in the \gls{katrin} experiment is \SI{10}{\m} long and has a diameter of  \SI{90}{\mm}.  The reference column density is \SI{5.0e17}{molecules\per\centi\metre\squared}. Superconducting solenoids generate a homogeneous magnetic field of up to \SI{3.6}{\tesla}, which adiabatically guides the \betaels{} towards the tube ends. The nominal magnetic field configuration of the \gls{katrin} experiment is set such that \betaels{} with starting angles of up to \SI{51}{\degree} with respect to the magnetic field are transmitted through the beam line \cite{KATRIN2005}.

\clearpage

\begin{figure}[ht]
    \centering
    \includegraphics[width=0.9\textwidth]{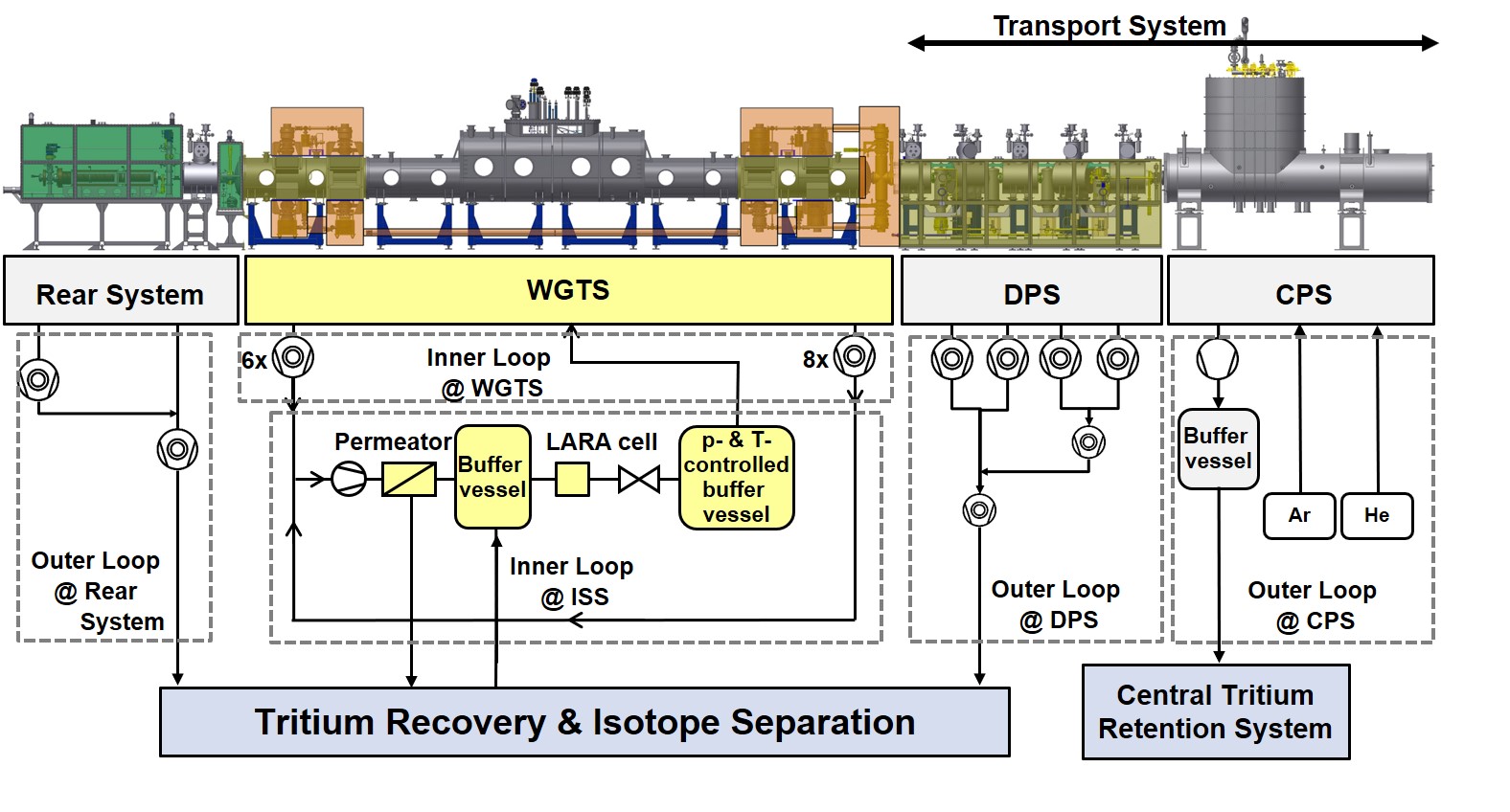}
    \caption{Scheme of the source and transport systems with tritium loops and the \gls{tlk} infrastructure. To achieve the required tritium column density, the tritium gas is operated in a closed Inner Loop (yellow parts). Tritium gas which diffuses out of both ends of the \gls{wgts} beam tube is pumped out by the pumping systems of the Outer Tritium Loop. All tritium loop systems are connected to the infrastructure systems of \gls{tlk}. All systems can be separated by gate valves. The Rear System contains the High Resolution Angular Selective Electron Gun which is i.a. used for energy loss measurements in the tritium source (\secref{sec:egun}).}
    \label{Figure:WGTS-with-Loops}
\end{figure}

A large amount of tritium on the gram scale is needed to operate the \gls{wgts}. For a given amount of tritium inside the source tube, there needs to be more than \num{e4} times as much tritium in the entire system. This is due to the fact that one has to operate dedicated loops and buffer volumes for tritium supply and tritium purification. 
Therefore, the use of the \gls{wgts} requires access to a laboratory with an adequate tritium infrastructure. For that reason the \gls{katrin} experiment was set up at the \gls{kit}, whose on-site \acrfull{tlk} is currently the only scientific laboratory equipped with a closed tritium cycle \cite{Doerr2005} and licensed to handle the required amount of tritium.
\figref{Figure:WGTS-with-Loops} gives an overview of how the source and transport systems are connected to the tritium loops and the \gls{tlk} infrastructure. For additional details, see \secref{sec:TritiumInnerLoopSystem}, \secref{sec:wgts_rear_wall}, \ref{Subsection:DifferentialPumpingSystem} and \ref{Subsection:CryogenicPumpingSystem}.

The high stability of the tritium isotopic composition and the column density of the \gls{wgts} are vital to the determination of the neutrino mass. These parameters are associated with several leading systematic uncertainties: activity fluctuations of the \gls{wgts}, energy-loss corrections due to scattering of \betaels{} with tritium molecules in the \gls{wgts}, and the final-state spectrum. The precise knowledge of these parameters is crucial to combining the data which will be taken over the extended measurement period of a few years \cite{KATRIN2005}. 
Table \ref{table:KATRINKeyParameters} summarizes the key parameters of the \gls{katrin} experiment, including the two key parameters of the \gls{wgts}: column density and tritium purity.
The stability requirement for the column density is \SI{< 0.1}{\percent\per\hour}. This leads to stringent stability limits for other operating parameters such as source temperature, injection rate and pumping speed.
The \gls{wgts} is planned to be used continuously for 10 weeks per measurement campaign and up to 3 campaigns per year.
The combination of such stringent stability and continuous usage requirements presents a great technical challenge. This section gives a prescription to how these challenges were met and overcome.

A detailed discussion of the \gls{wgts} operating parameters and their impact on the systematic uncertainties of the neutrino-mass measurement can be found in \cite{Babutzka2012}. 
The technical realization of the \gls{wgts} magnet cryostat system together with special tritium loops  is described in the following two subsections.

\clearpage

%% file: WGTSMagnet.tex
\subsection{
	WGTS magnet cryostat system}
\label{subsec:WGTS-magnet-cryostat-system}
\begin{figure}[!ht]
	\centering
	\includegraphics[width=\linewidth]{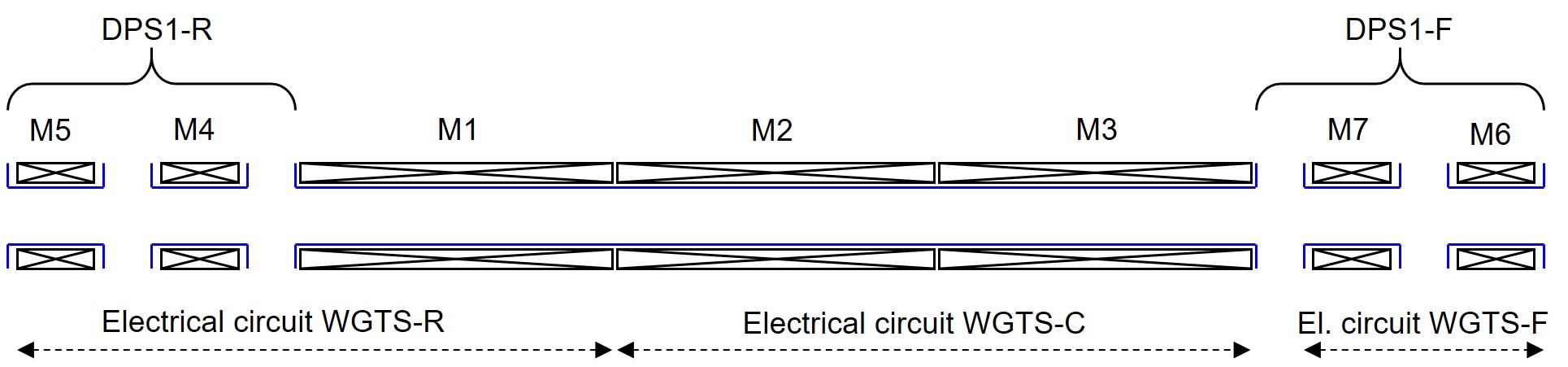}
	\caption{A sketch of the WGTS magnet cryostat. It consists of the central beam part with solenoids M1, M2 and M3, and a differential pumping section (DPS1-R/F) on both ends (solenoids M4-M7). The magnets are operated in 3 electrical circuits. Pump ports are located between M5 and M4, M4 and M1, M3 and M7, and M7 and M6. }
	\label{Figure:wgts_cad}
\end{figure}
The \gls{wgts} magnet cryostat mainly consists of the following subsystems (Figure~\ref{Figure:wgts_cad}):
\begin{itemize}
	\item The 10-m-long \gls{wgts} beam tube within magnets M1, M2 and M3
	(\figref{Figure:WGTSPrinciple}) encloses the windowless gaseous tritium source. The temperature of the tube is \SI{30}{K} at standard conditions. 
	\item The magnetic guiding system, with fields up to \SI{3.6}{T},  guides \betaels{} along the beam axis either downstream towards the spectrometers or upstream towards the \gls{wgts} rear wall.
	\item The differential pumping sections DPS1-R and DPS1-F with the
	corresponding beam tube elements and the pump ports at the ends of the \gls{wgts} beam tube reduce the tritium gas flow rate by a factor of about 200 and return the tritium to the injection system.
\end{itemize}

The magnet system consists of seven superconducting solenoids and four pairs of superconducting dipole magnets. Its setup is described in \secref{subsubsection:setup-magnet-system}. The commissioning results are reported in \secref{subsubsection:comm-cooling-system}. 
The requirements for temperature stabilization of the beam tube (\secref{Subsection:KATRINKeyParameters}) are quite demanding, and can only be met with a dedicated cooling system. This system and its commissioning results are described in \secref{subsubsection:cooling-system} and \secref{subsubsection:comm-cooling-system}.

\subsubsection{Magnet system setup}
\label{subsubsection:setup-magnet-system}
%
The \gls{wgts} magnet system is manufactured in one 16-m-long cryostat~\cite{Grohmann2008, Grohmann2009}, housing seven chambers of superconducting solenoids, a helium reservoir, beam tubes, and thermal shields. Seven solenoid modules with warm bores are installed in a straight line, surrounding five beam tube sections that are interconnected with four pumping ducts
(\figref{Figure:wgts_cad}). Three 3.3-m-long solenoid modules (M1, M2, and M3) are located at the central part of the \gls{wgts}, providing a homogeneous magnetic field along the 10-m-long central beam tube. Two 1-m-long solenoid modules are connected to the ends (M5 and M4 at the rear side, M7 and M6 at the front side). The solenoid modules are grouped in 3 electrical circuits: WGTS-R (M5, M4, and M1), WGTS-C (M2 and M3), and WGTS-F (M7 and M6) for driven-mode operation (operational scheme is explained in \cite{Arenz2018c}), each with its own \gls{psu}. In addition, four dipole coil pairs are installed for the purposes of beam alignment and calibration. Two dipole coil pairs (DRx and DRy) are wound on the rear end module M5 and two other pairs (DFx and DFy) on the front end module M6. These are used to deflect the guiding magnetic fields radially up to \SI{42}{\milli\metre} in the x- and y- directions relative to the z (axial) axis  \footnote{This is necessary for both beam alignment and systematic investigations, e.g. to investigate the density of the tritium source as a function of the radial position by means of electrons from an e-gun.}. 
	
The \gls{psu}s from FuG\footnote{FuG Elektronik GmbH, \url{https://www.fug-elektronik.de}} provide a current, which is stable to better than \SI{10}{ppm} per \SI{8}{\hour}, and which allows to meet the magnetic field stability requirement of better than \SI{0.03}{\percent\per month} at the source. The field stability of the magnets in WGTS-R and WGTS-C can be monitored by a closed-loop flux gate sensor in the \gls{psu}, while those in WGTS-F can be monitored via a precision resistor in the \gls{psu}. More details about the magnet system including the magnet safety system are described in~\cite{Gehring2006, Gil2018, Arenz2018c}. 
	
The superconducting coils of the \gls{wgts} are cooled in a liquid helium ({LH}e) bath at \SI{4.5}{\kelvin} and \SI{0.13}{\mega\pascal} that is provided in a \SI{1.5}{\metre\cubed} helium reservoir by Joule-Thompson expansion of supercritical helium~\cite{Grohmann2008}. The supercritical helium at \SI{5}{\kelvin}  and \SI{0.5}{\mega\pascal} is supplied by a LINDE\footnote{Linde Kryotechnik AG, \url{https://www.linde-kryotechnik.ch}} TCF~50 refrigerator \cite{Grohmann2010}. The cryostat including the magnet chambers and the He reservoir has a total {LH}e volume of about \SI{2.8}{\meter\cubed}. This volume is sufficient to keep the system at \SI{4.5}{\kelvin} for about \SI{24}{\hour} in case of an {LH}e supply interruption. The cryostat is equipped with a safety valve opening at \SI{0.2}{\mega\pascal} overpressure and a burst-disc opening at \SI{0.3}{\mega\pascal} overpressure for the \SI{4.5}{\kelvin} helium cooling system.

\subsubsection{Setup of the beam tube cooling system}
\label{subsubsection:cooling-system}
\begin{figure}
	\centering
	\includegraphics{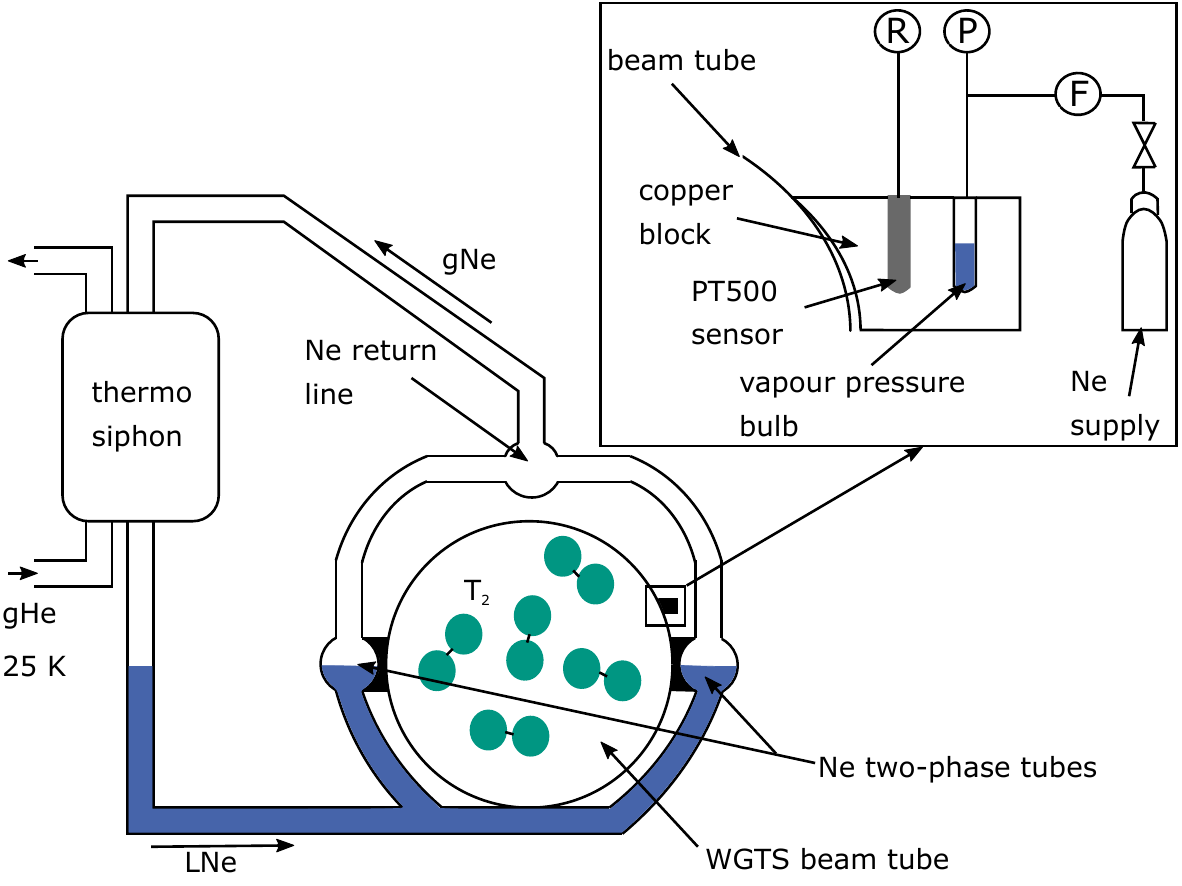}
	\caption{The two-phase cooling of \gls{wgts} beam tube. This figure presents the two-phase neon cooling of the \gls{wgts} central beam tube with the temperature monitoring and calibration system as described in the main text (R -- resistance, P -- pressure transducer, F -- flow meter). The abbreviation gHe stands for gaseous helium, gNe for gaseous neon and LNe for liquid neon. This figure is reproduced from~\cite{PhDSeitz2019} (adapted from~\cite{Grohmann2008}).}
	\label{Figure:wgts_cooling}
\end{figure}

In order to achieve the temperature stability requirement of \SI{0.1}{\percent} in the central beam tube of the \gls{wgts}~\citep{KATRIN2005}, a unique two-phase neon cooling system was developed~\cite{Grohmann2008, Grohmann2009}. Two 16-mm-diameter tubes are brazed to both sides of the central beam tube and are filled with two-phase neon (\figref{Figure:wgts_cooling}) with both liquid and vapor phase occupying about half of the cross-section. Neon was selected as it offers a suitable vapor pressure of \SI{2}{\bar} at \SI{30}{\kelvin}~\cite{Stull1947}. Other coolants allow for other beam-tube operation temperatures (\tabref{table:WGTS-modes}). 

The liquid neon evaporates due to heat input and diffuses towards the thermosiphon at one end of the cooling pipes. In a heat exchange with the \SI{25}{K} cooling circuit of gaseous helium, the neon is re-liquefied and flows back towards the cooling tubes. No mechanical pumping is required for this two-phase cooling system. The evaporation process of neon is controlled by four heating wires of up to \SI{2}{\watt} each. The heating power is adjusted according to the measured temperature fluctuations so that the resulting temperature stability is further improved.

\begin{table}[bt]
	\caption{Possible operation modes of \gls{wgts} magnet cryostat. By changing the coolant, the \gls{wgts} beam tube temperature can be operated at different temperature regimes. With a given coolant, the temperature can be fine-tuned by adjusting the pressure in the two-phase cooling system.}
	\begin{center}
		\begin{tabular}{lc}
			\hline
			\textbf{Coolant} & \textbf{Temperature Range for Operation}  \\
			\hline
			Neon & \SI{28}{K} - \SI{37}{K}\\
			Nitrogen & \SI{80}{K} - \SI{103}{K}\\
			Argon & \SI{90}{K} - \SI{115}{K}\\
			\hline
		\end{tabular}
	\end{center}
	\label{table:WGTS-modes}
\end{table}

The DPS1-R/F are cooled in separate cycles. For both inner parts (inside M4 and M7, Figure~\ref{Figure:wgts_cad}), a two-phase neon cooling system is applied, while the two outermost beam tube elements (inside M5 and M6) are cooled with gaseous nitrogen only. The gaseous neon is at a temperature of \SI{\sim 80}{\kelvin}. Its cooling cycle is coupled via a heat exchanger to the outer shield of the \gls{wgts}, which is cooled by liquid nitrogen~\citep{Grohmann2009}. A higher temperature and a worse temperature stability can be tolerated for these outer sections of the \gls{wgts}, as the tritium density is much lower than that in the central beam tube.

The pump ports at both ends of the central beam tube are each equipped with four \gls{tmp}, whereas each of the two pump ports in the DPS1-R/F has two \gls{tmp} (\secref{sec:innerloop-setup}). The pumps are operated at room temperature. In order to minimize the heat load from the pumps, the pumping ducts are lined with thermal radiation shield on their inner surface~\citep{Grohmann2009}. The concentric shields are blackened by Al$_2$O$_3$/TiO$_2$ and cooled with liquid nitrogen. The pumping chamber itself is cooled to \SI{\sim 30}{\kelvin} by the gaseous helium circuit.

For \kr{} operation (\secref{Subsubsection:GKrS}), the \gls{wgts} temperature has to be increased to about \SIrange{80}{100}{\kelvin} \footnote{The precise value is chosen in dependence on the experimental program.} to prevent krypton atoms from freezing out on the inner surface of the beam tube at partial pressures of \SI{<<1}{\milli\bar}~\cite{Leming1970}. The \gls{wgts} temperature is stabilized by replacing neon with argon~\cite{Grohmann2008, Itterbeek1964}. The temperature of the gaseous helium circuit is adjusted accordingly.

The temperature along the beam tube in the \gls{wgts} is monitored with a total of 52 platinum wire-wound glass sensors (Pt500 or Pt1000)~\citep{Grohmann2011}. Twenty-four Pt500 sensors are mounted on the central beam tube and two in each of the pump ports at the end of the central beam tube. 
The DPS1-R/F are equipped with ten Pt1000 sensors each, and two are further placed in each of the DPS1-R/F pump ports. The sensors are supplied with a current of \SI{500}{\micro\ampere} by a Keithley (model 6220)\footnote{Keithley Instruments, \url{https://www.tek.com/keithley}} DC current source. The voltage  is read out in a 4-wire sensing configuration by Gantner (Q.bloxx, model A107 and A105)\footnote{Gantner Instruments GmbH, \url{https://www.gantner-instruments.com}}. 

In the \gls{katrin} experiment, the relative temperature changes are monitored as well as the absolute temperature of the central beam tube \citep{Bodine2015}. This requires the calibration of the 28 Pt500 sensors, positioned at the central beam tube and in the pump ports~\citep{Grohmann2011}. The Pt500 sensors are each placed in a copper block brazed to the beam tube. The copper block contains a vapor pressure bulb, which is filled with two-phase neon during calibration, with  each the liquid and vapor phase occupying about half of the volume. The absolute temperature is derived from the saturation vapor pressure being read out with pressure transducers from Endress+Hauser\footnote{Type Cerabar S PMC71, Endress+Hauser GmbH+Co.KG, \url{https://www.de.endress.com}}. With this technique, the trueness of the temperature value of \SI{83}{\milli\kelvin} at \SI{30}{\kelvin} and \SI{163}{\milli\kelvin} at \SI{100}{\kelvin} are achieved for the Pt500 sensors~\cite{PhDSeitz2019}.

\subsubsection{Commissioning of the WGTS magnet cryostat system}
\label{subsubsection:comm-cooling-system}
After its delivery to KIT in 2015, the \gls{wgts} magnet cryostat system was positioned in the \gls{tlk} hall and connected to the cryogenic transfer lines and the electrical cabinets. An intensive commissioning campaign was conducted, including electrical tests and control system checks, and led to the  first cool-down to \SI{30}{\kelvin} in 2016. From that time until the end of 2018, the \gls{wgts} magnet cryostat was returned to room temperature once for half a year of maintenance. For the rest of the time, the \gls{wgts} magnet cryostat remained at \SI{30}{\kelvin} or \SI{100}{\kelvin}. The change of the operational temperature was done by thermal radiation only~\cite{Heizmann2017}: the two-phase tubes were evacuated and the gaseous helium circuit was set to a new temperature set-point. When the operational temperature was increased from \SI{30}{\kelvin} to \SI{100}{\kelvin}, thermeal radiation from the gaseous helium circuit at \SI{95}{\kelvin} heated up the beam tube until argon can be condensed in the two-phase tubes. This process took about two days~\cite{PhDSeitz2019}. In the case of cooling from \SI{100}{\kelvin} to \SI{30}{\kelvin}, thermal radiation from the beam tube was absorbed by the gaseous helium circuit at \SI{25}{\kelvin} until neon could be condensed in the two-phase cooling tubes. This took about three days~\cite{PhDSeitz2019}.

In the following, the main results of about three years of \gls{wgts} operation at \SI{30}{\kelvin} and \SI{100}{\kelvin} are summarized.
\begin{description}
\sisetup{detect-weight=true, detect-family=true}
\item[Temperature stability at \SI{30}{\kelvin}]
The temperature stability of the central beam tube at \SI{30}{\kelvin} is~\citep{PhDSeitz2019} (\figref{Figure:wgts_stab})
\begin{align}
\left|\frac{\Delta T}{T}\right|_{\mathrm{CB,30\,K}} & < \SI{5+-1e-5}{h^{-1}}.
\end{align}
%
%
For the DPS-1-R/F sensors, a temperature stability of
\begin{align}
\left|\frac{\Delta T}{T}\right|_{\mathrm{DPS1-R/F,30\,K}} & < \SI{8+-1e-5}{h^{-1}}
\end{align}
is found~\cite{PhDSeitz2019}. Both values safely met the requirements of $\left|\Delta T/T\right|_{\mathrm{CB}}<\SI{1e-3}{h^{-1}}$~\cite{KATRIN2005} and $\left|\Delta T/T\right|_{\mathrm{DPS1-R/F}} < \SI{1e-2}{h^{-1}}$~\cite{PhDSeitz2019}.
\begin{figure}
	\centering
	\includegraphics{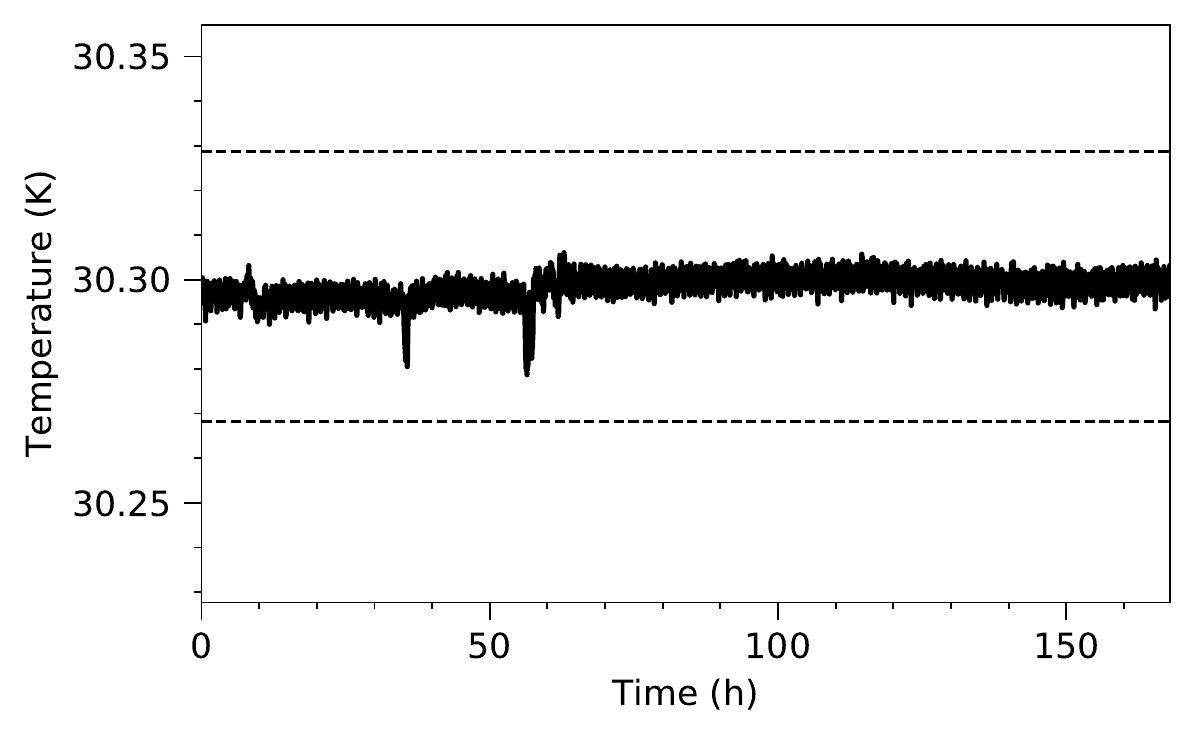}
	\caption{Temperature stability of the \gls{wgts} central beam tube at \SI{30}{\kelvin}. The temperature trend for one of the 24 Pt500 sensors is shown for one week of measurement time. The dashed lines mark the range with the specified value of \SI{+-0.1}{\percent}.}
	\label{Figure:wgts_stab}
\end{figure}
\item[Temperature stability at \SI{100}{\kelvin}]
The same temperature stability requirements as for \SI{30}{\kelvin} have to be fulfilled at \SI{100}{\kelvin}. For both the central beam tube and the DPS1-R/F, the measured stability is more than one order of magnitude better~\cite{PhDSeitz2019}:
\begin{align}
\left|\frac{\Delta T}{T}\right|_{\mathrm{CB,100\,K}} & < \SI{4.9+-0.4e-5}{\per\hour}, \\
\left|\frac{\Delta T}{T}\right|_{\mathrm{DPS1-R/F,100\,K}} & < \SI{5.43+-0.04e-4}{\per\hour}.
\end{align}
\begin{figure}
	\centering
	\includegraphics{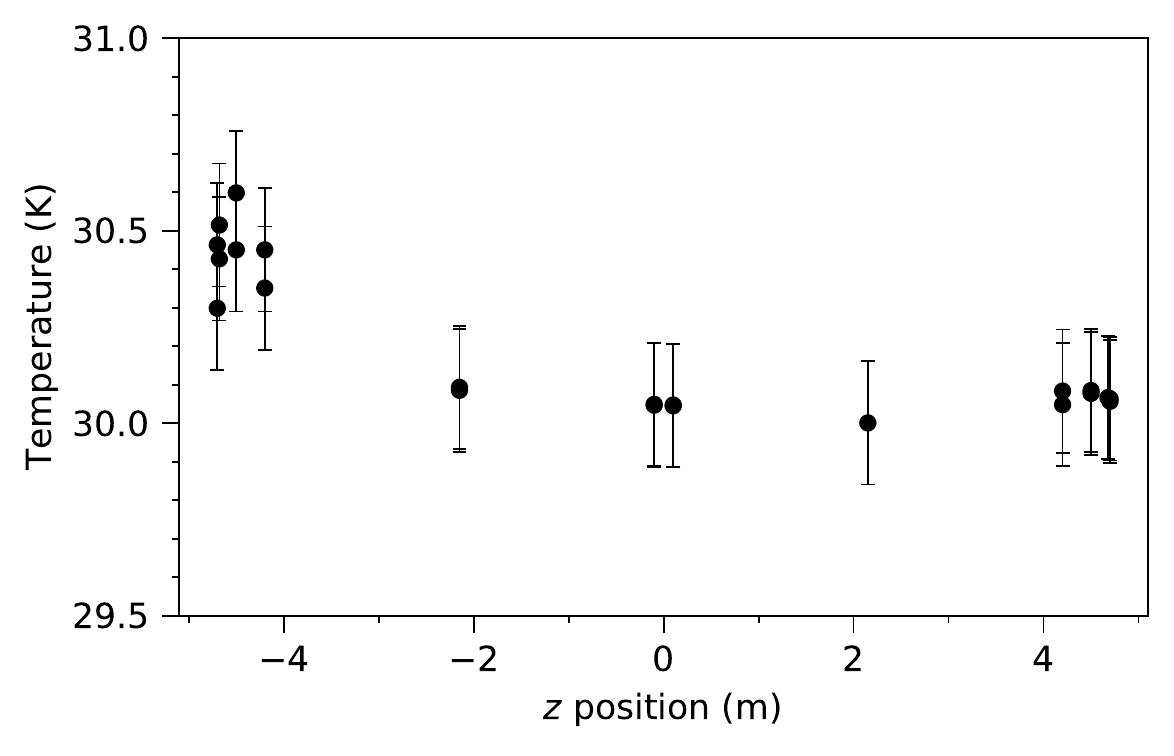}
	\caption{Temperature homogeneity of the \gls{wgts} central beam tube at \SI{30}{\kelvin}. The plot shows the mean temperature during one week of analyzed measurement time of the PT500 sensors at their position along the central \gls{wgts} beam tube. The errorbars are displaying the total uncertainty (including systematic contributions) of the measurement. The temperature shows an increase of \SI[separate-uncertainty=true]{594+-226}{\milli\kelvin} towards the rear end of the \gls{wgts} due to the heat load of the vapor pressure capillaries. Figure adapted from~\cite{PhDSeitz2019}.}
	\label{Figure:wgts_hom}
\end{figure}
\item[Temperature homogeneity] 
The temperature homogeneity  $\left|\Delta T/T\right|_{\mathrm{Hom}}$, or the temperature variation along the central beam tube of the \gls{wgts}, has to be better than $10^{-3}$ for both operating modes at \SI{30}{\kelvin} and \SI{100}{\kelvin}~\cite{KATRIN2005}. The first measurements with the \gls{wgts} demonstrator  showed a temperature inhomogeneity of $\SI{2.8e-2}{}$ at \SI{30}{\kelvin}, exceeding the requirement by more than one order of magnitude~\cite{Grohmann2013}\footnote{In order to test the novel cooling system of the tritium source beam tube, a twelve meter long test cryostat was built by the manufacturer and delivered to KIT in 2010. Apart from the magnets and the magnet cooling system, the so-called demonstrator contained all original components of the central \gls{wgts} magnet cryostat. After end of the tests the demonstrator was sent back to the manufacturer for final assembly of the complete \gls{wgts} magnet cryostat.}. The capillaries of the vapor pressure bulbs (\figref{Figure:wgts_cooling}) were identified as the reason for this result since they introduced additional heat to the system. After a re-arrangement of the capillaries inside the final \gls{wgts} cryostat, the inhomogeneity was reduced to~\cite{PhDSeitz2019} (Figure~\ref{Figure:wgts_hom})
\begin{align}
\left|\frac{\Delta T}{T}\right|_{\mathrm{Hom,30\,K}} = \SI{2.0+-0.8e-2}~. 
\end{align}
In contrast to the \gls{wgts} demonstrator design, the rear end of the central beam tube is at a higher temperature of about $\SI{0.6}{K}$. The consequences of this  higher temperature are small: since \betaels ~created in this region have to travel through the whole gas column and therefore have a higher probability to scatter and to lose energy, only a small fraction is used for the neutrino mass analysis. The very good temperature stability allows the consideration of the temperature inhomogeneity in the gas dynamics model of the tritium source~\cite{PhDKuckert2016, Kuckert2018, PhDHeizmann2019}. 

For \SI{100}{\kelvin} operation of the \gls{wgts}, the homogeneity requirement is nearly met~\cite{PhDSeitz2019}:
\begin{align}
\left|\frac{\Delta T}{T}\right|_{\mathrm{Hom,100\,K}} = \SI{1.0+-3.0e-3}~.
\end{align}
\item[Stand-alone time]
The magnets can be kept cold at \SI{4.5}{\kelvin} for about \SI{16}{\hour} after magnet cooling has been interrupted from the refrigerator at a {LH}e level of \SI{55}{\percent}. 
\item[Maximum operational currents] 
The \gls{wgts} magnets have reached the maximum design current of \SI{310}{\ampere} for the WGTS-R and WGTS-C, and \SI{209}{\ampere} for WGTS-F without training quench. The safety system that was developed for the driven mode of the magnets has been successfully tested~\cite{Gil2018}. The three dipole-coil pairs have been tested up to the current of \SI{110}{\ampere} needed to deflect the guiding magnetic field radially up to \SI{42}{\mm} for the design magnetic setting of the experiment. One dipole-coil pair (DFy) could not be operated because of a short circuit to ground under cold condition. In combination with the main solenoids at \SI{70}{\percent} of the design current, the two dipole coil pairs (DRx and DRy) at the rear side have been tested for the \SI{70}{\percent} \footnote{The ﬁelds were set to \SI{70}{\percent} of their design values in order to operate the magnets at a safe level without a quench risk of the complex systems.} setting, which is the nominal value for neutrino-mass operation mode (\tabref{table:KATRINKeyParameters}) \footnote{As already mentioned before, the dipole coils are only needed for systematic investigations. The unavailability of DFy can be compensated by adaptation of the measurement plan. }.  
\item[Current stability] 
The instabilities of the magnet currents at the standard setting (\SI{70}{\percent} of the design fields) are summarized in Table~\ref{tab:Magwgts}.
 In an \SIadj{8}{\hour} measurement, the data of the current sensors of the power supplies were taken for the calculation of the average value $I_\mathrm{av}$, the standard deviation $\sigma$ and their ratio as a measure of the relative fluctuation. These values are compared with $I_\mathrm{av, ext}$ and $\sigma_\mathrm{ext}$, calculated from external current sensor data.  The external sensors are installed on each circuit for magnet safety. The difference between $I_\mathrm{av}$ and $I_\mathrm{av, ext}$ is below \SI{0.05}{\percent}, and the relative fluctuation is in both cases within the specification of \SI{0.03}{\percent}.
\end{description}
\begin{table}[!ht]
	\caption{Instabilities of the magnet currents at the standard setting (\SI{70}{\percent} of the design fields). Eight hours of data were taken with the current sensors of the stable power supplies for calculating the average current  $I_\mathrm{av}$.  $I_\mathrm{av, ext}$ is calculated from the data of the external DC current transducer sensors$^a$  installed on each circuit for use in the magnet safety system. A small offset value at zero-current has been corrected for $I_\mathrm{av}$ and $I_\mathrm{av, ext}$. $\sigma$ and $\sigma_\mathrm{ext}$ are the standard deviations. }
	\label{tab:Magwgts}
	\vspace*{1ex}
	\begin{center}
	\begin{tabular}{lccc}
		\hline
		 & \textbf{WGTS-R} & \textbf{WGTS-C} & \textbf{WGTS-F}\\
		\hline
		$I_\mathrm{av}$ (\si{\ampere}) & \num{216.564} & \num{215.800} & \num{145.921} \\
		$\sigma$ (\si{\ampere}) & \num{2.8e-4} & \num{2.0e-5} & \num{2.0e-4} \\
		$\sigma/I_\mathrm{av}$ (\si{\percent}) & \num{1.3e-4} & \num{9.6e-6} & \num{1.3e-4} \\
		\hline
		$I_\mathrm{av, ext.}$ (\si{\ampere}) & \num{216.677} & \num{215.875} & \num{145.982} \\
		$\sigma_\mathrm{ext}$ (\si{\ampere}) & \num{3.0e-2} & \num{2.4e-2} & \num{2.5e-2} \\
		$\sigma_\mathrm{ext}/I_\mathrm{av, ext.}$ (\si{\percent}) & \num{1.4e-2} & \num{1.1e-2} & \num{1.4e-2} \\
		\hline
		\multicolumn{4}{l}{\scriptsize a) Type IT~400-S Ultrastab, LEM, \url{https://www.lem.com}}
	\end{tabular}
	\end{center}
\end{table}

\clearpage

%% file: WGTSInnerLoop.tex
\subsection{Tritium Loops}
\label{sec:TritiumInnerLoopSystem}

The required stability of the tritium column density and the purity of the tritium gas in the KATRIN source (\secref{subsec:WGTS-Principle}) 
can only be achieved by means of closed tritium loops and a dedicated tritium injection system. These parts of the \gls{katrin} experiment are described below.

\subsubsection{Tritium Loops Overview}\label{sec:Loop-overview}

\begin{figure}[!b]
	\centering
	\includegraphics[width=0.9\textwidth]{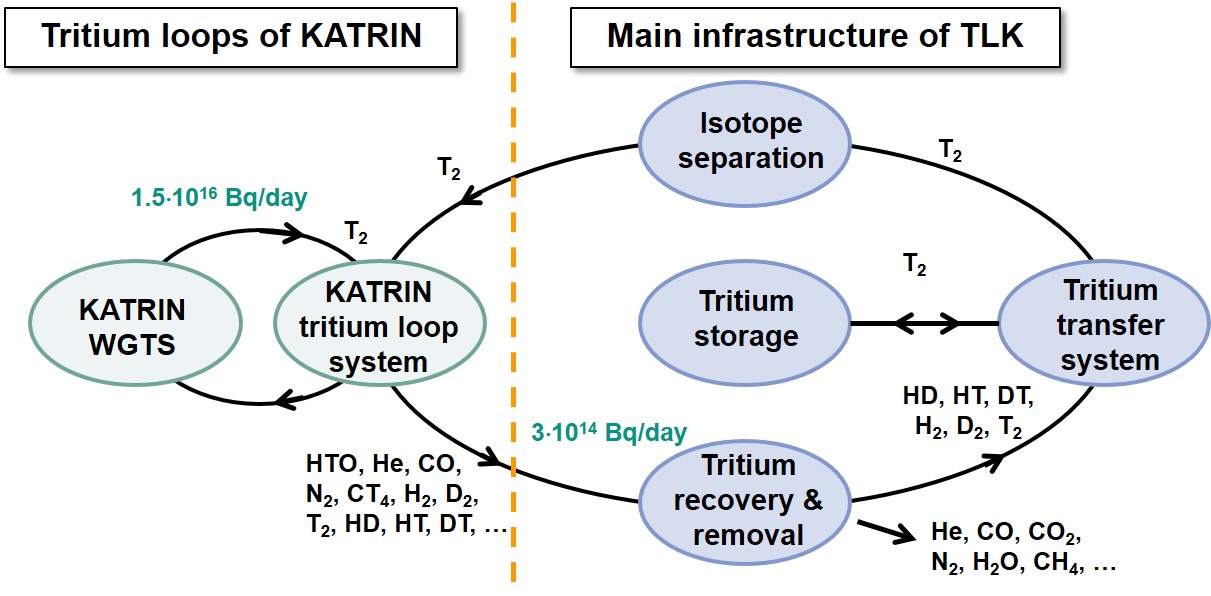}
	\caption{The tritium loop schematic (simplified) with focus on tritium paths. Displayed here is the \gls{il} (left, clockwise) and the Outer Loop (right, counterclockwise). The specific systems of the \gls{katrin} experiment are displayed in green, and the main infrastructure systems of \gls{tlk} in blue. The \gls{katrin} tritium loop system is part of both loops. Details are given in the main text.}
	\label{Figure:Loops_schematics}
\end{figure}

\begin{figure}[t]
	\centering
	\includegraphics[width=0.9\textwidth]{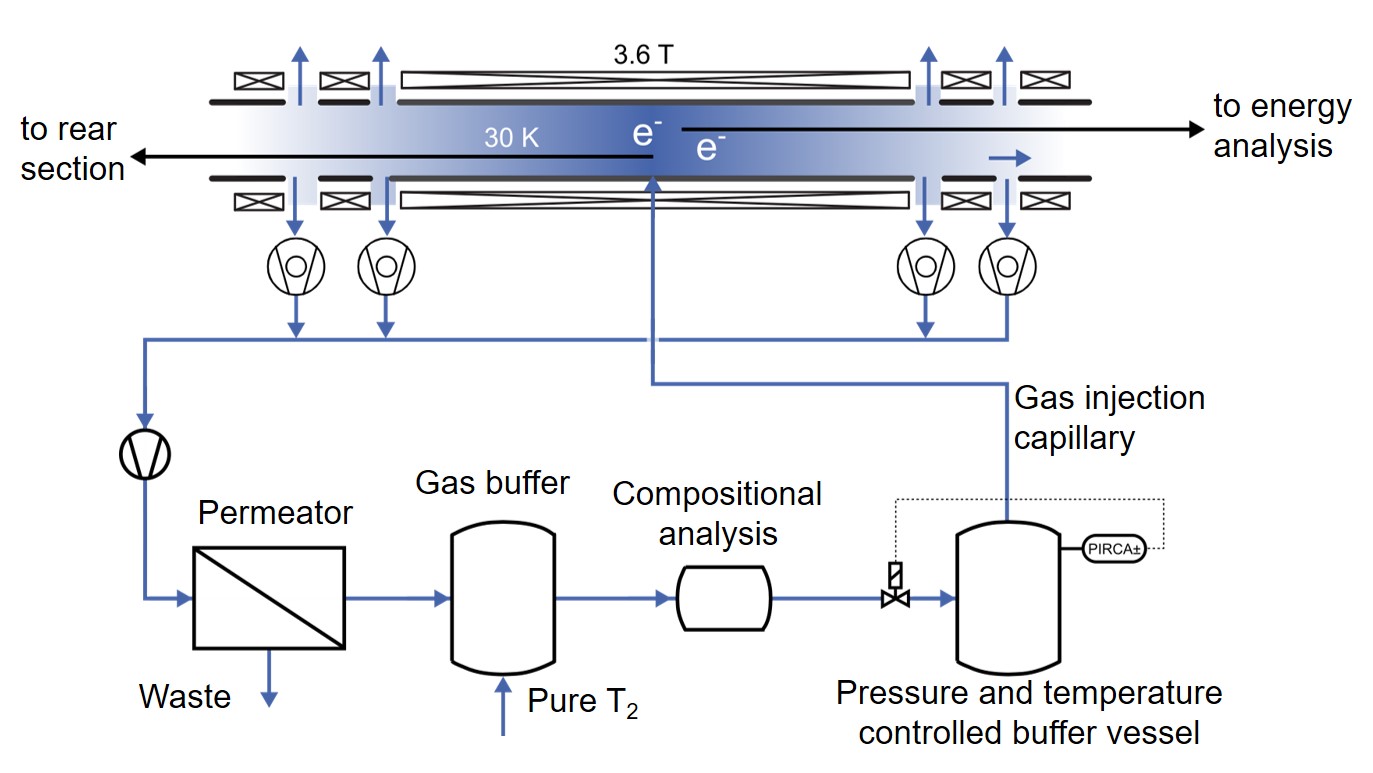}
	\caption{The schematics of the \gls{il} with stabilized injection of tritium gas into the beam tube.}
	\label{Figure:Principle-Inner-Loop}
\end{figure}

The tritium loop system consist of two interlinked loops, the \acrfull{il} and the \gls{ol}. A schematic is shown in \figref{Figure:Loops_schematics}. 
The loops have different functions and are operated differently:
\begin{description}
	\item[Inner Loop] The Inner Loop (\figref{Figure:Principle-Inner-Loop}) provides the tritium gas profile in the \gls{wgts} beam tube~\cite{Kuckert2018} by continuous and stable injection of high-purity tritium gas into the beam tube, and by continuous and stable pumping of tritium in the differential pumping sections DPS1-F/R (\secref{subsec:WGTS-magnet-cryostat-system}).  During the measurement phases, the tritium throughput is about \SI{40}{g} per day. A fraction of the circulated tritium (around \SI{1.2}{\%} of the flow rate) is continuously extracted from the \gls{il} and exchanged with pure tritium to keep the tritium concentration high. For detailed information, see~\cite{Kazachenko2008b}.  The monitoring of all hydrogen isotopologues is realized with a tritium-compatible in-line Laser-Raman system (\secref{SubSection:GasCompositionMonitoring}).
	\item[Outer Loop] 
	 The Outer Loop provides the \gls{katrin} experiment with pure tritium (purity above \SI{95}{\percent}), and collects and recycles degraded tritium gas (referred to as exhaust gas) originated from the \gls{il} and from the differential pumping systems at the \gls{dps} (\secref{Subsection:DifferentialPumpingSystem}) and the Rear System. The \gls{ol} is operated in batch-wise mode during normal working hours: the exhaust gas is stored in a buffer vessel and transferred in batches (up to \SI{20}{\litre}) to the \gls{tlk} infrastructure (\figref{Figure:Loops_schematics}) for hydrogen isotope recovery and impurity removal, and subsequently, for hydrogen-isotope separation. These specialized systems are distributed across the entire \gls{tlk} and connected with several hundred meters of piping. For detailed information on the \gls{ol} and on the \gls{tlk} infrastructure, see~\cite{Doerr2005} and~\cite{Welte17}.
\end{description}

\newpage
\subsubsection{KATRIN Tritium Loop System Setup}\label{sec:innerloop-setup}

\begin{figure}[ht]
	\centering
	\includegraphics[width=1.35\textwidth, angle=90]{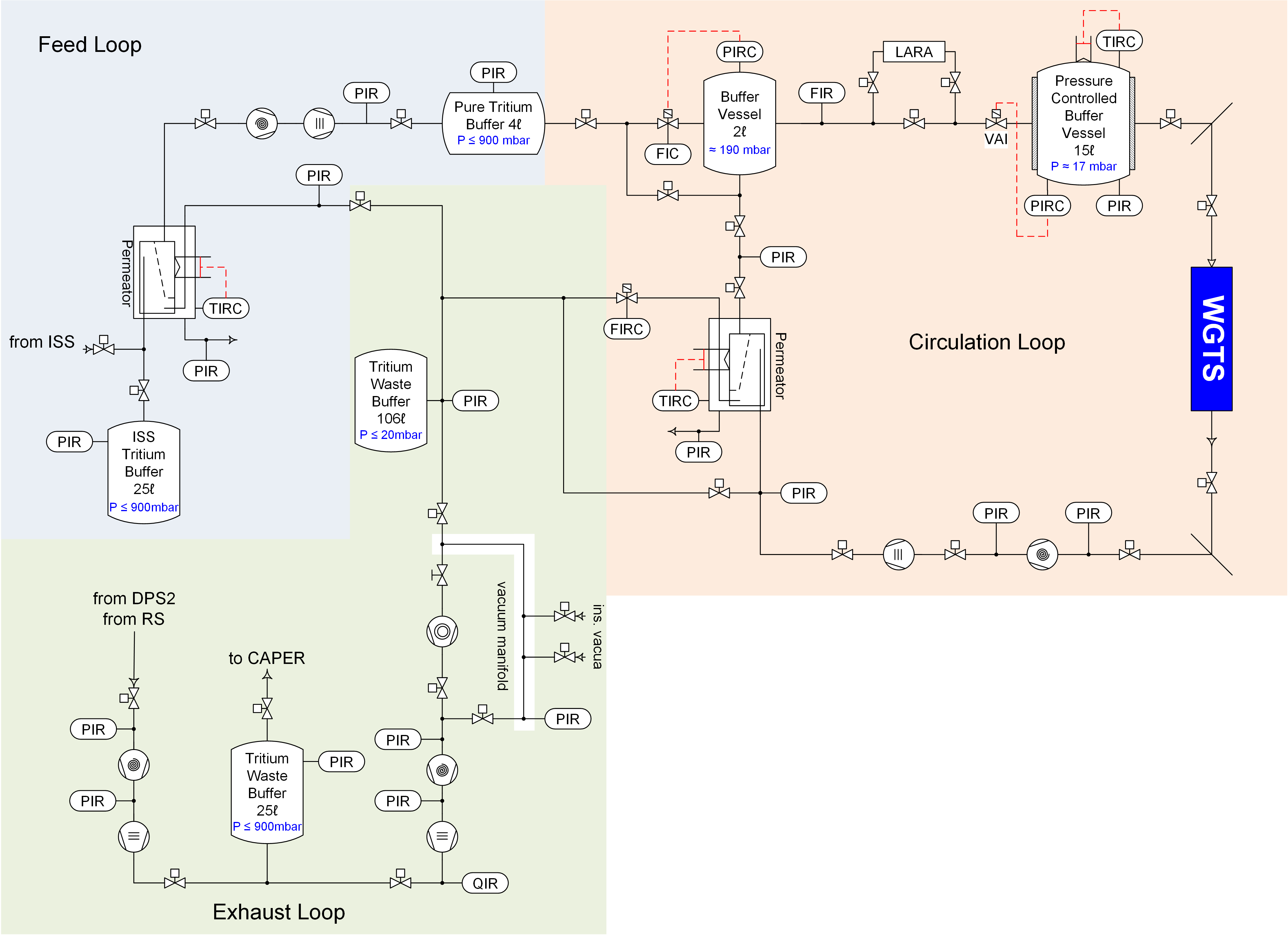}
	\caption{A simplified piping and instrumentation diagram of the main part of the \gls{katrin} Tritium Loop System. Displayed here is the Feed Loop (blue background), the Exhaust Loop (green) and the Circulation Loop (orange).  The front end \gls{tmp} systems at the \gls{wgts}, the \gls{dps} and the Rear System are depicted in a simplified scheme in \figref{Figure:WGTS-with-Loops}. }
	\label{Figure:loopinjection}
\end{figure}

The \gls{katrin} tritium loop system is part of both the \gls{il} and the \gls{ol} (\figref{Figure:Loops_schematics}). It includes all tritium systems necessary for stabilized  circulation of tritium, impurity separation via a permeator~\footnote{A permeator is a palladium-silver membrane filter, which is only permeable to hydrogen isotopes.}, intermediate tritium and exhaust gas storage,  gas transfer from the \gls{tlk} Isotope Separation System and to the \gls{tlk} Tritium Recovery \& Removal System.

The \gls{katrin} Tritium Loop System comprises (sub)loop systems, including:
\begin{description}
	\item [Circulation Loop] 
		The components of the Circulation Loop are placed in the \gls{iss} glove box (called Inner Loop@ISS) and in the \gls{wgts} glove boxes (called Inner Loop@WGTS), see \figref{Figure:WGTS-with-Loops}.
		The Circulation Loop (mainly orange part in \figref{Figure:loopinjection}) provides stabilized tritium gas injection into the \gls{wgts} and the recovery of the majority of the injected gas. It also keeps the tritium purity above \SI{95}{\percent} via constant purification of the circulating gas by means of a permeator (= removal of non-hydrogen-isotopologues) and by replacing a small fraction of the polluted tritium gas by pure \t2{}. The stabilization of tritium gas injection is maintained using a temperature- and pressure-stabilized buffer vessel, which is connected to the \gls{wgts} beam tube via a capillary. The capillary (\SI{2.1}{\milli\metre} inner diameter, \SI{5}{\metre} length) is thermally coupled directly to the beam tube. Therefore, its temperature and conductance are defined by the beam tube temperature. 
		
		The sub-system Inner Loop@WGTS collects tritium at both ends of the \gls{wgts} beam tube (at DPS1-F/R, see \secref{subsec:WGTS-magnet-cryostat-system}) and sends it back into the Inner Loop@ISS. This part of the Tritium Loop System mainly consists of 14 \gls{tmp}s of type Leybold Turbovac MAG W 2800 in the first stage and four \gls{tmp}s of type Pfeiffer HiPace300 in the second stage (\figref{Figure:WGTS-with-Loops}).The combination of Circulation Loop and the \gls{wgts} beam tube forms the \gls{il} of the \gls{katrin} experiment.
 	\item [Feed Loop] 
	  	The Feed Loop (blue part in \figref{Figure:loopinjection}) supplies the Circulation Loop with pure tritium gas from the \gls{iss}. Its \SIadj{25}{\litre} buffer vessel, filled batch-wise 1-2 times a week, provides an intermediate storage of pure tritium from the \gls{iss}.  
	\item [Exhaust Loop]
	    The Exhaust Loop (green part in \figref{Figure:loopinjection}) collects gas that has to be removed from the Circulation Loop and transfers it to the \gls{tlk} infrastructure for purification and isotope separation. This gas consists of tritium, deuterium, and protium \footnote{Protium is the hydrogen isotope with one proton and one electron.}, as well as impurities, including tritiated methane, helium, nitrogen, and carbon monoxide. The exhaust gas originates either from the Circulation Loop (separated by the permeator) or the pumping sections at the \gls{dps}, the \gls{cps} and the Rear System.  The Exhaust Loop comprises a \SIadj{25}{\litre} buffer vessel for intermediate storage of the exhaust gas. The exhaust gas is sent batch-wise to the \gls{tlk} infrastructure (\gls{ol}) 2-3 times a week.
	\item [Outer Loop@DPS] 
		The Outer Loop@DPS collects tritium gas from the second differential pumping section (DPS2-F, see \secref{Subsection:DifferentialPumpingSystem}) and transfers it to the Exhaust Loop. This part contains four \gls{tmp}s in the first stage and 2 \gls{tmp}s in the second stage (\figref{Figure:WGTS-with-Loops}).
	\item [Outer Loop@CPS]
		The Outer Loop@CPS connects the argon frost pump of the \gls{cps} to the Central Tritium Retention System of \gls{tlk}. In addition, it includes the system for argon frost preparation and helium purging (\secref{Subsection:CryogenicPumpingSystem}).
	\item [Outer Loop@Rear System]
		This system collects tritium gas that is not removed by DPS1-R, and transfers it to the Exhaust Loop.
\end{description}

The combination of these subsystems with the main infrastructure systems of the \gls{tlk} forms the \gls{ol} (\figref{Figure:Loops_schematics}). 

\clearpage

To meet the functionality requirements of the complex Loop system distributed over \SI{30}{\metre} along the beam line at \gls{wgts}, \gls{dps}, \gls{cps} and \gls{rs} (each with its individual glove box system), a large variety of valves, sensors, actuators, pumps, and analytical systems is necessary. Custom-made pipes connect sensors, valves and other components with all-metal sealing. An integral He-leak rate of \SI{<e-9}{\milli\bar\litre\per\second} at each flange connection was ascertained to ensure safe operation while tritium is circulating. 

\begin{figure}[t]
	\centering
	\includegraphics[width=0.8\textwidth]{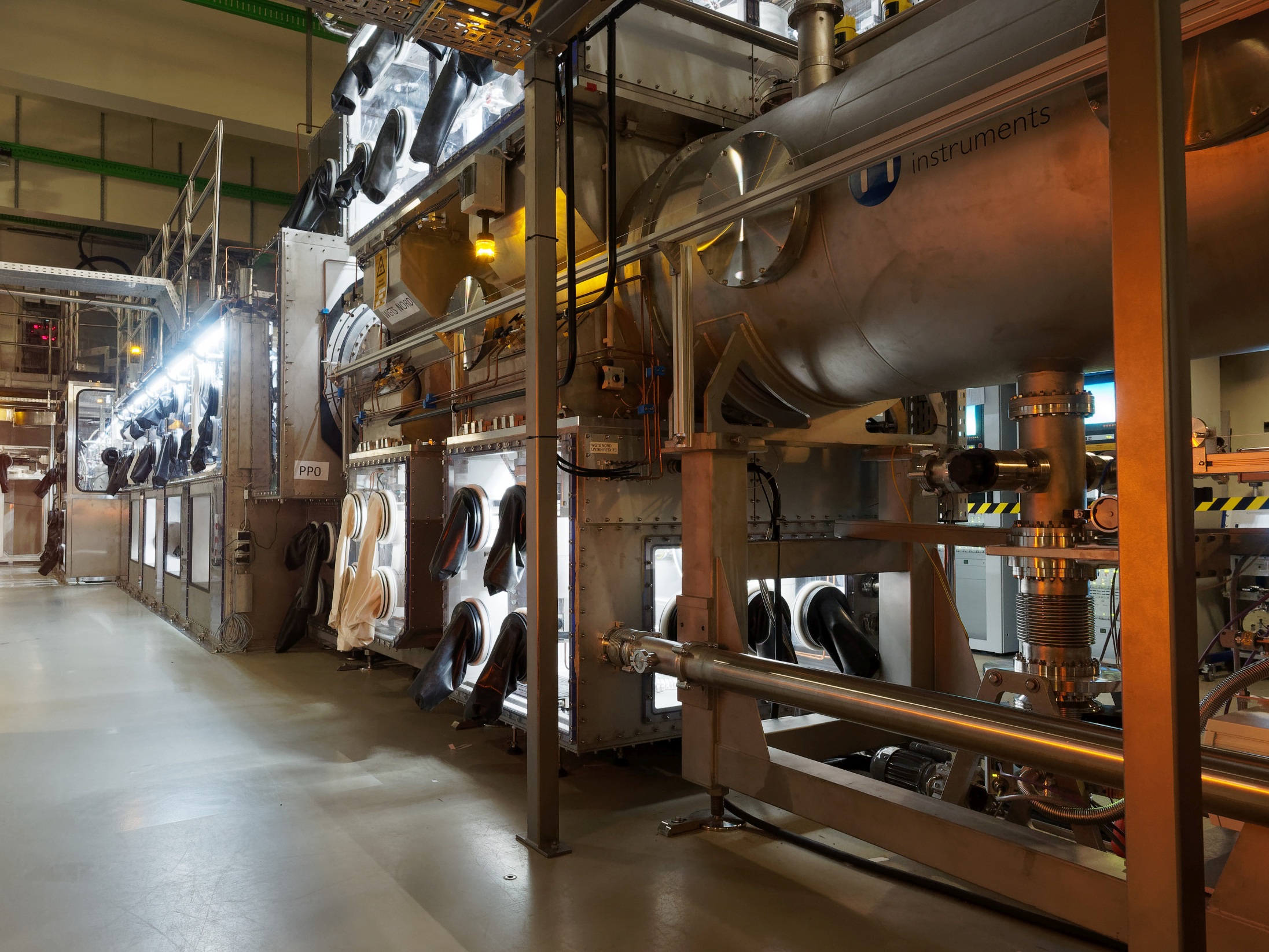}
	\caption{A View along the beam line. Shown on the right side is a part of the \gls{wgts} magnet cryostat with the glove boxes around the DPS1-F pumps (top and bottom). To the left follows the glovebox system around \gls{dps} and \gls{cps}.}
	\label{Figure:2ndContainment}
\end{figure}

The design and setup of all tritium systems were done in compliance with the administrative and technical framework of the \gls{tlk}, which includes stringent requirements on quality assurance and safety. An introduction to the framework is given in~\cite{Welte2015}. \figref{Figure:2ndContainment} shows a segment of the \gls{katrin} beam line with the glovebox systems inside \gls{tlk}.

\subsubsection{Commissioning Results}
\label{sec:commissioning-results}
This subsection describes the commissioning results of the \gls{il} system with respect to the stability of gas injection. 

Before the \gls{wgts} magnet cryostat was available, a part of the \gls{il} system was commissioned via a bypass tube, mimicking the \gls{wgts} beam tube~\cite{Priester2015}. The gas processing performance of the \gls{tlk} infrastructure was tested with comparable tritium throughputs as needed for \gls{katrin} operation~\cite{Welte17}.

\begin{figure}[t]
    \centering
    \includegraphics[width=0.8\textwidth]{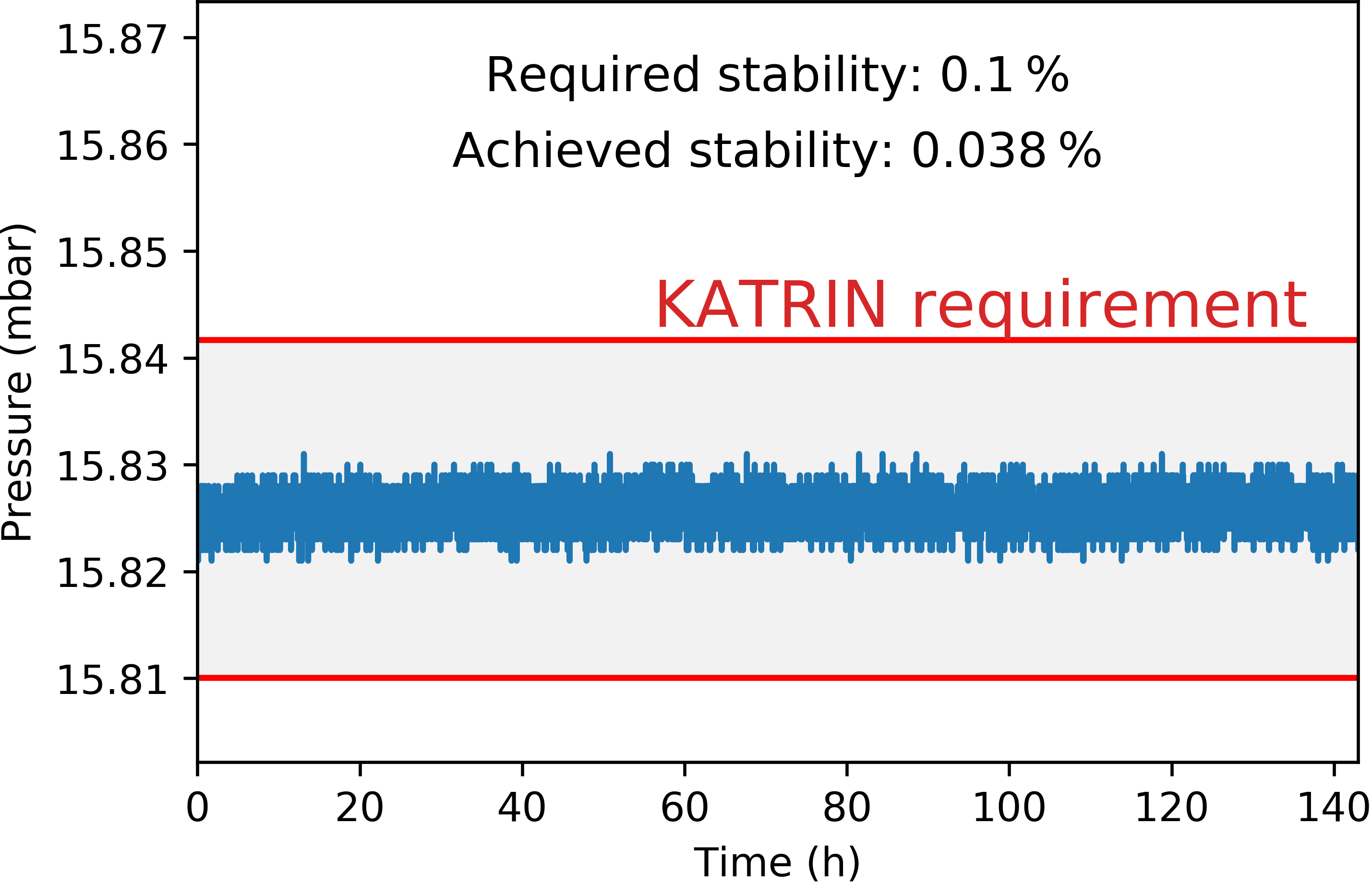}
    \caption{Achieved pressure stability. Shown is the pressure in the buffer vessel responsible for stabilizing the injection pressure (blue) together with the \SI{0.1}{\percent} stability requirement (red). The relation between buffer vessel pressure and injection gas flow rate is discussed further down in the text.}
    \label{Figure:loopstability}
\end{figure}

During the inactive commissioning of the whole Loop system, deuterium was circulated through the \gls{wgts}, \gls{dps}, and \gls{cps} in the same manner as tritium would be during neutrino-mass measurements. With deuterium at the \gls{wgts} beam tube temperature of \SI{30}{\kelvin} and a throughput of \SI{26}{\gram\per\day}, a stability of the injection pressure of \SI{0.038}{\percent} could be reached, far exceeding the requirement of \SI{0.1}{\percent}, as can be seen in \figref{Figure:loopstability}.

\begin{figure}[b]
	\centering
	\includegraphics[width=0.8\textwidth]{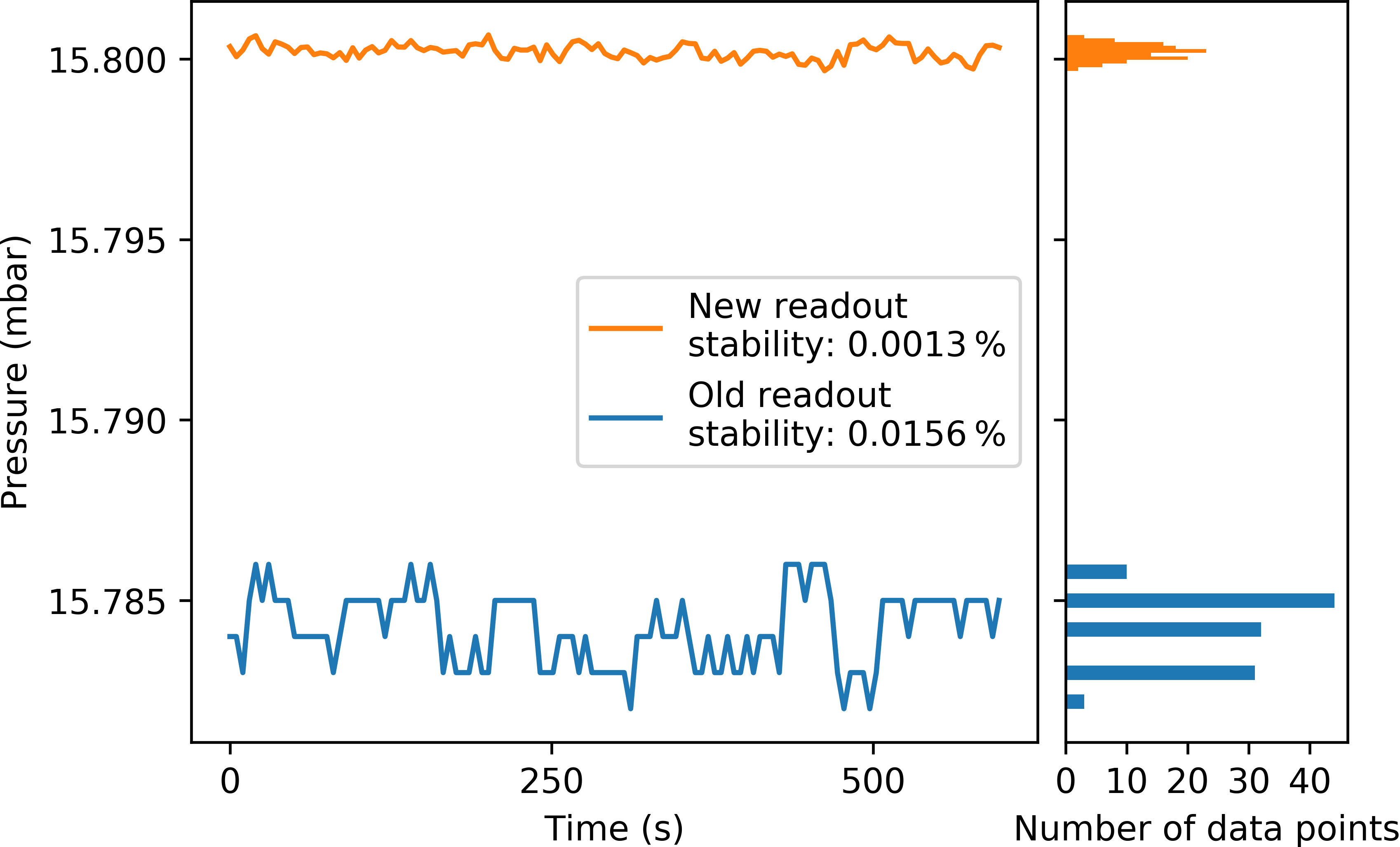}
	\caption{The improvement in pressure transducer readout by deploying a high-resolution analog-to-digital converter in the control circuits. Shown here is a comparison between the old and the new readout stability for the pressure in the buffer vessel responsible for stabilizing the injection pressure.
	}
	\label{Figure:loopadc}
\end{figure}

The pressure fluctuations shown in \figref{Figure:loopstability} were found to be caused by a too-coarse digitization of the pressure transducer signal, used in the regulation system. During stable operation, the signal only flips between two digitization steps, causing the regulation system to lag slightly behind the real pressure change. This has imprinted a periodic fluctuation in the signal. This proved to be an issue, especially for set points at low presure. The control circuit was subsequently improved by adding a dedicated high-resolution analog-to-digital converter. The final pressure stability improvement is demonstrated in \figref{Figure:loopadc}. A stability of \SI{0.0013}{\percent} was  achieved at the pressure set point used in nominal measurement conditions. Furthermore, it is possible to stabilize the injection pressure at the level of \SI{0.1}{\percent} for pressures as low as \SI{0.2}{\milli\bar}.

\begin{figure}[!ht]
    \centering
    \includegraphics[width=0.8\textwidth]{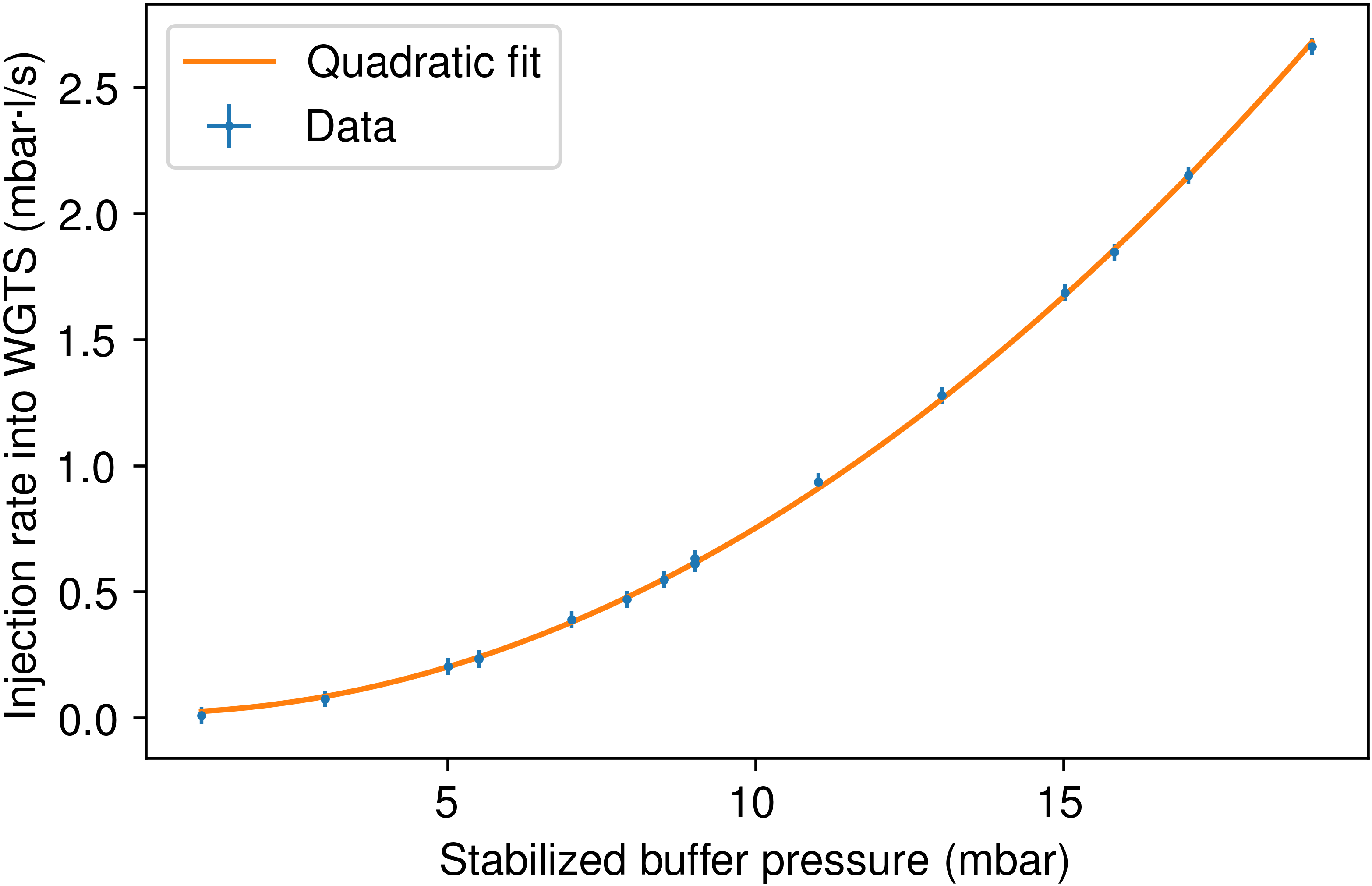}
    \caption{Characterization of the injection flow rate dependence on the buffer vessel pressure. Tritium gas is injected from the stabilized buffer vessel over the injection capillary into the middle of the \gls{wgts}. The injection rate through the capillary is proportional to its conductance and the pressure drop over its length. The pressure at the entrance of the capillary is in the mbar region, and is less than \SI{e-2}{\milli\bar} at its exit. The conductance in viscous flow is proportional to the mean pressure. With a negligible pressure at the outlet of the capillary, a quadratic dependence of the injection flow rate on the pressure of the stabilized buffer is found.
    }
    \label{Figure:Loops_injrate}
\end{figure}

The relation between the buffer vessel pressure and the injection gas flow rate has been characterized in the range from \SIrange{1}{19}{\milli\bar}. The gas flow inside the injection capillary is expected to be predominantly viscous. In this flow regime, the flow $q = C(p) \cdot p$ depends approximately quadratically on the buffer vessel pressure, as the conductance $C(p)$ is approximately linear in pressure $p$. The observed relation in \figref{Figure:Loops_injrate} agrees well with this expectation.

\clearpage

%% file: WGTSRearWall.tex
\subsection{WGTS Rear Wall}
\label{sec:wgts_rear_wall}

\subsubsection{Potential Definition}
The starting potential of the \betael{} is provided by the cold and strongly magnetized plasma in the \gls{wgts}.
In order to accurately measure the \betael{} energy, a homogeneous starting potential distribution needs to be established over the whole magnetic flux tube volume in the \gls{wgts}.
Spatial inhomogeneities eventually occur due to the formation of local space charges in the plasma. 

The \acrfull{rw} is an important tool for determining and manipulating the plasma potential distribution. 
It is a gold-coated stainless steel disk with an outer diameter of \SI{145}{\milli \meter} in the rear system. 
The magnetic field along the \gls{wgts} confines \betael{} as well as low energy ions and electrons from the plasma on the magnetic field lines which terminate on the \gls{rw} surface. Therefore, a good conductance between the \gls{rw} and the plasma is assumed.
The surface potential is defined by the sum of the gold work function and the bias voltage of up to $\pm$\SI{500}{\volt} which can be applied.

A picoammeter is connected to the \gls{rw} to measure the current flowing through the disk towards ground. 
The \gls{rw} can be baked out at temperatures of about \SI{120}{\celsius}.  
In the center of the disk, a \SI{5}{\milli \meter} hole allows for the transmission of the \gls{egun} beam (\secref{sec:egun}). 
The \SI{134}{\tesla\centi\meter\squared} flux tube is completely mapped onto the \gls{rw} such that \betael{}s, secondary electrons and ions produced within the \gls{wgts} hit its surface where they are absorbed and neutralized.

Dedicated simulations show that both the \gls{rw} and the \gls{wgts} beam tube surface potential establish the major boundary conditions for the plasma potential distribution within the source system \cite{PhDKuckert2016}. 
These simulations indicate that there is a \gls{rw} bias voltage at which the work function differences between the beam tube and \gls{rw} are compensated, and the spatial inhomogeneity of the plasma potential is minimized. This optimal point can be determined experimentally by measuring the $\beta$-rate at various \gls{rw} bias voltages. The optimal bias voltage produces a radially homogeneous endpoint $E_0$ distribution. This procedure can also be performed by measuring the radial variation in line position of the $L_3$ line of meta-stable \kr{} in the presence of tritium gas.
In this configuration, the radial variations of the plasma potential are minimized, providing optimum starting conditions for the \betaels{}.

\subsubsection{Rear Wall Setup}

\begin{figure}[!ht]
 \centering
 \includegraphics[width=.6\textwidth]{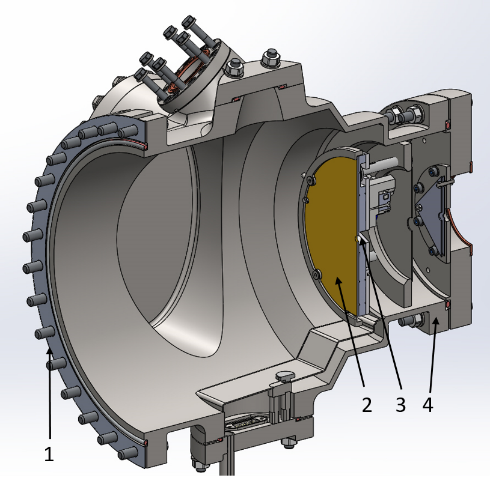}
 \caption{Cross section of the entire integrated \gls{rw} unit. (1) and (4) Flanges of the \gls{rw} chamber to the \gls{wgts} magnet cryostat and Rear System, respectively. (2) The \gls{rw} itself. (3) The central hole for \gls{egun} beam transmission.}
 \label{fig:rw_1}
\end{figure}

\figref{fig:rw_1} shows the cross section of the \gls{rw} unit. The \gls{rw} is mounted on ceramic rods for electrical isolation. The \gls{rw} surface itself is gold; it has a homogeneous and constant work function to guarantee a constant surface potential. Tests of gold surfaces in air and in vacuum showed a work function of about \SI{4.2}{\electronvolt}, which is almost \SIadj{1}{eV} off from the ideal work function value. This can be attributed to surface impurities, and is still acceptable for the \gls{katrin} experiment. The average work function spread across the surface is around \SI{20}{\milli\electronvolt} \footnote{While mono-crystalline gold would be a more ideal surface, it is infeasible to grow a surface of the size required.}. The \gls{rw} can be mounted and exchanged manually. For this purpose, the vacuum chamber in front of the \gls{rw} has a \SI{200}{\milli\metre}-diameter flange. Another flange (not shown here), which faces the \gls{rw} at an incident angle of \SI{55}{\degree}, carries a UHV-proof, \SI{137}{\milli\metre}-diameter quartz window. It allows the illumination of the \gls{rw} with light in the far UV (down to \SI{200}{\nano\metre}) in order to feed the \t2{} plasma with additional photoelectrons. This measure enhances the conductivity and maintains the quasi-neutrality of the plasma. Due to the very compact construction of the \gls{rw} chamber, it was milled out of a solid stainless steel block.

\subsubsection{Rear Wall Illumination}
\label{sssec:rw_uv}
\begin{figure}[!ht]
 \centering
 \includegraphics[width=.7\textwidth]{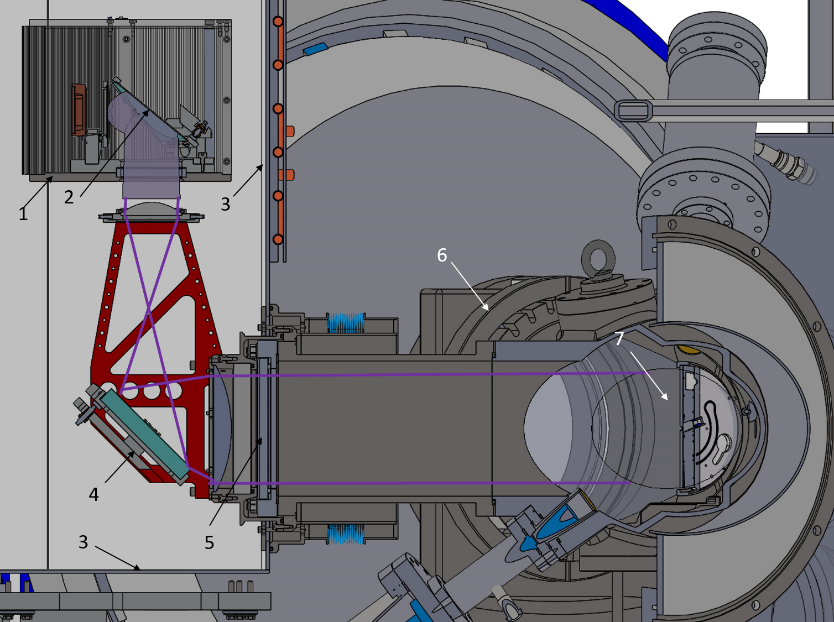}
 \caption{Setup for UV-illumination of the \gls{rw}. (1) Steel box housing the Cermax\textsuperscript{\textregistered} lamp. (2) Beam splitting mirror. (3) Recess of second containment, giving direct access to Cermax\textsuperscript{\textregistered} lamp and optics. (4) Adjustable mirror. (5) UHV-proof quartz window. (6) Flange and valve at \gls{wgts} entrance. (7) The \gls{rw} itself. See text for the working principles of this UV-illumination setup.}
 \label{fig:rw_2}
\end{figure}

\figref{fig:rw_2} shows the optical setup for the UV-illumination of the \gls{rw}. It had to fit inside the narrow channel between the superconducting magnet of the Rear System and the entrance of the \gls{wgts} magnet cryostat. The steel box in the upper right houses a short-arc Cermax\textsuperscript{\textregistered} xenon lamp\footnote{Type Cermax PE1000DUV (lifetime \SI{1000}{\hour}), Perkin Elmer, \url{https://www.perkinelmer.com}} and shields any disturbing stray magnetic field of the order of \si{\milli\tesla}. The lamp yields a UV power of $\approx \SI{0.3}{\watt}$ in the  wavelength range of $\lambda =$~\SIrange{200}{266}{\nano\metre} for producing photoelectrons from the \gls{rw}. Watercooling removes \SI{1}{\kilo\watt} of excess heat from the lamp. One of the two beam-splitting mirrors reflects only the UV part of the spectrum. Two lenses, positioned with a focal point between them, enlarge the approximately parallel beam from the lamp to the size of the \gls{rw}. A mirror adjusts the beam direction. A shallow recess in the second containment gives access to the lamp and the optical components for replacement and adjustment.

\begin{figure}[t]
 \centering
 \includegraphics[width=.7\textwidth]{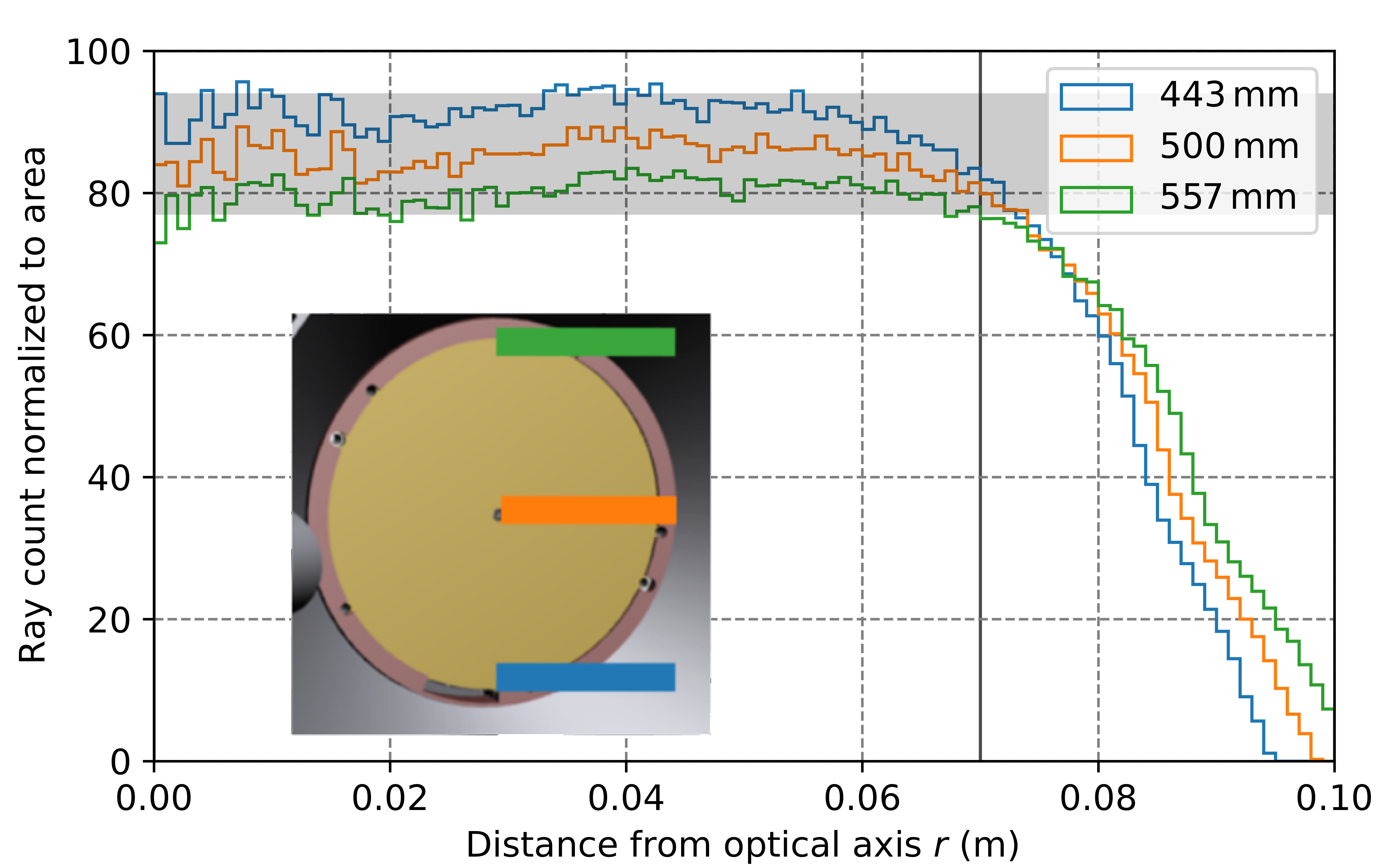}
 \caption{Simulation of the light distribution on the \gls{rw}. The intensities were sampled along three lines: closest at 443 mm (blue), center at 500 mm (orange) and furthest at 557 mm (green). The vertical, solid gray line shows the maximal radius $R=\SI{7}{\centi\meter}$ of the \gls{rw}, while the shaded area indicates the homogeneity requirement of $\pm 10\%$ centered around the mean of the data up to $R$. Since the \gls{rw} is at an angle with respect to the optical axes, a small gradient from the near end to the far end cannot be avoided.}
 \label{fig:rw_3}
\end{figure}

It is not trivial to achieve homogeneous illumination of the \gls{rw}. The arc of the Cermax\textsuperscript{\textregistered} lamp is placed in the focus of a parabolic mirror. In the far field, the radial intensity profile of the beam decreases towards the rim and is similar to a Gaussian. Its transformation to a homogeneous profile is enabled by projecting the sharp intermediate focus of the arc image between the two lenses. The requisite condition is that the angular emittance at any point of this image reaches the solid angle of the full light cone. The final configuration was optimized by simulations with the optics program COMSOL\footnote{COMSOL Multiphysics including Ray Optics package, \url{https://www.comsol.com}.}. The results shown in \figref{fig:rw_3} yield the specified homogeneity of \SI{+-10}{\percent} (gray horizontal band) over the entire \gls{rw} up to its maximal radius of \SI{7}{\centi\meter} (marked with a solid line). 

%% file: TransportSystem.tex
\section{Transport System}
\label{sec:transport_system}

The Transport System adiabatically transports \betaels{} from the source to the Spectrometer System and in parallel reduces the tritium flow rate into the spectrometer system by more than 12 orders of magnitude. It consists of the Differential Pumping System and the Cryogenic Pumping System. Both systems are described in the following two subsections.

\input{TransportSystemDPS}

\input{TransportSystemCPS}

\clearpage

%% file: TransportSystemDPS.tex
\subsection{Differential Pumping System}
\label{Subsection:DifferentialPumpingSystem}

The Differential Pumping System (\gls{dps}) is located between the \gls{wgts} (\secref{sec:source_system}) and the Cryogenic Pumping System \gls{cps} (\secref{Subsection:CryogenicPumpingSystem}), as can be seen in~\figref{Figure:WGTS-with-Loops}. It is the second differential pumping stage (historically called DPS2-F, since it comes after DPS1-F) to reduce the tritium flow rate in the beam tube in the downstream direction to the spectrometers. The first differential pumping stage is part of the \gls{wgts} magnet-cryostat, which includes pump ports (\secref{subsec:WGTS-magnet-cryostat-system}) and the tritium loops system (\secref{sec:innerloop-setup}).

\begin{figure}[b]
	\centering
	\includegraphics[width=0.9\textwidth]{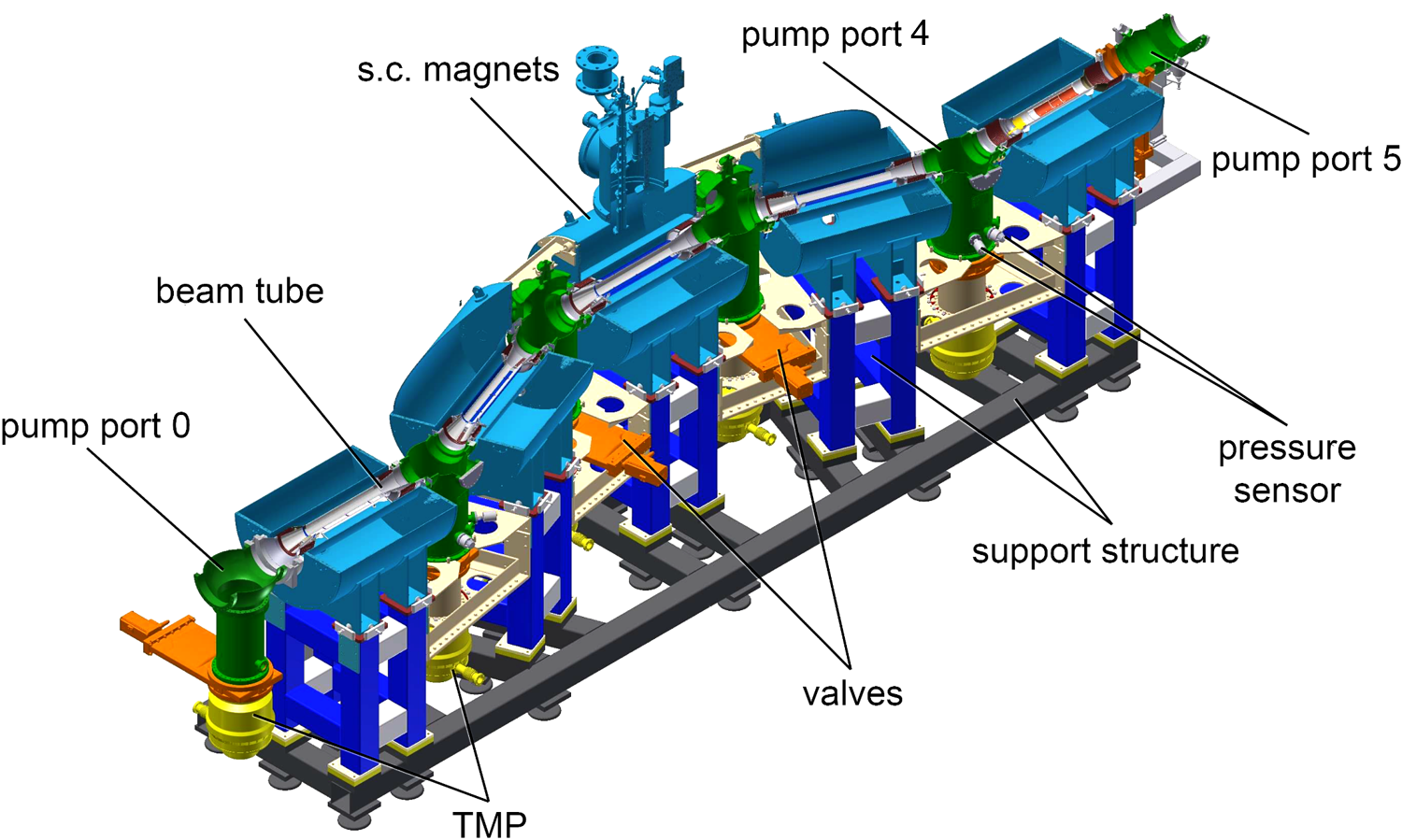}
	\caption{CAD drawing of the \gls{dps} . Five super-conducting (s.c.) magnets (light blue) surround the beam tube at room temperature. The gas is pumped by \gls{tmp}s connected to the pump ports P0 to P5 (green). The figure is adapted from \cite{Friedel2019}.}
	\label{Figure:DPS-System}
\end{figure}

The \gls{dps} was the first complex magnet cryostat delivered and commissioned in the \gls{katrin} experiment \cite{KATRIN2005,Lukic2012}. After a quench during the commissioning in 2011, one of the safety diodes at the superconducting magnet coils was damaged. Safety diode replacement was considered cost- and time-prohibitive, since the diodes were installed deep inside the magnet cryostat. Instead, a new, simplified design of the \gls{dps} was proposed (\figref{Figure:DPS-System}) based on 5 custom-designed single superconducting magnet cryostat systems (\secref{SubSubSection:DPSMagnetSystem}), a beam tube at room temperature (\secref{SubSubSection:DPSPumpingSystem}), and an additional pump port, called PP0 (\figref{Figure:DPS-vacuum-flow-diagram}). This new design also featured an access port in the cryostat, which allowed the cold bypass safety diodes to be more easily replaced, should they be damaged.

\subsubsection{DPS Principle and Basic Requirements}

The basic requirements for the \gls{dps} are:
\begin{itemize}
	\item To guide the \betaels{} adiabatically from the \gls{wgts} magnet cryostat to the \gls{cps} magnet cryostat.
	\item To reduce the neutral tritium flow rate (about \SI{1}{\milli\bar\litre\per\second}) in the direction of the \gls{cps} by at least seven orders of magnitude~\cite{KATRIN2005}. This is done in combination with the first differential pumping stage DPS1-F, as shown in \figref{Figure:WGTS-with-Loops} and \figref{Figure:wgts_cad}.
	\item To block and remove magnetically guided tritium ions from the source. 
\end{itemize}

\begin{figure}[t]
	\centering
	\includegraphics[width=0.9\textwidth]{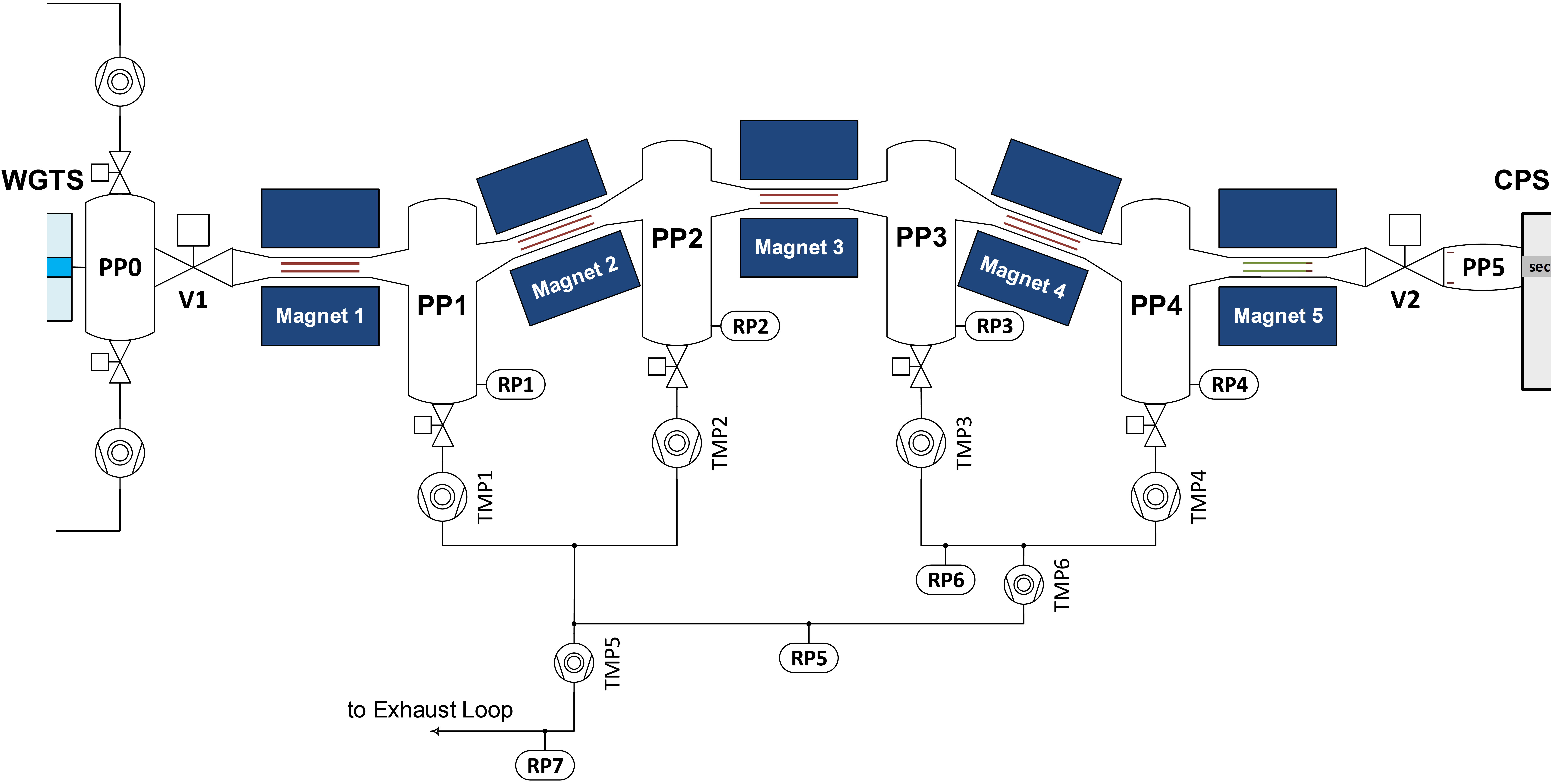}
	\caption{Schematic of the second differential pumping stage. The pumping system is part of the tritium loops@DPS (\secref{sec:innerloop-setup}). V: gate valve, PP: pump port, TMP: turbomolecular pump, RP: pressure gauge. This simplified schematic does not show all pressure gauges. More information about the pumping system is given in \secref{SubSubSection:DPSPumpingSystem}.}
	\label{Figure:DPS-vacuum-flow-diagram}
\end{figure}

The adiabatic guidance of the \betaels{} is done with a magnet system consisting of 5 single superconducting magnets (\secref{SubSubSection:DPSMagnetSystem}).
While the reduction of the neutral tritium flow rate is achieved by the method of differential pumping (\secref{SubSubSection:DPSPumpingSystem}), the reduction of the ions is not possible with that method. This is because ions follow magnetic field lines and cannot be pumped out through pump ports which are  positioned outside of the field lines.
To overcome this issue and to avoid spectrometer contamination by magnetically-guided tritium ions from the source, a dedicated ring and dipole electrode system was developed and installed inside of the \gls{dps} beam tube. A detailed discussion of ion blocking and removal is given in \secref{SubSection:IonBlocking}. 

\figref{Figure:DPS-vacuum-flow-diagram} shows a simplified vacuum flow diagram of the \gls{dps} system. Along the beam line, it consists of a sequence of beam tube elements and pump ports to which \gls{tmp}s are connected. Differential pumping \cite{War90}  is achieved by the conductance difference of the beam tube elements with respect to the effective pumping speed of the pump ports, which is about ten times higher. The beam tube elements are configured as a chicane: they are tilted by \SI{20}{\degree} to each other in horizontal plane. This geometry prevents the beaming effect by eliminating a direct line-of-sight from the \gls{wgts} to the spectrometers. This measure enhances the removal of neutral gas.

\subsubsection{Magnet System}
\label{SubSubSection:DPSMagnetSystem}

The company Cryomagnetics, Inc.\footnote{Cryomagnetics, Inc., Oak Ridge, TN, USA, \url{https://www.cryomagnetics.com}.} built six similar magnets for the Source and Transport System, each with a maximum magnetic field of \SI{5}{\tesla}. Five of these magnets are used in the \gls{dps}, while the sixth is used in the Rear System (\figref{Figure:RearSystem}). The magnets are cooled by liquid helium which is continuously recondensed by a cryocooler system~\cite{Arenz2018c}.

\begin{table}[!ht]
	\caption{Main technical requirements for the magnets of \gls{dps}.}
	\begin{center}
		\begin{tabular}{p{8cm}cp{6cm}} 
			\hline
			\textbf{Subject} &~& \textbf{Requirement} \\
			\hline
			Magnetic flux density at the center of the beam tubes &~& \SI{>5}{\tesla} (nominal field)\\[0.5ex]
			Magnetic magnetic flux density inside the pump ports within transported magnetic flux & & \SI{>1}{\tesla} (at nominal field)$^\textrm{a}$ \\[0.5ex]
			Magnetic field homogeneity at the center of the beam tubes & & 
			better than \SI{100}{ppm} over a  length of \SI{100}{\milli\meter} along the axis$^\textrm{b}$\\[0.5ex]
			Magnetic field decay in persistent current mode & & \SI{<0.1}{\percent\per month} \\[0.5ex]
			Coaxiality between the solenoid and the warm bore& & \SI{<1}{\milli\meter}\\[0.5ex]
			Neighboring magnetic fields && Magnets must withstand forces due to neighboring magnet systems at full magnetic field\\[0.5ex]
			Operation without refilling of liquid helium && \SI{9}{months}\\
			\hline
			\multicolumn{3}{l}{\scriptsize a) The field needs to be at least \SI{1}{T} to ensure adiabatic transport of decay electrons into the spectrometer.}\\
			\multicolumn{3}{l}{\scriptsize b) Specific to \gls{fticr} functionality. }\\
		\end{tabular}
	\end{center}
	\label{table:DPSrequirements}
\end{table}

The main technical requirements for the magnets of the \gls{dps} are summarized in \tabref{table:DPSrequirements}. Each of the \gls{dps} magnets contains a main solenoid with 2 correction coils at each end. The magnets are enclosed in liquid helium bath cryostats with a recondenser unit cooled by a cold head. The systems can run for at least 9 months without the need of a liquid helium refill. A heater in the liquid helium vessel, which is regulated by pressure measurement in the helium gas phase, keeps the pressure above atmospheric. The magnets are nominally operated in persistent current mode. Magnet coils are protected against damage from overheating in case of a quench by freewheeling cold bypass diodes in the liquid helium vessel. A quench relief valve and a rupture disk are mounted in parallel on each liquid helium vessel to release the helium gas rapidly in the case of a quench. All these safety components, as well as the liquid helium refill nozzle and the hand valve to the insulation vacuum, are placed outside the second containment glovebox of the \gls{dps} (see \secref{SubSubSection:DPSBeamtube}). 

For a more detailed description, including the most important technical parameters, refer to \cite{Arenz2018c}.

\subsubsection{Beam Tube, Pump Ports and Glove Box}
\label{SubSubSection:DPSBeamtube}

For historical reasons, the section between the \gls{wgts} magnet cryostat and the \gls{dps} system is called the Pump Port Element 0, although it is connected to the \gls{wgts} beam tube vacuum. The same holds for the Pump Port Element 5 between \gls{dps} and \gls{cps}, which is connected to the \gls{cps} beam tube vacuum (\figref{Figure:DPS-vacuum-flow-diagram}). 

The \gls{dps} magnets are mounted on a support frame custom made by the KIT main workshop (Technik-Haus). The frame is made of non-magnetic stainless steel. It is designed to support the weights of the magnets and to withstand the magnetic forces between the magnet systems. The positions of the magnets can be adjusted  horizontally. The vertical positions can be changed with metal shims, which also allow a moderate tilting of the magnets. 

According to the \gls{tlk} regulation of safe tritium handling, all primary systems need to be contained in secondary containments (usually gloveboxes) \cite{Doerr2005,Welte2015}. For the \gls{dps}, a custom-made glovebox was designed and built by the company GS\footnote{GS GLOVEBOX Systemtechnik GmbH, Germany, \url{https://http://www.glovebox-systemtechnik.de}}. Specific requirements for the \gls{dps} call for all cryogenic liquids and compressed gases of the magnets, as well as access to their valves and accessories, to be outside the secondary containment. Hence, the treadable slab of the glovebox closes around the magnet domes with rubber liners. The safety outlet of the insulation vacua of the magnets are led out through the second containment via formed bellows and are closed with dropout plates. PP0 and PP5 are housed in the appendices of the \gls{dps} glovebox, which also provide the closing of the secondary containment to the \gls{wgts} magnet cryostat and the \gls{cps} magnet cryostat.
The maximum allowed leak rate for each glovebox is \SI{0.1}{vol\percent\per\hour}.

\subsubsection{Pumping System}
\label{SubSubSection:DPSPumpingSystem}

Two \gls{tmp}s of type Turbovac MAG W 2800\footnote{Leybold GmbH, Germany, \url{https://www.leybold.com}} are attached to PP0.
Four Turbovac MAG W 2800 \gls{tmp}s are connected to PP1 - PP4, with one per pump port. The sequential positioning of the \gls{tmp}s leads to differential pumping of the tritium gas along the beam line. In order to reach the desired compression factor, the gas from the outlet of the \gls{tmp}s at PP3 and PP4 is compressed via a HiPace300 \gls{tmp}\footnote{Pfeiffer Vacuum GmbH, Germany, \url{https://www.pfeiffer-vacuum.com}}. Then the gas is piped, together with the outlet of the \gls{tmp}s from PP1 and PP2, to a second HiPace300 \gls{tmp}, which pumps the gas to the Exhaust Loop of the Tritium Loop System (see \secref{sec:innerloop-setup} and \figref{Figure:DPS-vacuum-flow-diagram}). The pressure inside the \gls{dps} beam tube is monitored with both a Capacitance Diaphragm Vacuum Gauge (Baratron\textsuperscript{\textregistered})\footnote{MKS Instruments, USA \url{https://www.mksinst.com}} and a cold cathode gauge at each of PP1 and PP2. PP3 and PP4 are each equipped with a cold cathode and an extractor gauge, and, in case of PP4, with a quadrupole mass spectrometer. A Baratron\textsuperscript{\textregistered} pressure gauge measures the pressure at the outlet of the PP3 and PP4 \gls{tmp}s (RP6 in \figref{Figure:DPS-vacuum-flow-diagram}). Another Baratron\textsuperscript{\textregistered} pressure gauge is placed between the first HiPace 300 \gls{tmp} and the second HiPace 300 \gls{tmp} (RP5 in \figref{Figure:DPS-vacuum-flow-diagram}).
The \gls{tmp}s and the cold cathode vacuum gauges are shielded with soft iron against the stray field of the \gls{dps} magnets.
Additionally, a new fixture was developed for the Leybold MAG W2800 \gls{tmp}s, which holds the torque momentum in case of a rotor crash~\cite{SturmPopov2018}. 
Because all components used in the tritium-containing primary system of the \gls{dps} need to be fully tritium compatible, they must all be metal-sealed to the outside and cannot contain any halogenic materials. While the \gls{tmp}s are not technically fully tritium compatible, the \gls{tlk} made an exception based on the results of a study on the usability of MAG W2800 \gls{tmp}s for tritium pumping~\cite{Priester2013}.

Similar to the \gls{bte} and the \gls{ppe}, the pumping system is classified as a primary system according to the Technical Conditions of Delivery and Acceptance of \gls{tlk} \cite{Doerr2005,Welte2015}.
Therefore, the maximum allowed helium leak rate is \SI{1e-9}{\milli\bar\litre\per\second} for a single flange connection and \SI{1e-8}{\milli\bar\litre\per\second} for the total \gls{dps} beam tube system.

\subsubsection{Commissioning Results}

\paragraph{Magnet tests}

The five \gls{dps} magnets (and the \gls{rs} magnet) met most requirements in tests performed at the company prior to delivery to KIT. These included magnetic force tests in triplet arrangement with forced quenches to simulate the extreme conditions in the final setup. Three out of six magnets met the stringent requirements of field homogeneity at the center of the solenoid necessary to house the \gls{fticr} in the magnet bore. One of these magnets was then placed on position five (between PP4 and V2, \figref{Figure:DPS-vacuum-flow-diagram}) for the \gls{fticr} (\secref{sec:ionblocking}). In 2016, the RS magnet showed an excessive field drift caused by wire damage. It was successfully repaired later at the manufacturer site in USA. For further details, see \cite{Arenz2018c}.

\paragraph{Influence of magnetic field on the beam tube pressure sensors}

Due to the shielding of the relevant pressure gauges with soft iron material, the influence of the magnetic field on the pressure measurement inside the \gls{dps} beam tube is \SI{<=10}{\percent}, which is negligible for the \gls{katrin} experimental operation.
	
\paragraph{Vacuum tests}

After the commissioning of the \gls{dps} beam tube and the \gls{ol} at the \gls{dps} (but prior to start of the circulation tests with deuterium), a final pressure of about \SI{e-9}{\milli\bar} was reached in the \gls{dps} beam tube. This subsystem cannot be baked, so this is the best value.
Helium leak tests of the \gls{dps} beam tube showed that leak rate of the  flanges exceeded the limit of \SI{1e-9}{\milli\bar\litre\per\second} for tritium containing systems at \gls{tlk}. All three flanges were QDS\textsuperscript{\textregistered} connections with HELICOFLEX\textsuperscript{\textregistered} seals in the beam tube:

\begin{itemize}
	\item Connection between \gls{bte}1 and PP1: \SI{3e-8}{\milli\bar\litre\per\second},
	\item Connection between \gls{bte}1 and PP2: \SI{3e-8}{\milli\bar\litre\per\second},
	\item Connection between \gls{bte}2 and PP2: \SI{1.5e-8}{\milli\bar\litre\per\second}.
\end{itemize}

The \gls{tlk} formally accepted these leak rates since the tritium pressure in this part of the \gls{katrin} setup is always below \SI{1}{\milli\bar}. The effective tritium leak rate into the second containment is therefore at least three orders of magnitude smaller than the measured helium leak rate.

\paragraph{Tritium reduction factor measurements}

During the \gls{firsttritium} measurement in Spring 2018 
\cite{Aker2020_firstOperationTritium}, the tritium reduction factor of the \gls{wgts} (in the DPS1-F section) and the combined tritium reduction factor of the \gls{wgts} and \gls{dps} (in the combined DPS1-F and DPS2-F sections) were measured. From these measurements, the tritium reduction factor of the \gls{dps} was derived. The gas composition used for these measurements was 1\% DT in D$_2$, corresponding to 0.5\% of the nominal source activity~\cite{Aker2020_firstOperationTritium}.

In the \gls{wgts} tritium reduction factor measurement, the gas was circulated through the \gls{il} and the \gls{dps}. The valve V2 (\figref{Figure:DPS-vacuum-flow-diagram}) to the \gls{cps} was closed and the Outer Loop@DPS (\secref{sec:innerloop-setup}) was not connected to the Exhaust Loop. A comparison of the injection rate into the \gls{wgts} beam tube with the pressure rise in the Outer Loop@DPS (sensor RP7, \figref{Figure:DPS-vacuum-flow-diagram}) gave a \gls{wgts} reduction factor of \SI{3.5e+3}{} with a 5\% uncertainty.
In the determination of the combined reduction factor of the \gls{wgts} and \gls{dps}, the integral tritium flow into the \gls{dps} during \gls{firsttritium} was calculated. Based on flow sensor readings and \gls{lara} measurements, this was about  \SI{3e+14}{Bq}. After the end of the \gls{firsttritium} measurement, the tritium accumulated on the Ar frost inside the \gls{cps} (\secref{Subsection:CryogenicPumpingSystem}) was purged into a buffer vessel. The activity was determined with a liquid scintillation counter to be \SI{2.1e+7}{Bq}. 

The combined DT reduction factor for the \gls{wgts} and \gls{dps} is thus about \SI{1.4e+7}{} with an uncertainty of 10\%, while the reduction factor for the \gls{dps} alone is about \SI{4e+3}{}. For a gas composition of almost pure tritium (98\%, similar to regular neutrino-mass mode operation), an even higher reduction factor is achieved as the average molecular mass of T$_2$ is higher than those of DT and D$_2$. Repeating the discussed methodology but using data of the first two neutrino mass measurement campaigns, the combined reduction factor for pure (98\%) tritium was determined to be \SI{9.6+-1.0e+7}{} at nominal column density~\cite{marsteller2020neutral}.

Thus, the performance of the combined differential pumping stages of the \gls{wgts} and \gls{dps} does surpass the required value for tritium flow rate reduction into direction of \gls{cps}, and therefore enables an enlarged time interval for the necessary \gls{cps} regeneration (\secref{Subsubsection:ArgonPump}).

%% file: TransportSystemCPS.tex
\subsection{Cryogenic Pumping System}
\label{Subsection:CryogenicPumpingSystem}

\subsubsection{CPS Principle and Basic Requirements}

The \gls{cps} is the last part of the transport section.
It guides \betaels{} adiabatically towards the spectrometer section, while reducing the tritium flow rate by at least seven orders of magnitude to achieve the required overall tritium flow rate reduction factor of at least \SI{1e+14}{}. Upon exiting the downstream end of the \gls{cps}, a tritium flow rate of not more than \SI{1e-14}{\milli\bar\litre\per\second} is allowed into the \gls{ps}.

Following the tritium flow rate reduction by seven orders of magnitude in the previous stage, additional \gls{tmp}s are no longer sufficient to achieve the required additional factor of \num{e7} due to back-diffusion in the pumps. 
Since the integral amount of tritium leaving the \gls{dps} during one 60-day measurement campaign is about \SI{0.5}{cm^3}, a sorption pump integrated into the beam tube was chosen as the workhorse in the next pumping stage. 

The main requirements for such a pump are a high pumping speed for tritium per surface unit (specific pumping speed), long-term tritium retention at operating conditions, and an easy and complete removal of the tritium from the beam tube surface at stand-by conditions (regeneration phase).

The pump of choice given these requirements is a cryo-sorption pump with a condensed argon layer, prepared on the surface of the gold-plated inner beam tube wall.
The main advantage of condensed gas layers on a gold surface compared to solid adsorbents is that one can easily remove the layer together with any adsorbed tritium, minimizing the residual tritium contamination. 

An overview of the whole \gls{cps} magnet cryostat system is given in the next section. The argon cryopump is described in detail in \secref{Subsubsection:ArgonPump}.

\subsubsection{Magnet-Cryostat Setup}

\begin{figure}[th]
	\centering
	\includegraphics[width=\linewidth]{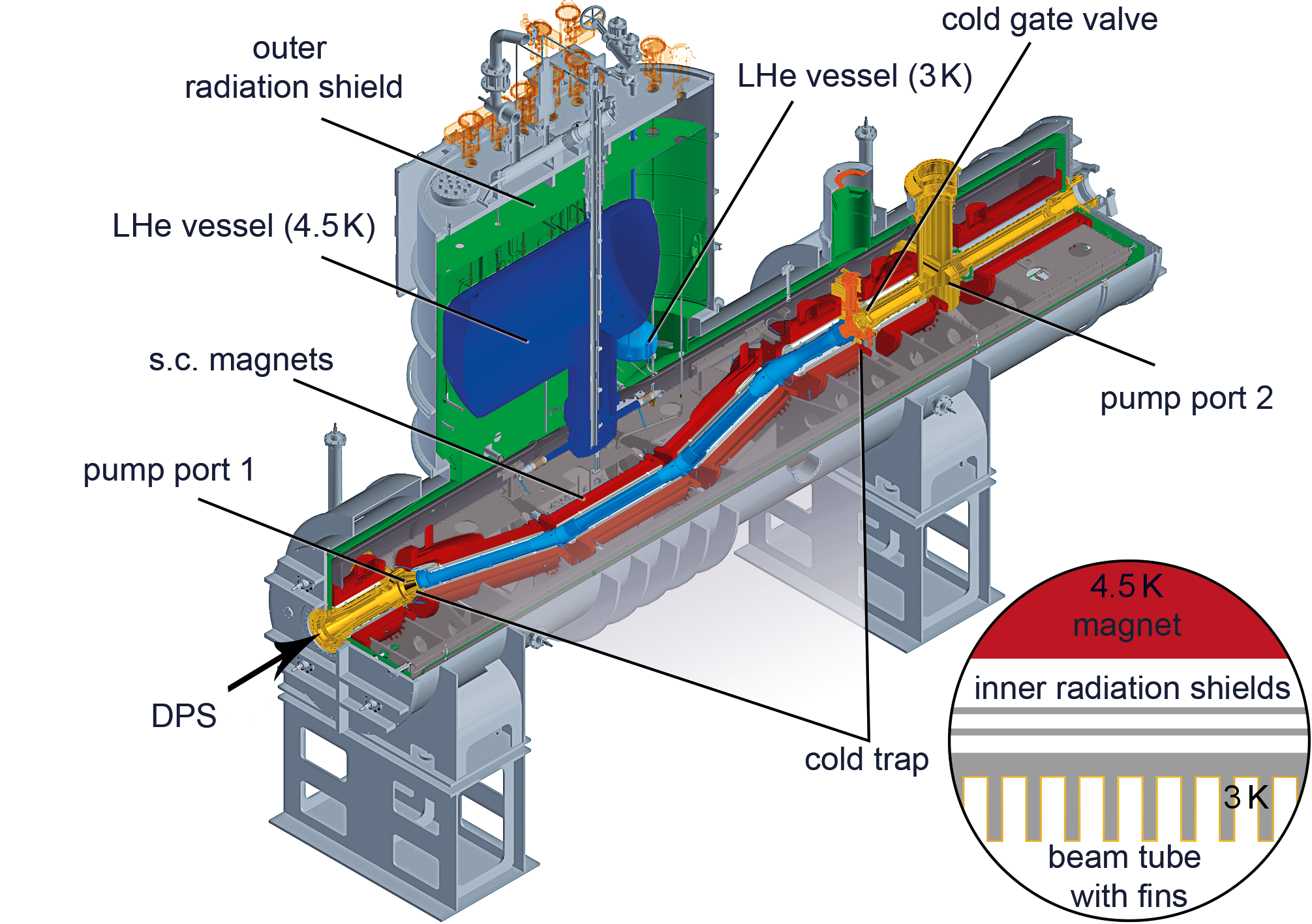}
	\caption{CAD drawing of the \gls{cps} - 7 superconducting magnets (red) surround the gold-plated beam tube. For the cooling of the magnets and beam tube, a \SI{1300}{\litre} liquid helium vessel supplies a reservoir of \SI{4.5}{\kelvin} helium. The cold trap (blue) is highlighted between pump port 1 and the cold gate valve. A detailed description of the other labeled components is given in the text. The figure is adapted from~\cite{Friedel2019}.}
	\label{Figure:CPS_cad}
\end{figure}

The \gls{cps} was built by ASG Superconductors\footnote{ASG Superconductors S.p.A., Italy, \url{https://www.asgsuperconductors.com}} and has a mass of about \SI{12}{\tonne}.
The magnet cryostat is manufactured in one \SIadj{7}{\meter}-long and \SIadj{4}{\meter}-high system~\cite{Gil2010}, housing 7 chambers of superconducting solenoids on one cold support structure, liquid helium reservoirs, 7 beam tubes, thermal shields, and 2 pump ports (\figref{Figure:CPS_cad}). 
Pump port 2 contains a horizontal port, where the \gls{fbm} can be inserted (\secref{Subsubsection:ForwardBeamMonitor}), as well as a vertical one for the \gls{ckrs} (\secref{Subsubsection:CKrS}).
The chicane in sections 2-4 is tilted by \SI{15}{\degree} (\figref{Figure:CPS_cad}) relative to the \gls{ms} axis, thereby increasing the probability of neutral gases hitting the wall.

Each of the 7 beam tube sections are surrounded by a superconducting solenoid. Three magnet modules are assembled with short separation distance from each other because of the 15-degree chicane. A total of 14 tie-rods are holding about 5~tonnes of the cold mass.  

The superconducting coils of the \gls{cps} are cooled in a liquid helium bath at \SI{4.5}{\kelvin} and \SI{1.3}{\bar} provided in a \SI{1300}{\ell} helium reservoir. The cryostat, including the magnet chambers and the helium reservoir, has a volume of about \SI{1600}{\ell} for the total liquid helium inventory. The cryostat is equipped with a safety valve rated at \SI{2}{\bar} overpressure and a burst disk rated at \SI{3}{\bar} overpressure for the \SI{4.5}{\kelvin} helium cooling system according to the European pressure device standard.

The solenoid modules are electrically connected in one current circuit and are operated in driven mode with a power supply. A stable power supply from Bruker GmbH provides a current stability of better than \SI{100}{ppm} per \SI{8}{\hour} in order to meet the required magnetic field stability better than \SI{0.1}{\percent\per month}. The magnetic field stability of the magnets is monitored by a closed-loop flux gate sensor of the power supply. The \gls{cps} magnets were commissioned with a magnet safety system consisting of a network of 15 active quench detection systems \cite{Gil2017}. More details about the magnet system, including the magnet safety system, are described in~\cite{Arenz2018c}.

The main part of the beam tube, consisting of sections 2 - 5, is a cold trap cooled to \SI{3}{\kelvin} which functions as the above-mentioned sorption pump.
For this purpose, a \SIadj{3}{\kelvin} beam tube cooling system  with two separate cooling circuits is installed. In the primary cooling circuit of the cooling system, liquid helium is filled from the \SIadj{4.5}{\kelvin} reservoir into the \SIadj{3}{\kelvin} vessel, which is connected to an external pumping station through a \SI{15}{\meter}-long cryogenic pumping line. The cold helium in the small \SIadj{13}{\litre} vessel is pumped to a saturation pressure of \SI{0.3}{\bar} in order to establish \SI{3}{\kelvin} inside. Before reaching the external pumping station, the cold helium is warmed up to room temperature through the long pumping line and 3-stage heat exchangers. In the secondary cooling circuit of the \SI{3}{\kelvin} beam tube cooling system, helium between \SI{2.5}{\bar} and \SI{4.5}{\bar} with a rate of about \SI{0.5}{\gram\per\second} from the refrigerator flows through a heat exchanger loop that is assembled in the \SIadj{3}{\kelvin} vessel and through the cooling tubes brazed on  beam tube sections 2 - 5, and finally to the \SI{4.5}{\kelvin} reservoir. 
The other beam tube sections are connected to a liquid nitrogen cooling circuit.

In order to minimize thermal heat load, a radiation shield cooled by nitrogen (green in \figref{Figure:CPS_cad}) surrounds the beam tube and magnets.
Between the magnetic coils and the beam tube, there is a second radiation shield (light brown in \figref{Figure:CPS_cad}) cooled down with helium, which allows for beam tube heating up to \SI{80}{\kelvin} for the regeneration of the cold trap. The regeneration process can in principle be done during magnetic field operation. Nevertheless, for safety reasons the regeneration is done with discharged magnets. 

Downstream from the cold trap, a gate valve (V3) operated at a temperature below \SI{10}{\kelvin} (cold gate valve, \figref{Figure:CPS_cad} and \figref{Figure:CPS-vacuum-flow-diagram}) can be closed within a few seconds in case of a cryogenic failure in order to prevent unintended release of the adsorbed tritium towards the spectrometers.

\subsubsection{Argon Cryo Pump}
\label{Subsubsection:ArgonPump}

\begin{figure}[tb]
	\centering
	\includegraphics[width=\linewidth]{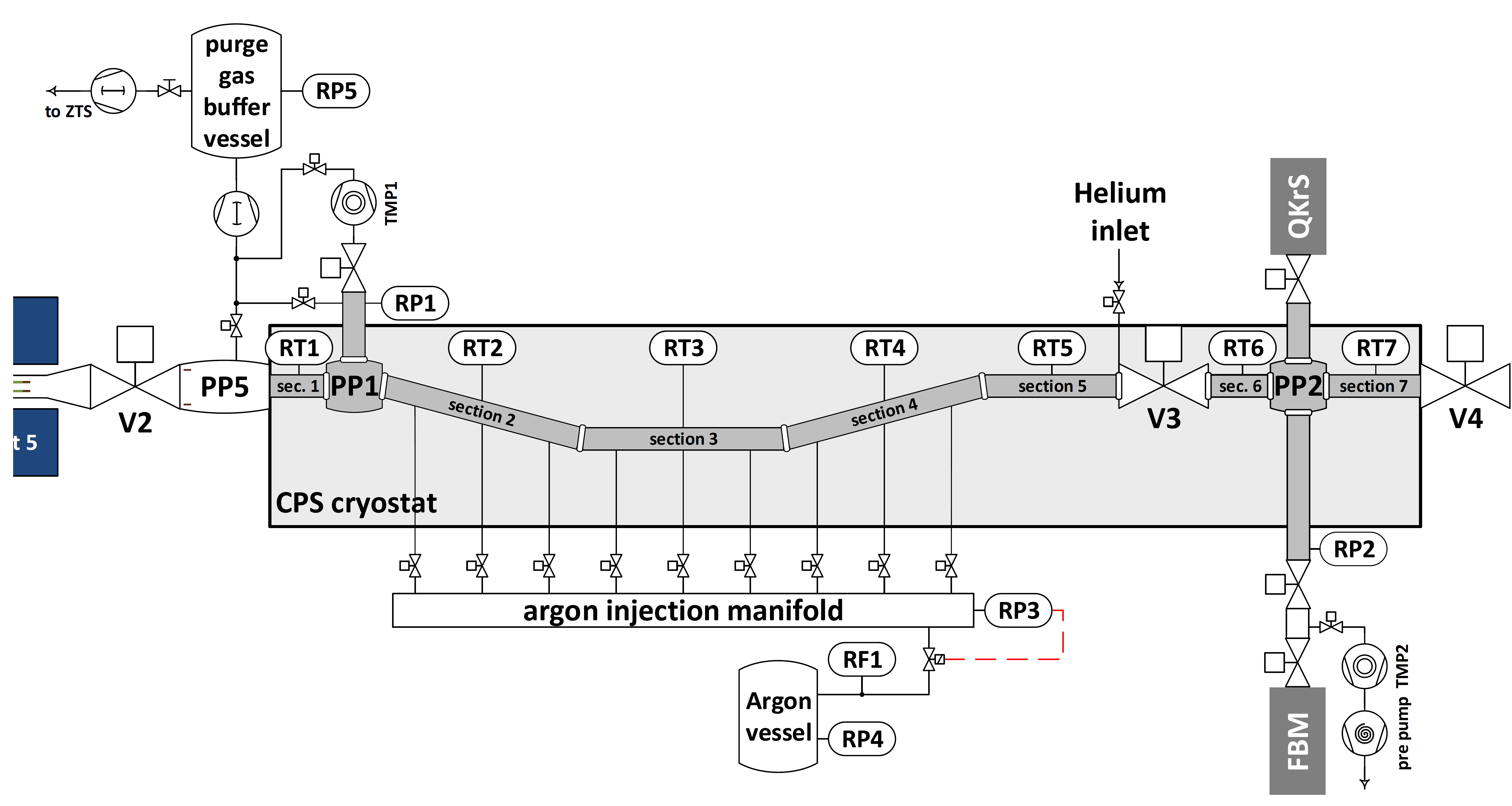}
	\caption{Schematic of the argon pump flow diagram. Shown are the main parts needed for argon pump preparation and regeneration. RT stands for  temperature sensor, RP for pressure sensor.}
	\label{Figure:CPS-vacuum-flow-diagram}
\end{figure}

The argon pump has the three operation modes~\cite{Gil2010}: preparation mode, standard operation mode, and regeneration mode. Each of these modes are operated at different temperatures.

\begin{description}
	\item [Preparation of argon pump]
During the preparation of the argon frost layer, the beam tube sections 2-5 are stabilized to \SI{6}{\kelvin}, providing optimal conditions for porosity and therefore large adsorption ability of the argon frost layer~\cite{Nepijko2005}. To increase the surface of the cryopump and therefore the pumping capacity by a factor of three, 90 2-mm-long fins are mounted on the inner beam tube surface of the beam tube elements 2-4 \cite{Gil2010}.
The argon is injected via 9 capillaries, 3 for each beam tube section 2-4. In order to avoid freezing of the argon inside the capillaries the capillaries are heated up to 75K. An argon inlet system controls the flow rate into the \gls{cps} (\figref{Figure:CPS-vacuum-flow-diagram}).
Its main parts are a \SIadj{4}{\litre} argon buffer vessel and a manifold, which is pressure stabilized and distributes argon into the 9 capillaries. The pressure stabilization is realized by a control valve, which is regulated by the pressure in the manifold. 
Combining the readout of the buffer vessel pressure together with its volume reveals the inserted argon amount $pV$. The necessary amount of argon for the preparation of the argon frost pump is about \SI{6}{\bar\litre}. The \SIadj{4}{\litre} argon buffer vessel has therefore to be refilled during pump preparation.
\item [Standard operation of argon pump] 
With the built geometry, a sticking coefficient $\alpha=0.7$ can be estimated for tritium molecules hitting the cold trap~\cite{Friedel2019}. 
Adsorbed tritium molecules can desorb again, either prompted by thermal desorption or induced by \betaels{}.
Since the second effect can not be prevented, the first one has to be minimized.
The mean sojourn time $\tau_{\mathrm{des}}$, including only thermal desorption, on the argon frost layer is given by~\cite{Jousten2008}
	\begin{equation}
	\label{eq:tau_des}
    \tau_{\mathrm{des}}=\tau_0\cdot\exp \left ( \frac{E_{\mathrm{des}}}{RT} \right) \ ,
    \end{equation}
    where $\tau_0$ is a material and gas specific time constant~\cite{Boer1956},
    $E_{\mathrm{des}}$ is the desorption energy for one mole of adsorbed gas, and ${R=\SI{8.314}{\joule\mole\per\kelvin}}$ the molar gas constant.
    Therefore, it is necessary to keep the temperature low in order to increase the desorption time.
    As $E_{\mathrm{des}}$ for tritium on an argon frost layer is not known with sufficient accuracy \cite{Kazachenko2008a}, the temperature of the cold trap was chosen conservatively to be \SI{3}{\kelvin}.
\item [Regeneration of argon pump] The regeneration of the argon pump for the removal of tritium and argon is done by purging with helium gas and warming up the beam tube to 80K. For radiation safety reasons the regeneration needs to be conducted before an accumulated activity of no more than \SI{1}{Curie}  (=\SI{3.7e+10}{Bq})~\cite{KATRIN2005,Friedel2019}.
Under standard measurement conditions, this limit is reached only after more than one year. Hence, the regeneration of the argon pump is performed at least once a year during regular \gls{katrin} maintenance phases.
For the regeneration process the cold gate valve must be closed.
While purging helium into \gls{bte}5 (\figref{Figure:CPS-vacuum-flow-diagram}), the cold trap is warmed to \SI{80}{\kelvin} within \SI{2.5}{\hour}.
At the same time the gas is pumped via \gls{dps}-PP5 and \gls{cps}-PP1.
After bringing back the cold trap temperature to standard conditions, a new argon frost layer can be generated.
\end{description}

\subsubsection{Commissioning Results with  the magnet-cryostat system}

The first cooldown of the \gls{cps} was performed in May 2016~\cite{Roettele2017}.
\begin{description}
	\item[Stand-alone time] The \gls{cps} magnets can be kept cold at \SI{4.5}{\kelvin} for about \SI{16}{\hour}, when the magnet cooling has been interrupted from the refrigerator at a liquid helium level of \SI{65}{\percent}. 
	\item[Reaching the design current] The \gls{cps} magnets had a training quench at \SI{194}{\ampere} (\SI{97}{\percent} of the maximum design current of \SI{200}{\ampere}) during the first ramping~\cite{Gil2017}. After the training quench, there were two additional (unexpected) quenches at \SI{167}{\ampere}. The following investigations showed that the magnetic field of the neighboring magnet (the first \gls{ps} magnet, PS1) is necessary to stabilize the \gls{cps} magnetic field: Once the neighboring magnet PS1 have been charged to its full current, the \gls{cps} magnets could be charged up at least to \SI{90}{\percent} of the maximum design current. 
	\item[Current stability] The instability of the magnet current at the nominal setting (\SI{70}{\percent} of the design fields) has been checked, as performed for the \gls{wgts} in \tabref{tab:Magwgts}. The relative current fluctuation is within the specification of $\sigma/I_{av} = \SI{4.5e-4}{\percent}$  with $I_{av} = \SI{140.040}{\ampere}$ and $\sigma = \SI{6.3e-4}{\ampere}$. 
\end{description}

\subsubsection{Commissioning results with the argon frost pump}
Argon frost preparation and purging with helium was successfully done for the first time in December 2017.

In each beam tube section of the cold trap, the temperature is monitored by 3 rhodium-iron sensors with 50 mK accuracy. During the first activation of the \SIadj{3}{K} cooling system, the measured temperatures on the beam tube did not reach the expected \SI{3}{K}, but varied between \SI{3.4}{K} and \SI{6.2}{K} \cite{Roettele2017}. 

In order to investigate the origin of the temperature discrepancy, the heat transfer module of the commercial simulation program COMSOL Multiphysics\textsuperscript{\textregistered}\footnote{COMSOL Multiphysics GmbH, Germany, \url{https://www.comsol.de}} was used with a finite-element-method simulation. The results of these simulations are discussed in detail in \cite{Friedel2019}. Except for some warmer spots, such as the bolts connecting  the warmer inner radiation shield to the beam tube sections, most of the beam line areas, in particular those with the fins, are in a temperature range between 3 - 4 K. The simulated temperature profile was used further for the simulations of the reduction factor.
The outcomes showed that, despite of the locally higher beam tube temperature, the reduction factor of the \gls{cps} exceeds the required factor of \num{e7}~\cite{Friedel2019} by at least four orders of magnitude.
This positive result is due to the safety margin (the conservative value of 3 K, a third beam tube element with argon frost) which was implemented during the design phase of the \gls{cps}.

In order to verify the simulation and to test the performance of the cryo pump, a dedicated inlet system for deuterium was installed temporarily at the \gls{dps}-PP5, i.e. at the inlet of the cold trap.
Within this setup, it was possible to control and vary the inlet flow into the \gls{cps} by three orders of magnitude in the molecular flow range using an orifice behind a leak valve.
Combining the measured pressure at \gls{cps} PP1 and PP2 with a MolFlow+ simulation\footnote{CERN, 
	\url{https://molflow.web.cern.ch}} \cite{Molflow}, the reduction factor for deuterium could be deduced~\cite{Friedel2019, PhDRoettele2019}.

\begin{figure}[]
	\centering
	\includegraphics[width=\linewidth]{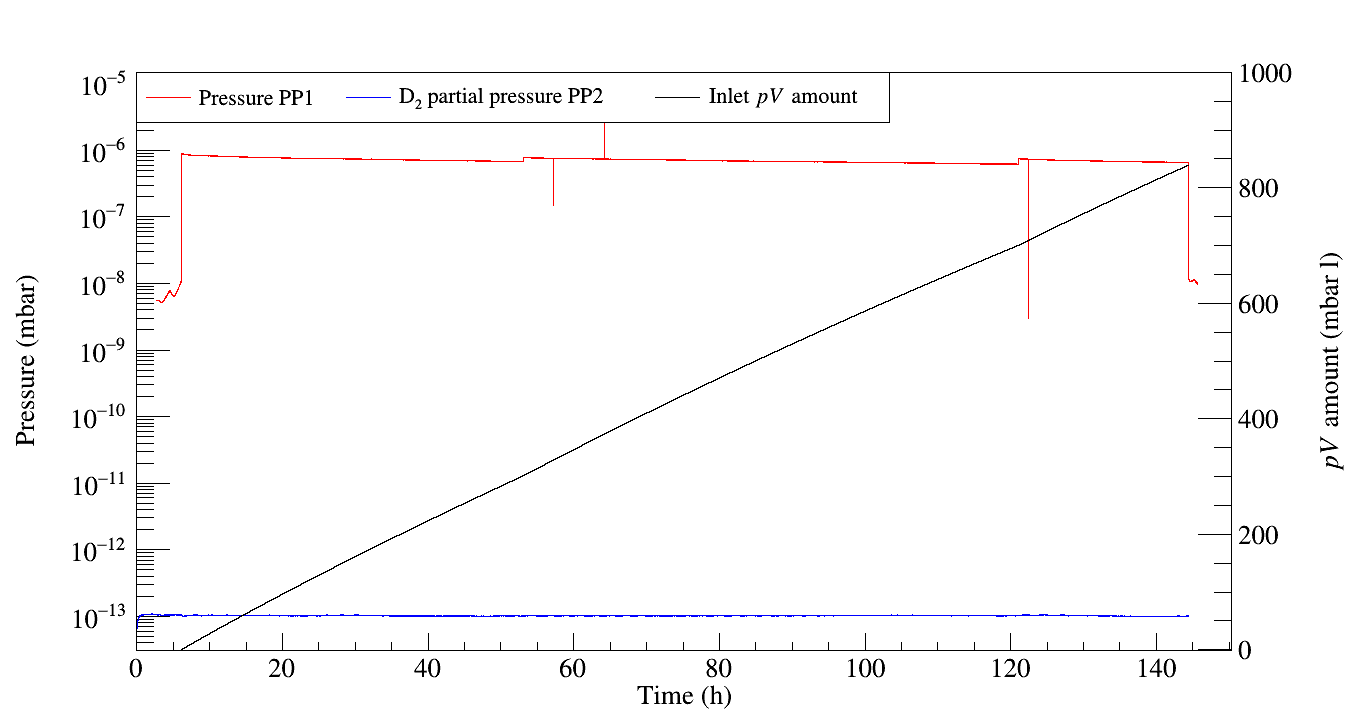}
	\caption{Reduction factor measurement of \gls{cps} - The (partial) pressure at PP1 and PP2 are plotted over the measurement time. Additionally on the right $y$-axis the inserted $pV$ amount of deuterium is shown. The figure is adapted from~\cite{PhDRoettele2019}. }
	\label{Figure:CPS_reduction}
\end{figure}

In \figref{Figure:CPS_reduction}, the result of the measurement is shown.
After opening the valve at DPS-PP5 (at about \SI{6}{\hour}) the pressure at \gls{cps}-PP1 increased by about two orders of magnitude, while the partial pressure of deuterium at PP2 was not effected.
The two small stepwise rises at about \SI{55}{\hour} and \SI{120}{\hour} in the red curve correspond to a slight re-adjustment of the leak valve.
The three spikes in the PP1 pressure were induced by small fluctuations of the \gls{cps} nitrogen cooling influencing the pressure sensor for a short time.
After around \SI{145}{\hour} the deuterium injection was stopped.
During the whole injection time, no increase in the partial pressure at \gls{cps}-PP2 was seen.
Due to the sensitivity of the residual gas analyzer,
 only a lower limit of the reduction factor $R$ can be stated.
With the exact geometry of the pressure sensors at room temperature taken into account, the experiment established (\figref{Figure:CPS_reduction}) a lower limit 
of~\cite{PhDRoettele2019}:

\begin{equation}
	R\geq 10^{8}  .
\end{equation} 

This detailed analysis includes a conservative estimate of the uncertainty for both the readout of the cold cathode at \gls{cps}-PP1 and the residual gas analyzer at \gls{cps}-PP2. The injected $pV$ amount of \SI{850}{\milli\bar\litre} equals the operation of the cold trap of approximately 6000 days in standard \gls{katrin} operation.
Therefore, the capacity of the cold trap was also confirmed.
A more detailed analysis of additional measurements can be found in~\cite{PhDRoettele2019}.

%% file: SpectrometerSystem.tex
\section{Spectrometer System}
\label{sec:spectrometer_system}

The \gls{katrin} spectrometer system consists of three \acrfull{mace} filter spectrometers, a \acrfull{ps} and \acrfull{ms}          
in the main beam line as well as a \acrfull{mos} in a parallel setup.
Common to all spectrometers is to fulfill the requirements on vacuum, electrostatic potential, and magnetic guiding field in order to be operated as \gls{mace} filters.
At the same time their individual configuration and application are important in complementing each other in the \gls{katrin} setup.

\input{SpectrometerSystemPrinciple}

\clearpage

\input{SpectrometerSystemVacuum}
\clearpage

\input{SpectrometerSystemHV}
\clearpage

\input{SpectrometerSystemMagnet}

\clearpage

\input{SpectrometerSystemBackgroundMitigation}
\clearpage

%% file: SpectrometerSystemPrinciple.tex
\subsection{Spectrometer Principle and Basic Requirements}
\label{subsec:specFilterPrinciple}

\begin{figure}[!t]
	\begin{center}
	\includegraphics[width=0.95\textwidth]{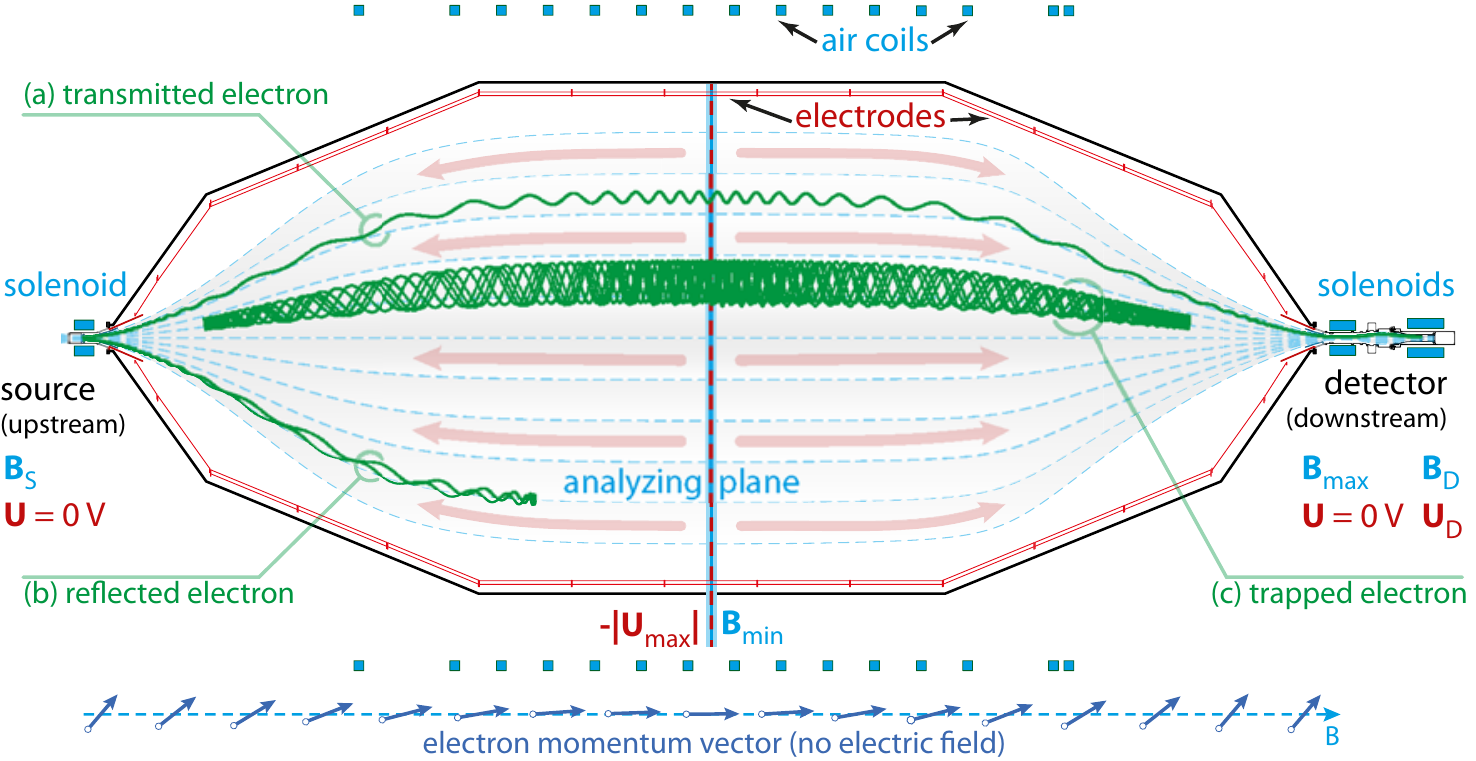}
	\end{center}
	\caption{The \gls{mace} filter principle.
	Superconducting solenoid magnets provide the magnetic guiding field.
	Wire electrodes together with the vacuum vessel on high voltage create the electrostatic retarding potential.
	Electrons emitted from the source are being collimated magnetically under energy conservation, while the retarding potential slows them down and analyzes their kinetic energy as an integrating high-pass filter.
Electrons with sufficient energy to pass the filter are transmitted and counted by the detector (track a); those with insufficient kinetic energy are reflected (track b).
Electrons originated from inside the spectrometer volume may remain trapped due to the magnetic bottle effect at both ends (track c).    
At the bottom the magnetic adiabatic collimation is represented by electron momentum vectors without retardation. 
}
	\label{fig:mace_2019}
\end{figure}

The \gls{mace} filter technique is based on the collimation of isotropically emitted electrons by the inverse magnetic mirror effect~\cite{Beamson1980,Lobashev1985,Picard1992}.
It applies magnetic adiabatic collimation in combination with an electrostatic energy filter for energy selection.
Its basic configuration is shown in \figref{fig:mace_2019}.
The technique provides superior energy resolution in the \si{eV} range at electron energies of several \si{keV} and a high luminosity for signal electrons.
It is thus an excellent choice for the \gls{katrin} experiment~\cite{Drexlin2013}.

Signal \betaels\ are emitted in a region of high magnetic field $B_\mathrm{S}$ in the source.
The magnetic field is produced by the superconducting solenoids at the spectrometer entrance and exit.
It guides the electrons towards the center of the spectrometer as it drops by several orders of magnitude.
Maintaining adiabatic particle transport with full energy conservation, the magnetic gradient force transforms the cyclotron energy of the isotropically emitted electrons into the longitudinal component.
In this case, the orbital magnetic moment $\mu$ of the electron's cyclotron motion and accordingly it's angular momentum $\vec l$ around a magnetic field line are conserved,
\begin{equation}
    \label{eq:magneticmoment}
    \mu = | \vec \mu | = \frac{\rm e}{2 m_{\rm e}} |\vec l| = \frac{E_\perp}{B}={\rm constant}
    \,,
\end{equation}
in a non-relativistic approximation.
Here $E_\perp$ and $B$ denote the transversal component of the electron energy and the magnetic field, respectively.
The transversal kinetic energy can be written in terms of kinetic energy $E$ and pitch angle $\theta = \angle(\vec{p}, \vec{B})$ between electron momentum $\vec{p}$ and magnetic field $\vec{B}$, which results in $E_\perp = E \cdot \sin^2\theta$.

The fundamental requirement of the \gls{mace} filter is to provide adiabatic propagation of the signal electrons by maintaining sufficiently small magnetic field gradients and by initiating electrostatic retardation.
In a region of small field gradients, the momentum transformation \eqref{eq:magneticmoment} occurs over a distance of several meters, requiring an overall length of \SI{23}{\metre} of the \gls{katrin} \acrfull{ms}.

The transversal kinetic energy reaches its minimum in the so-called analyzing plane, where the magnetic field reaches a small value in the \SI{0.5}{\milli\tesla} range $(B_\mathrm{min} \ll B_\mathrm{max})$.
Concurrently, the transversal energy is transformed into longitudinal energy due to energy conservation, which is then analyzed by the maximum retarding potential $|U_\mathrm{max}|$ in the analyzing plane.

Electrons are transmitted if their longitudinal kinetic energy is larger than the filter energy, $E_\parallel = E \cdot \cos^2\theta > q U_\mathrm{max}$, where $q$ is the electron charge.
In this respect the \gls{mace} filter acts as a high-energy-pass filter for electrons.
The finite width $\frac{\Delta E}{E}$ is essentially the remaining fraction of the perpendicular energy component that is not completely transformed into the longitudinal part.
The energy resolution of such a configuration for an isotropically emitting electron source is given by the ratio~\cite{Otten2008}
\begin{equation}
    \label{eq:energyresolution}
	\frac{\Delta E}{E} = \frac{B_{\rm min}}{B_{\rm max}}
    \,.
\end{equation}

The \gls{katrin} \gls{ms} achieves an unprecedented energy resolution of $\Delta E = E \cdot \SI{0.3}{\milli\tesla} / \SI{6}{\tesla} \approx \SI{1}{\electronvolt}$ at $E = \SI{18.6}{\kilo\electronvolt} \approx E_\mathrm{max}$ by design.
As described in \secref{Subsection:KATRINKeyParameters}, the design resolution can be adjusted on purpose. 
The energy resolution of a \gls{mace} filter must not be confused with other particle detectors, defined by the width of the energy distribution of the detected events.
For the \gls{mace} filter, the knowledge of the width and the shape of the transmission function of the high-energy-pass filter is crucial. 
Instead of introducing an energy error, the width of the transmission function, when fully characterised by electron gun measurements, does not restrict us to resolve even smaller structures in the spectrum.

The overall transmission characteristics needs to take the magnetic field at the source position into account.
The acceptance angle of transmitted electrons relative to the magnetic field direction depends on the magnetic field ratio between that of the source position $B_\mathrm{S}$ and at the largest magnetic field $B_\mathrm{max}$ region in the beam line.
Referring to the magnetic bottle effect\footnote{This magnetic field configuration is also referred to as Mirror Trap in magnetic plasma refinement studies.}, the maximum acceptance angle $\theta_\mathrm{max}$ can be derived as
\begin{equation}
    \label{eq:thetaMax}
	\theta_\mathrm{max} = \arcsin \left ( \sqrt{\frac{B_\mathrm{S}}{B_\mathrm{max}}} \right )
    \,.
\end{equation}

In the standard magnetic field configuration of the \gls{katrin} experiment the maximum acceptance angle is $\theta_\mathrm{max} \approx 51 ^{\circ}$.
A side effect of this design is trapping of electrons by the magnetic bottle effect.
The electrons originating from the low magnetic field region in the center of the spectrometer are reflected by the increasing magnetic field at the spectrometer entrance and exit, see track (c) in \figref{fig:mace_2019}.
A hybrid trap for electrons, formed by the retarding potential on one side and the magnetic bottle on the other is not shown in \figref{fig:mace_2019}, but is a combination of tracks (b) and (c).
For positively charged particles or ions, the retarding potential itself is the trap.
For the stable low-background operation of a \gls{mace} filter it is crucial to prevent ions from entering the main volume as well as the creation of electrons within.  

The $\upbeta$-spectrum can be measured in integral form by varying the retarding potential, or more precisely by varying the potential difference between the source and the spectrometer. 
The requirements of the \gls{mace} filter are the basis of the electromagnetic and mechanical design of the \gls{katrin} experiment~\cite{Glueck2013}.
A detailed discussion is available in~\cite{PhDWandkowsky2013}.

The key requirements and design criteria for the spectrometers are the vacuum quality, high voltage and magnetic field stability, and optimal electron transport from the source to detector.
They are summarized here:

\begin{itemize}
	
\item The vacuum systems of the spectrometers have been designed to reach a pressure in the lower \SI{e-11}{\milli\bar} regime
(see \secref{sec:SDS_vacuum}).
The main	 reason for this challenging requirement is the reduction of the background originated from inside the spectrometer (see \secref{subsec:background}). 

\item The electrostatic retarding potential defines the energy filter.
A high-precision high-voltage supply, distribution, and monitoring system has been developed to provide the retarding potential with a relative precision in the ppm ($10^{-6}$) range for voltages of down to \SI{-35}{\kilo\volt} (see \secref{sec:SDS_HV}).
The motivation for this is to limit the energy-scale-related contribution to the systematic uncertainty budgeted for the neutrino mass analysis.
		
\item The magnetic field in the analyzing plane $B_\mathrm{min}$ defines the energy resolution of the spectrometer for a given $B_\mathrm{max}$ and is a key parameter for the neutrino mass analysis.
Due to the finite spatial resolution of the pixelated detector wafer~\cite{Amsbaugh2015}, a nearly homogeneous magnetic field across the analyzing plane is required to minimize systematic uncertainties (see \secref{subsection:MSmagnetsystem}).

\item The electromagnetic configuration must ensure that the transversal-to-longitudinal energy transformation is complete before the electrons reach the analyzing plane, where the electrostatic retarding potential is maximum.
The magnetic field generated by the Air Coil Systems (see \secref{subsubsection:aircoilsystem}) as well as the retarding potential offsets provided by the inner electrode system (see \secref{sec:SDS_IE}) are fine-tuned to fulfill this requirement.

\end{itemize}


\subsubsection{Main Spectrometer related background}
\label{subsec:background}
 
The majority of the background events observed in the \gls{katrin} detector are produced in the spectrometer section.
Since these electrons originate from inside the spectrometer, they appear in the same energy window as the filtered signal electron of the $\upbeta$-spectrum. 
There are four main possible sources of background, which are described here.
Three of these sources are suppressed by appropriate countermeasures, leaving one dominating source of background. 
 
\paragraph{Electrons from the spectrometer walls} 
Backgrounds from the inner spectrometer surface are induced by secondary electrons created by cosmic muons and ambient gamma radiation from building walls, concrete, or the vessel material itself.
The primary mechanism to reduce this background is magnetic shielding by the Lorentz force~\cite{PhDWandkowsky2013}.
This magnetic shielding effect, however, requires an excellent axial symmetry of the electromagnetic fields~\cite{Glueck2013};
otherwise, surface-generated electrons would quickly drift into the inner magnetic flux tube volume and become a background source.
The dominant non-axial magnetic field component is the earth magnetic field.
Other sources include stray fields from magnetized steel bars and rods in the building materials of the spectrometer hall.
In addition to magnetic shielding, the inner surface of the spectrometer is equipped with a wire electrode (see \secref{sec:SDS_IE}), which can be charged with a more negative voltage than the vessel wall.
Low energy electrons from the wall are reflected back in this configuration.
Dedicated investigations showed that the background caused by electrons from the spectrometer walls is not dominant and is effectively mitigated by the measures mentioned here \cite{Altenmueller2019a, Altenmueller2019b}.

\paragraph{Decay of Radon atoms in the volume} 
Another source of primary electrons is radon decay; in particular, short-lived radon isotopes such as $^{219}$Rn ($t_{1/2} = \SI{3.9}{\second}$) and $^{220}$Rn ($t_{1/2} = \SI{56}{\second}$), which can decay before they are pumped out \cite{Drexlin2017, Wolf2017}.
Following the $\alpha$-decay of a radon isotope, these background electrons can be produced by various processes \cite{Wandkowsky2013}.
Effective measures to reduce this background component have been investigated in prototype measurements with the \acrfull{ps} \cite{Fraenkle2011, Goerhardt2018} that led to the design of the cryogenic baffles described in \secref{subsec:MS_vacuum}.  
At the nominal pressure of \SI{e-11}{\milli\bar}, there is sufficient time to empty the trap regularly with appropriate measures, such as magnetic or electric pulsing \cite{Arenz2018d}, before too many secondary electrons are produced. 

\paragraph{Magnetic trapping conditions} 
Electrons produced in a region with low magnetic field, enclosed by strong magnetic fields on both sides, can be trapped by the magnetic mirror effect.
At the low pressure of \SI{e-11}{\milli\bar} in the spectrometers, a primary electron can ionize residual gas molecules over a period of several hours \cite{Wandkowsky2013}, until it has lost enough energy to leave the trap.
Since about half of the low-energy secondary electrons are accelerated by the electrostatic field of the \gls{mace} filter towards the detector, a single radioactive decay with a primary keV-energy electron can increase the background rate considerably \cite{Mertens2013}.
The number of secondary electrons produced depends on the primary energy.
They are correlated in time, which lead to a non-Poissonian distribution of the background rate.  

Neutral atoms and molecules are not deterred from entering the volume of the sensitive flux tube\footnote{The sensitive volume of the magnetic flux tube refers to the region where the magnetic field guides electrons to the focal plane detector.}.
Thus, gaseous radioactive atoms, such as tritium from the \gls{wgts}, can enter the volume to produce primary electrons and a multitude of secondary electrons. Due to the long half-life of tritium, a large effective pumping speed helps to remove the tritium molecules before they decay. 

\paragraph{Ionization of Rydberg atoms}
The inner surface of the Main Spectrometer was exposed to ambient air during the construction phase.
This exposure caused a small activity (order of  \SI{1}{\becquerel\per\metre\squared}) of surface-implanted \textsuperscript{210}Pb from the decay chain of \textsuperscript{222}Rn.
Residual gas atoms (predominantly hydrogen) in Rydberg states are produced in sputtering processes at the inner spectrometer surface by \textsuperscript{206}Pb recoil ions following the $\alpha$-decay of \textsuperscript{210}Po.
The Rydberg atoms are able to enter the magnetic flux tube where a small fraction is ionized by black-body radiation. The resulting sub-\si{\electronvolt} scale electrons are accelerated by the retarding voltage as they leave the spectrometer and contribute to the background at the detector.

\subsubsection{Penning trap between the Pre- and Main Spectrometers} 
\label{subsec:PS_MS_Penning_Trap}

When combining two \gls{mace} filter spectrometers in series, a Penning trap can be formed between their analysis regions.
In case of the \gls{ps} and \gls{ms}, this happens due to the superposition of negative electric retarding potentials and the magnetic guiding field.
Measurements showed that Penning discharges can occur at a pressure above \SI{5e-10}{\milli\bar}, leading to a high background rate.
At the nominal pressure of \SI{e-11}{\milli\bar}, the tandem spectrometer setup can run for hours without a Penning discharge.
In addition, a wire \cite{aker2019suppression} is moved in regular time intervals across the high-field region close to the magnet between the \gls{ps} and \gls{ms}.
This wire catches the stored electrons before they can trigger a Penning discharge (see \secref{subsubsection:penningwiper}).

%% file: SpectrometerSystemVacuum.tex
\subsection{The Spectrometer Vacuum System \label{sec:SDS_vacuum}}
\label{subsec:MS_vacuum}

The \gls{uhv} systems of the spectrometers have two main functions.
The first is to maintain a low enough pressure that allows \betaels{} to pass both spectrometers undisturbed.
The second function is to prevent the production of backgrounds in situ.
While the first condition can be easily met at \SI{e-7}{\milli\bar}, the second one requires a pressure in the lower \SI{e-11}{\milli\bar} regime. The pressure inside the spectrometers affects mostly the storage time of trapped particles, which lose energy through collisions with residual gas molecules.
Secondary electrons, produced in this process, add to the background rate seen at the detector.
Without countermeasures, they represent the dominant background of the experiment.

Both spectrometer vacuum systems meet the stringent requirement of a \gls{uhv} in the low \SI{e-11}{\milli\bar} regime by employing a combination of cascaded \glspl{tmp} and \gls{neg} pumps. The \gls{neg} pumps are made of a zirconium-vanadium-iron alloy\footnote{SAES Group, Type St707, \url{https://www.saesgetters.com/st707-strips}}, which has been found later to be the dominant source of radon-induced background in the spectrometers \cite{Fraenkle2011}. As demonstrated in measurements \cite{Goerhardt2018, Wolf2017} and by extensive simulations \cite{Drexlin2017}, the radon influx into the sensitive volume of the magnetic flux tube can be sufficiently reduced by \gls{ln2} cooled baffles, mounted between the \gls{neg} pumps and the inner spectrometer volume. This countermeasure has been implemented.

The pumping systems have been specified under the assumption of a pressure-dominating hydrogen outgassing rate of the stainless steel surfaces in the order of \SI{e-12}{\milli\bar\litre\per\second\per\centi\metre\squared}.
The stainless steel alloy (316LN) used in the construction of the vacuum vessels was chosen for its low magnetic permeability ($\mu_r \approx 1.005$), its mechanical properties, and its low radioactivity. The walls of the spectrometer vessels have been designed to be operated in a temperature range of \SIrange{-20}{350}{\celsius}. The upper value was defined by the required temperature for the thermal activation of the \gls{neg} pumps, since the initial design intended to heat the getter material through thermal radiation from the walls. The temperature of the spectrometers is regulated by a combination of electrical heating tapes and a thermal fluid pumped through channels welded to the outer surface of the walls. The force exerted by the atmospheric pressure and the reduced tensile strength at higher temperatures led to a wall thickness of \SI{10}{\milli\metre} for the \gls{ps} and \SI{32}{\milli\metre} for the \gls{ms}.
In the following sections, we give a short overview of the vacuum systems of both spectrometers and update the descriptions since the older publications on their design and performance \cite{Arenz2016, Wolf2009, Luo2007}. 

\subsubsection{Vacuum System of the Pre-Spectrometer \label{sec:PS_vacuum}}

\begin{figure}[!t]
    \centering
    \includegraphics[width=0.6\textwidth]{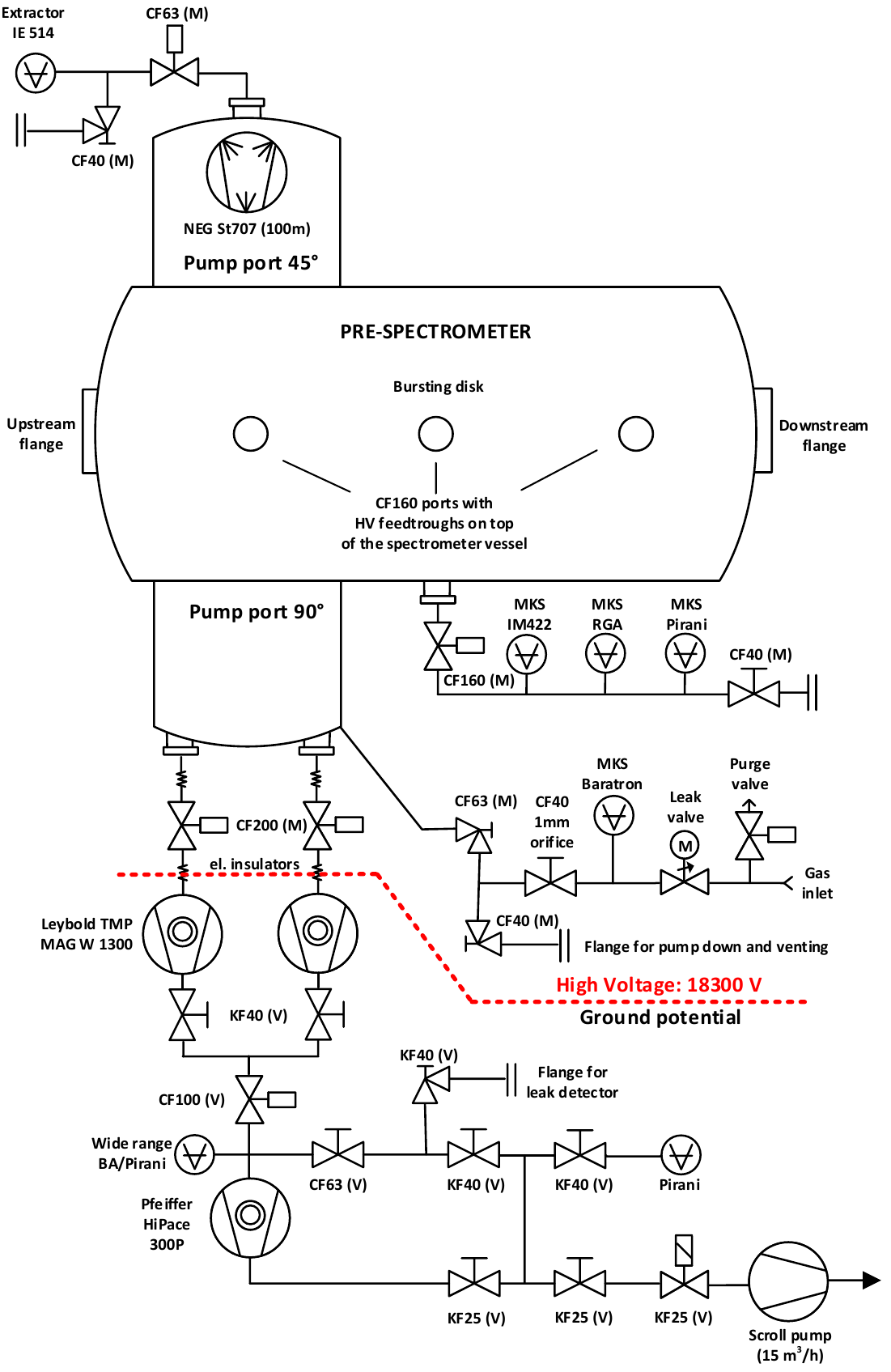}
    \caption{Schematics of the vacuum system of the Pre-Spectrometer. 
    }
    \label{Figure:pre-spec-vacuum}
\end{figure}

The vacuum vessel of the \gls{ps} is \SI{3.38}{\metre} long and has an inner diameter of \SI{1.68}{\metre}.
It has a volume of \SI{7.6}{\metre\cubed} and an inner surface area of \SI{25.6}{\metre\squared}.
The inner electrode system, also made of 316LN stainless steel, adds another \SI{13}{\metre\squared} to the surface area.
On top of the \gls{ps}, three DN160 ports with CF flanges accommodate the high voltage feedthroughs, which connect to the inner electrode system.
For safety reasons a bursting disk, which breaks at an over-pressure of \SI{0.5}{\bar}, is added to one of the ports.

At both ends are conical ceramic insulators attached to DN500 flanges, which connect the vessel (operated down to \SI{-35}{\kilo\volt}) 
to the central beam line (ground potential).
At the upstream end, a DN200 all-metal gate valve connects the \gls{cps} to a short section of the beam line, passing through the warm bore of the first superconducting spectrometer magnet.
At the downstream end, a flapper valve is integrated into the beam line inside the warm bore of the second spectrometer magnet, connecting the conical insulators of the \gls{ps} and the \gls{ms}.
The flapper valve, designed and built at the University of Washington, is described in \cite{Arenz2016}. 
  
The \acrfull{ps} served as a prototype for the \acrfull{ms} \cite{Bornschein2006}, initially.
After successfully testing the vacuum performance and the \gls{mace} filter properties of the \gls{ps}, an unexpected source of background was discovered -- magnetically trapped keV-energy electrons from $^{219}$Rn decays \cite{PhDFraenkle2010}. It was known before that the short-lived $^{219}$Rn isotope is released from the \gls{neg} material.
However, the trapping of the primary electron for several hours and the subsequent increase of the background rate by secondary electrons was not expected in the original design \cite{KATRIN2005}.
This initiated the test of a \gls{ln2}-cooled copper baffle mounted between the \gls{neg} pump and the inner volume \cite{Goerhardt2018}.
The successful reduction of the radon-related background led to the implementation of the cryogenic baffles for the \gls{ms} \gls{neg} pumps.
Since secondary electrons from trapped electrons in the \gls{ps} would not be able to overcome the retarding potential of the \gls{ms} in the final setup, the current setup of the \gls{ps} vacuum system does not have a cryogenic baffle, which would otherwise reduce the effective pumping speed of the \gls{neg} pump for tritium and hydrogen.  

The \gls{ps} vacuum system has been described in the original \gls{tdr} in 2005 \cite{KATRIN2005}.
\figref{Figure:pre-spec-vacuum} shows the updated flow diagram of the vacuum system.
The general layout is still close to the original design with the following changes:

\begin{itemize}
\item Some of the vacuum gauges have been relocated to different flanges. All vacuum gauges are mounted behind valves, allowing the removal of a sensor without venting the spectrometer. 

\item The \gls{neg} pump in the \SI{45}{\degree} pump port has been upgraded from \SI{90}{\metre} to \SI{100}{\metre} of \gls{neg} strips.
The original St707 \gls{neg} strips have been replaced by a new low-activity batch with reduced radon emanation \cite{Fraenkle2011}, which is also used in the Main Spectrometer \gls{neg} pumps.

\item An automated gas inlet system was installed.
It is used for regular in-situ calibrations of the vacuum gauges, and for special measurements when the spectrometer is operated at an elevated pressure (up to \SI{e-8}{\milli\bar} by injecting clean Ar gas) to investigate the effects of ions from the source and transport section, as well as radon emanation from the \gls{neg} pump.
The main components of the system are an electrically controlled leak valve, a capacitive gauge\footnote{MKS Instruments, Baratron 626, \url{https://www.mksinst.com/c/capacitance-manometers}} (\SI{0.1}{torr}), and a \SIadj{1}{\milli\metre} calibrated orifice used to monitor the gas flow into the vacuum vessel.

\end{itemize}

\subsubsection{Vacuum System of the Main Spectrometer \label{sec:MS_vacuum}}

\begin{figure}[!t]
    \centering
    \includegraphics[angle = -90, width=\textwidth]{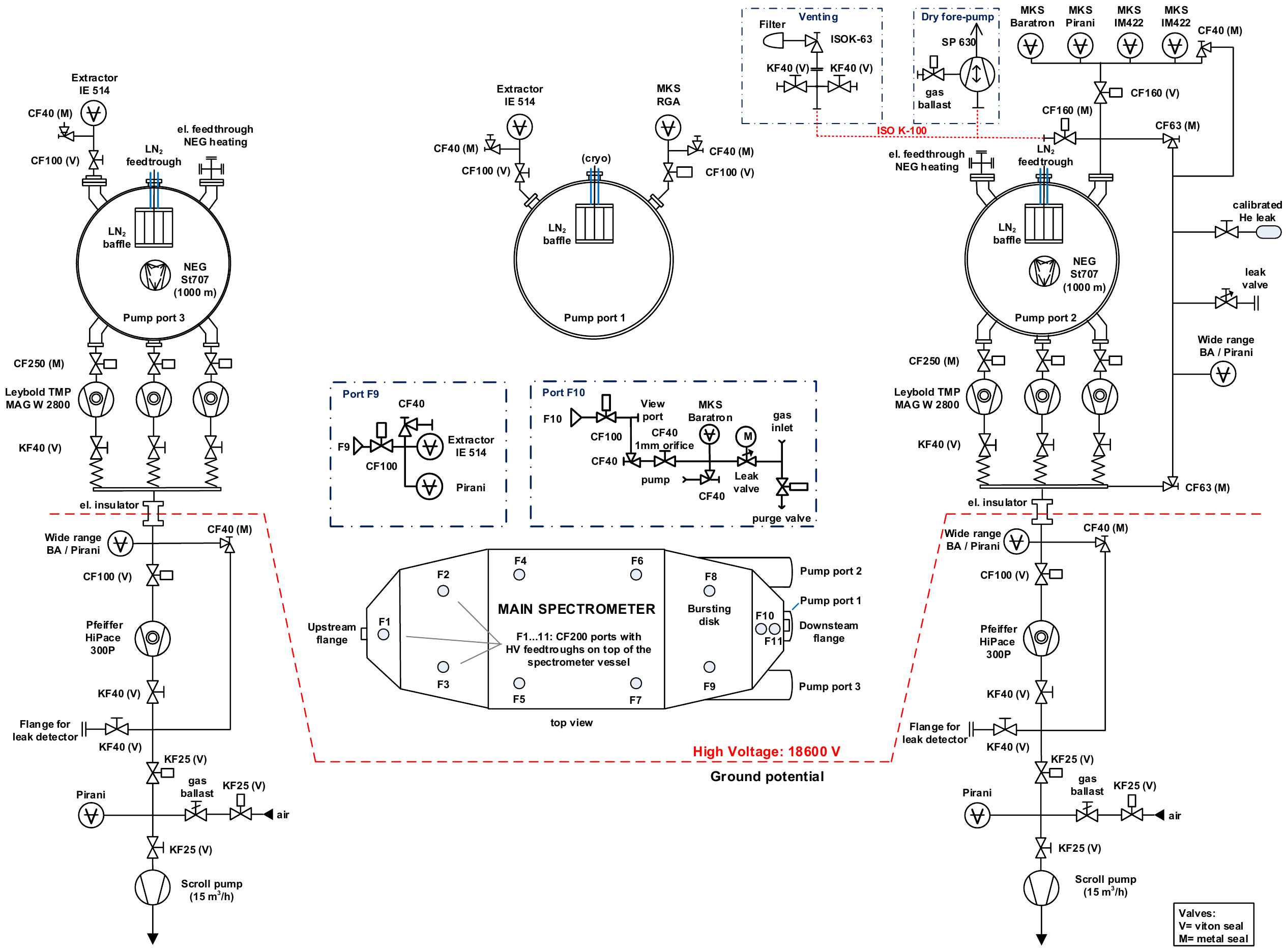}
    \caption{Schematics of the vacuum system of the Main Spectrometer. 
    }
    \label{Figure:main-spec-vacuum}
\end{figure}

The vacuum vessel of the \gls{ms} is \SI{23.23}{\metre} long and has an inner diameter of \SI{9.8}{\metre}. It has a volume of \SI{1240}{\metre\cubed} and an inner surface area of \SI{690}{\metre\squared}.
The inner electrode system including all wires, also made of 316LN stainless steel, adds another \SI{532}{\metre\squared} to the surface area.
On the upper half of the \gls{ms}, eleven DN200 ports with CF flanges accommodate high voltage feedthroughs, which connect to the inner electrode system.
For safety reasons a bursting disk, which breaks at an over-pressure of \SI{0.5}{\bar}, is added at port F8 (see \figref{Figure:main-spec-vacuum}).
The ports are also used for a pressure gauge (F9), a gas inlet system and a sapphire window for a UV source (F10). 

Like the \gls{ps}, conical ceramic insulators are attached to DN500 flanges at each end, connecting the vessel (operated down to \SI{-35}{\kilo\volt})
to the central beam line (ground potential).
At the upstream end, the aforementioned flapper valve connects the \gls{ms} to the \gls{ps}, passing through the warm bore of the second spectrometer magnet.
At the downstream end, a similar flapper valve passes through the warm bore of the Pinch magnet and connects the \gls{ms} to the detector system.

A detailed description of the \gls{ms} vacuum system and its commissioning has been published in \cite{Arenz2016}. \figref{Figure:main-spec-vacuum} shows the updated flow diagram of the vacuum system. The general layout is still close to the version described in \cite{Arenz2016} with the following changes:

\begin{figure}[!t]
    \centering
    \includegraphics[width=0.8\textwidth]{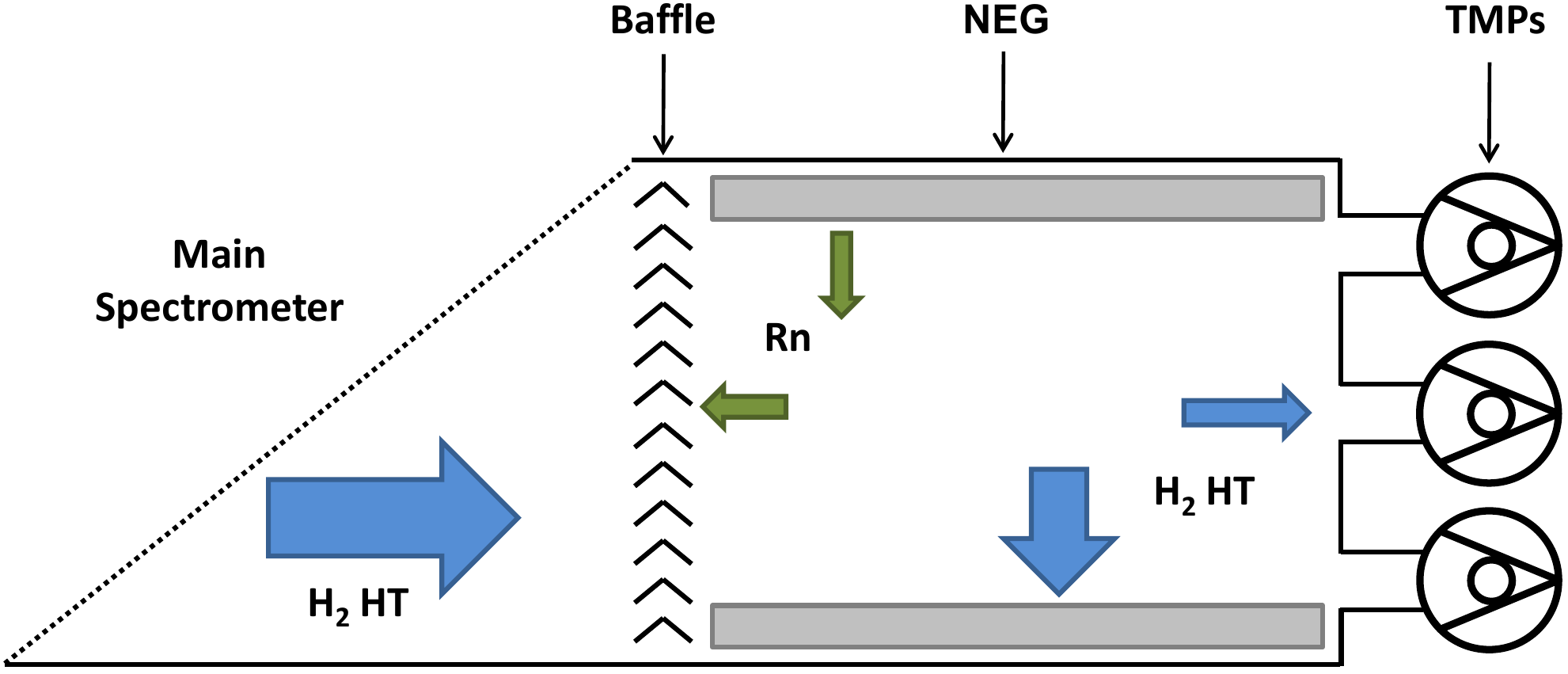}
    \caption{This drawing shows a cut through a \SIadj{1.7}{\metre} diameter pump port of the \gls{ms}. The cryogenic baffle separates the inner volume from the \acrfull{neg} pumps. 
    Residual gases from inside the \gls{ms} can pass the baffle and are being pumped by the \gls{neg} pump.
    Rn gas (also $\mathrm{H_2O}$) sticks on the cold surface and cannot pass into the \gls{ms}. 
    Noble gases, which cannot be pumped by the \gls{neg} material, are pumped out by \acrfullpl{tmp}. 
    Figure reproduced from \cite{PhDKosmider2012}. See also \cite{PhDGoerhardt2014} for details. 
    }
    \label{Figure:main-spec-pumpport}
\end{figure}

\begin{itemize}
   \item The \gls{neg} pumps have to be activated at a temperature of at least \SI{350}{\celsius} (\SI{24}{\hour}). In the initial test in the empty vacuum vessel, the required temperature was reached by the heating system.
   However, short-circuits appeared in the inner electrode system during the baking of the spectrometer at temperatures above \SI{200}{\celsius}. Since this temperature is sufficient for the thermal conditioning of stainless steel for \gls{uhv} operation, it was decided to reconfigure the \gls{neg} pumps for electrical heating.
   In the new design, the \gls{neg} pumps can be activated at temperatures above \SI{350}{\celsius}, which also reduces the required activation time.

   \item In the \gls{ms}, the \gls{neg} pumps are the main source of radon-induced background events. Therefore, only two out of three possible \gls{neg} pumps were installed (see \figref{Figure:main-spec-pumpport}). In combination with cryogenic baffles between the \gls{neg} pumps and the inner volume, the radon-induced background has been reduced to an acceptable level \cite{Wolf2017}. These measures reduced the effective pumping speed of \SI{e6}{\litre\per\second} for H$_2$ in the initial design to \SI{2.5e5}{\litre\per\second}. When the outgassing rate of the stainless steel is reduced by operating the \gls{ms} at a temperature of \SI{10}{\celsius}, this pumping speed is still sufficient to reach a pressure of low \SI{e-11}{\milli\bar}.

   \item For the in-situ calibration of vacuum gauges and special measurements at elevated pressure (up to \SI{e-8}{\milli\bar}), a gas inlet system, similar to the \gls{ps} gas inlet, was installed.
   It allows the regulation and monitoring of gas flow with an electrically controlled leak valve.
   Its capacitance gauge also measures the pressure behind a \SIadj{1}{\milli\metre} diameter calibrated orifice.

\end{itemize}  

After completing the baking sequence in May 2017, the electrical activation of the \gls{neg} pumps, and the cooling of the cryogenic baffles to liquid nitrogen temperature, the requirements for the \gls{mace} filter conditions were met by achieving an ultimate pressure of \SI{e-11}{\milli\bar}\footnote{Pressure gauges use nitrogen calibration. Value has to be multiplied by a gas correction factor of 2.3 for a hydrogen-dominated vacuum, no other gas correction factors have been applied.} in the \gls{ms}, operated at a temperature of \SI{9.5}{\celsius}.

%% file: SpectrometerSystemHV.tex
\subsection{High Voltage System} 
\label{sec:SDS_HV}

\begin{figure}[!t]
	\centering
	\includegraphics[width=1.0\textwidth]{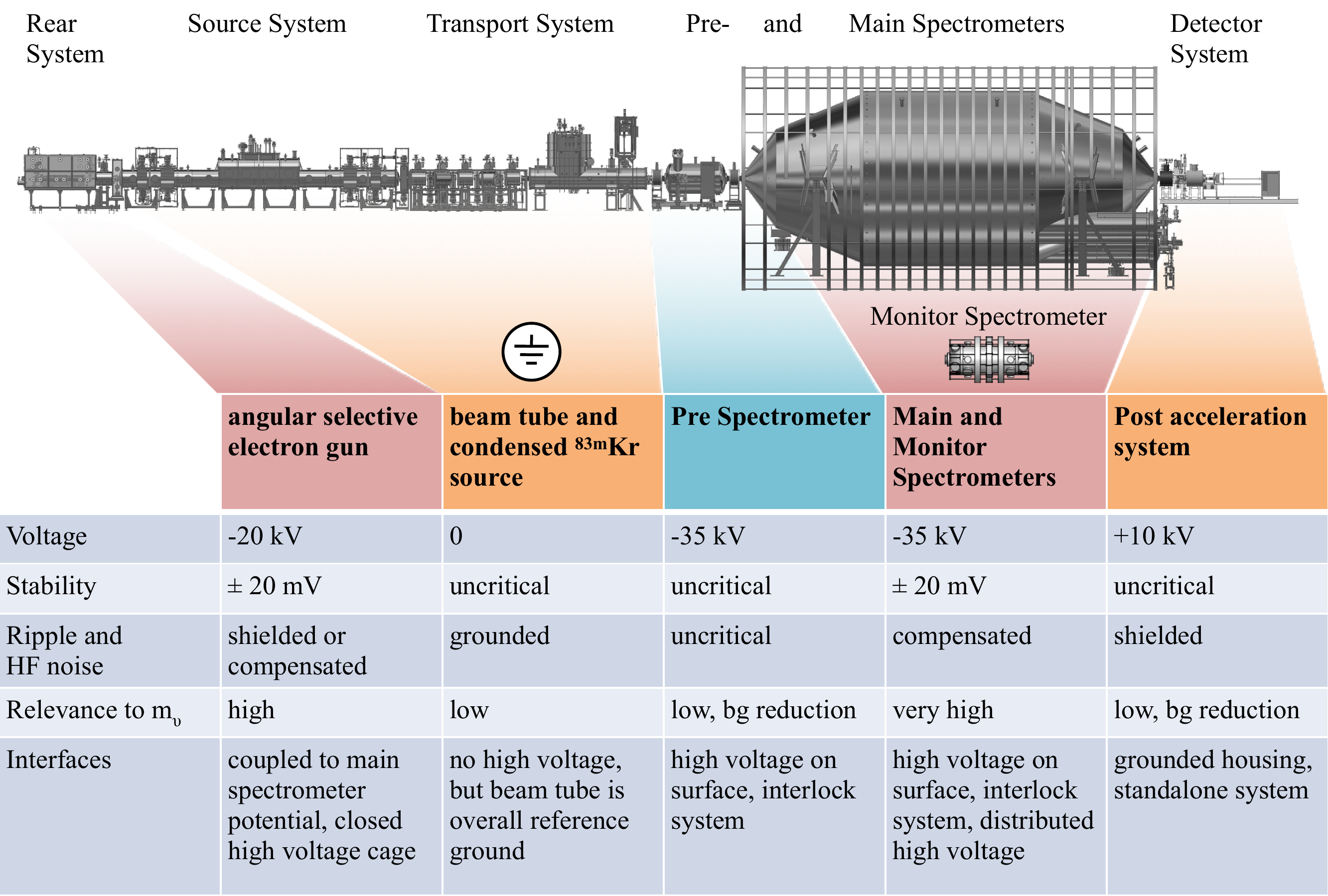}
	\caption{Overview of the high-voltage related parts of the \gls{katrin} experiment from the angular selective electron gun at the upstream end to the spectrometers and the detector system at the downstream end of the beam line.
	The source system is grounded and therefore represents the overall reference ground.
	The requirements of the \acrfull{ps} are moderate since it only filters the low energy part of the electron spectrum.
	The focus is on the \acrfull{ms} with its precision retarding potential, which can be connected to the angular selective electron gun for precision transmission studies, and to the \acrfull{mos} for long term energy scale reference.
	The positive post-acceleration potential of the detector system is independent and does not influence the spectroscopic precision of the \gls{mace} filter.  
	}	
\label{fig:HV-overview}
\end{figure}

The effective retarding potential for the \betaels\ is the difference between the electric potential at the point of the tritium decay within the \acrfull{wgts} and the electric potential which the electrons experience within the \acrfull{ms}.
In \figref{fig:HV-overview}, an overview of all high voltage related parts in the \gls{katrin} beam line is shown.
The actual source tube of the \gls{wgts} acts as the reference ground for the entire measurement of the retarding potential in the \gls{ms}, because this is the potential difference the electrons experience in a \gls{mace} filter setup.
All parts of the transport system are also connected to the source tube reference ground.

\subsubsection{High Voltage at the Pre-Spectrometer}

The \acrfull{ps} has only moderate requirements on high-voltage precision, but needs to be operated down to \SI{-35}{\kilo\volt} in order to filter the low energy part of the \betaspec{} or during calibration with the \kr\ sources.
The high-voltage setup of the \gls{ps} is similar to that of the \gls{ms}.
The primary high voltage is on the vacuum vessel and the inner electrodes are used for adjustments.
Due to its purpose of coarse filtering of low energy \betaels{}, the stability and noise requirements are uncritical, thus off-the-shelf equipment is sufficient.
The primary high voltage for the vessel is provided by a standard \SI{35}{\kilo \volt} high-voltage power supply\footnote{FuG-Elektronik HCN 140-35000, \url{https://www.fug-elektronik.de}}.
Floating on top of the primary high-voltage source, two \SI{5}{\kilo \volt} offset supplies\footnote{ISEG-HV NHQ 226L \url{https://iseg-hv.com/files/media/iseg_datasheet_NHQ_en_22.pdf}} with two channels each provide potentials for upstream, downstream, and central inner electrodes (see \cite{PhDGoerhardt2014} for details).
The primary high voltage is monitored by a standard voltage divider\footnote{Julee Research Labs KV-50 (now Ohm-Labs), \url{http://www.ohm-labs.com}}, but is not relevant for precision monitoring nor data analysis.
In addition, the current of the downstream electrode of the pre-spectrometer is monitored for the detection of ions, see \secref{sec:ion-detection} and \figref{Figure:IonsPS}.

\subsubsection{High Voltage at the Main Spectrometer}

When the \gls{wgts} potential is stable, the precision of the \gls{katrin} \gls{mace} filters depends primarily on the stability of the retardation potential \cite{Kaspar2004}.
The surface in the \acrfull{ms} where the electric retardation is maximum and the magnetic field is minimum defines the analyzing plane\footnote{Effectively the analyzing plane is the region of minimal longitudinal energy of the electrons. This coincides with the location of the maximum electric potential, if the adiabatic collimation, i.e. the adiabatic conversion of transverse energy into longitudinal energy in the spectrometer is done correctly.}.
In an ideal, symmetric \gls{mace} filter like the \gls{katrin} \gls{ms}, the analyzing plane is a flat plane in the center of the spectrometer perpendicular to the beam axis.

The acceptable relative deviation of the retardation potential in the analyzing plane from its intended value is in the ppm (\num{e-6}) range.
Furthermore, the stability of the retarding potential has to be assured from very short time intervals (sub-$\upmu$s) up to the duration of a whole measurement campaign (60 days).

The electric potential at each point of the analyzing plane is a superposition of the potentials of the vessel itself, all internal electrodes, as well as the grounded beam tubes at either end of the \gls{ms}.

\begin{figure}[!t]
	\centering
	\includegraphics[width=1.0\textwidth]{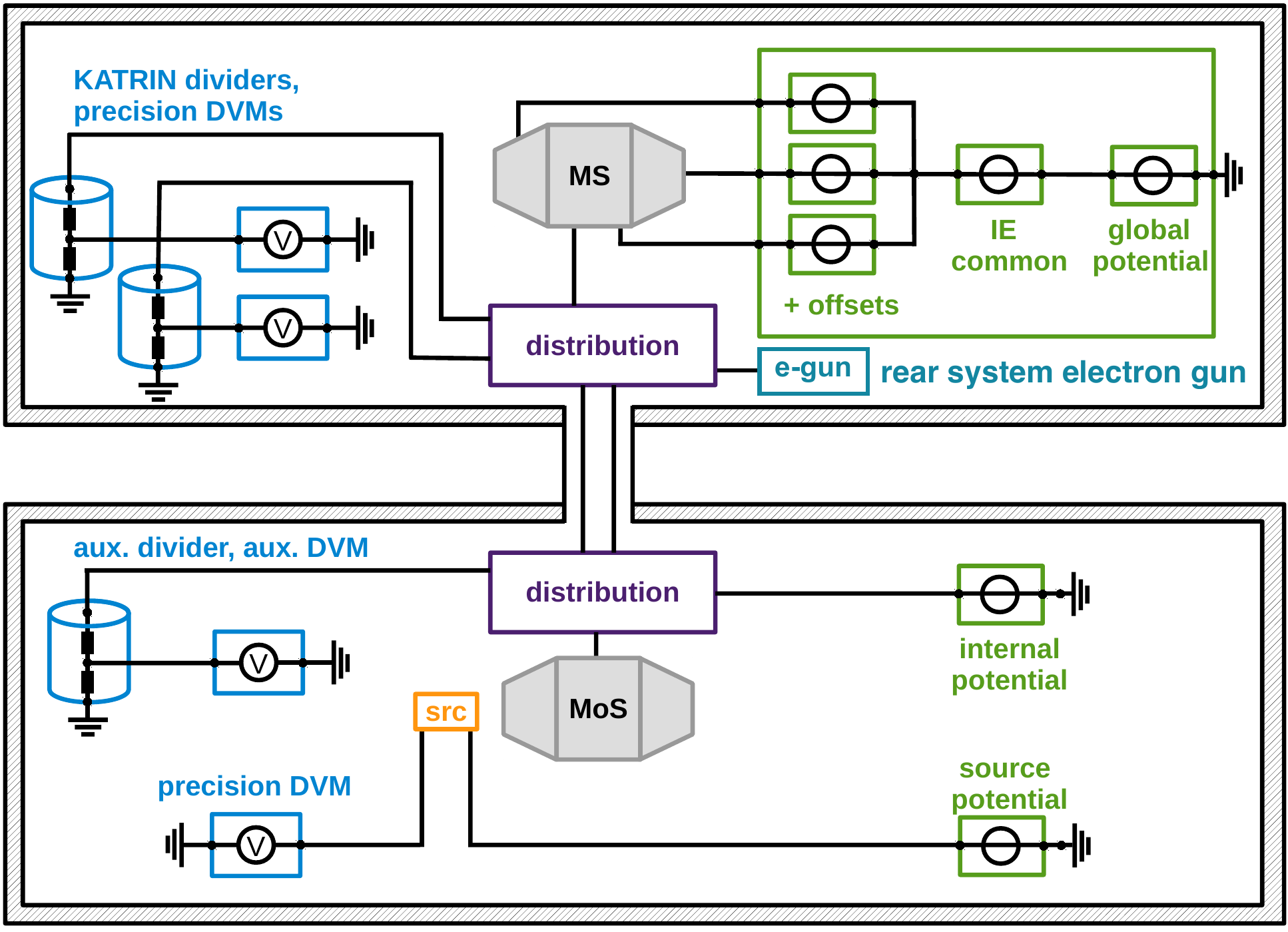}
	\caption{Distribution of precision high voltage among \acrfull{ms}, \acrfull{mos}, voltage dividers, and electron gun. The electron gun and the \acrshort{mos} can be operated in stand-alone mode. All voltages and voltage dividers are monitored by \acrfullpl{dvm}.
	}	
\label{fig:HV-distribution}
\end{figure}

In order to perform precision studies of the electron transmission properties, the potential of the \gls{ms} can be distributed to the electron gun at the far upstream end of the beam line.
The \gls{mos} can also be connected to the \gls{ms} potential for monitoring the energy scale stability (\secref{SubSection:MonitorSpectrometer}).
The entire high-voltage distribution and monitoring scheme is shown in \figref{fig:HV-distribution} and will be discussed in the following sections.

\subsubsection{Inner Electrodes of the Main Spectrometer \label{sec:SDS_IE}}

The retarding potential of the \gls{ms} is mainly established by applying a negative high voltage to the spectrometer vessel.
In order to fine tune the shape of the electrical field inside, the inner surface of the \gls{ms} is covered with an electrically isolated (partly dual-layer) wire electrode system \cite{VALERIUS2010291}.
An additional function of the wire electrode system is to prevent electrons from the vessel walls, e.g. created by cosmic muons or radioactive decays, from entering the \gls{ms} volume.
Therefore, this \acrfull{ie} is put on a slightly more negative potential than the vessel.
Otherwise those electrons from the walls could potentially find their way to the detector, contributing to the background of the experiment. 

\begin{figure}[!t]
	\centering
	\includegraphics[width=1.0\textwidth]{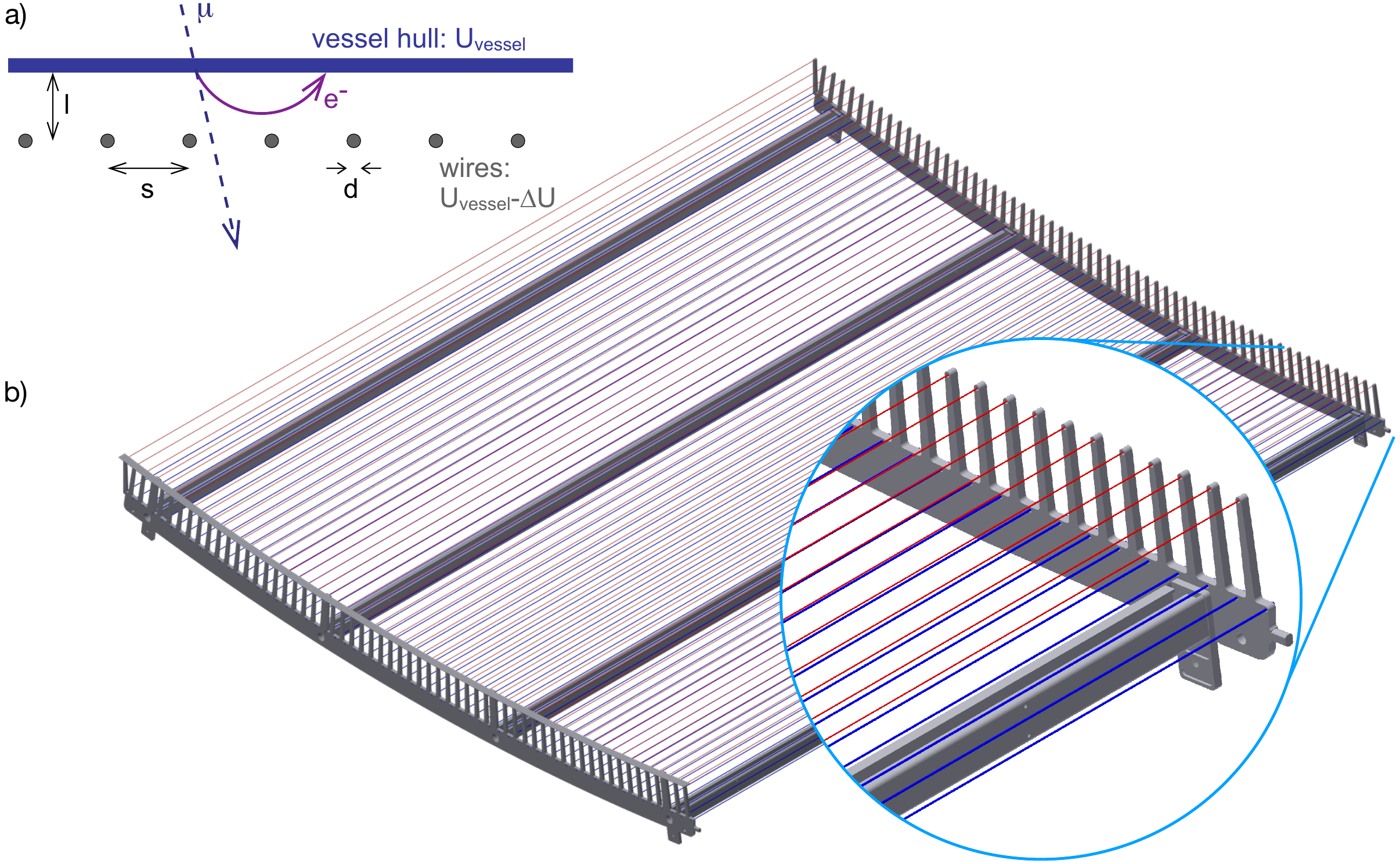}
	\caption{Wire electrode of the \gls{ms}. a) Working principle of the system for the one-wire layer case. Secondary electrons, which are created by cosmic muons (or radioactivity like $\alpha$- or $\beta$-decay or $\gamma$-radiation) hitting the vessel, are reflected due to a negative potential applied to the wires at a distance $l$ (here \SI{15}{cm}) to the wall. Each wire has a diameter $d$ and a distance $s$ (here typically \SI{2.5}{\cm}) to its neighbor. b) Drawing of a $\SI{2}{\meter} \times \SI{1.5}{\meter}$ large segment of the central dual-layer wire electrode system. Wire diameter $d$ is \SI{200}{\micro\meter} for the red and \SI{300}{\micro\meter} for the blue wire. For visibility the wire diameter is enlarged in this drawing. 
	}	
\label{fig:wire_electrode_schematics}
\end{figure}

The working principle and a segment of the \gls{ie} are shown schematically in \figref{fig:wire_electrode_schematics}. 
The geometrical cross section of the wires for the interaction with muons or gammas are negligible compared to the vessels wall.  The wire diameter $d$ and inter-wire spacing $s$ were chosen such that secondary electron emission from the wires is about 2 orders of magnitude smaller than that from the vessel.
The secondary electrons have energies of typically a few eV up to a few tens of eV, which can be reflected back to the vessel wall effectively by a wire voltage of ${\mathcal O}(-100)\,\mathrm{V}$ relative to the vessel voltage. It should be noted that the \gls{ie} is not the only shield against secondary electrons from the vessel entering the magnetic flux tube. The dominating shielding is provided by the axial magnetic field of the air coil system (\secref{subsubsection:aircoilsystem}). 

The requirement of a small geometrical cross section for background reasons precipitates the dependence of the electric potential inside the spectrometer on the electric potential of the wire electrodes $U_\mathrm{wire}$ and of the vessel potential $U_\mathrm{vessel}$.
For a simple geometry shown in \figref{fig:wire_electrode_schematics}, the electric potential far away from the wire (i.e. at least a few wire distances $s$) can be described by an effective potential $U_\mathrm{eff}$:
\begin{equation}
	U_\mathrm{eff} = U_\mathrm{wire} + \frac{U_\mathrm{vessel} - U_\mathrm{wire}}{S}~.
\label{eqn:wire_eletrode_eff_pot}
\end{equation}
The screening factor $S$ is given by geometry:
\begin{equation}
  S = 1 + \frac{2 \pi l}{s \cdot \ln{\frac{s}{\pi d}}}~.
\label{eqn:wire_eletrode_screening_factor}
\end{equation}

\eqnref{eqn:wire_eletrode_eff_pot} and \eqnref{eqn:wire_eletrode_screening_factor} show that the distance between the wires and the  wire layers have to be kept with high precision.
Therefore, a wire position and tension measurement system has been developed and set up in the clean rooms during production and installation \cite{Prall2010,PhDPrall2011,PhDHillen2011}. 

For a complex geometry like the \gls{katrin} \gls{ms} with its \gls{ie}, the spatial dependence of the electric potential was calculated with a dedicated boundary element method \cite{Furse2017}. 

\begin{figure}[!!!!b]
	\centering
	\includegraphics[width=0.8\textwidth]{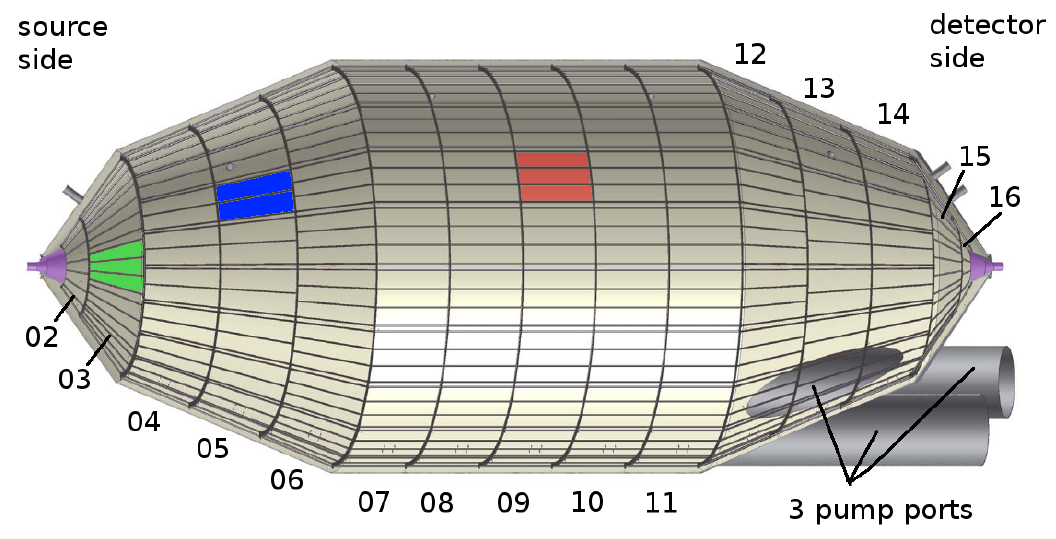}
	\caption{	
Overview of the ring structure of the \gls{ms} \gls{ie} with numbering scheme and 248 segments covering the inner surface of the \gls{ms} vessel (from \cite{PhDZacher2015}).
Rings 2 and 3 cover the upstream steep cone (module highlighted in green).
Rings 4, 5, and 6 cover the upstream flat cone (module highlighted in blue).
Rings 7 to 11 cover the cyclindrical part in the center (module highlighted in red).
From rings 12 on this scheme is continued for the downstream flat and steep cones.
Not numbered are rings 1 and 17, which correspond to the solid electrodes at the upstream entrance and downstream exit of the \gls{ms} (highlighted in purple).
    }
	\label{fig:wire_electrode_overview}
\end{figure}

The \gls{ie} comprises 248 segments arranged in 15 rings (see \figref{fig:wire_electrode_overview}).
Each ring can be set on a different potential to fine-tune the electric potential. An exception is in the 5 rings of the central, cylindrical part of the \gls{ms}; they are short-circuited in order to maintain a homogeneous electric retardation potential in the analysis plane. 
In addition, each ring is divided along the vertical axis into an eastern and a western half ring, which can be temporarily switched to a different potential in order to eject stored particles by $\vec E \times \vec B$ drift (see \secref{subsection:MSbackgroundMitigation}). 

The 5 central rings as well as the three rings in the flat conical parts at both ends of the \gls{ms} feature two wire layers for exceptional background shielding.
However, the two layers in some parts of the \gls{ie} were short-circuited during the baking of the vessel at \SI{300}{\celsius}.
The whole \gls{ie} is now operated as a single wire layer system (see \secref{sec:MS_vacuum}).
In total, 11 different voltages need to be supplied for the wire electrode system, i.e. 22 different voltages in the electric dipole mode. 
Furthermore, individual potentials are applied to the two metal electrodes at both ends of the \gls{ms} vessel.
These electrodes are needed to avoid a local Penning trap near the edges of the vessel. 
\tabref{tab:values_wire_electrodes} exhibits the typical electrical potential values of the wire electrode system in standard (single-wire layer) operation.

\begin{figure}[!!!!b]
	\centering
	\includegraphics[width=0.8\textwidth]{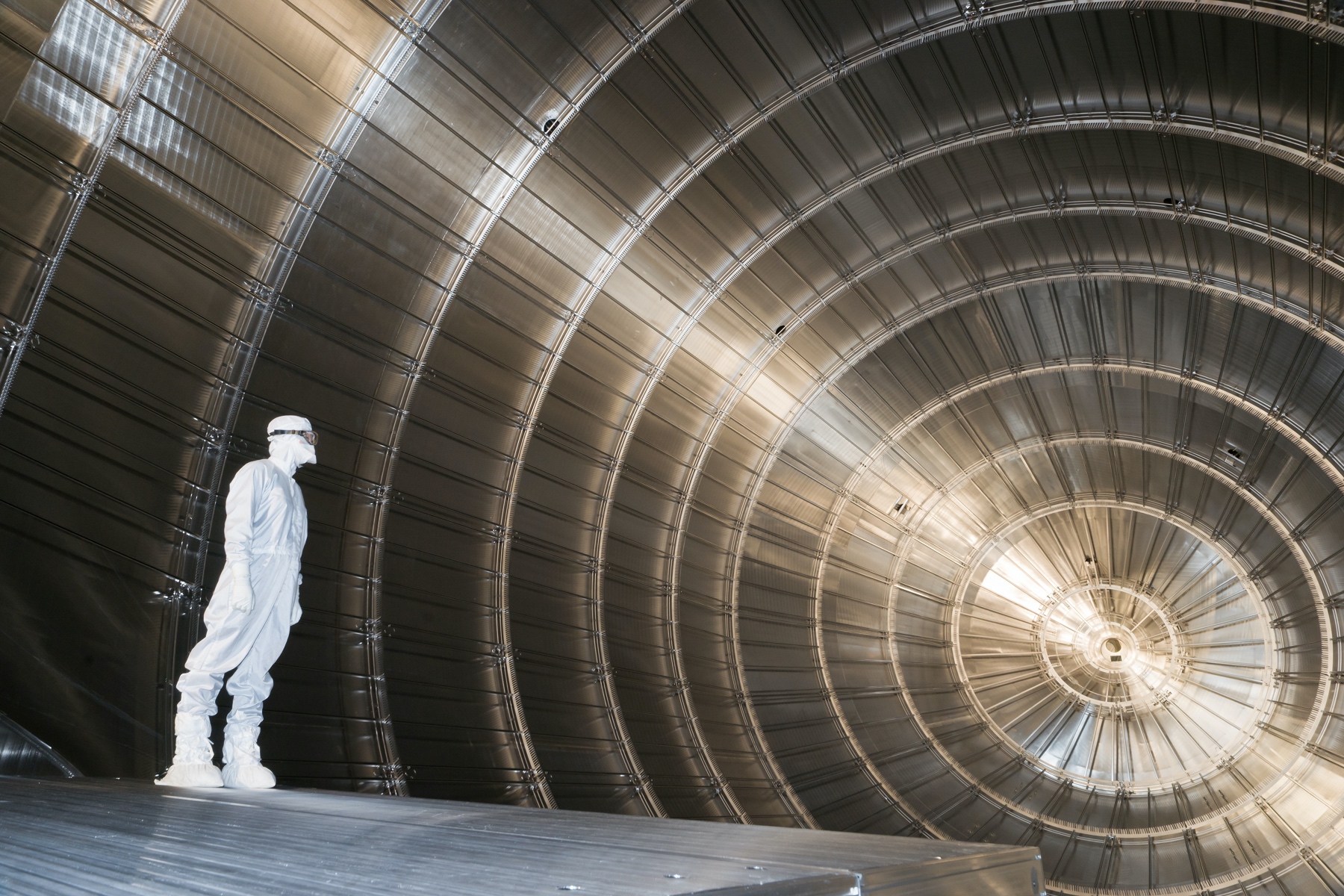}
	\includegraphics[width=0.8\textwidth]{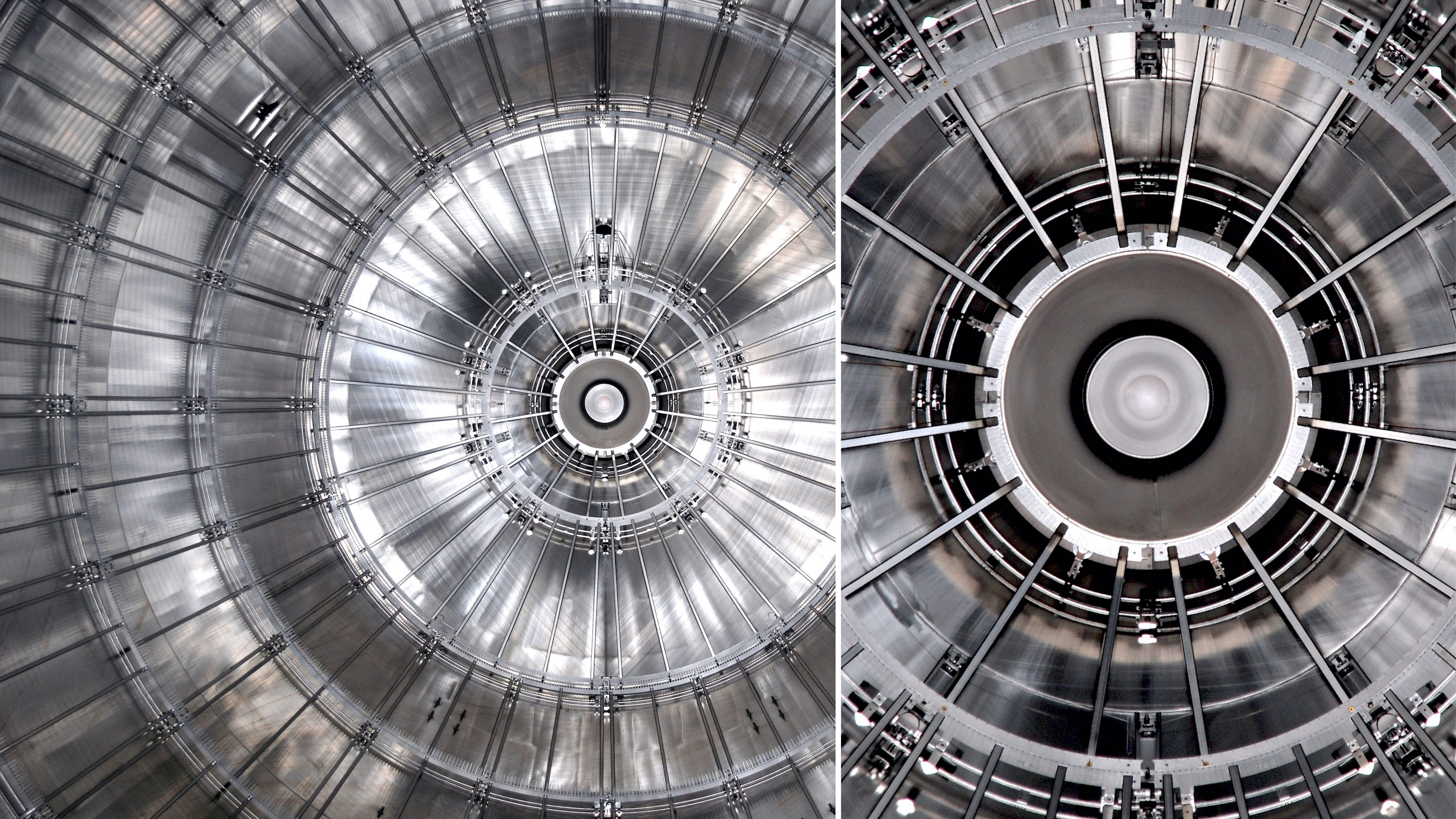}
	\caption{Upper photo from inside the main spectrometer during mounting of the wire electrode system under clean room conditions. Photo provided by M. Zacher. The lower left photo shows the finalized installation with grounded entry electrode in the center made of aluminum (insulator not visible due to perspective), surrounded by the full metal anti-Penning electrode (ring 1) made of titanium, followed by rings 2 and 3 of the steep cone as well as rings 4, 5 and 6 of the flat cone, all made of stainless steel identical to the vessel. A detailed view of the entry electrode construction can be seen in the lower right photo. Both photos provided by T. Th\"ummler. 
	}
	\label{fig:wire_electrode_photo}
\end{figure}

\figref{fig:wire_electrode_photo} shows a photo from inside the main spectrometer during mounting of the wire electrode system under clean room conditions.

\begin{table}[!ht]
	\caption{Electrical potential $\Delta U$ with respect to the \gls{ms} vessel potential $U_\mathrm{veseel}$ on the \gls{ie} rings in single-wire layer operation mode. Both (up- and downstream) full metal anti-Penning electrodes (rings 1 and 17) are usually shorted to the \gls{ms} vessel potential and not listed here. Refer to \figref{fig:wire_electrode_overview} for geometrical overview.
	}
	\begin{center}
	\begin{tabular}{ccc}
	\hline
	\textbf{Location} & \textbf{Ring} & \textbf{Voltage $\Delta U$ (V)} \\ 
	\hline 
	upstream steep cone & 2 & -160 \\ 
	upstream steep cone & 3 & -160 \\ 
	upstream flat cone & 4 & -200\\ 
	upstream flat cone & 5 & -200\\ 
	upstream flat cone & 6 & -200\\ 
	\hline
	cylindrical part & 7 -- 11 & -200 \\ 
	\hline
	downstream flat cone & 12 & -200 \\ 
	downstream flat cone & 13 & -200\\ 
	downstream flat cone & 14 & -200\\ 
	downstream steep cone & 15 & -160\\ 
	downstream steep cone & 16 & -160\\ 
	\hline
	\end{tabular}
	\end{center}
	\label{tab:values_wire_electrodes}
\end{table}

\clearpage

\subsubsection{High Voltage Distribution and Monitoring}
\label{subsec:HV-Distribution-Monitoring}

The \gls{ms} vessel itself is charged up to almost the full negative voltage of the filter potential.
This avoids the technical challenge related to large voltage differences of the \gls{ie} to the \gls{ms} vessel.
The inner wire electrodes are only a little more negative (see \tabref{tab:values_wire_electrodes}).
With the vessel at a potential different from the ground potential, insulating parts within the beam tube are used to separate the DC voltage of the vessel (down to \SI{-35}{\kilo\volt}) from the grounded parts of the beam tube, e.g. the grounded electrodes (see \figref{fig:wire_electrode_photo}).

\begin{figure}[!t]
	\centering
	\includegraphics[width=1.0\textwidth]{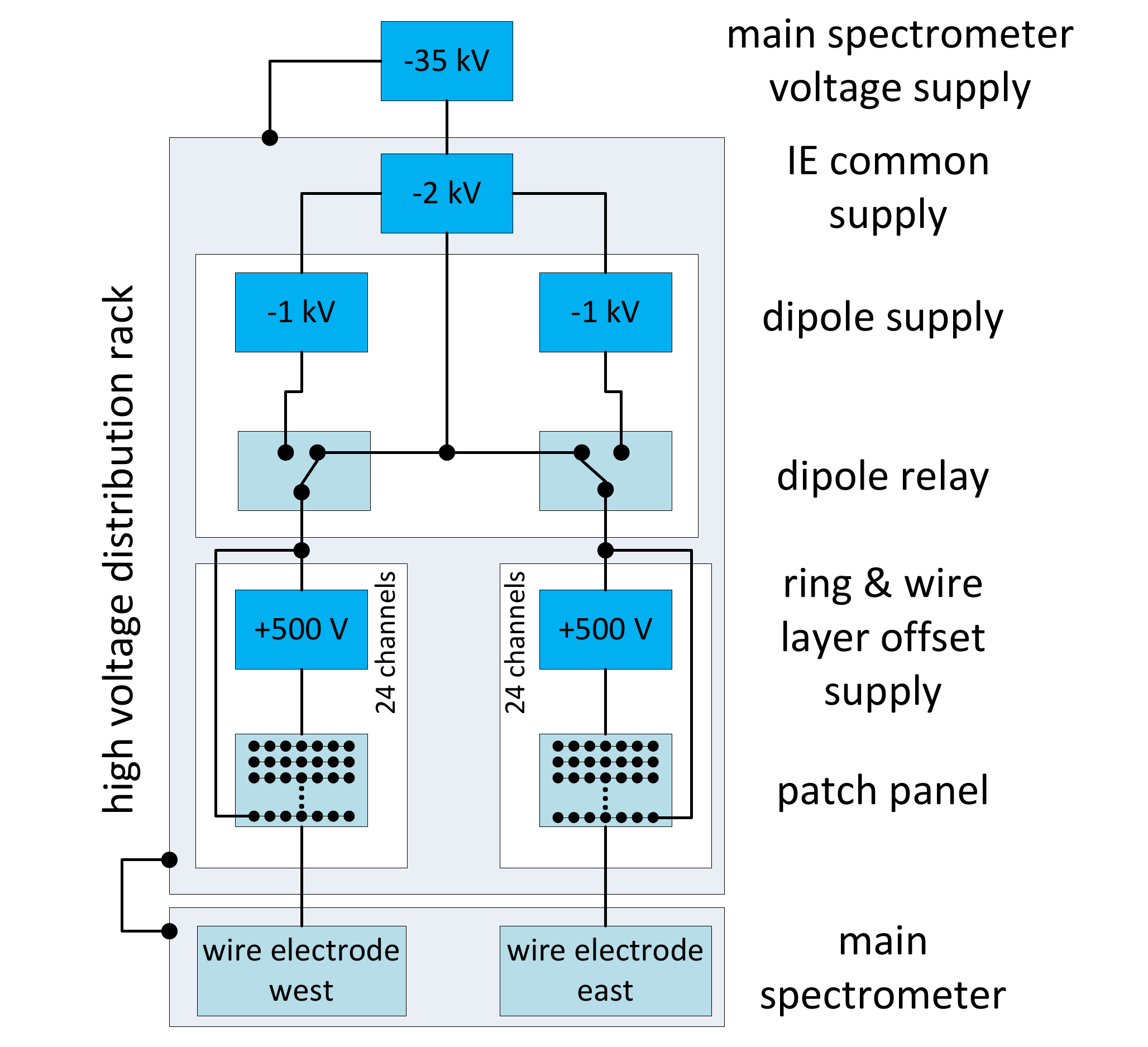}
	\caption{Simplified schematic overview of the HV system of the \gls{ms}.
	The main HV power supply delivers a voltage down to \SI{-35}{\kilo\volt}, which is distributed to the \gls{ms} vessel and the HV distribution rack. Here, an additional so-called \gls{ie} common HV power supply is used to deliver a negative offset potential to the \gls{ie}.
	With dipole power supplies and relays, each one of the inner electrode dipoles (east and west) can be loaded with fast switching negative voltages for the removal of stored charged particles.
	For the 22 isolated elements of the inner electrode of each dipole (11 in single wire layer mode), 24 channels of positive offset power supplies are available. The distribution is done with the patch panel.
	The \gls{ie} common voltage is also connected to the patch panel to allow for the direct connection of the \gls{ie}, e. g. for the central electrodes, to this potential. Not shown here is the precision monitoring of the \gls{ie} common HV relative to the ground potential with the help of custom-made ppm-precision HV dividers read out by 8.5-digit \glspl{dvm}  and, redundantly, the \gls{mos}. Figure reproduced from \cite{PhDRest2019}.	
	}
	\label{fig:hv_system_ms_overview}
\end{figure}

The main spectrometer high-voltage system consists of several voltage supplies, high-voltage dividers, and digital voltmeters to create and measure the retarding potential of the \gls{ms}.
The overall distribution scheme between the \gls{ms}, \gls{mos}, voltage dividers, and electron gun is shown in \figref{fig:HV-distribution}.
\figref{fig:hv_system_ms_overview} focusses on the distribution in the \gls{ms} and its \gls{ie}.
The key units are:

\begin{itemize}
	\item \textbf{The main HV power supply} delivers the basic voltage of \SI{-18.4}{\kilo\volt} (down to \SI{-35}{\kilo\volt}) to the vessel of the \gls{ms} and provides the reference potential of the HV distribution rack, where all other power supplies are located.
	In order to reduce the high-frequency (HF) noise and AC fluctuations of the HV, an active post-regulation system is used (see last paragraph of \secref{sec:post_regulation}). This primary potential can also be routed to the angular selective electron gun at the upstream end of the beamline (see \secref{sec:egun}).
	\item \textbf{The \gls{ie} common HV power supply} delivers a negative offset of usually \SI{-200}{\volt} (max. \SI{-2}{\kilo\volt}) to the whole \acrfull{ie}, resulting in an absolute retarding potential of \SI{-18.6}{\kilo\volt}. Since the analyzing plane is determined by the most negative potential in the center of the \gls{ms}, the inner wire layer of the central cylindrical part of the \gls{ie} is hard-wired to this offset potential.
	\item \textbf{Two dipole power supplies} in combination with a relay each create a fast switching ($\mathcal{O}(\mathrm{ms})$) dipole voltage of down to \SI{-1}{\kilo\volt} for one side or the whole of the \gls{ie} (east and/or west). This feature can be used to remove background caused by magnetically trapped charged particles in the \gls{ms} with an $\vec E \times \vec B$ drift \cite{PhDHilk2016}. The dipole power supplies are  disabled (short-circuited by relays) during single tritium runs and can be activated to get rid of the stored particles between runs.
	\item \textbf{Multiple offset power supplies} deliver positive voltages of up to \SI{500}{\volt} on top of the \gls{ie} common potential for single channels of the wire electrode system.
	The configuration can be arranged by the so-called patch panels. Each of these matrices (one for the east and one for the west dipole) with 23 rows and 23 columns connects 22 of the 24 offset power supplies with the 22 isolated parts of the wire electrode (or 11 respectively, in single wire layer mode, see \secref{sec:SDS_IE}) or directly to the \gls{ie} common potential.
	By this arrangement, all elements of the \gls{ie} can be operated at individual or common potentials and any arbitrary configuration.
\item \textbf{The precision measurement of the retarding potential} is performed with custom-made ppm-precision HV dividers K35 \cite{Thuemmler2009} and K65 \cite{Bauer2013a} and two 8.5-digit \glspl{dvm}\footnote{Fluke 8508A, \url{https://us.flukecal.com}} (see \figref{fig:HV-distribution}). 
The high-voltage dividers have been calibrated at PTB Braunschweig, the German national metrology institute, and a novel in-house calibration method has been successfully developed and tested \cite{Rest_2019}.
The offset values and gain correction factors of the \glspl{dvm} are determined twice per week based on a set of  \SI{10}{\volt} reference standards\footnote{Fluke 732B, \url{https://us.flukecal.com}}.
For traceability, a subset of the reference standard units is calibrated once per year at an external calibration lab based on the Josephson standard.
Additionally, this voltage is also distributed to the \gls{mos}, which runs in parallel to the tritium measurement with a $^\mathrm{83m}$Kr source in order to monitor the HV.\\	
\end{itemize}

The precision HV dividers provide a ppm-precise monitoring of the high voltage. However, in order to simplify the analysis and for practical reasons it is also desired that not only the monitoring but also the set points and absolute value of the applied HV can be provided with ppm-precision.
For the \gls{ie} common and offset power supplies, the same level of precision is not required, since their outputs are in the range of a few \SI{100}{\volt}.
Such devices with uncertainties in the \SI{10}{\milli\volt} range or better are commercially available.
For the main HV power supply this requirement is challenging, since the absolute value is much higher.
A custom-design precision high-voltage power supply has been developed (HCP 70M-35000)\footnote{FuG-Elektronik, \url{https://www.fug-elektronik.de}} to provide voltages down to \SI{-35}{\kilo\volt} with a stability of 2\,ppm over \SI{8}{\hour}.
When the post-regulation system is running, the stability requirement for the HV power supply is less stringent, since the HV is regulated by an active control circuit (see last paragraph of \secref{sec:post_regulation}).\\
The HV system of the \gls{ms} has been tested and operated successfully in two commissioning phases of the spectrometer and detector section as well as during the first measurement campaign with $^\mathrm{83m}$Kr \cite{Arenz2018b} and tritium \cite{Aker2020_firstOperationTritium}.

\paragraph{Post-regulation}
\label{sec:post_regulation}

The \gls{ms} provides no electrostatic shielding and therefore acts as a large antenna for HF interference.
Due to the large capacitive coupling between the \gls{ie} and the \gls{ms} vessel (many nF) and the impossibility to connect the \gls{ie} in such  a way that they form their own Faraday cage from all other electrodes, it must be assured that an effective Faraday cage is formed by the grounded  beam tube parts in conjunction with the \gls{ms} vessel.

\begin{figure}[!t]
	\centering
	\includegraphics[width=0.9\textwidth]{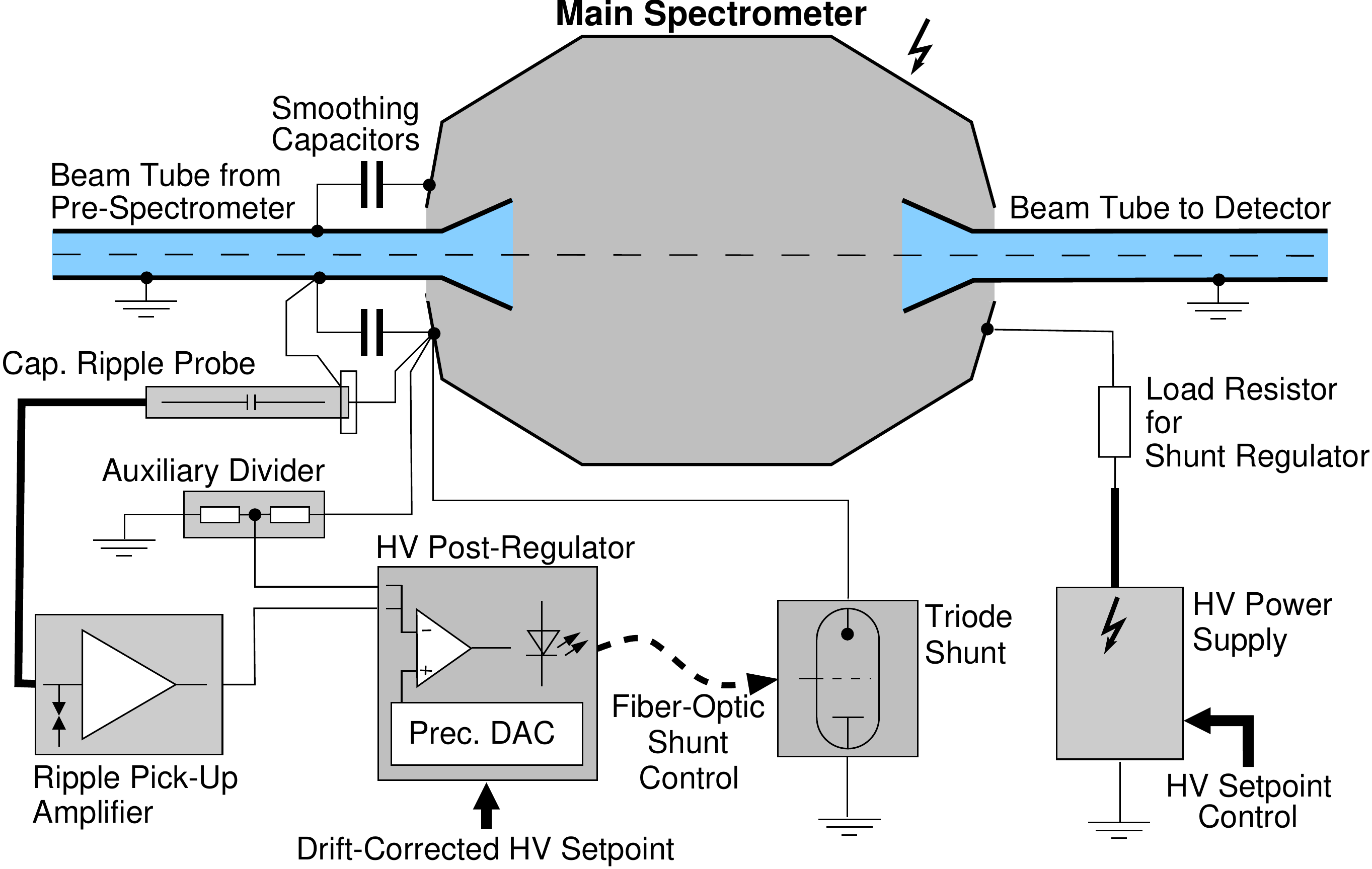}
	\caption{Overview of the concept of the post regulation system with capacitive pick-up probe, smoothing capacitors, regulator and feedback loop.
	}	
\label{fig:post-regulation}
\end{figure}

To build the Faraday cage from the beam tube interrupted by insulators without galvanically reconnecting it to the vessel, the insulators must be made ineffective for any AC voltages forming across them. 
The HV power-supply units that feed the \gls{katrin} \gls{ms} are not able to actualize this.
This is due to the parasitic inductance of the connection line from the supply unit to the spectrometer and the output impedance of the supply unit that is not low enough in some frequency ranges (such as \SI{1}{\Hz} to \SI{1}{\kilo\Hz}).
Creating a low AC impedance across the insulators by simply attaching large capacitors would require very bulky devices that in turn have large parasitic inductances and render them ineffective at higher frequencies (MHz).
Moreover, capacitors that large would store a lot of energy that might result in a highly destructive electromagnetic transient in case of an accidental HV discharge event.

The post-regulation system (see \figref{fig:post-regulation}) implements an active compensation loop to remove high frequency noise from the precision retarding potential.
A capacitive pick-up probe detects the ripples on the high-voltage potential, which is then amplified and inverted.
The inverted ripple is then fed into the regulator which drives a triode shunt to re-apply the potential correction to the vessel potential.
The stability of the regulation is maintained by referring to an auxiliary divider and to precision set points from the \acrfull{scs}.

\clearpage

%% file: SpectrometerSystemMagnet.tex
\subsection{Magnet system}
\label{subsection:MSmagnetsystem}

The magnetic field inside the \acrfull{ms} is dominated by the fringe fields of the superconducting magnets of the \acrfull{ps} and the detector system.
The magnetic field at the analyzing plane of the \gls{ms} is further optimized with the normal-conducting air coil system (\secref{subsubsection:aircoilsystem}) surrounding the \gls{ms}.
It can provide a magnetic field strength of up to \SI{2}{\milli\tesla}.

\subsubsection{Beam line magnets}
\label{subsubsection:beamlinemagnets}
The guiding magnetic field of the beam line in the spectrometer system is defined by four strong superconducting magnets~\cite{Arenz2018c}: two in the \gls{ps} and two in the detector system. 
\begin{itemize}
	\item Two magnets of the \gls{ps}: They are two \SI{4.5}{\tesla} cryogen-free conduction cooled superconducting magnets installed between the entrance of the \gls{ps} and the \gls{cps} called \gls{ps1} and between the exit of the \gls{ps} and the \gls{ms} called \gls{ps2}.
	 A two-stage Gifford-McMahon (GM) cryocooler\footnote{Sumitomo RDK-415D \url{https://www.sumitomocorp.com/}} is adapted for cooling of the magnet.
	 The magnets are operated in driven mode at a nominal field of \SI{3.1}{\tesla}.
	 The air-cooled power supplies\footnote{FuG-Elektronik NTS~800-5, \url{https://www.fug-elektronik.de}} provide a sufficient current stability of \SI{<+-100}{ppm} per \SI{8}{\hour}.
	 The current stabilities of the power supplies are monitored by DC current transducers\footnote{DCCT, IT~200-S Ultrastab manufactured by LEM, \url{https://www.lem.com/}}.
	 The stability of the \gls{ps} magnet currents at their nominal settings (\SI{70}{\percent} of the design fields) are summarized in table~\ref{tabMagps}.
	\item Two magnets of the detector system: The magnets of the detector system comprise the \gls{pch} and the \gls{det}~\cite{Amsbaugh2015}.
	The \SI{6}{\tesla} \gls{pch} provides the highest magnetic field for the experiment (nominal operation at \SI{4.2}{\tesla}), while the \gls{det} delivers a \SI{3.6}{\tesla} field (nominal operation at \SI{2.5}{\tesla}), matching the field at the source.
	To achieve the requirement of high field stability \SI{0.03}{\percent\per month}, the magnets are designed for persistent current mode operation, like the \gls{rs} and the \gls{dps} magnets, which have persistent switch heaters inside the liquid helium bath. The magnets are cooled in a liquid helium bath at a small over-pressure of about \SI{4.8}{\kilo\pascal} by recondensing boiling helium with a two-stage pulse-tube cryocooler\footnote{Cryomech PT410, \url{https://www.cryomech.com}}. The magnetic field drift of the magnets were checked during the commissioning of the magnets with a NMR probe; they are well within the specification~\cite{Amsbaugh2015, Arenz2018c}.
\end{itemize}

\begin{table}[!ht]
	\caption{The stability of the \gls{ps} magnet currents at their nominal settings (\SI{70}{\percent} of the design fields).
	Eight hours of data from the external flux gate current sensors were used for the calculation of the average current $I_\mathrm{av, ext.}$ and its standard deviation $\sigma$\,(A).
	A small offset value at zero-current has been corrected for $I_\mathrm{av, ext.}$.
	}
	\label{tabMagps}
	\vspace*{1ex}
	\begin{center}
	\begin{tabular}{ccc}
	\hline
	 & \textbf{PS1} & \textbf{PS2} \\
	\hline
	$I_\mathrm{av, ext.}$ (A) & \num{109.459} & \num{108.806} \\
	$\sigma$ (A) & \num{3.8e-4} & \num{3.2e-3} \\
	$\sigma/I_\mathrm{av, ext.}$ ($\%$) & \num{3.5e-4} & \num{3.0e-3} \\
	\hline
	\end{tabular}
	\end{center}
\end{table}

The beam line magnetic fields in the spectrometer system will be set in standard nominal configuration to analyze the tritium \betaspec.
In addition, they can be configured either symmetrically or asymmetrically to the analyzing plane for background investigations and the alignment of the \gls{ms}.
In the symmetric field configuration, the magnetic fields in the \gls{ms} will be set symmetrically from the analyzing plane either by setting the \gls{ps2} and the \gls{pch} at the same field or by using the air coils. For the asymmetric field configuration in the \gls{ms}, either the two \gls{ps} magnets are turned off or the air coils are configured individually for the investigation of background electrons from the \gls{ms} vessel wall~\cite{Altenmueller2018}. 

\subsubsection{Air coil system}
\label{subsubsection:aircoilsystem}
\glsreset{lfcs}
\glsreset{emcs}

The air coil system aims to provide optimal transmission properties for signal electrons and to reduce background electrons by efficient magnetic shielding.
Optimal \gls{mace} filter characteristics in the \gls{ms} can be achieved by operating the air coils in addition to the beam line magnets, as investigated in~\cite{Glueck2013}. 

The air coil system is designed to fine-tune the magnetic field in a range of up to \SI{2}{\milli\tesla} in the analyzing plane.
There are two large-volume air coil systems mounted on a cylindrical holding structure around the \gls{ms} vessel: the \gls{lfcs} and the \gls{emcs}.
The \gls{lfcs} comprises 14 \SIadj{12.6}{\metre}-diameter normal-conducting solenoids installed around the longitudinal axis of the \SIadj{23.23}{\metre}-long \gls{ms} vessel.
The \gls{emcs} consists of two sets of dipole coils installed around the x- and y-axes of the \gls{ms} to compensate for the earth magnetic field in the \gls{ms}.
Each of the 16 air coils can be individually charged by its associated power supply\footnote{SM-3000 series, Delta Elektronika, \url{https://www.delta-elektronika.nl}} for fine tuning of the magnetic field in the analyzing plane.
The details of the air coil systems are reported in~\cite{Glueck2013, PhDErhard2016, Erhard2018}. 

\begin{figure}
\center
\includegraphics[width=.8\linewidth]{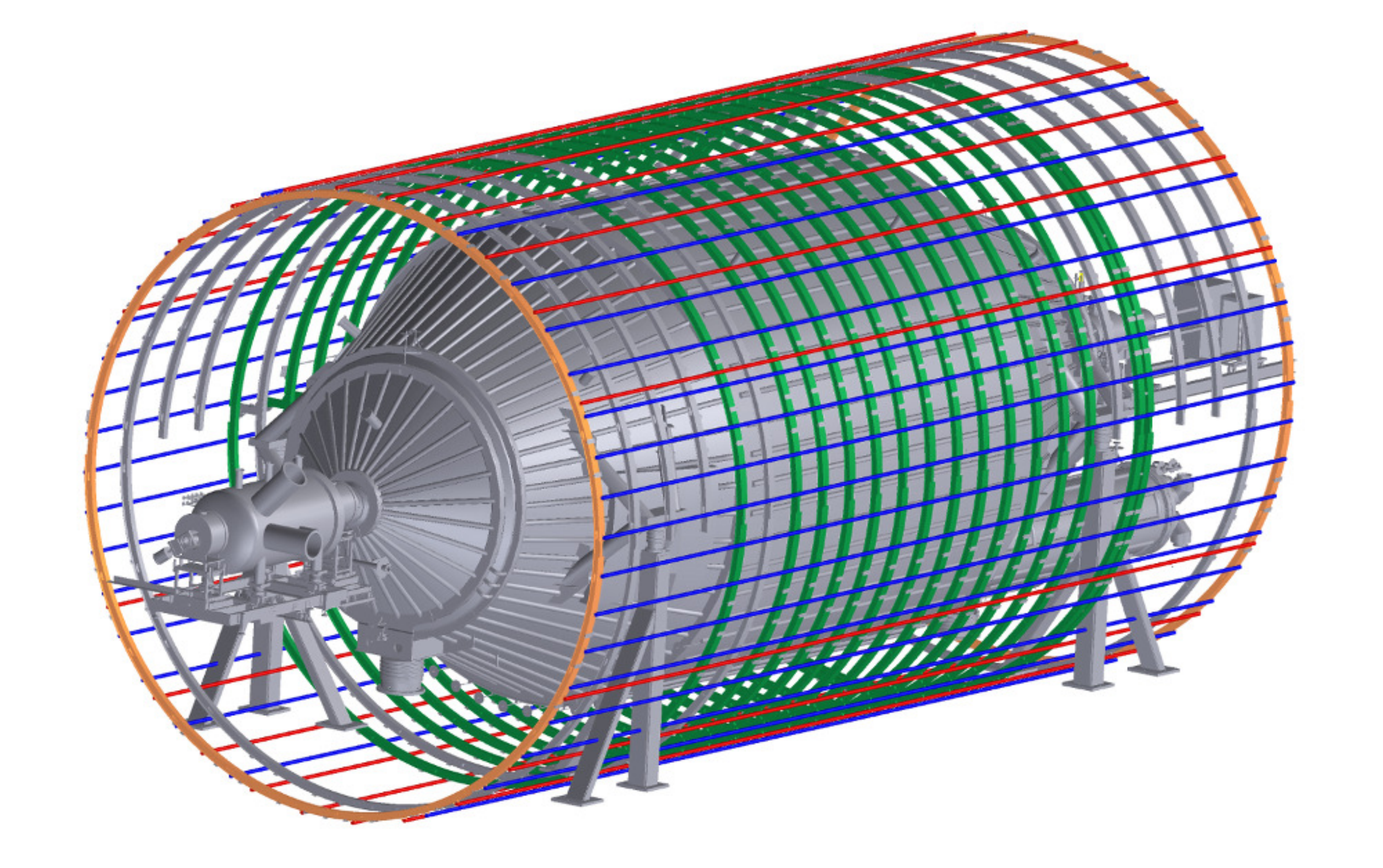} 
\caption{A schematic view of the spectrometer system with the air coil systems around the \gls{ms}: 14 \gls{lfcs} coils (green circles) and 2 \gls{emcs} wire loops (blue/red straight lines) correct and fine-tune the magnetic field in the \gls{ms}. All coils can be operated individually. Figure reproduced from \cite{Glueck2013}. CC-BY 3.0.
 }
\label{Figure:MSaircoils}
\end{figure}

The current stability of the 16~individual power supplies of the air coils is monitored by the same type of DCCT as the \gls{ps} magnets.
A relative uncertainty of $\sigma/I < 10^{-3}$ is reached for the current monitoring of the 16 power supplies.
The details of the magnetic field measurements and analysis are described in  \secref{SubSection:SpectrometerMagneticFieldMonitoringSystem} and \cite{PhDErhard2016, Letnev2018}.

Fine-tuning of the individual \gls{lfcs} currents allows to optimize the homogeneity of the magnetic fields in the \gls{ms} for tritium runs.
The flexibility of adjusting the magnetic field also allows the investigation of backgrounds from the \gls{ms} vessel wall or its \gls{ie}.
By increasing the magnetic field in the \gls{ms}, the transported magnetic flux volume can be reduced, which would help reduce the background rate during the standard operation.
In addition, a special magnetic-pulsing mode (\secref{subsubsection:magneticpulse}) can effectively remove stored electrons from the transported flux-tube.
For this purpose the nominal currents of the 14~ \gls{lfcs} coils can be individually inverted.
For certain background investigations, a static current inversion is applied to the \gls{lfcs}, providing field lines that connect sections of the vessel wall with the detector.
Along these field lines, background electrons from the vessel walls, otherwise magnetically shielded from the detector, can be directly detected.

%% file: SpectrometerSystemBackgroundMitigation.tex
\subsection{Background Mitigation}
\label{subsection:MSbackgroundMitigation}

A low background rate is vital for the success of the \gls{katrin} experiment.
Different active methods are available to reduce the background.

\subsubsection{Magnetic Pulse}
\label{subsubsection:magneticpulse}
The air coil system at the \gls{ms} (\secref{subsubsection:aircoilsystem}) is equipped with the so-called flip-boxes that can invert the current direction in each coil individually. At typical coil currents of up to \SI{120}{\ampere}, the inversion takes less than \SI{0.1}{\second}.
By inverting multiple \gls{lfcs} coils simultaneously, the magnetic guiding field inside the \gls{ms} is inverted. This process takes place within one second, after which the currents are flipped back to normal. This magnetic pulse deforms the magnetic field lines so that stored electrons are guided towards the \gls{ms} vessel walls, where they are removed.
In addition, a radial drift is induced by the magnetic field change that supports the electron removal.
Tests of this system have been carried out during several commissioning phases.
The test results indicate the system's effective removal of a large fraction of stored electrons from the flux volume; the details can be found in \cite{Arenz2018d}.

\subsubsection{Penning Wiper}
\label{subsubsection:penningwiper}
Due to the superposition of the negative electric potentials of the \gls{ps} and the \gls{ms}, with the magnetic field of the \gls{ps2}, a Penning trap for negatively charged particles is formed in the inter-spectrometer space.
If an electron passing through this region loses kinetic energy via synchrotron radiation, or elastic or inelastic scattering on residual gas particles, it can be trapped.
Also secondary electrons can populate the trap, even without any energy loss.
As a consequence, this trap accumulates electrons and can lead to the creation of additional background.
Once the accumulated electron density is high enough to trigger residual gas ionization processes and discharges, it interrupts measurements and can damage hardware downstream in the spectrometer and detector systems.
These problems do not allow for operating both spectrometers at high voltages at the same time unless there is a reliable countermeasure to limit the population in the Penning trap. 

\begin{figure}
	\center
	\includegraphics[width=1\linewidth]{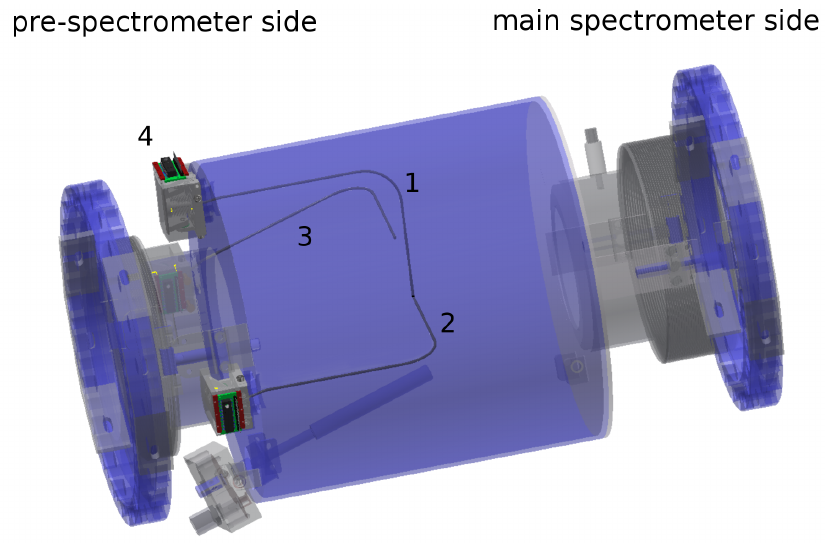} 
	\caption{A schematic view of the three electron catchers (1),(2), and (3) inside the valve of the \gls{ps2} magnet; (4) is a catcher's microcontroller.
	The valve connects to the \gls{ps} exit flange on the upstream end (left) and  to the \gls{ms} entry flange on the downstream end (right).
	Figure reproduced from \cite{aker2019suppression}.
	 }
	\label{Figure:penningWiperCAD}
\end{figure}

As a countermeasure, the mechanical removal of trapped electrons was developed and tested.
This is based on the fact that the trapped electrons will hit a wire anywhere in the flux tube within less than \SI{\sim  1}{\milli\second} due to the magnetron rotation around the symmetry axis \cite{Beck2010}.
Finally, three electron "catchers" were implemented at different positions inside the valve of the \gls{ps2} magnet (see \figref{Figure:penningWiperCAD}) \cite{aker2019suppression}.
An electron catcher is a \SIadj{2}{\milli\metre}-diameter Inconel rod that can be moved in or out of the magnetic flux tube by means of an external pneumatic muscle driven by pressurized air connected to the catcher.
The rod is large enough to reach the central axis of the beam line so that all stored electrons in magnetron motion are intercepted.
The ability to move the catcher is important to prevent unnecessary losses in statistics due to shadowing of detector pixels if the catcher were fixed inside the flux tube.
Three catchers are installed for redundancy as the number of operation cycles is limited to \num{1.5} million each\footnote{The only moving parts are custom made membrane bellows by company COMVAT, \url{https://www.comvat.com}.}.
The electron catchers can stop or prevent discharges and empty the trap when needed, e.g. in pressure conditions up to few \SI{e-9}{\milli\bar} during background studies.
They can be controlled by the \gls{orca} software manually or automatically in run scripts (see \secref{Subsection:experimental_control}).
The actuation of the electron catchers when the detector rate surpasses a predefined safety threshold (during a Penning discharge) has been implemented in a detector safety script.
Moreover, each catcher is equipped with a photoelectric sensor that detects its position and indicates when the catcher moves in or out of the flux tube.
This signal is recorded by the \acrfull{daq} system together with the \acrfull{fpd} data for offline analysis.

\begin{figure}
	\center
	\includegraphics[width=1\linewidth]{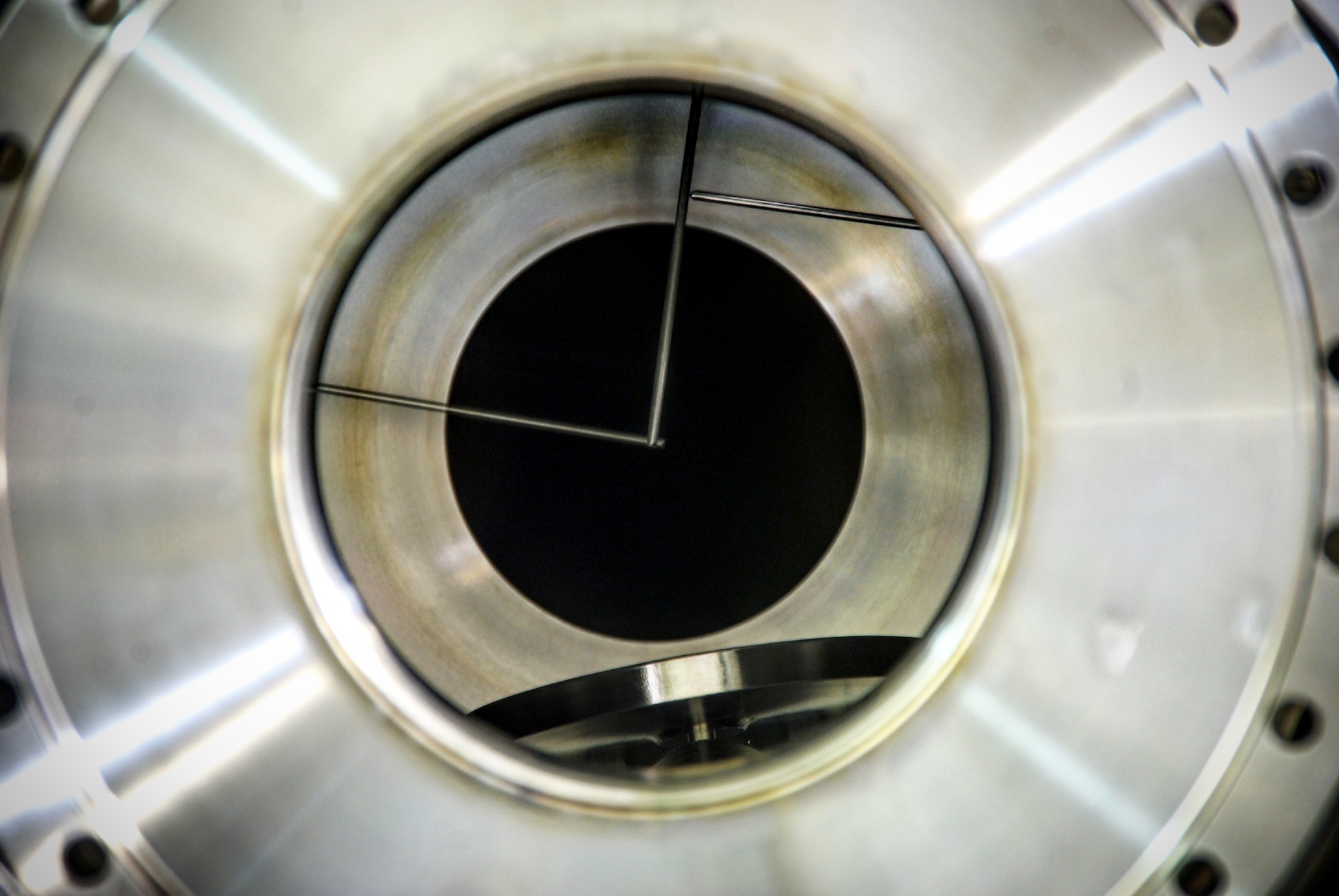}
	\caption{Photo of the electron catchers inside the valve: two of them (upper and left) are inserted into the flux tube, one (right) is retracted. Figure reproduced from \cite{aker2019suppression}. 
	}
	\label{Figure:penningWiperPhoto}
\end{figure}

\subsubsection{Electric Dipole}
\label{subsubsection:electricdipole}

\begin{figure}
	\centering
	\subfloat[]{{\includegraphics[width=0.4\linewidth]{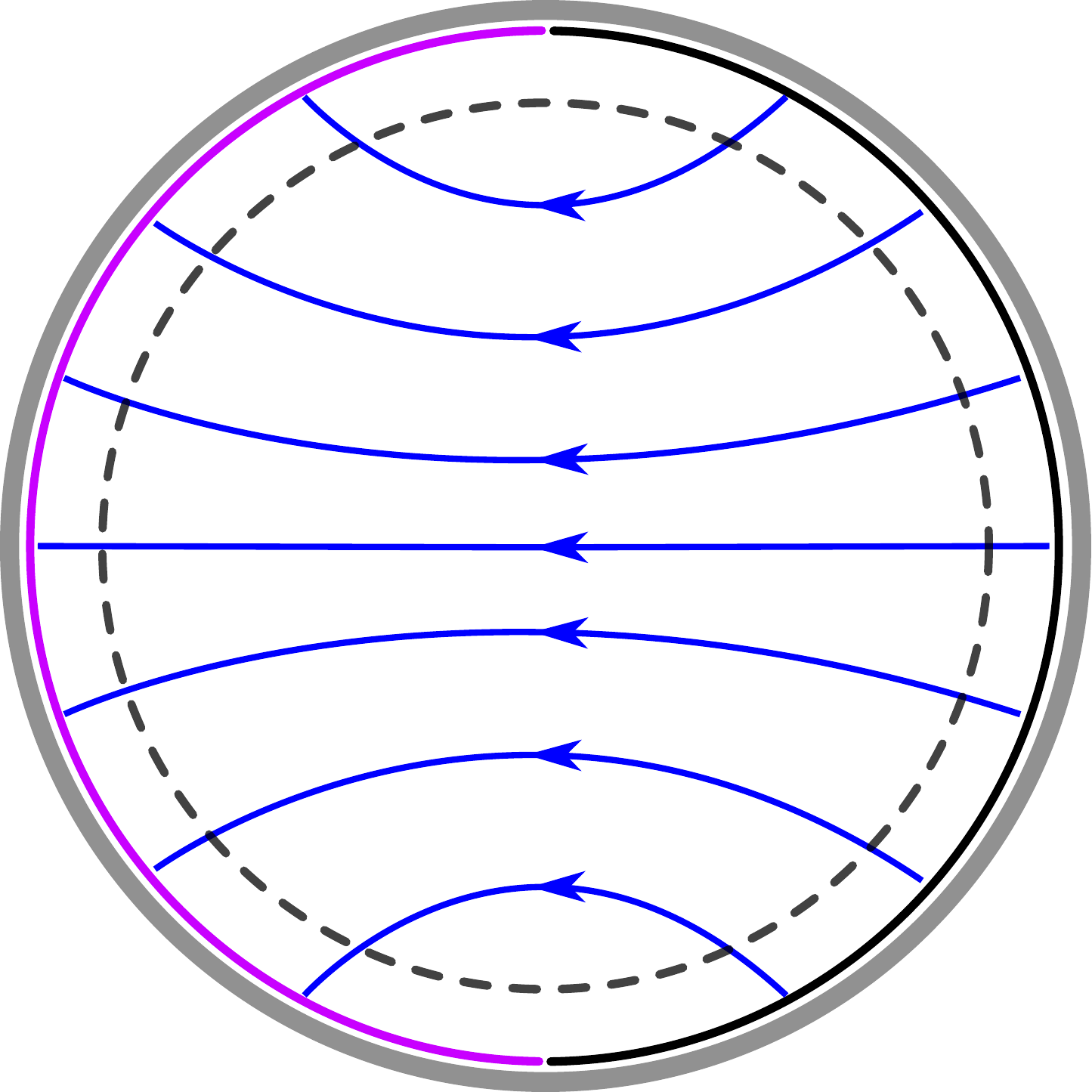}}}
	\qquad
	\subfloat[]{{\includegraphics[width=0.4\linewidth]{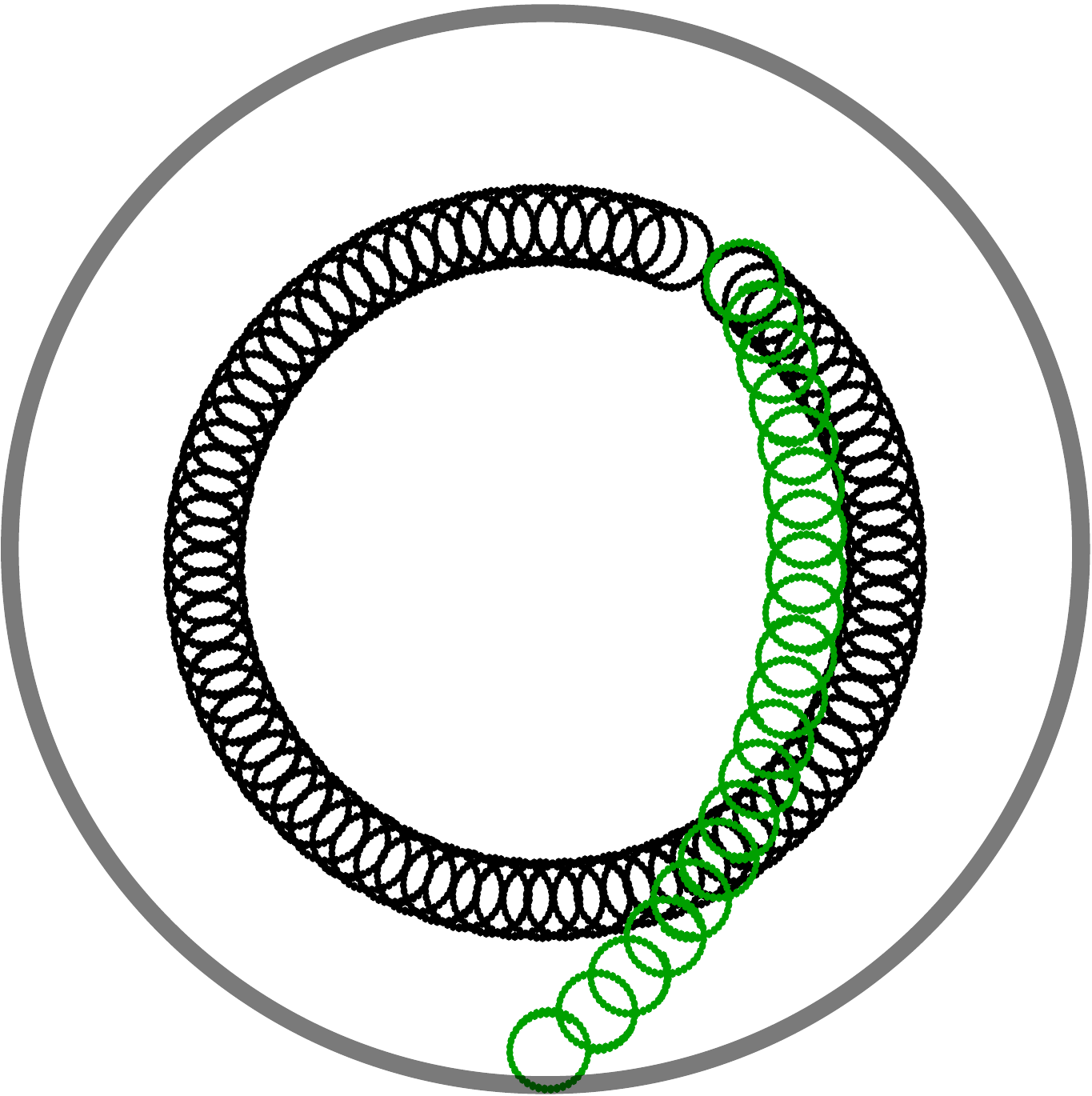}}}
	\caption{Electron removal with electric dipole: a) Visualization of the \gls{ms} dipole field. The dashed line indicates the flux tube. b) Trajectory of a stored electron (black) and of removal through the application of a dipole pulse (green). Figures adopted from \cite{PhDWandkowsky2013} 
	}
	\label{fig:dipole}
\end{figure}

As mentioned in \secref{sec:spectrometer_system}, electrons created inside the \gls{ms} volume can be trapped within the high magnetic fields at both ends of the vessel.
In the \gls{ms}'s ultrahigh vacuum ($\approx\SI{e-11}{\milli\bar}$), these electrons can be stored up to several hours, potentially inducing additional background mainly through the ionization of residual gas.
To mitigate this process, the \gls{ms} can be operated in dipole mode, described in \secref{sec:SDS_HV}.
In this configuration, different potentials on the western and eastern part of the \gls{ie} form an electric dipole field in the spectrometer (see \figref{fig:dipole} a).
Through the combination with the guiding magnetic field, an $E \times B$ drift is induced, driving stored electrons towards the vessel walls (see \figref{fig:dipole} b).
Since this drift is superimposed with the magnetron motion of the stored particle, this action is effective only when it is accomplished within one magnetron turn.
Therefore, the efficiency of this method depends on the electric dipole field strength and the kinetic energy of the stored electron, limiting its application to removing electrons up to a few \si{\kilo\electronvolt}.
For efficient removal, a short dipole pulse of about one second is sufficient.
The electric dipole mode, static or pulsed, and its on-off time is controlled by the \gls{orca} software.
The electric dipole cannot be applied during standard data taking, because the field would distort signal electron transmission.
However, its short operation time allows its operation during changes of the retardation potential in a tritium run.
For analysis purpose, the dipole status is stored together with the \gls{fpd} data. Besides background mitigation, the electric dipole field can also be used for background electron investigations (e.g. electron energy spectrum measurement with various vessel potentials).
More details about the electric dipole can be found in \cite{PhDWandkowsky2013, PhDHilk2016}. 

\subsubsection{UV irradiation system -- the Light Hammer}
\label{subsubsection:lighthammer}

The Rydberg background depends on the condition of the inner surfaces of the \gls{ms}.
In order to have the option to improve the surface conditions using UV light, a microwave-powered UV irradiator\footnote{Light Hammer MK6, manufactured by Heraeus Noblelight, \url{https://www.heraeus.com/de/hng/home_hng/home_noblelight.html}} was installed at port F10 of the \gls{ms} (see \figref{Figure:main-spec-vacuum}).
The interior of the \gls{ms} is illuminated by shining the UV light through a \SIadj{2}{\milli\meter}-thick sapphire window on a 63CF flange.
Background measurements, performed before and after a three-day irradiation of the \gls{ms}, did not show a significant change in the background for a baked \gls{ms}.
Further investigations will follow, but as of now the impact on the Rydberg background component is inconclusive. 

%% file: DetectorSystem.tex
\section{Detector System}
\label{Section:DetectorSystem}
\label{sec:detector_system}
\glsreset{fpd}

The \gls{fpd} system \cite{Amsbaugh2015} for the \gls{katrin} experiment consists of a multi-pixel silicon \pin -diode array, custom readout electronics, two superconducting solenoid magnets (see \secref{subsubsection:beamlinemagnets}), an ultra high-vacuum system, a high-vacuum system, calibration and monitoring devices, a scintillating veto, and a custom data-acquisition system. It is designed to detect the low-energy \betaels{} selected by the \gls{katrin} main spectrometer.  

\input{DetectorSystemPrinciple}

\input{DetectorSystemOverview}

\input{DetectorSystemFPD}

\input{DetectorSystemDAQ}

\input{DetectorSystemCalibration}

\input{DetectorSystemVeto}

\clearpage

%% file: DetectorSystemPrinciple.tex
\subsection{Detector Principle and Basic Requirements}
\label{Subsection:FPDPrincipleAndBasicRequirements}

The main objective of the \gls{fpd} system is to detect \betaels{} transmitted by the main spectrometer inside the nominal magnetic flux tube of \SI{134}{\tesla\centi\metre\squared}. This requires two superconducting magnets to adiabatically guide the \betaels{} from the spectrometer to the detector wafer. The \betaels{} are accelerated by a \gls{pae} at a typical voltage of \SI{10}{\kilo\volt} prior to reaching the wafer in order to reduce the backscattering probability and to shift the signal peak into a region of lower intrinsic background. The \betaels{} are detected with a monolithic 148 pixel \pin -diode array on a single silicon wafer. Each pixel has an area of \SI{44}{\square\milli\metre} and a design capacitance of \SI{8.2}{\pico\farad}. Since the \gls{fpd} system beam pipe couples to the main spectrometer, it must be maintained at a pressure on the order of \SI{1e-9}{\milli\bar} or below.

%% file: DetectorSystemOverview.tex
\subsection{Focal Plane Detector System Overview}
\label{Subsection:FPDSystemOverview}

\begin{figure}[ht]
    \centering
    \includegraphics[width=1.0\textwidth]{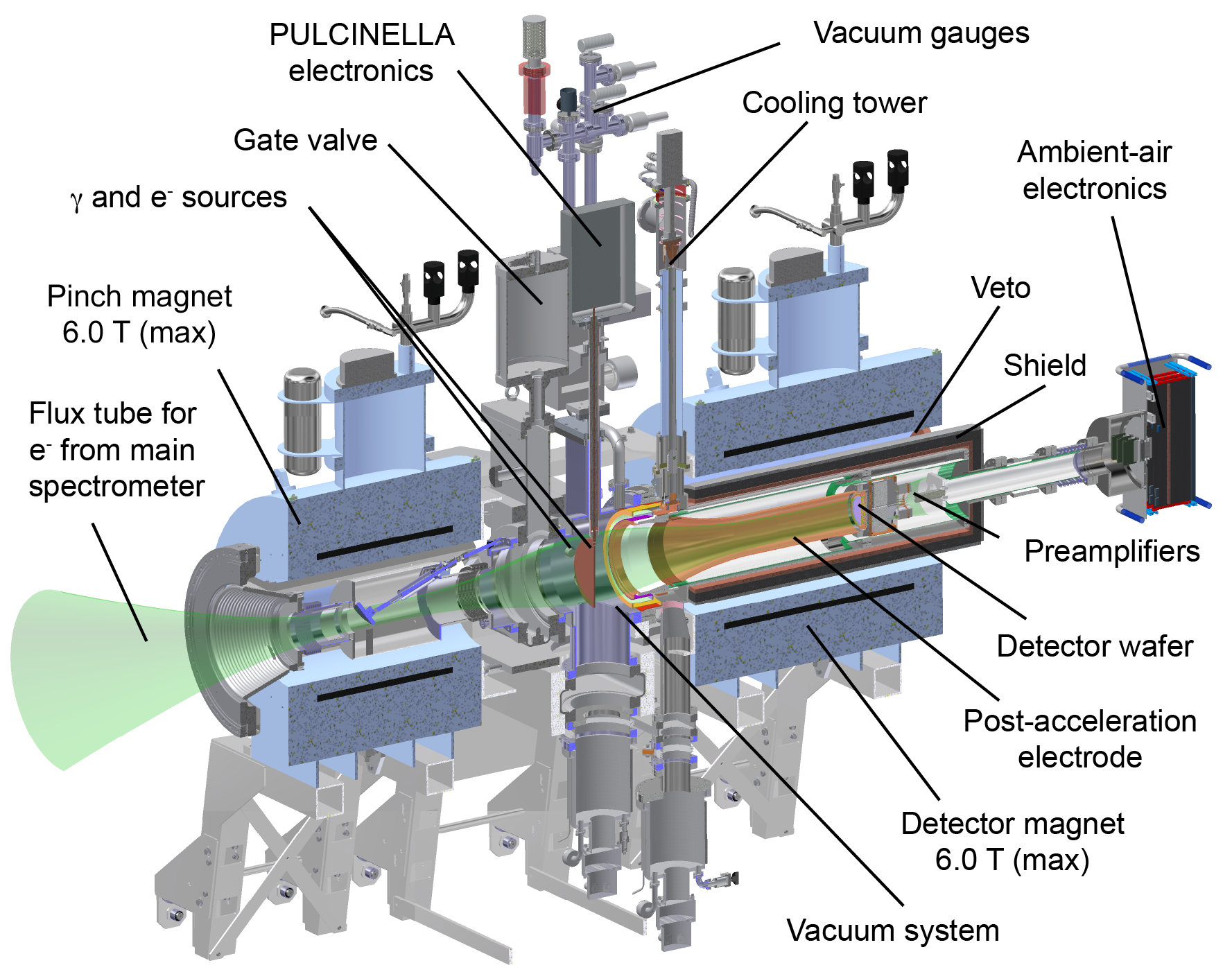}
    \caption{The primary components of the \gls{fpd} system. The main spectrometer is located to the left side of the figure. This figure is adapted from \cite{Amsbaugh2015}.}
    \label{Figure:FPDSystemOverview}
\end{figure}

An overview of the detector system is provided in \figref{Figure:FPDSystemOverview}. \betaels{} transmitted by the main spectrometer are adiabatically guided through the bore of the pinch magnet and past a gate valve that separates the main spectrometer vacuum from the \gls{fpd} system vacuum. The \betaels{} strike the multi-pixel silicon \pin -detector (see \secref{Subsection:FPD}) located inside the bore of the detector magnet. The signal of the \betaels{} is processed by custom readout electronics and DAQ system (see \secref{Subsection:FPDElectronicsAndDAQ}). An electron source and a $\gamma$-emitter, located between the two magnets, serve as calibration sources (see \secref{Subsection:FPDCalibrationAndPAE}). A shield and a veto system (see \secref{Subsection:FPDPassiveShieldAndVetoSystem}) are installed inside the bore of the detector magnet, reducing backgrounds in the detector.

%% file: DetectorSystemFPD.tex
\subsection{Focal Plane Detector}
\label{Subsection:FPD}

\begin{figure}[!ht]
    \centering
    \includegraphics[width=1.0\textwidth]{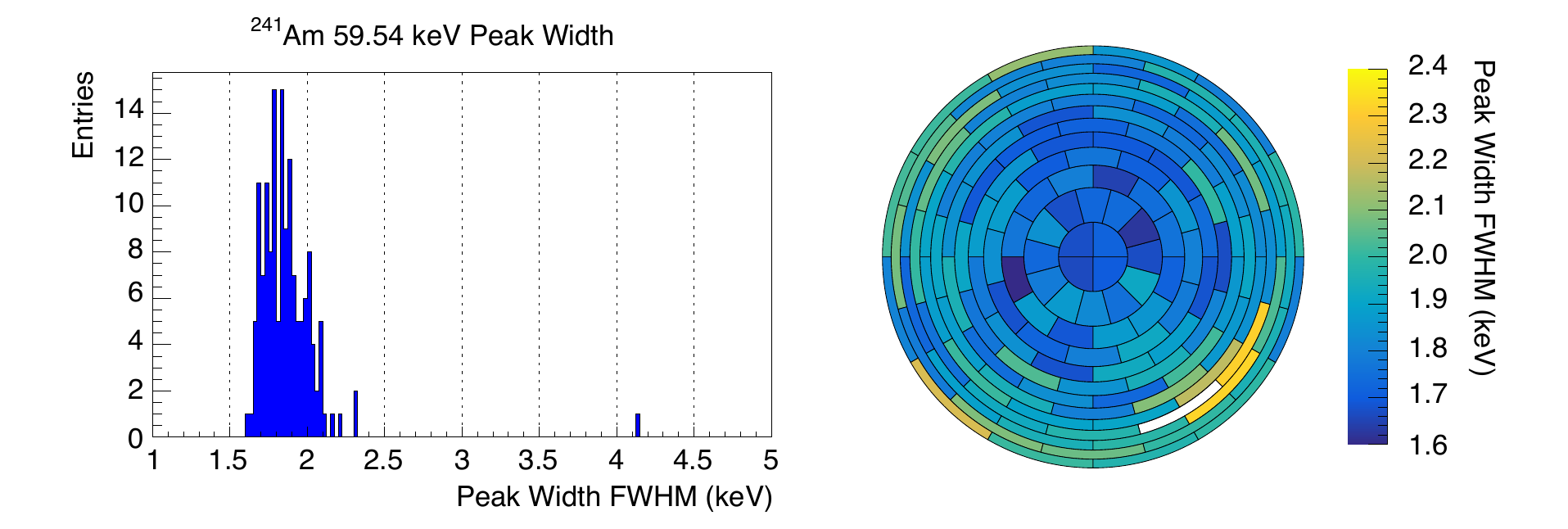}
    \caption{Energy resolution of the \gls{fpd} system with the wafer installed in July 2018. The resolution is determined from the full width at half maximum (FWHM) of the \SI{59.54}{\kilo\electronvolt} line of the \am241{} calibration source. The mean value of the distribution is \SI{1.85}{\kilo\electronvolt}, the sigma value is \SI{0.13}{\kilo\electronvolt}. One of the pixels has a much worse energy resolution of about \SI{4.1}{\kilo\electronvolt} compared to the other pixels. This pixel, marked in white in the right plot, is typically excluded for analysis.}
    \label{Figure:FPDEnergyResolution}
\end{figure}

The central piece of the detector system is a monolithic 148-pixel \pin -diode array on a single silicon wafer with a thickness of \SI{503}{\micro\metre}, a diameter of \SI{125}{\milli\metre}, and a specified dead layer thickness of \SI{100}{\nano\metre}. It has a sensitive area of \SI{90}{\milli\metre} in diameter which is segmented in pixels of equal area as shown in \figref{Figure:FPDEnergyResolution}. The wafer is operated at a bias voltage of \SI{120}{\volt} and is mounted on a feedthrough flange \cite{VanDevender2012}. The electrical connection with each pixel is made by a spring-loaded Interconnect Devices pogo pin.

Three different wafers were installed at the detector system over the course of the \gls{katrin} commissioning measurements. Two pixels of the initially installed wafer could not be used for data taking due to an electrical short between them. This wafer was replaced in May 2014. The replacement wafer was damaged presumably by ion bombardment during a discharge of the Penning trap located between the \gls{ps} and \gls{ms} (see section \ref{subsubsection:penningwiper}). The damaged wafer was replaced by another wafer in July 2018. The performance of this wafer with regard to energy resolution is shown in \figref{Figure:FPDEnergyResolution}.
Typically not all pixels are used for the neutrino mass measurement campaigns. The exact pixel selection varies between measurement campaigns and depends on pixel characteristics such as noise and alignment between \gls{wgts} and \gls{fpd}.

%% file: DetectorSystemDAQ.tex

\subsection{Readout Electronics and Data Acquisition}
\label{Subsection:FPDElectronicsAndDAQ}

\begin{figure}
\includegraphics[width=\textwidth]{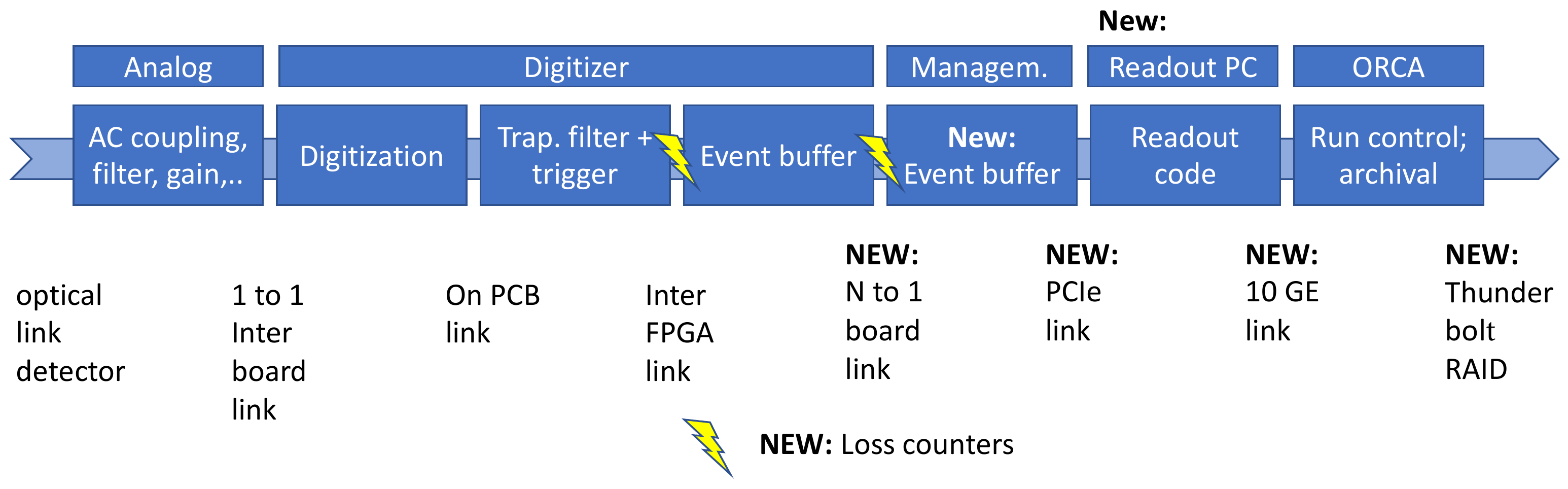}
\caption{Improved signal chain for DAQ rates of several \si{\mega cps}. From left to right the hardware components (top) are listed with their functional blocks (middle). At the bottom is the description of the link connecting the blocks.}
\label{fig:daqChain}
\end{figure}

The \gls{katrin} DAQ system needs to cover a wide range of event rates, from single-channel rates of order \SI{1}{cps} to \SI{100}{\kilo cps}, and order \SI{10}{cps} to above \SI{1}{\mega cps} total rate over the whole detector \cite{KATRIN2005}. To meet these requirements, a modular, programmable custom DAQ system for multi-channel systems up to nearly 500 channels has been developed.
The basic design has been used for the commissioning of the \gls{katrin} \gls{fpd} system and is described in \cite{Amsbaugh2015}. Since then, its performance for the highest rates has been optimized. The signal chain consists of analog filters, digitization, digital signal processing, readout by a Linux PC and management, monitoring, and control with the \gls{orca} software \cite{Howe2004} . 
The development of the \gls{katrin} DAQ system has benefited from the electronics being programmable. Recently for the highest event rates of about \SI{100}{\kilo cps} per channel, the DAQ layout has been redesigned once again. The embedded Linux readout PC has been replaced by a PCI to PCI Express (PCIe) bridge and an external readout system.
The modernized PCIe link is able to transfer up to \SI{250}{\mega\byte\per\second} and thus exceeds the maximal data rate of the backplane in the order of \SI{160}{\mega\byte\per\second}. In order to utilize the improved link, the data from the digitizer cards are transferred to the central management card and shipped in larger blocks of up to \SI{32}{\kilo\byte} of event data.
For this purpose, we were able to make use of an unused peer-to-peer connection between the digitizer cards and the central management card to collect event data.
Only the transfer of large enough blocks ensures a reasonable data transfer rate. In event-by-event readout mode, a maximal data rate of \SI{125}{\mega\byte\per\second} is achieved with the new readout concept.
With an event size of 24 Bytes in event-by-event mode, a total event rate of about \SI{5}{\mega cps} can be reached. In order to detect an overflow of the event buffers in the DAQ electronics at the highest rates, event loss counters have been added (see \figref{fig:daqChain}).

Finally, the link between the readout PC and \gls{orca} has been upgraded to \SI{10}{\giga\bit} Ethernet. The complete readout software has been designed in ways that no bottlenecks occur during the whole readout chain. The readout rate is purely defined by the performance of the DAQ electronics and thus the whole DAQ system runs very stably.

\subsubsection{Digital filters}

\begin{figure}
\begin{minipage}{0.33\textwidth}
\includegraphics[width=\textwidth]{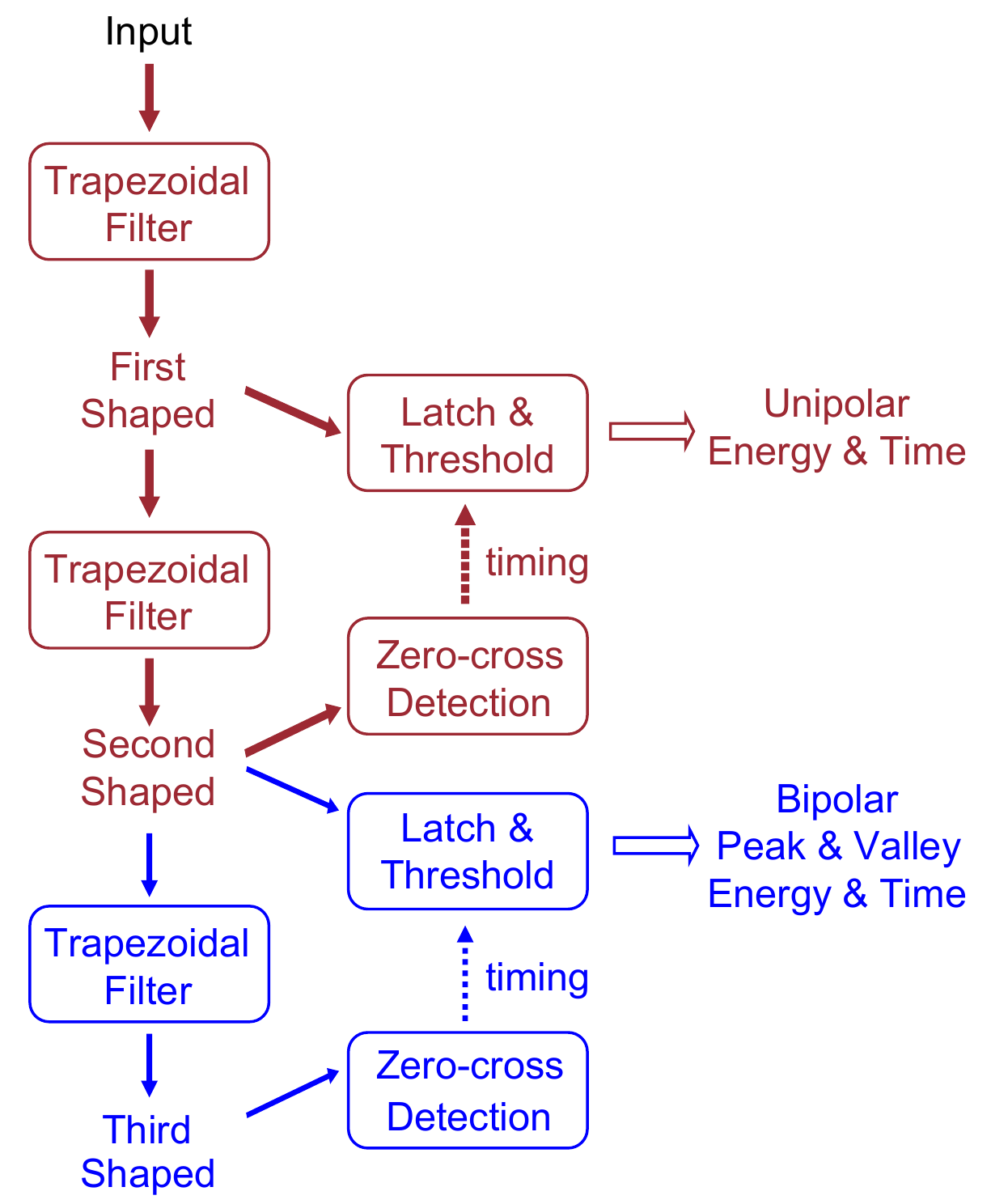} 
\end{minipage}
\begin{minipage}{0.33\textwidth}
\includegraphics[width=\textwidth]{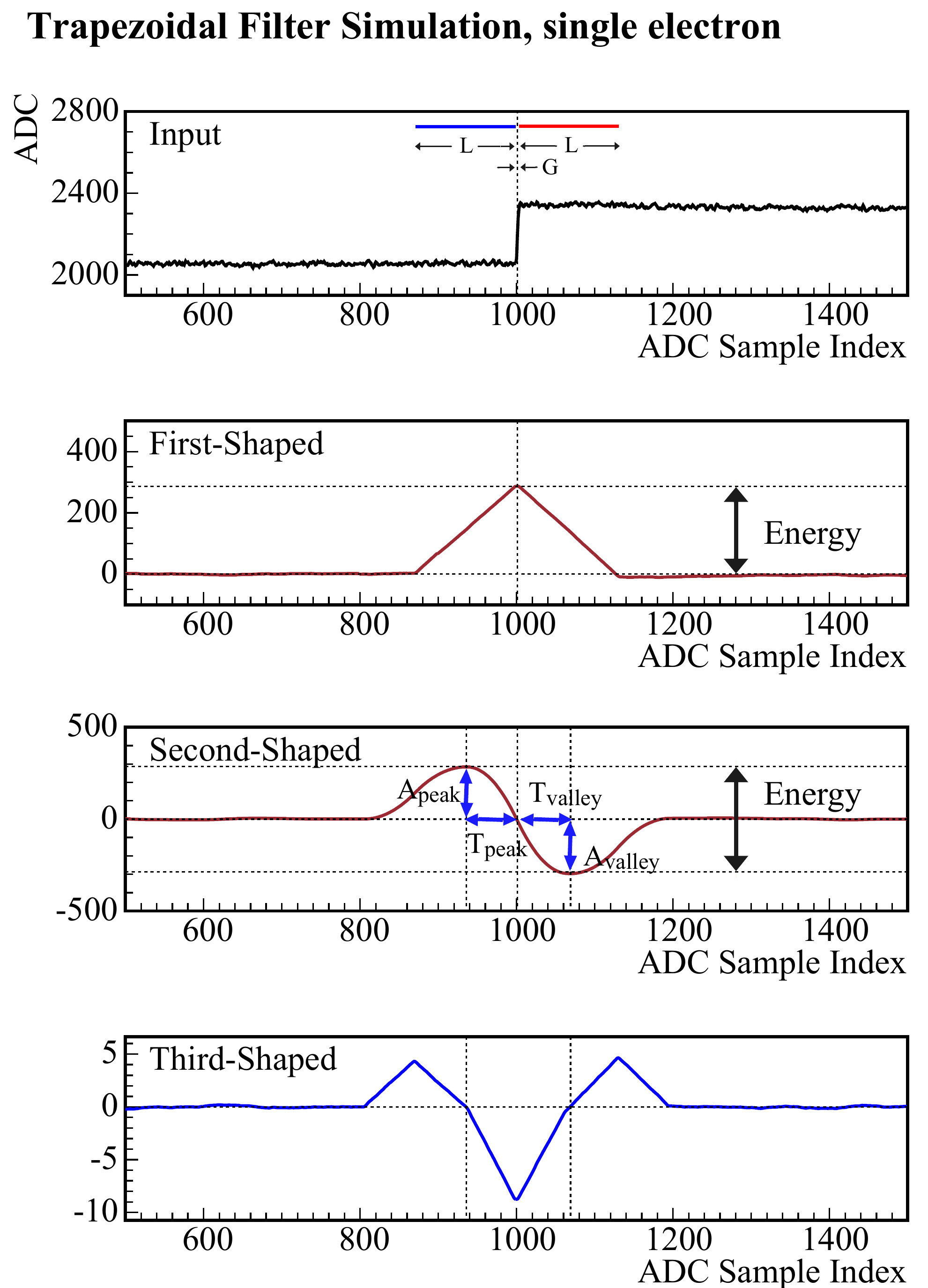} 
\end{minipage}
\begin{minipage}{0.33\textwidth}
\includegraphics[width=\textwidth]{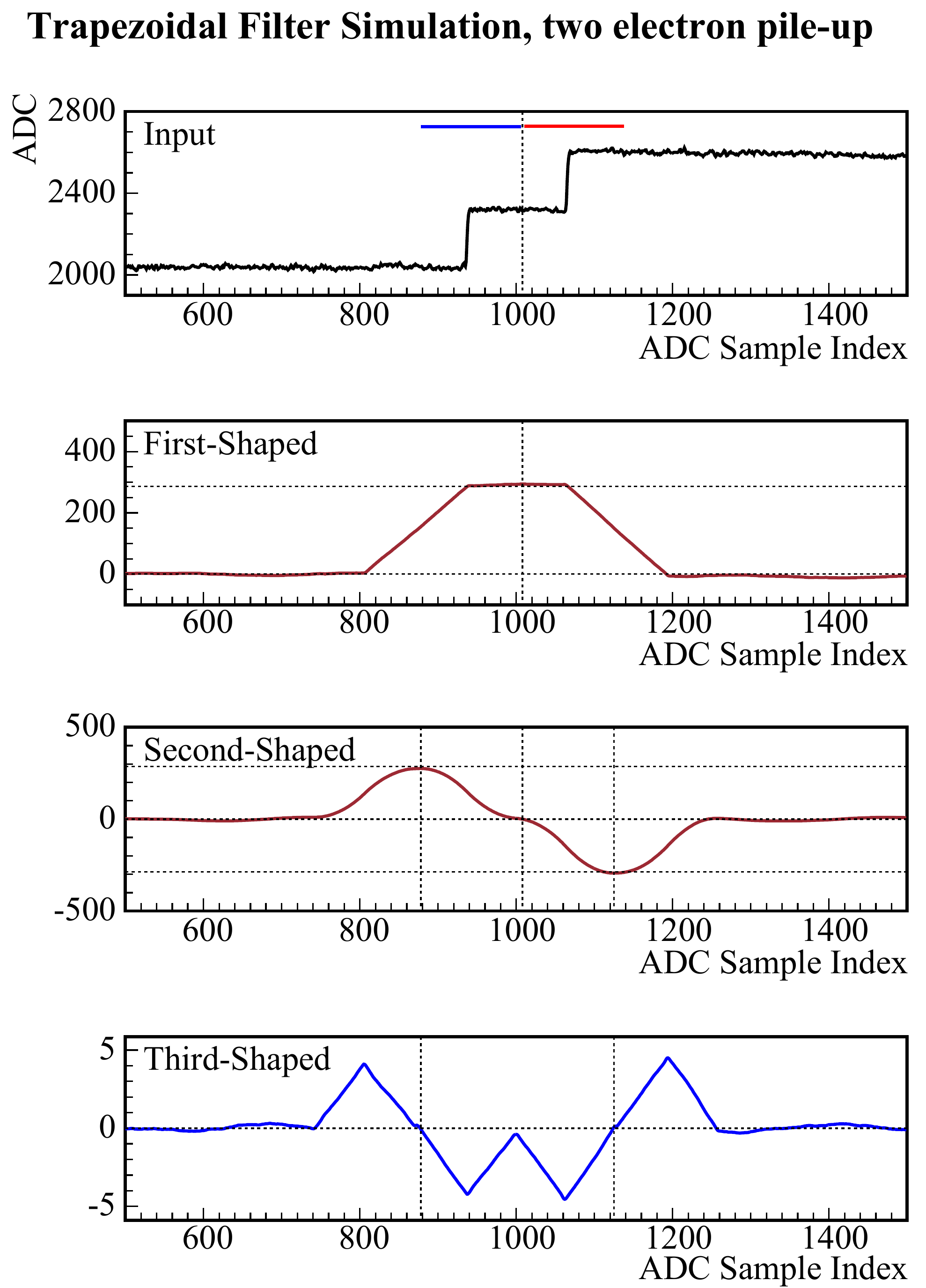}
\end{minipage}
\caption{Outline of the digital filters (left) and simulation of the filter behavior, for a single-electron event (middle) and two-electron pile-up event (right). In the outline, the components shown in blue indicate the extension to the filter to accommodate higher data taking rates. The filter latencies are subtracted in the simulation traces for illustration purposes.}
\label{fig:daqFilter}
\end{figure}

Charge deposited in the wafer pixels is first processed by a charge-integrating preamplifier. The output of the preamplifier to a typical incident electron with \SI{28}{\kilo\electronvolt} energy has a step-like shape with a rise-time of \SI{\sim 100}{\nano\second} and a long discharging time of \SI{1}{\milli\second}. The trace is digitized at \SI{20}{\mega\Hz} and then processed by a chain of trapezoidal filters \cite{Jor94a, Jor94b} for triggering and energy estimation (\figref{fig:daqFilter}).

The trapezoidal filter is a digital filter that converts the step shape input into a trapezoidal shape output. The FPGA implements the filter as an infinite impulse response filter defined as
\begin{align}
  y_i = y_{i-1} + (x_i - x_{i-L}) - (x_{i-(L+G)} - x_{i-(2L+G)})
\end{align}
where $x_i$ and $y_i$ are the $i$-th sample values of the input and output, respectively, and $L$ and $G$ are adjustable parameters called the shaping length and gap length. 
The filter can be equivalently described by a finite impulse response filter, providing a better picture for these parameters:
\begin{align}
  y_i = \sum_{j=1}^{L} x_{i+j-L} - \sum_{j=1}^{L} x_{i+j-(2L+G)}
\end{align}
here the filter consists of two integration windows of length $L$ separated by $G$ (in \figref{fig:daqFilter} indicated by the red and blue bars), and the output is the difference between the two integrations.
For a step-like input, the height of the output trapezoid scales to the height of the input step, providing an energy estimation, if the rise-time is fully contained within the gap length. For a time-scale longer than the shaping length, the filter can be viewed as a differentiator, to be used for peak detection.

The system was originally designed with two stages of the trapezoidal filters, where the first filter provides energy estimation and the second filter determines the timing. The shaping length and the gap length are programmable, and for the first filter, the shaping length is adjusted to optimize the energy resolution of the wafers, typically ranging between \SI{1.6}{\micro\second} and \SI{6.4}{\micro\second}, and the gap length is chosen at \SI{200}{\nano\second} to fully contain the rise time. The second filter reads the output of the first filter and detects the peak position by the positive-to-negative zero-crossing of its output. The shaping length of the second filter is set to be half the length of the first filter, in order to make it work as a differentiator for this time scale. The gap length of the second filter is set to zero.

On every zero-crossing, after adjusting the filter latency, the output of the first filter is compared with a programmable threshold, and if it is above the threshold, a trigger is issued. On every trigger, an event data packet is created with the height of the first filter output as its energy and the time of the second filter zero-crossing as its time.

This two stage system provides optimal energy resolution at low rates. At high rates, where the time between events becomes shorter than the preamplifier discharging time constant, the discharging tail from a previous electron affects the energy estimation of the following electron (tail pile-up). If two electrons arrive within the shaping length, only one trigger will be issued for two incident electrons (peak pile-up), causing inaccurate electron counting; in particular, if two electrons with the same energy arrive in the same interval as the shaping length, the generated events will just look like a single electron event of the one-electron energy, instead of sum of the two electrons, providing no way to discriminate between single and double occupancy events (\figref{fig:daqFilter}, middle and right). More generally, the recorded energy of two-electron peak pile-up events vary depending on the time between the two electrons.

In order to mitigate the tail pile-up effects and to detect the peak pile-up events, while still keeping the optimal energy resolution of the two-stage system, an extension was made to extract more information from the shapes. The output of the second filter, originally used for triggering via zero-cross detection, has a peak and valley associated with each incident step, where the amplitudes are proportional to the input step height, providing another energy estimation. Unlike the trapezoidal shape from the first filter, this peak-and-valley shape is bipolar and therefore the tail pile-up effects are largely suppressed (mathematically, the first order of the exponential discharging curve is eliminated by the differentiation), at the cost of increased noise. The time from the peak to the valley has information about the rise-time within the shaping window, providing a way to discriminate peak pile-up events. 

To record these information in the output of the second filter, another trapezoidal filter, with a shaping length \SI{50}{\nano\second} and zero gap length was appended to the two stage chain. On zero-crossing of the third filter, the amplitudes and timestamps of the peak and valley, $A_{\mathrm{peak}}$, $T_{\mathrm{peak}}$, $A_{\mathrm{valley}}$, and $T_{\mathrm{valley}}$, are recorded for each trigger, in addition to the (unipolar-shaping) energy and time-of-event from the two-stage part. Simulation studies predict that this upgrade will extend our rate limit from \SIrange{\sim 1}{\sim 100}{\kilo cps \per pixel} for \SI{28.6}{\kilo\electronvolt} mono-energetic electrons from the e-gun operating in continuous emission mode. The upgrade has been successfully used for the e-gun operating in pulse mode with varying multiplicities per pulse and varying spread in arrival times; this resulted in challenging peak pile-up situations at \SI{\sim 10}{\kilo cps \per pixel}, but use of an algorithm, utilizing the bipolar shaper information, accurate counting with less than 0.1\% correction errors was achieved.



\subsubsection{Run modes and performance}

The DAQ system provides several data acquisition modes in order to cover a wide range of event rates.
Table \ref{table:DAQModes} summarizes the parameters recorded in the available run modes. Trace and energy modes are event-by-event or so-called list modes, while histogram and counting modes calculate integral parameters on-the-fly in the electronics and ship data every \SI{1}{\second}. The trace mode is limited to a total event rate of about \SI{10}{\kilo cps}. The list mode reaches the highest data rates of \SI{125}{\mega\byte\per\second} and is able to handle a total event rate up to about \SI{5}{\mega cps}. The histogram and count rate modes are only limited by the digital filter parameters. Still in these modes, the rates should be limited to about \SI{100}{\kilo cps} per channel in order to get precise estimates for the particle energies.

A large number of events within a short time interval might cause overflows in internal buffers in the DAQ chain. In order to detect and correct the recorded values, loss counters have been implemented in all data transfer stages in the electronics.

The DAQ system records all events in a round-robin fashion from all the digitizer boards. Thus, the exact time order is not guaranteed within the electronics when the events are acquired in the list mode. In order to ensure the fastest readout possible, \gls{orca} records the data as provided by the electronics. Afterward, an independent process checks data quality, sorts all events by time order, and stores the data in the \gls{katrin} ROOT format for further analysis (see section \ref{Subsection:data_analysis}).

\begin{table}[!ht]
	\caption{ Data-record characteristics for each DAQ mode. Trace and energy mode provide event-by-event data. Some of the information is not processed in the DAQ electronics but can be calculated from the data via software.}
	\begin{center}
	\begin{tabular}{lcccc}
	\hline
	\textbf{DAQ mode} 		& \textbf{Trace} 	& \textbf{Energy} 	& \textbf{Histogram} 	& \textbf{Counting} \\
	\hline
	Time stamp, sec & yes   	& yes  		& yes     		& yes \\
	Time stamp, subsec & yes 	& yes  		& no   			& no \\
	Channel map 	& yes   	& yes    	& yes       	& yes \\
	Event identifier & yes 		& yes 		& no        	& no \\
	Trigger rate 	& yes  		& yes    	& yes       	& yes \\
	ADC trace 		& yes 		& no	 	& no        	& no \\
	Energy per event & yes  	& yes 		& no      		& no \\
	Energy histogram & software 	& software  	& yes      		& no \\
	Bipolar energy & software 		& yes 		& no 			& no \\
	Pileup parameter & software 	& yes 		& no 			& no \\
	\hline
	Record size per event & \SI{4124}{\byte\per event} & \SI{24}{\byte\per event} & - & - \\
	Record size per second & -  & -	& \SI{4144}{\byte\per channel} & \SI{192}{\byte\per card} \\
	Data rate @ \SI{1}{\kilo\hertz} &  \SI{3.9}{\mega\byte\per\second} & \SI{23.4}{\kilo\byte\per\second} & \SI{599}{\kilo\byte\per\second}	& \SI{1.3}{\kilo\byte\per\second} \\
	Data rate @ \SI{1}{\mega\hertz} & - & \SI{22}{\mega\byte\per\second}  & \SI{599}{\kilo\byte\per\second} & \SI{1.3}{\kilo\byte\per\second}\\
	\hline
	Max. acq. rate 	& \SI{e4}{events \per\second} & \SI{5e6}{events \per\second}	& \multicolumn{2}{c}{limited by filter} \\
	\hline
	\end{tabular}
	\end{center}
	\label{table:DAQModes}
\end{table}

\subsubsection{Grid synchronization signal}

The \gls{katrin} experiment, which uses an integrating measurement method, must know its voltages to a precision \SI{<=60}{\milli\volt} (at \SI{18.6}{\kilo\electronvolt}). One hurdle to getting these very precise voltage values (on the order of \SI{10}{ppm}) is the voltage ripples within the mains power source. This necessitates the design of a grid synchronization box, which keeps track of the phase within the \SI{50}{\hertz} frequency of the mains power signal during a data event trigger. \cite{gridSynchLT}

The box takes the mains power as an input, and outputs a synchronization pulse at the start of each new mains power period. When used as a DAQ trigger, this output signal allows \gls{katrin} to analyze data at the same phase within the mains voltage ripple. The output signal is emitted as an optical pulse to take advantage of the same data acquisition chain as the other detector readout channels.

\begin{figure}
\includegraphics[width=\textwidth]{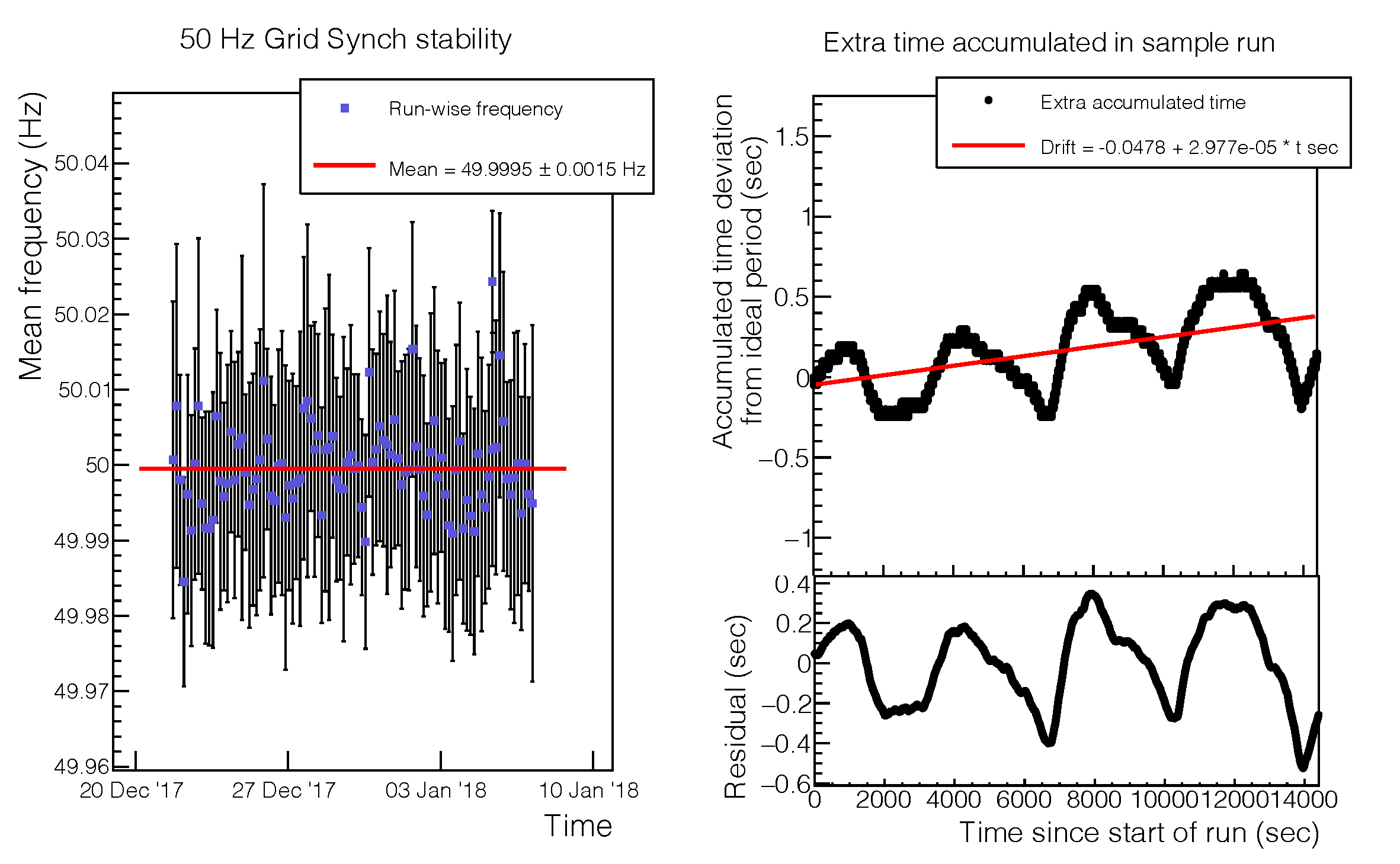}
\caption{Left: frequency stability of the mains power over the course of 100 runs, fit with a zeroth-order polynomial. Right: a more detailed view of a single sample run, showing fluctuations and drifts in mains frequency. The first-order polynomial fit's extracted slope gives an estimate of the drift at the run-level. The periodic nature is expected: it's a direct result of the power company compensating to keep the signal it delivers at \SI{50}{\hertz}.}
\label{Figure:gridSynchFreqStability}
\end{figure}

\figref{Figure:gridSynchFreqStability} demonstrates that the mains frequency remains quite close to the expected \SI{50}{\hertz} (left plot), with a fluctuation on the \SI{0.01}{\percent} level and negligible drift over long periods of time (right plot). These fluctuations and drifts are within acceptable limits for the experiment. The box's timing resolution is approximately \SI{50}{\micro\second}, and the 100 consecutive 4-hour long runs used in this analysis came from a long-term background measurement campaign between December 2017 and January 2018.

%% file: DetectorSystemCalibration.tex
\subsection{Calibration and Post-Acceleration System}
\label{Subsection:FPDCalibrationAndPAE}

Two different calibration sources are used at the detector system. An encapsulated \am241{} $\gamma$-source with an activity of about \SI{18}{\mega\becquerel} can be moved into the detector line of sight without breaking the vacuum. This source provides an absolute energy-scale calibration independent of dead layer effects.

A disk made of titanium allows the creation of electrons via the photoelectric effect by illuminating the disk with ultraviolet light from an array of six UV LEDs (wavelength \SI{255}{\nano\metre}). The photoelectrons are adiabatically guided by the magnetic field of the detector system to the wafer and their energy can be adjusted from \SIrange{0}{25}{\kilo\electronvolt} by applying a negative voltage to the titanium disk. The disk can be moved into the flux tube without breaking the vacuum.

The \gls{pulcinella} \cite{Amsbaugh2015} is connected to the titanium disk and is designed to measure picoamp-scale currents with \SI{3}{\percent} accuracy. \gls{pulcinella} can be used to determine the absolute efficiency of the detector by comparing the measured photocurrent leaving the disk with the recorded event rate. \gls{pulcinella} was used as a Faraday cup for measuring electron or ion currents from different sources further upstream in the \gls{katrin} beamline during the \gls{katrin} commissioning measurements.

The detector wafer and electronics are mounted at the downstream end of a trumpet-shaped, copper \gls{pae}. The \gls{pae}  is operated at a voltage of \SI{10}{\kilo\volt} and accelerates the \betaels{} in order to reduce the backscattering probability and to shift the signal peak into a region of lower intrinsic background. The \gls{pae} separates the vacuum of the \gls{katrin} beamline from an intermediate vacuum system which houses the detector electronics.

%% file: DetectorSystemVeto.tex
\subsection{Passive Shield and Veto System}
\label{Subsection:FPDPassiveShieldAndVetoSystem}

The \gls{fpd} system is equipped with a passive shield which consists of two nested cylindrical shells: a \SI{3}{\centi\metre} thickness of lead that reduces the $\gamma$-background and an inner \SI{1.27}{\centi\metre} thickness of oxygen-free, high-conductivity copper to block lead X-rays. The passive shield is surrounded by a plastic scintillator based veto system for muon tagging. 

The veto system described in \cite{Amsbaugh2015} suffered from a poor signal-to-noise ratio in spite of active cooling. This resulted in a large data set and inconvenient complexity during normal operation. This veto system has been upgraded for higher detection efficiency, better stability, reduced data size, improved operability, and better accessibility for maintenance and common repairs.

The new veto system was designed for higher light output. The scintillating panel thickness is doubled to \SI{20}{\milli\meter}. \gls{sipm}s\footnote{Hamamatsu Photonics, S13081-050CS, \url{https://www.hamamatsu.com/us/en/index.html}} are directly mounted to the \gls{wls} fibers using adapters developed by the T2K experiment \cite{Amaudruz2012}. The \gls{sipm} are mounted in a housing box without active cooling. The temperature of each box is monitored by a RTD-100 sensor. The new system consists of eight staves and two end-cap scintillating panels\footnote{Saint-Gobain Crystals, BC-408, \url{https://www.crystals.saint-gobain.com}}. Each panel is wrapped with a high-efficiency reflector\footnote{3M, Vikuiti Enhanced Specular Reflector, \url{https://www.3m.com}}. For each panel, four \SI{1}{\milli\meter} diameter \gls{wls} fibers\footnote{Kuraray, Y-11(200)M, \url{http://kuraraypsf.jp/psf/ws.html}} are installed in U-shaped grooves. The fibers fit loosely in the grooves allowing them to be replaced in the event of damage.

Custom electronics modules were developed to supply \gls{sipm} bias voltages and to amplify \gls{sipm} signals before the \gls{flt} modules in the detector readout system. All the potentiometers are replaced with digital-to-analog converters which can be configured through a \gls{usb} interface. The \gls{usb} interface also provides access to the temperature readings and output voltage readback. A \gls{sbc} which is connected to the \gls{usb} interface and located outside the strong magnetic field region, provides monitoring and controls over \gls{http} with RESTful API together with web-browser GUI. The \gls{sbc} also implements calibration sequence logic and bias-voltage control based on the temperature reading.

The new veto system produces \SI{\sim 50}{{photoelectrons (p.e.)}\per \gls{sipm}} per cosmic-ray muon (peak value) even with the loose grooves, which is to be compared with \SIrange{3}{5}{{p.e.}\per \gls{sipm}} per muon of the old system. This high light output allows us to set the detection threshold high at around \SI{10}{{p.e.}} for a \SI{>99.7}{\percent} per-panel muon detection efficiency. The high detection efficiency provides stability against gain fluctuation, typically caused by temperature drift. At this threshold, the contamination of the muon signal by \gls{sipm} thermal events is far below the percent level, even without cooling. The coincidence logic used in the old veto is no longer necessary. In the nominal configuration, each stave panel observes \SI{\sim 20}{cps \per \gls{sipm}} muon events and \SI{20}{cps \per \gls{sipm}} environmental gamma events, where the rates depend on geometry, with negligible contribution from \gls{sipm} thermal events. The environmental gamma events can be removed in the offline analysis by requiring inter-panel coincidence.

%% file: CalibrationAndMonitoringSystem.tex
\section{Calibration And Monitoring Systems}
\label{sec:calibration_and_monitoring_systems}


\input{CalibrationAndMonitoringSystemOverview}

\input{CalibrationAndMonitoringSystemGasComposition}

\input{CalibrationAndMonitoringSystemActivity}

\input{CalibrationAndMonitoringSystemIonRetention}

\input{CalibrationAndMonitoringSystemMagneticField}

\input{CalibrationAndMonitoringSystemMonitorSpectrometer}

\input{CalibrationAndMonitoringSystemElectronSources}

\clearpage

%% file: CalibrationAndMonitoringSystemOverview.tex
\subsection{Overview}

In the \gls{katrin} experiment, numerous interplaying parameters and processes contribute to the overall performance of the experiment, which needs to operate in a reliable, stable and reproducible way over a measurement period of several years. Thus, it is paramount to (i) precisely monitor key parameters and processes that affect the determination of the effective electron antineutrino mass; and to (ii) calibrate both these key parameters and the overall system behavior. The crucial monitoring and calibration steps in the chain of processes, from the generation of \betaels{} through transporting them to the \gls{ms} for energy analysis, are discussed in detail here, highlighting the performance of the monitoring and calibration systems during the early \gls{katrin} commissioning campaigns. 

The two key parameters relevant to the generation of \betaels{} in the \gls{wgts} are the composition of the injected tritium gas and the column density. The injected tritium gas is not pure \t2{} as it contains dynamically varying impurities of DT and HT, and is monitored by \gls{lara} (\secref{SubSection:GasCompositionMonitoring}). The column density, which must be tracked because the energy of the electrons exiting the source is affected due to scattering with gas molecules inside the source, is monitored by multiple systems: by \gls{bixs} for \betaels{} moving  upstream and by the \gls{fbm} (\secref{Subsection:ActivityMonitoring}) for \betaels{} moving downstream. In tritium \betadec{}, a range of positively and negatively charged atomic and molecular ions are generated, which would adversely affect the measurement background if they were to reach the \gls{ms}. The methodologies for detecting, blocking and removing these ions are described in \secref{SubSection:IonBlocking}.

The energy analysis of the \betaels{} is performed in the \gls{ms}, where the performance and energy resolution critically depend on precise knowledge of the magnetic and electric fields inside it. The necessary input data for electromagnetic calculations are provided by two magnetic sensor networks: a spatially fixed and a mobile one, both of which are described in \secref{SubSection:SpectrometerMagneticFieldMonitoringSystem}. An ultra-stable high voltage reference and divider setup, together with \rb{}/\kr{} electron energy data from a galvanically-connected \gls{mos} (\secref{SubSection:MonitorSpectrometer}) allow a high accuracy determination of the  retarding potential of the \gls{mace} filter and a continuous check of its stability at the ppm level.

Besides these monitoring tasks, an additional important issue is the calibration of the \gls{ms} electron energy scale. The zero-point energy and the absolute energy scale are not fixed \emph{a priori}, but instead depend on a range of intrinsic and extrinsic quantities. Specifically, an \gls{egun} with well-defined electron beam characteristics (point-like source, narrow energy and tunable angular distributions) is deployed to investigate system alignment and a range of effects influencing \betael{} trajectories (\secref{sec:egun}). Two \rb{}/\kr{} sources (one condensed, one gaseous) are used for energy calibration of the \gls{katrin} system and the study of plasma effects in the \gls{wgts}, using their (quasi) mono-energetic conversion electrons (\secref{Subsubsection:CKrS} and \secref{Subsubsection:GKrS}).

%% file: CalibrationAndMonitoringSystemGasComposition.tex
\clearpage
\subsection{Gas Composition Monitoring}
\label{SubSection:GasCompositionMonitoring}

The determination of the neutrino mass is associated with the measurement of \betaels{} from the decay of a tritium atom within the \gls{wgts}. The caveat here is that tritium is predominantly present in diatomic molecular form. The tritium gas in the \gls{wgts} is expected to contain some impurities. Though these might only be minute, they can become noticeable in high-precision measurements such as \gls{katrin}. Sources of impurities include deuterium, which is always encountered in the reservoir gas, and hydrogen, which can enter the main gas stream via outgassing from the vessels' steel walls and connecting tubing, as well as other tritiated molecules, further discussed in this section. Additionally, by incorporating a permeator in the gas circulation loop, all six possible hydrogen isotopologues become present in the circulating gas. It is therefore paramount that the gas composition be monitored at all times.

\subsubsection{Measurement Principle and Requirements} 

The fluctuations of the tritium gas composition during a spectral scan will manifest themselves as non-negligible effects on the \gls{katrin} neutrino mass analysis. Accurate knowledge of the relative concentration of the tritiated isotopologues \t2{}, DT and HT is of high importance because they appear as a weight in the sum of the individual \betaspec{} of the three isotopologues, which are slightly different due to their respective \gls{fsd}. Therefore, the development of precise methods to monitor multi-component gas composition simultaneously and in real-time is crucial. 

\glsreset{lara} 

\glsfirst{lara} was selected as the monitoring method of choice, as it is a non-invasive and fast in-line measurement technique. It also provides quantitative concentration data simultaneously for all gas constituents.

Three \gls{lara}-measured quantities are required for the neutrino mass analysis and for feedback to the \gls{katrin} tritium loop operation: the hydrogen isotopologue concentrations, $c_i$; the tritium purity, $\epsilon_{\rm{T}}$ (\eqnref{eq:epsT}); and the HT/DT ratio, $\kappa$ (\eqnref{eq:kappa}). The first is predominantly utilized in \gls{katrin} loop operation control, while the latter two are required during the data analysis phase.

\subsubsection{Laser Raman spectroscopy system} 

\begin{figure}[!ht]
	\centering
		\includegraphics[width=\textwidth]{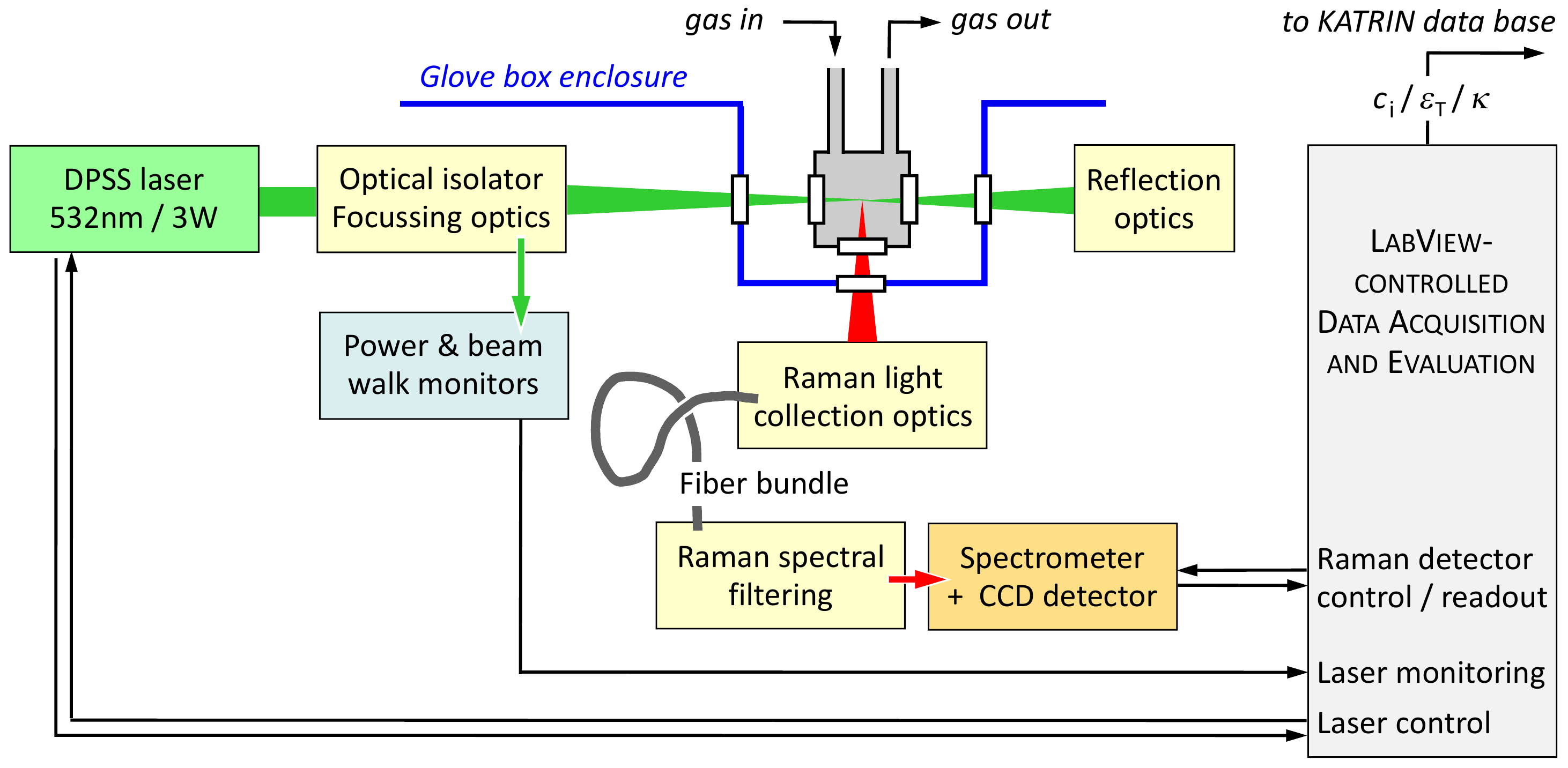}
	\caption{Concept of the \gls{lara} system for real-time, in-line monitoring of the tritium gas composition.}
	\label{fig:LARA-Figure-1_rev1c}
\end{figure}

The implementation of the \gls{lara} monitoring system is shown schematically in \figref{fig:LARA-Figure-1_rev1c}. The setup closely follows the proposal in the original \gls{tdr} \cite{KATRIN2005}. Additional details of the system can be found in \cite{Sturm2010paper}. Other technical and procedural improvements over the years since these two publications are summarized below. 

A continuous wave green ($\lambda_{\rm{L}}= \SI{532}{\nano\meter}$) laser \footnote{Laser Quantum, \emph{Finesse} DPSS Nd:YVO$_4$, \url{https://www.laserquantum.com/products/detail.cfm?id=33}}, operated at an output power of $P_{\rm{L}}= \SI{3}{\watt}$, is utilized. The laser light is guided, polarized and focussed into a Raman gas cell. In order to increase the Raman light intensity, the laser beam is back-reflected through the cell, effectively doubling the Raman response yield. Note that the back-reflected laser light is directed out of the beam axis via a Faraday isolator for laser power monitoring and for preventing potential laser beam walk during long-term operation.

The Raman gas cell itself is a tritium-compatible design, as described in \cite{Taylor2001}. The cell is in-line mounted in the Inner Loop between the two main tritium circulation buffer vessels (\figref{Figure:loopinjection}). As the cell is situated within the secondary glove box enclosure, access to laser radiation and Raman light is through anti-reflection coated windows in a bespoke appendix extension of the glove box. The Raman cell can be temporarily exchanged with a calibration cell, which incorporates a Raman intensity standard \footnote{NIST, SRM 2242, \url{https://www-s.nist.gov/m-srmors/view_detail.cfm?srm=2242}} for \emph{in situ} absolute light intensity calibration. For a detailed description, refer to \cite{Schloesser2013c,Schloesser2015}.

The Raman light from the (nearly cylindrical) excitation volume is collected at right angle to the laser excitation axis (\SI{90}{\degree} scattering geometry) and focussed onto a "slit-to-slit" optical fiber bundle. Before entering the spectrometer \footnote{Princeton Instruments, Acton HTS, \url{https://www.princetoninstruments.com/wp-content/uploads/2020/08/SpectraPro-HRS_Datasheet_07102020.pdf}} for Raman spectral analysis, the radiation is cleaned by removing any residual laser light via a long-pass filter \footnote{Semrock, \emph{RazorEdge \SI{532}{\nano\meter}}, \url{https://www.semrock.com/FilterDetails.aspx?id=LP03-532RE-25}}. The spectrometer range and resolution are fixed such that all six hydrogen isotopologues can be resolved and recorded simultaneously by the high-sensitivity CCD-array detector \footnote{Princeton Instruments, \emph{Pixis 2KB}, \url{https://www.princetoninstruments.com/wp-content/uploads/2020/04/PIXIS_2K_datasheet.pdf}}, as demonstrated in \cite{Sturm2010paper}.

System control, operation parameter recording, Raman spectrum acquisition, and real-time data treatment and analysis are integrated in a dedicated LabVIEW program \cite{James2013}. The analysis outputs for $c_i$, $\epsilon_{\rm{T}}$, and $\kappa$ are extracted from the spectra and transmitted to the \gls{katrin} slow control database.

\begin{figure}[!ht]
	\centering
		\includegraphics[width=\textwidth]{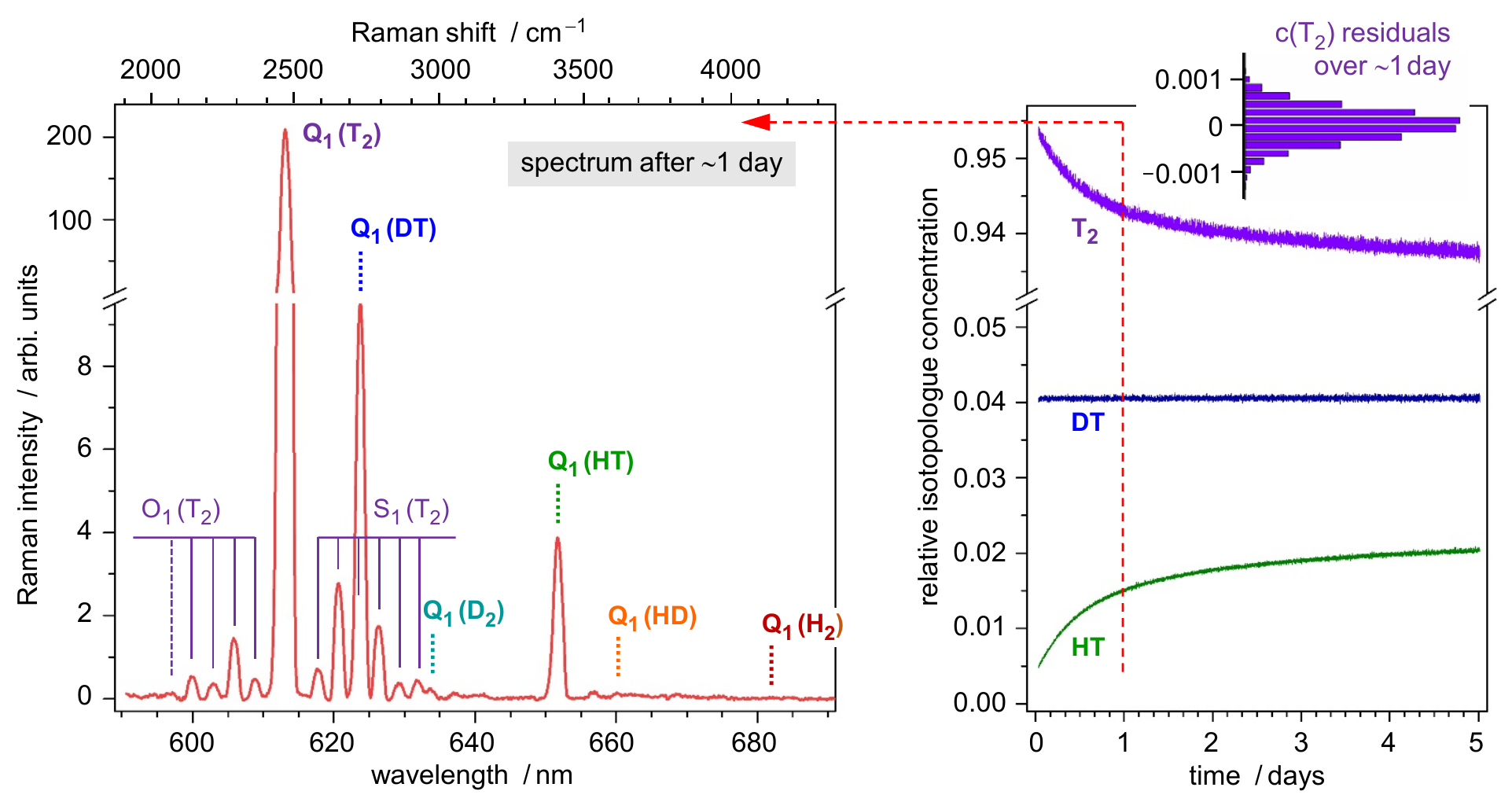}
	\caption{Representative \gls{lara} data, recorded continuously during the May/June 2013 LOOPINO test runs. \textbf{Left:} a Raman spectrum with annotated isotopologue branches. \textbf{Right:} temporal variation of the relative concentrations $c_{\rm{XT}}$ for the three tritium isotopologues \t2, DT and HT. \textbf{Inset:} distribution of the concentration residuals for \t2, over the period of $\approx1$ day, relative to the running mean. }
	\label{fig:LARA-Figure-2_rev2b}
\end{figure}

The Raman spectral data is accumulated for \SIrange{20}{30}{\second}. 
Selected sample data from a test setup ("LOOPINO") of the actual tritium loop \cite{Fischer2011} operated between 2010-2013 are shown in \figref{fig:LARA-Figure-2_rev2b}. This demonstrates the \gls{lara} system performance over extended periods of time, when operating under \gls{katrin} run conditions. An extended analysis of the \gls{lara} system's monitoring capabilities over multiple measurement periods has recently been published~\cite{lara2020}.

The concentration data shown in the figure clearly reveals the reliability and reproducibility of the \gls{lara} monitoring system, from the recorded raw data, through the customized automated analysis chain, to the concentration output. In the \figref{fig:LARA-Figure-2_rev2b} inset, a histogram of the residuals (relative to the running mean) for the \t2 concentration highlights the excellent reliability of the gas circulation system and its monitoring. The histogram data covers a period of about one day, corresponding to around 3000 individual measurements, and yields a measurement precision of $\sigma \approx 3\times10^{-4}$. This is well within the \gls{katrin} requirements of $\sigma = 1\times10^{-3}$. The concentration data $c_i$ and the derived quantities $\epsilon_{\rm{T}}$ and $\kappa$ are calculated from the Q-branch Raman intensities, measured on an individual spectra basis. Note that the relative concentrations change over time (towards equilibrium); this behavior can be attributed to expected contamination by hydrogen exchange reactions with the stainless steel vessel walls and the loop tubing.

\subsubsection{Performance of the Raman setup} 

Since its first implementation\cite{KATRIN2005}, the \gls{lara} monitoring system has been continuously tested and improved through a number of measurement campaigns. These include the \gls{tilo} test run using only non-tritium isotopologues \cite{Lewis2008}, and the LOOPINO tritium test runs in 2010 \cite{Fischer2011} and 2013 (presented here). 

The first \gls{lara} setup had provided satisfactory results as early as 2005, but improvements since then showed that the \gls{katrin} requirements could be met and in some cases exceeded. These improvements include: reducing the acquisition time by nearly two orders of magnitude, down to \SI{20}{\second}; upgrading components (excitation laser and CCD light detector); integrating a double-pass configuration, nearly doubling the Raman light intensity; implementing a fully quantitative intensity calibration \cite{Schloesser2013c}; and developing a custom data acquisition and evaluation software package \cite{James2013}.

The \gls{katrin} requirements with respect to tritium gas composition monitoring, as well as the actual performance data of the \gls{lara} system, are shown in \tabref{tab:LARA-table}.

\begin{table}[!ht]
\caption{\gls{katrin} requirements and achieved experimental realization of the tritium gas composition. The "TDR 2004" and "Revised 2015" requirements are found in \cite{KATRIN2005} and \cite{Bodine2015}, respectively. The achievements are derived from \gls{lara} data obtained between 2013 and 2015 in the LOOPINO configuration \cite{Fischer2011} and the systematic investigation of the calibration uncertainty \cite{Schloesser2015}. Precision and trueness are stated relative to their value ( $|\Delta X/X|$ ).}
\begin{center}
\label{tab:LARA-table}
	\begin{tabular}{llccccc}
		\toprule
		                     &    & \multicolumn{2}{c}{\textbf{Requirements}} & \multicolumn{3}{c}{\textbf{Achievements}}\\
			                 &    &  \textbf{TDR 2004} & \textbf{Revised 2015} & \multicolumn{3}{c}{\textbf{since 2013}} \\
			                 \midrule
		$\epsilon_{\rm{T}}$  &    & \num{>0.95} & \num{>0.95} & \multicolumn{3}{c}{\num{>0.93}} \\
		\ Precision  &  (\num{e-3}) & \num{2} & \num{1} & \multicolumn{3}{c}{\num{<0.01}} \\
		\ Trueness & (\num{e-2})   & - & \num{3} & \multicolumn{3}{c}{\num{0.18}} \\
		\hline
		$\kappa$  &    & - & \num{0.1} & \multicolumn{3}{c}{\num{<0.1}} \\
		\ Precision  & (\num{e-3}) & - & - & \multicolumn{3}{c}{\num{0.02}} \\
		\ Trueness & (\num{e-2})   & - & \num{10} & \multicolumn{3}{c}{\num{6.0}} \\
		\hline
		&&&& $\rm{T}_2$ & $\rm{DT}$ & $\rm{HT}$\\
		$c_i$ &    & - & - & \num{>0.93} & \num{\sim 0.04} & \num{<0.02} \\
		\ Precision  & (\num{e-3}) & - & - & \num{0.32} & \num{4.81} & \num{7.79} \\
		\ Trueness & (\num{e-2})   & - & - & \num{0.4} & \num{6.4} & \num{6.1} \\
		\bottomrule
	\end{tabular}
\end{center}
\end{table}

Overall, the \gls{lara} monitoring system is capable of providing the required information on $c_i$, $\epsilon_{\rm{T}}$ and $\kappa$ every \SI{30}{\second}. The tabulated example demonstrates that the requirements are met. Further improvement on trueness values are expected once new data for the theoretical Raman transition matrix elements which incorporate proper error estimates are available. 

%% file: CalibrationAndMonitoringSystemActivity.tex
\subsection{Activity Monitoring}
\label{Subsection:ActivityMonitoring}


The neutrino mass measurement depends on the accurate description of inelastic scattering of the electrons with the gas molecules inside the source. This description is strongly influenced by two key experimental parameters, column density and tritium purity.

The column density, $\rho d$, represents the number of molecules within the flux tube volume, and can be obtained by combining an \emph{in situ} measurement of the tritium purity with an activity (decay rate) measurement. The count rate of \betaels{} from the source, $S$, as measured by activity detectors scales as

\begin{equation}
    S = C \cdot \epsilon_{T} \cdot \rho d
    \label{Equation:ColumnDensity}
\end{equation}

where $C$ is a proportionality constant encompassing experimental properties such as detector efficiency and acceptance. Small fluctuations of the source parameters lead to changes in the shape of the differential \betaspec{}. Fluctuations in the column density are expected to be in the \SI{0.1}{\percent} range. Given the targeted sensitivity for the neutrino mass measurement, column density and tritium purity should not exceed an uncertainty of $\delta m^{2}_{\nu} =$ \SI{7.5e-3}{\square\electronvolt} in the neutrino mass analysis.

There are two activity monitoring systems in the \gls{katrin} experiment which measure the count rate of \betaels{} from the tritium decay in the \gls{wgts}. These monitoring systems provide information on fluctuations in the \gls{wgts} activity on a timescale of minutes, and can be used in conjunction with the measured tritium purity to monitor the column density with \SI{0.1}{\percent} precision.

One of these activity detectors, \gls{bixs}, uses X-ray detection to measure the current-induced bremsstrahlung of \betaels{} hitting the \gls{rw}. The other activity detector, the \gls{fbm}, is a moveable detector which makes measurements of the relative electron flux downstream of the transport section.

\subsubsection{BIXS monitor} 
\label{Subsubsection:BIXSmonitor}

\begin{figure}[!ht]
    \centering
    \includegraphics[width=.7\textwidth]{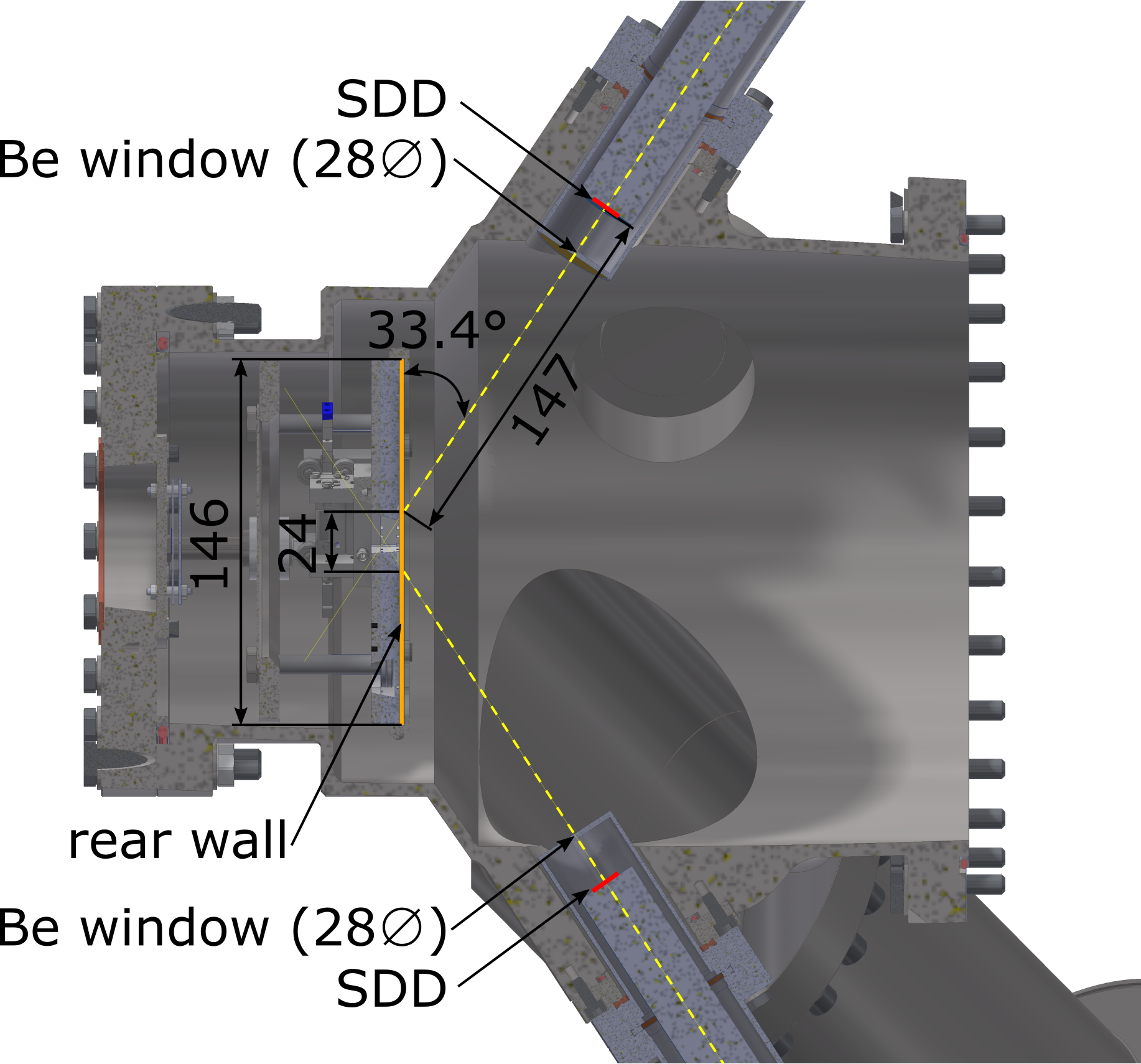}
    \caption{Sectional drawing of the \gls{rw} vessel. The \gls{rw} and the two \gls{sdd}s are highlighted. Mechanical dimensions are given in mm and degrees.} 
    \label{Figure:BIXS-RWV-CAD}
\end{figure}

The \gls{bixs} system \cite{Babutzka2012} is part of the \gls{cms}, and is used for continuous tritium activity monitoring of the \gls{wgts}. It consists of the \gls{rscm}, the \gls{rw} vessel with X-ray detectors, and the \gls{rw}. A sectional drawing of the \gls{rw} vessel with the detectors is shown in \figref{Figure:BIXS-RWV-CAD}.

The \betael{} flux coming from the \gls{wgts} is compressed by the \gls{rscm} to fit the \gls{rw} vessel and the gold-coated \gls{rw} surface. The outer diameter of the \gls{rw} is \SI{14.6}{\cm} and the magnetic field strength at the \gls{rw} position is approximately \SI{1.6}{\tesla}. Half of the produced \betaels{} inside the \gls{wgts} are emitted in the upstream direction, and are directly guided to the \gls{rw}. The other half is emitted in the downstream direction towards the spectrometer. The \betaels{} with a kinetic energy less than the retarding potential are reflected back to the \gls{rw}, either by the magnetic mirror effect or by the electrostatic potential of the spectrometers. The resulting total \betael{} flux on the \gls{rw} surface is approximately \SI{1e11}{e^-\per\second}.

The X-ray radiation, generated during the absorption process of the \betaels{} in the gold coating of the \gls{rw}, is monitored by two \glspl{sdd} \footnote{KETEK, AXAS-M, \url{https://www.ketek.net/wp-content/uploads/2017/09/KETEK-AXAS-M-Product-Information.pdf}}. Each of these detectors has an active area of \SI{92}{\milli\metre\squared}, with an energy resolution of \SI{160}{\electronvolt} (FWHM) at \SI{5.9}{\kilo\electronvolt}. A \SI{250}{\micro\metre}-thick, \SI{28}{\milli\metre}-diameter beryllium (Be) window is mounted in front of each \gls{sdd} to prevent tritium contamination. On the \gls{rw}-facing side, the Be windows are sputter-coated with \SI{100}{\nano\metre} of gold to reduce tritium adsorption. Two digital pulse processors \footnote{AMPTEK, DP5, \url{https://www.amptek.com/-/media/ametekamptek/documents/products/dp5.pdf?dmc=1&la=en&revision=030fc2ce-402d-4848-a12b-9a4f93e64af5}} are used for data acquisition. The intrinsic \gls{bixs} detector background of both \gls{sdd}s is \textless \SI{3.0}{\milli cps}, with a low-energy threshold of \SI{800}{\electronvolt}.  

\begin{figure}[!ht]
    \centering
    \includegraphics[width=.8\textwidth]{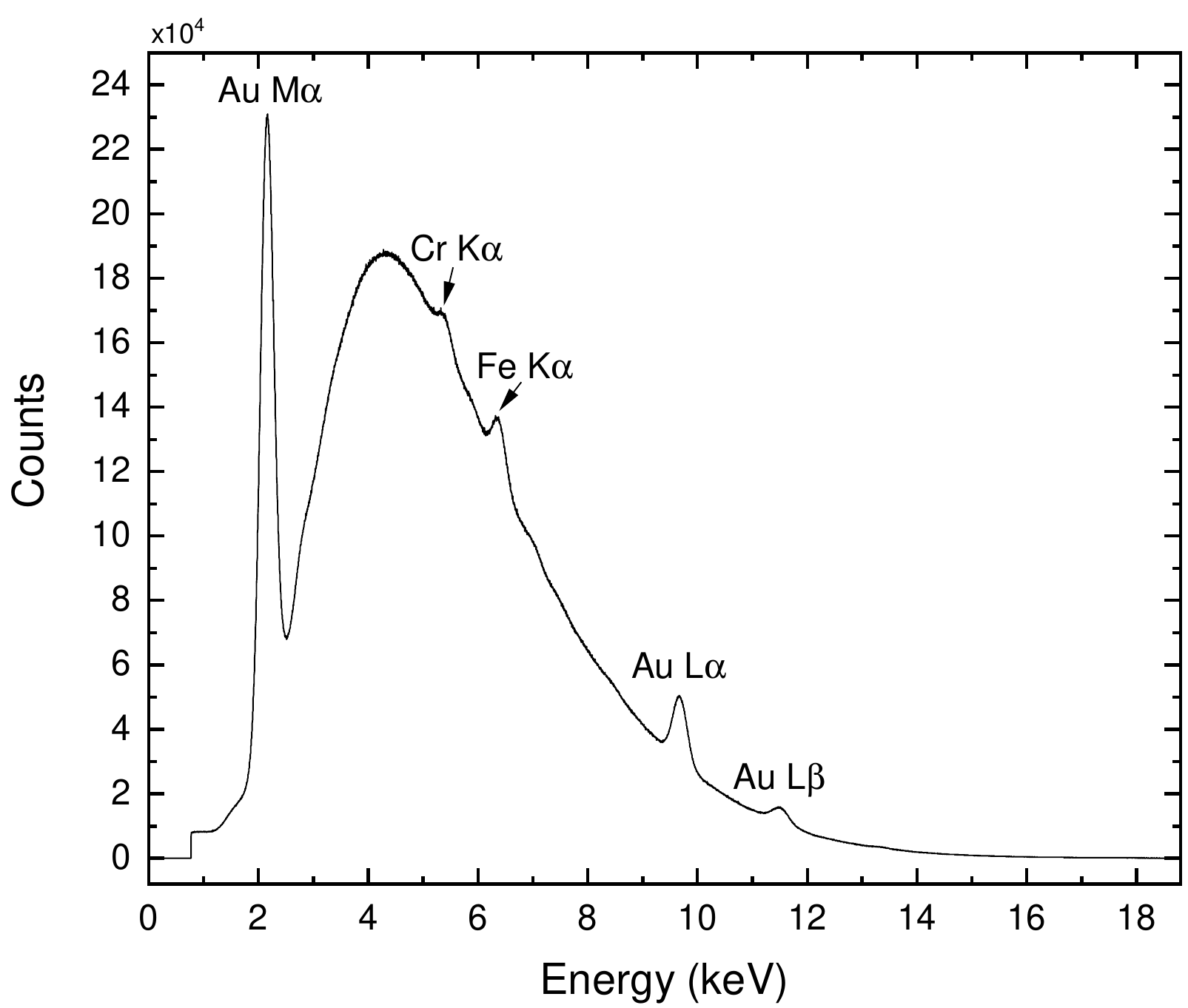}
    \caption{An example spectrum of the \gls{bixs} system. The continuous bremsstrahlung background is superimposed by fluorescence lines. Detectable fluorescence radiation is emitted by the gold coating on the \gls{rw} and the dominant constituents of the stainless steel system, iron and chromium. For details about the notation and transition energies, see \cite{XRD09}. }
    \label{Figure:BIXS-01P}
\end{figure}

In a recent commissioning experiment \cite{Rollig2013,Bornschein2017b}, the \gls{bixs} system demonstrated the ability to monitor the activity of a gaseous tritium source at the \SI{0.1}{\percent} level. During a tritium measurement campaign in March 2019, the \gls{bixs} system operated for the first time. A \gls{bixs} spectrum is shown in \figref{Figure:BIXS-01P}. The prominent fluorescence lines can be used for \emph{in situ} energy calibration. For stability monitoring, the integral count rate is used. The combined count rate of both \gls{bixs} detector systems at full column density is approximately \SI{11.3}{\kilo cps}. This allows stability monitoring on the \SI{0.1}{\percent} level in measurement cycles of duration \SI{100}{\second}. 

\subsubsection{Forward Beam Monitor} 
\label{Subsubsection:ForwardBeamMonitor}

The \gls{fbm} \cite{PhDEllinger2019,beglarian2021forward,Ellinger2017,Babutzka2012,PhDSchmitt2008} is used for monitoring the relative intensity of the electron flux. It is located in the transport section, mounted between the last two superconducting solenoids in \gls{pp2} of the \gls{cps} (\secref{Subsection:CryogenicPumpingSystem}). At this location inside the flux tube at the \gls{cps}, the \betael{} flux density is approximately \SI{e6}{\per\second\per\milli\metre\squared}, and the magnetic field is approximately \SI{0.84}{\tesla} (at center).

\begin{figure}[!ht]
    \centering
    \includegraphics[width=0.8\textwidth]{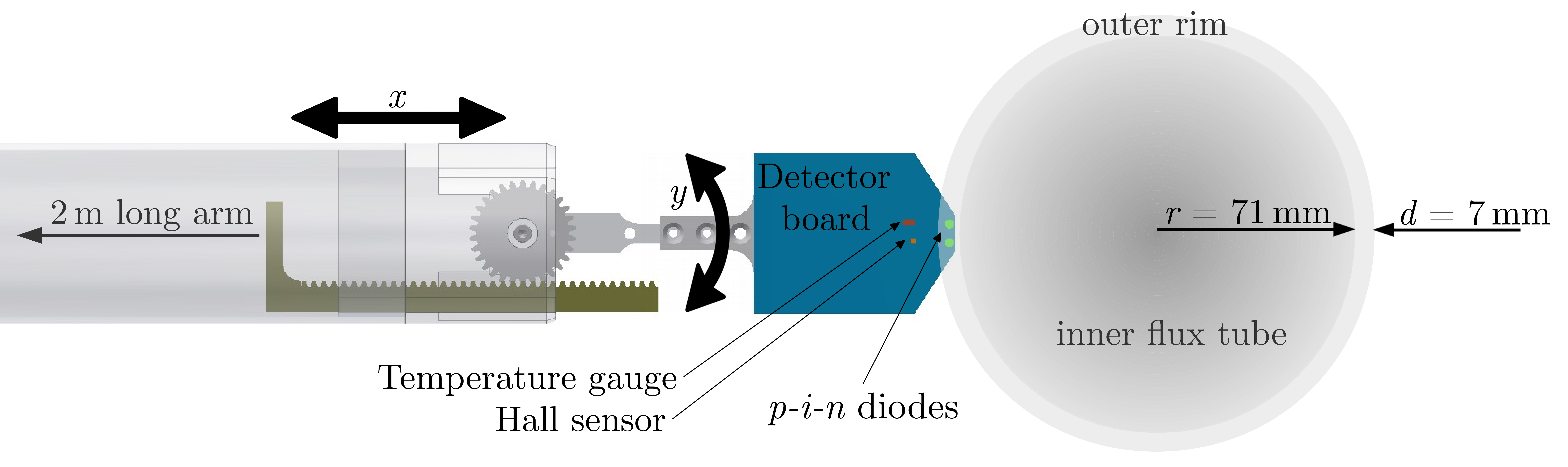}
    \caption{The \gls{fbm} front end. When the \gls{fbm} is in standard monitoring position during neutrino mass measurements, it is located in the outer rim of the flux tube where the source activity is continuously monitored without shadowing the \gls{fpd}. }
    \label{Figure:FBM}
\end{figure}

A vacuum manipulator with a \SI{2}{\metre}-long bellow enables the \gls{fbm} detector board to be inserted directly into the flux tube through \gls{pp2} of the \gls{cps}. Two independent motion systems drive the \gls{fbm}: the first system moves the detector-mounted end ("front end") linearly along the (horizontal) $x$-direction and the second system performs rotational movement. The \gls{fbm} can be moved throughout the cross section of the flux tube with a precision of better than \SI{0.1}{\milli\metre} relative to its starting position. The \gls{fbm} front end mechanics and the monitoring position are illustrated in \figref{Figure:FBM}.

The \gls{fbm} detector board is equipped with two silicon \pin{} diodes: one of type S9055 and size \SI{0.031}{\milli \meter \squared} \footnote{Hamamatsu, S9055, \url{https://www.hamamatsu.com/resources/pdf/ssd/s9055_series_kpin1065e.pdf}} for Channel 1, and one of type S5973 and size \SI{0.12}{\milli \meter \squared} \footnote{Hamamatsu, S5973, \url{https://www.hamamatsu.com/resources/pdf/ssd/s5971_etc_kpin1025e.pdf}}  for Channel 2 \cite{PhDEllinger2019}.
Both these diode channels are read out by two data acquisition processors \footnote{AMPTEK, PX5, \url{https://www.amptek.com/-/media/ametekamptek/documents/products/px5.pdf?dmc=1&la=en&revision=531ccb5b-9802-455e-8969-b5c88436e2b5}} \footnote{AMPTEK, DP5, \url{https://www.amptek.com/-/media/ametekamptek/documents/products/dp5.pdf?dmc=1&la=en&revision=030fc2ce-402d-4848-a12b-9a4f93e64af5}}. The \gls{fbm} can measure the \betael{} flux with a precision of \SI{0.1}{\percent} in less than \SI{60}{\second}. The detector board also includes a temperature gauge and a Hall sensor for additional monitoring of the flux tube properties. These measurement devices are read out continuously during neutrino mass measurements. During dedicated commissioning and calibration runs, the \gls{fbm} can map out the entire cross section of the flux tube.

\begin{figure}[!ht]
    \centering
    \includegraphics[width=0.7\textwidth]{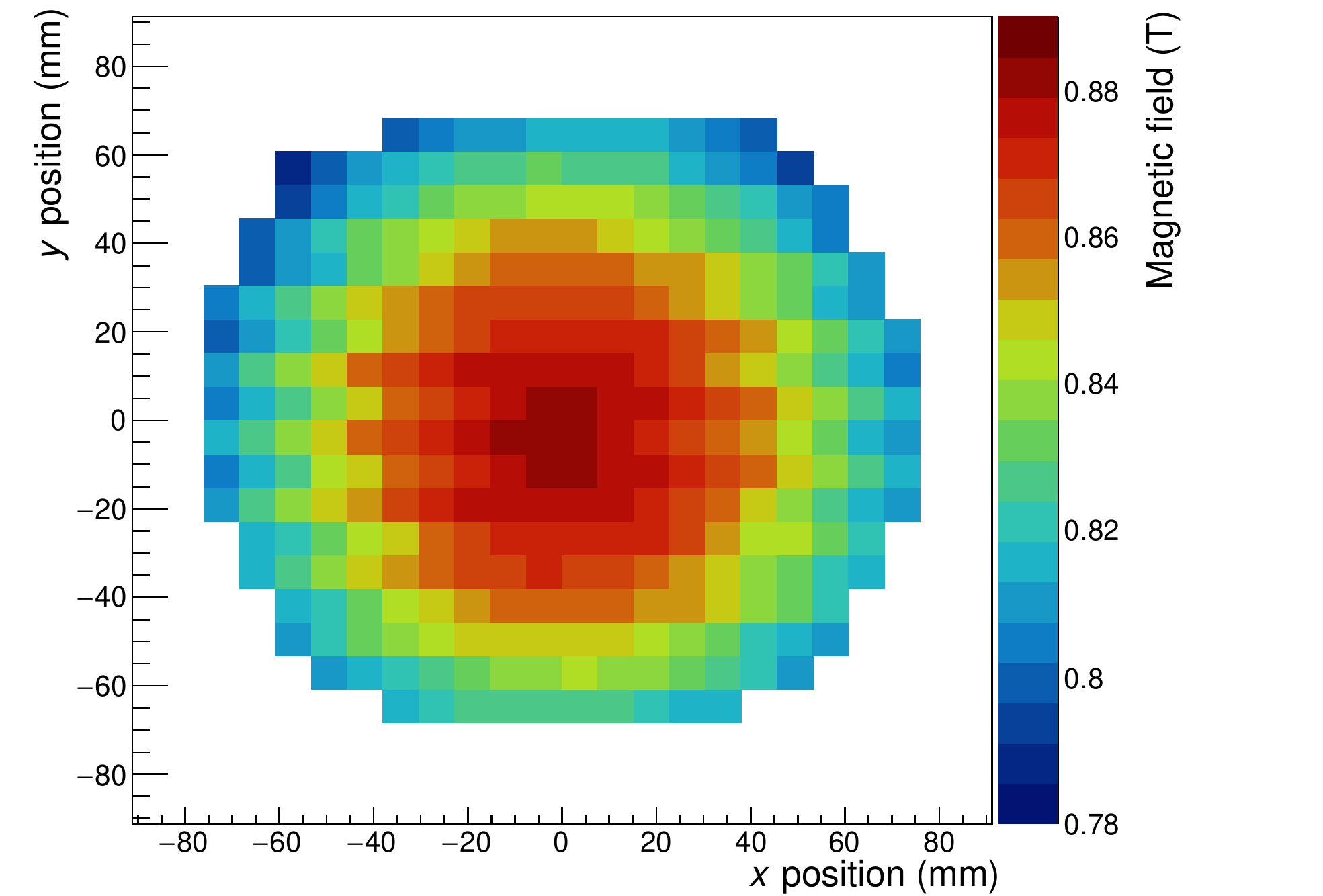}
    \caption{Cross section of the magnetic flux tube in \gls{cps} \gls{pp2}, as measured by the Hall sensor on the \gls{fbm} detector board. }
    \label{Figure:FBMmagfield}
\end{figure}

One such commissioning scan of the magnetic flux tube was performed in August 2018. This scan comprises \num{304} measurement locations within the flux tube in a grid pattern with \SI{7.6}{\milli\metre} spacing. Magnetic field and temperature data were collected for \SI{10}{\second} at each measurement point. The resulting magnetic field strength, corrected by temperature dependency, is shown in \figref{Figure:FBMmagfield}. This measured magnetic field map agrees with results from simulations, and confirms that the \betaels{} coming from the tritium source follow the magnetic flux tube, as expected. 

\begin{figure}[!ht]
    \centering
    \subfloat[\label{Subfigure:FBMcalibration}]
    {\includegraphics[width=0.49\textwidth]{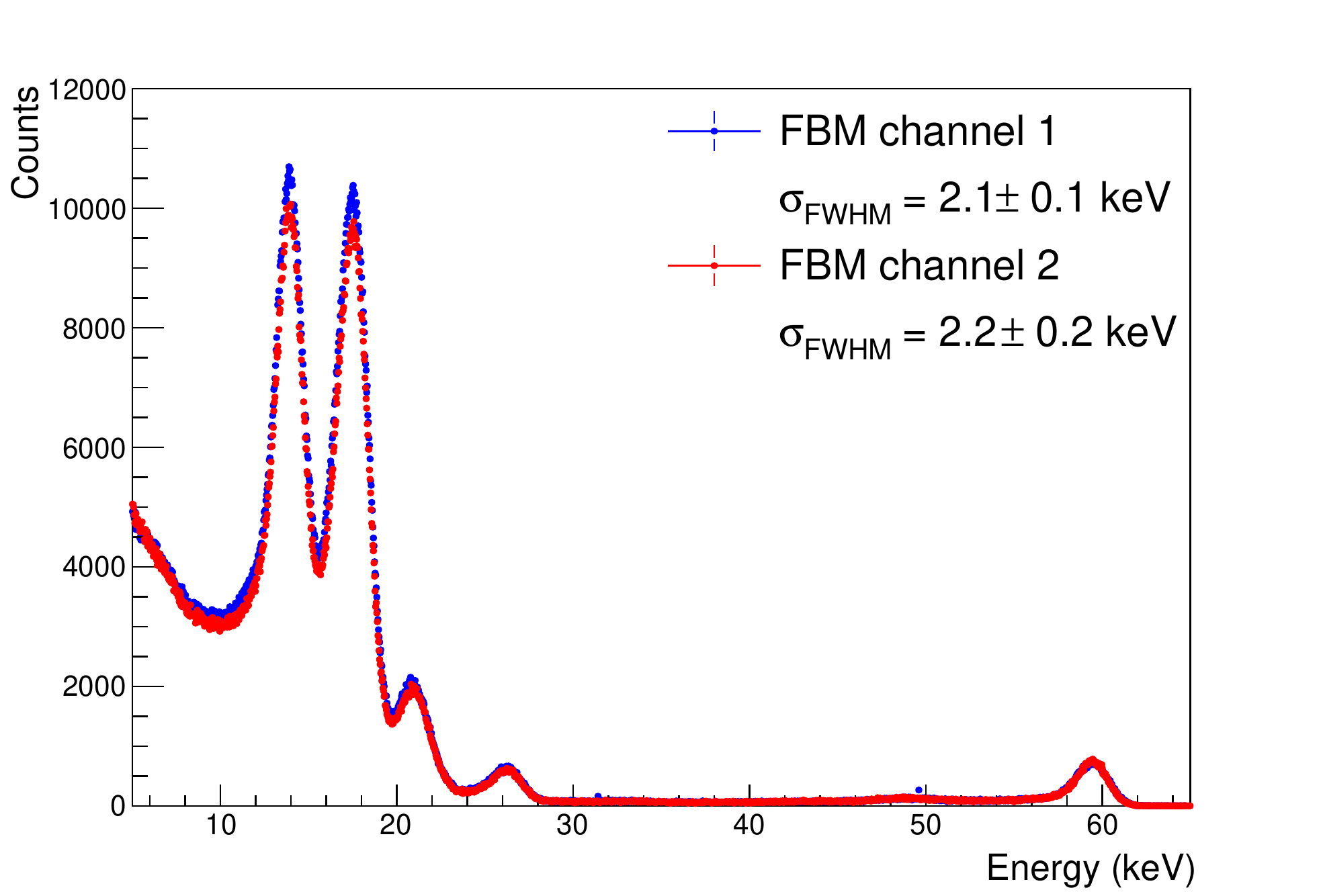}}
    \hspace{0.2cm}
    \subfloat[\label{Subfigure:FBMspectrum}]
    {\includegraphics[width=0.49\textwidth]{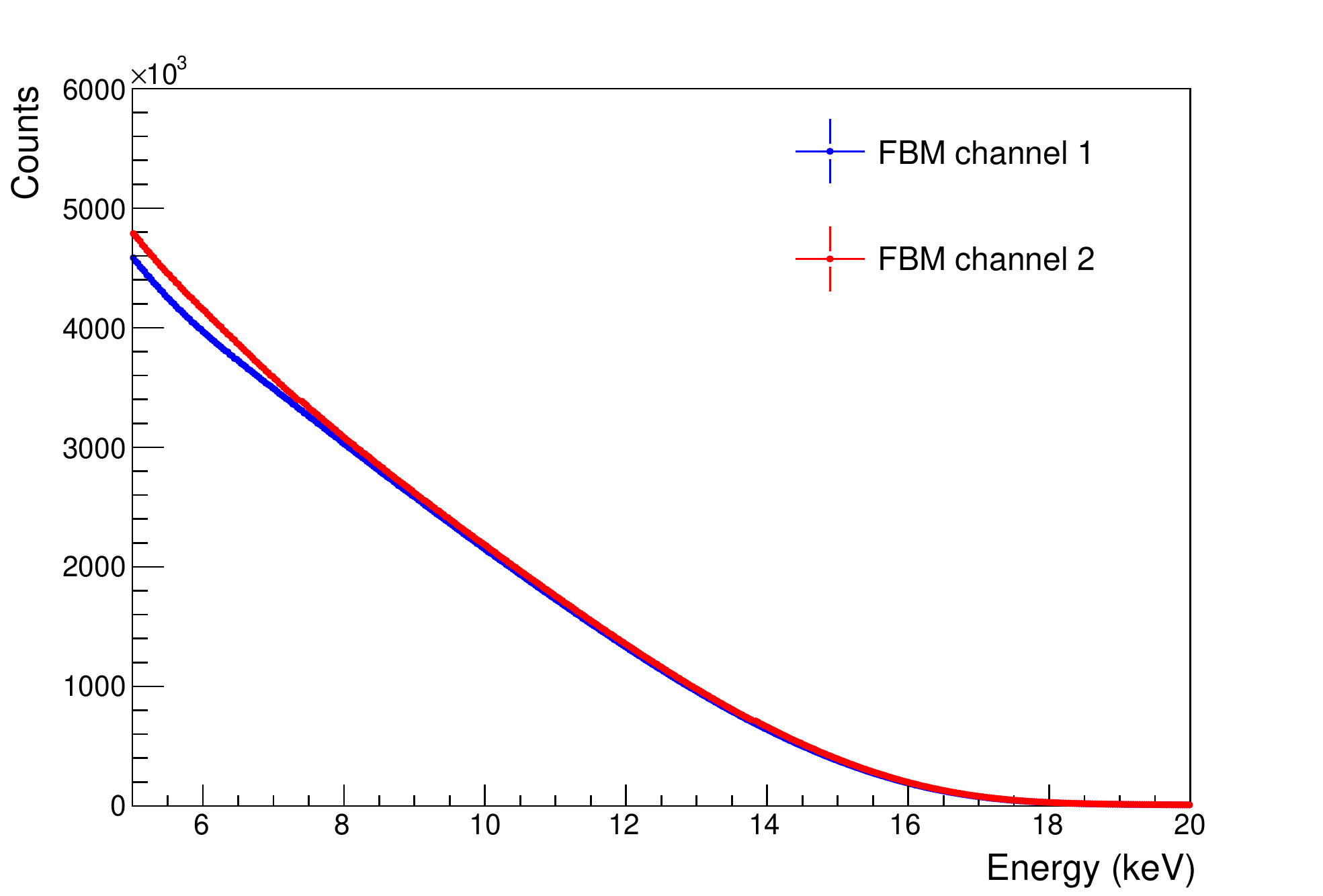}}
    \caption{\gls{fbm} spectra from \am241{} calibration and \gls{firsttritium}. (a) \am241{} spectra, the known peak at \SI{59.54}{\kilo\electronvolt} is fit to obtain the \pin{} diode energy resolution. (b) Tritium spectra from channel \num{1} and channel \num{2} \pin{} diodes obtained during the \gls{firsttritium} campaign.} 
\end{figure}

The energy resolution of the \pin{} diodes was measured during \gls{fbm} commissioning in April 2018. This was done using an \am241{} calibration source (see \figref{Subfigure:FBMcalibration}). Using the known peak at \SI{59.54}{\kilo\electronvolt}, the \gls{fbm} energy resolution was found to be $\sigma_{\mathrm{FWHM}} =$ \SI{2.1\pm0.1}{\kilo\electronvolt} for channel \num{1} and $\sigma_{\mathrm{FWHM}} =$ \SI{2.2\pm0.2}{\kilo\electronvolt} for channel \num{2}, where each channel corresponds to a \pin{} diode. 

Initial results from the first tritium campaign in June 2018 are shown in \figref{Subfigure:FBMspectrum}. The differential tritium spectrum is recorded and an energy threshold of  $E_{\mathrm{th}} \approx$ \SI{5}{\kilo\electronvolt} was set to remove background for subsequent analysis. The commissioning results and first tritium spectra demonstrate that the \gls{fbm} is fully operational and can be used for continuous monitoring of the \betael{} count rate from the source.

%% file: CalibrationAndMonitoringSystemIonRetention.tex
\subsection{Ion creation, blocking, and removal}
\label{SubSection:IonBlocking}

Tritium ions are created in the \gls{wgts} as a consequence of tritium \betadec{}. In order to prevent harmful background and systematic effects to the neutrino mass measurements, ion blocking, removal and detection are vital \cite{PhDKlein2018}.

\subsubsection{Tritium ion creation and induced background} 

Tritium ions originate in the \gls{wgts} via tritium \betadec{} at a rate of \SI{1e11}{ions\per\second} under nominal conditions. Additionally, because each \betael{} ionizes an average of 36~more tritium molecules, the total ion creation rate is \SI{4e12}{ions\per\second}.

These ions undergo transformation by scattering with neutral tritium molecules, resulting in mostly T$_3^+$ ions. T$^-$ ions are also created by dissociative attachment of secondary electrons to tritium molecules.
Recombination of positive ions with negative ions and secondary electrons creates a positive ion flux of \SI{2e11}{positive~ions\per\second} between the \gls{wgts} and the \gls{dps}; the negative ion flux in this region is significantly smaller. Nearly all ions have thermal energies of a few \SI{}{\milli\electronvolt}, corresponding to \SI{30}{\kelvin}. Only a small fraction of ions from molecular dissociation processes retain energies of several \SI{}{\electronvolt}.

Unlike neutral tritium, tritium ions cannot be pumped out of the beamline in the transport section; instead, they follow magnetic field lines and will induce background if they reach the spectrometers. Inside the \gls{ps}, these ions are accelerated by the negative high voltage to \SI{}{\kilo\electronvolt}-range energies and implanted into the spectrometer vessel walls. 
There is the danger that some tritium could reemerge as neutral gas, diffuse into the \gls{ms}, and create background by decaying there. 
They are considered a background here because tritium decaying in the \gls{ms} yields \betaels{} created under conditions different than those in the \gls{wgts}.
Any neutral gas flux into the \gls{ps} will reach the \gls{ms} with about \SI{3}{\percent} probability.
According to simulations, less \SI{1e-3}{} of the ions entering the \gls{ps} is expected to be transmitted into the \gls{ms} instead of being implanted. 
Additonally, in the \gls{ms}, the \SI{}{\kilo\electronvolt} ions could create background by ionization of residual gas.
In order to restrict the combined background from both mechanisms to less than \SI{0.2}{\milli cps}, the upper limit on the ion flux into the \gls{ps} is set to \SI{1e4}{ions\per\second}.
This is significantly smaller than the radiation protection limit of less than \SI{2e8}{tritium~ions\per\second} into the \gls{ps} over the course of three years of \gls{katrin} operation.
\par

\begin{figure}[!ht]
    \centering
    \includegraphics[width=\textwidth]{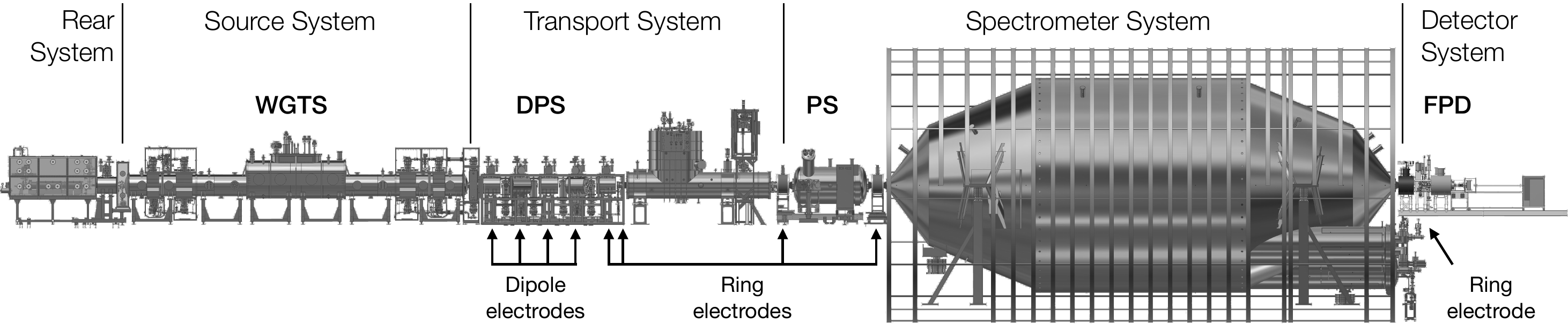}
    \caption{Position of ring and dipole electrodes along the beamline. There is no electrode in the \acrfull{wgts}. Four dipole electrodes are installed at the \acrfull{dps}. Five ring electrodes (split-rings) are installed at the exit of the \gls{dps} as well as in the \acrfull{ps}  and \acrfull{fpd} magnets. }
    \label{Figure:Ionsbeamlineoverview}
\end{figure}

\subsubsection{Ion blocking and removal with ring and dipole electrodes} 
\label{sec:ionblocking}

\begin{figure}[!ht]
    \centering
    \includegraphics[width=\textwidth]{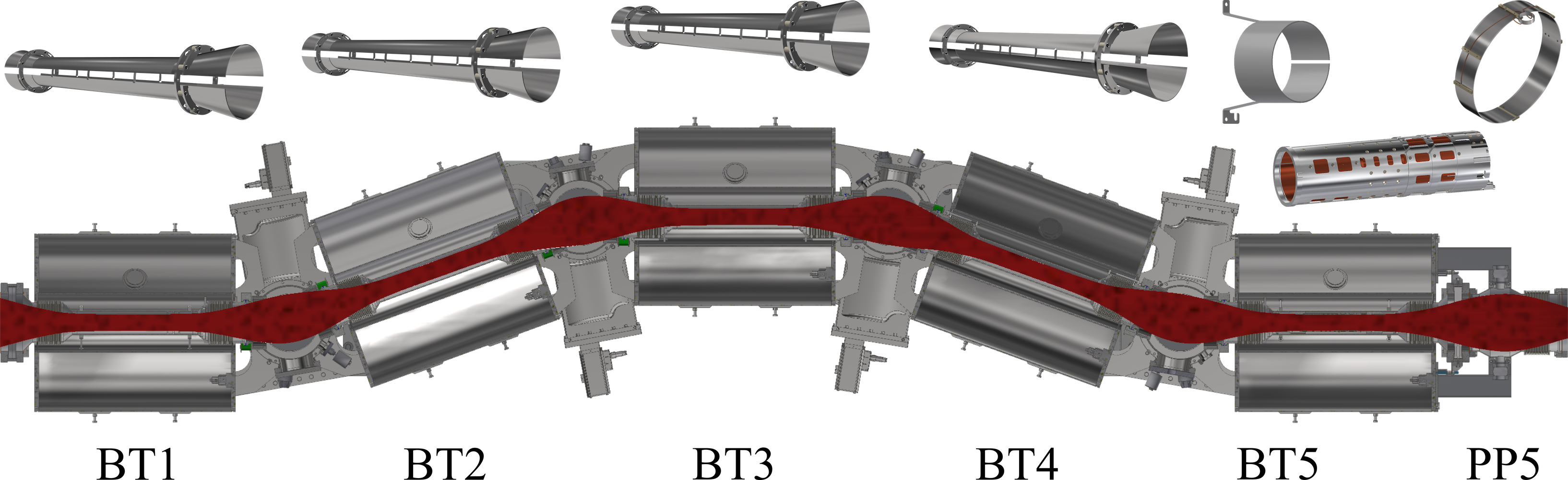}
    \caption{A CAD-drawing of the entire \gls{dps} is shown, with perspective drawings of the various kinds of electrodes directly above their respective locations. There are four dipole electrodes located in \gls{bt1} to \gls{bt4}. Two ring electrodes sit in \gls{bt5} and \gls{pp5}. In between them, the cylindrically-shaped \gls{fticr} mass spectrometry unit is installed.  The magnetic flux tube is indicated in red. }
    \label{Figure:IonsDPSoverview}
\end{figure}

Five ring electrodes are installed along the \gls{katrin} beamline in order to stop positive ions with a positive electrostatic potential (see \figref{Figure:Ionsbeamlineoverview} and \figref{Figure:IonsDPSoverview}). 
Their nominal voltages, ranging from \SIrange{+5}{+200}{\volt}, have been optimized during commissioning measurements. 
Each ring electrode has a split to prevent the creation of an induced electric current, should a superconducting magnet quench. 
Note that the ring electrode in \gls{dps} \gls{pp5} is considerably larger in size than the other four ring electrodes in order to accommodate the larger flux tube at that position. The blocked positive ions are stored in the beamline due to the gas flow from the \gls{wgts}.
As a consequence, the plasma density will increase and so will the probability of plasma instabilities which could change the energy of the \betaels{}. For this reason, the positive ions need to be removed. To this end, three dipole electrodes in the \gls{dps} create a dipole potential and remove the ions via $E\times B$ drift.
The transversal drift causes ions to be neutralized on the metallic lobes at the side of the dipole electrodes (\figref{Figure:IonsDPSoverview}), leaving only neutral tritium to be pumped off. 
The lobes are welded to the upper electrode in the first three \gls{dps} beamtubes, and to the lower electrode in \gls{bt4}. 
The voltages of the dipole electrodes in the first three \gls{dps} beamtubes have been optimized to values ranging from \SIrange{-5}{-85}{\volt}. 

The dipole electrode in \gls{bt4} of the \gls{dps} creates a dipole potential defined by applying \SI{+20}{\volt} to the upper and \SI{+25}{\volt} to the lower electrode. This removes secondary electrons which are stored in the beamline between the \gls{dps} and \gls{ps}. These stored secondary electrons could otherwise accumulate and neutralize the positive potentials of the ring electrodes. Inside all \gls{dps} beamtubes, cables connecting the ring and dipole electrodes are covered with a conductive shield to prevent charging up and a consequent creation of blocking potentials.

\glsreset{pp5} 
\glsreset{bt4} 

\subsubsection{Ion detection} 
\label{sec:ion-detection}

Several detectors are used to detect the presence of ions inside the beamline. They monitor the residual ion flux into the spectrometers and are used to assess systematic effects by ions and plasma in the source section.

\begin{figure}[!ht]
    \centering
    \includegraphics[width=\textwidth]{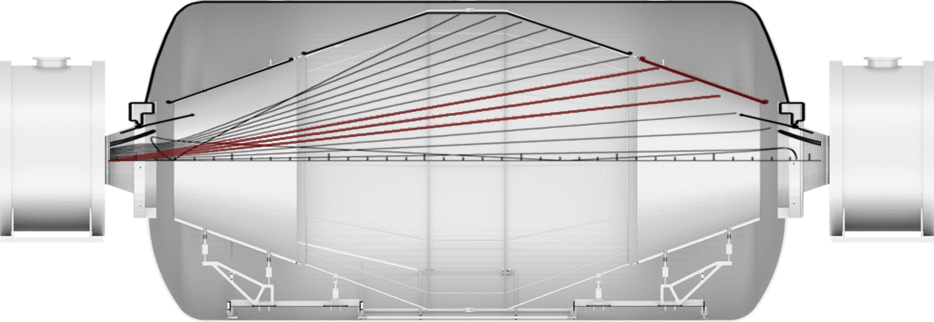}
    \caption{Simulated trajectories of T$_3^+$ ions in the \gls{ps}. Red: \SI{20}{\percent} of the ions reach the downstream cone electrode of the \gls{ps} (at \SI{-19}{\kilo \volt}).}
    \label{Figure:IonsPS}
\end{figure}

When the positive ions enter the \gls{ps} at negative high voltage, they are accelerated onto the downstream cone electrode (\figref{Figure:IonsPS}). The ions neutralize and eject secondary electrons, which creates a current which can be measured with an ammeter \footnote{Keithley 6514, \url{https://www.tek.com/keithley-low-level-sensitive-and-specialty-instruments/keithley-high-resistance-low-current-electrom}} inside the voltage supply of the cone electrode.
Calibration measurements \cite{PhDFriedel2020} showed that a current of \SI{10}{\femto \ampere} corresponds to \SI{5(1)e4}{ions \per \second} , which can be distinguished from a $3~\sigma$ background current fluctuation after \SI{2}{\hour} of measurement. 
As a safety precaution, closure of the valve between the \gls{cps} and \gls{ps} is triggered if three consecutive current measurement samples (each with a sampling rate of about \SI{1}{\hertz}) exceed a threshold of \SI{5}{\pico\ampere}; this corresponds to a rate of \SI{3e7}{ions\per\second} into the \gls{ps}.

The \gls{ps}-based ion detection methods can be calibrated using the \gls{pulcinella} disk, which is situated at the front of the \gls{fpd}  (\secref{Subsection:FPDCalibrationAndPAE}).
Because the \gls{pulcinella} disk is moved into the flux tube during calibration, which obscures the \gls{fpd}, no neutrino mass measurements can be done simultaneously.
The neutralization current is measured with an ADC \footnote{Texas Instruments, DDC-114, \url{https://www.ti.com/lit/ds/symlink/ddc114.pdf?ts=1608072005155&ref_url=https\%253A\%252F\%252Fwww.ti.com\%252Fproduct\%252FDDC114}}, which achieves an accuracy of about \SI{3}{\percent} in the \SI{}{\pico\ampere} range \cite{Amsbaugh2015, PhDMartin2017}.

In order to investigate plasma related effects in the source and transport section, several more current measurement techniques are available.
The neutralization current of ions, \betaels{} and secondary electrons can be measured on the \gls{rw}, as well as on the four dipole electrodes in the \gls{dps}.
For current measurements at the \gls{rw}, an ammeter \footnote{Keithley, 6487, \url{https://de.tek.com/datasheet/series-6400-picoammeters/model-6487-picoammeter-voltage-source}} with a sensitivity of \SI{0.4}{\pico \ampere} is used. The four \gls{dps} dipole electrode currents are measured using an ammeter \footnote{RBD, 9103, \url{https://rbdinstruments.com/products/files/9103-picoammeter.pdf}} which is sensitive down to \SI{0.5}{\pico\ampere}.
For these dipole electrodes, the ammeters are placed between the voltage supply and the lobe-bearing dipole halves, as shown in \figref{Figure:IonsDPSoverview}.
A neutralization current at the dipole electrodes of \SI{5}{\pico\ampere} (\SI{3e7}{ions\per\second}) can be distinguished from a $3\sigma$ background fluctuation, given an ammeter acquisition time of \SI{400}{\milli\second} and neglecting the uncertainty on the background.

An \gls{fticr} unit \cite{UbietoDiaz2009, PhDUbietoDiaz2011, Heck2014} (\figref{Figure:IonsDPSoverview}) for the distinction of different ion species is located in the \gls{dps}. 
Two endcaps create a Penning trap for ions, which are then excited into coherent motion so that their induced image current can be measured. 
Through Fourier transformation, the $e/m$-ratios can be determined~\cite{Blaum2006}, which allows for identification of the original ion.

This unit was not used in the measurement campaigns of 2017 - 2019.

%% file: CalibrationAndMonitoringSystemMagneticField.tex
\subsection{Spectrometer Magnetic Field Monitoring System}
\label{SubSection:SpectrometerMagneticFieldMonitoringSystem}

The magnetic field strength in the analyzing plane is one of the key operational parameters in the \gls{katrin} experiment; therefore thorough monitoring is essential for a successful neutrino mass analysis. Because it's not possible to make direct measurements of magnetic fields inside the \gls{ms} vessel, \gls{katrin} relies on electromagnetic calculations \cite{Osipowicz2012} and a precise model \cite{PhDErhard2016, Furse2017}.

The input data for the calculations and the model come from two sensor networks: one stationary and one mobile monitoring system. The stationary magnetic monitoring system is mounted directly on the \gls{ms} vessel, which allows for continuous monitoring as close as spatially possible to the magnetic flux tube. The mobile magnetic monitoring system consists of several robots which circulate along the outside of the \gls{ms} vessel hull, in order to map the magnetic field over a large volume \cite{PhDErhard2016}.

\subsubsection{Stationary Magnetic Monitoring System}

\begin{figure}[!ht]
	\centering
	\includegraphics[width=1.0\textwidth]{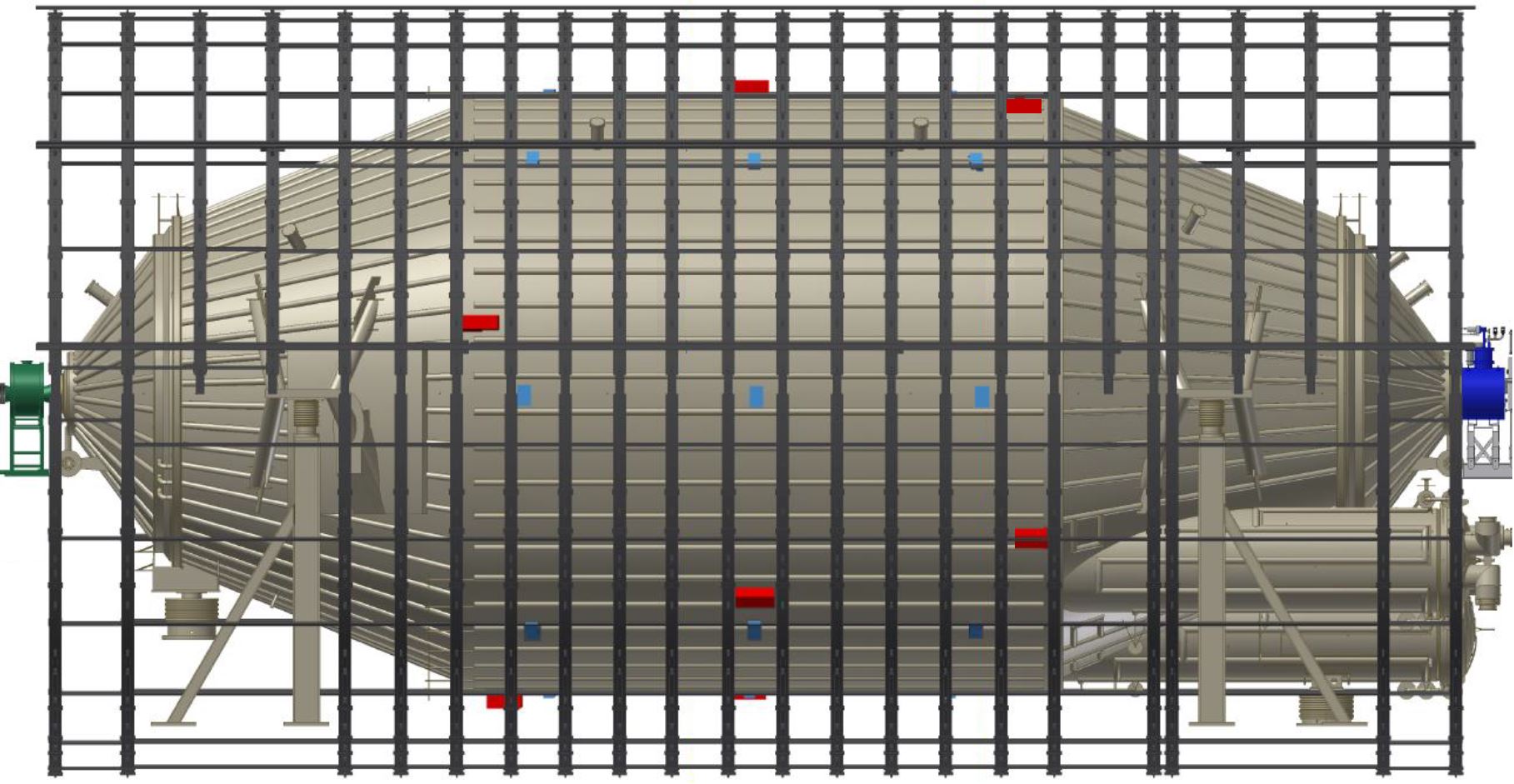}
	\caption{The \gls{ms} vessel and air coil system, as seen from the east side. The red boxes indicate the positions of the Bartington sensor units. The locations the Philips sensor units are marked with blue boxes. Figure adapted from \cite{PhDErhard2016}. }
	\label{fig:MainSpecMagnetometer}
\end{figure}

The stationary magnetic monitoring system consists of 38 sensor units located on the outer vessel surface in the central part of the \gls{ms} (\figref{fig:MainSpecMagnetometer}).
Each unit contains a triaxial magnetometer with custom electronics.
The units are divided into two categories based on the magnetometer type used.
Twenty-four of the 38 sensor units utilize magneto-resistance magnetic field sensors by Philips Semiconductors\footnote{NXP Semiconductors (formerly Philips Semiconductors), types KMZ10B and KMZ20M, \url{www.nxp.com}}, and 14 are based on high-precision magnetometers by Bartington Instruments\footnote{Bartington Instruments, type Mag-03MSB1000, \url{www.bartington.com}}.

The Philips magnetic field sensor units are distributed on three rings on the \gls{ms} vessel. Within each ring, the units share the same axial position but differ in azimuthal position. Each of the three rings accommodates eight sensor units. The axial positions of the magnetometer rings with respect to the analyzing plane are \SI{+-3.6}{\meter} and \SI{0}{\meter}. The units on the central magnetometer ring use magnetic field sensors of type KMZ10B, the other units type KMZ20M. These two Philips sensor unit types are designed for low-field measurements in the range of 0.1 to approximately \SI{2}{\milli\tesla} \cite{Philips_KMZ20M}. The sensor units have temperature sensors incorporated, which allow for calibration of the measured flux to within \SI{5}{\percent} systematic uncertainty \cite{PhDErhard2016}.

The Bartington sensor units are mounted on steel cable rings wrapped around the \gls{ms} vessel hull. The three magnetometer rings are mounted at axial positions of \SIlist{-4.5;-0.14;4.3}{\meter} relative to the analyzing plane. The central ring houses six sensor units, the outer two rings house four units each. The triaxial flux gate sensors on each unit can measure very low magnetic fields from \SI{<50}{\nano\tesla} to \SI{1}{\milli\tesla}. The sensors are manufactured with a linearity uncertainty of \SI{<0.0015}{\percent}, a negligible temperature dependence of \SI{200}{ppm\per\celsius} and an orthogonality uncertainty of \SI{<0.5}{\degree}. The relative uncertainty estimation on the measured magnetic flux is \SI{<0.5}{\percent} \cite{Bartington_Mag03}. In addition to the magnetometer, each Bartington sensor unit contains an inclinometer\footnote{{Murata Electronics Oy}, {SCA121T dual axis inclinometer}, \url{https://www.murata.com/en-eu/products/sensor}} and a laser-based position measurement system. These auxiliary devices provide accurate orientation (\SI{+-0.23}{\degree}) and positioning measurements (\SI{+-1}{\milli\meter}) \cite{MasterAntoni2013}.

\subsubsection{Mobile Magnetic Monitoring System}

\begin{figure}[!ht]
    \centering
    \includegraphics[width=0.7\textwidth]{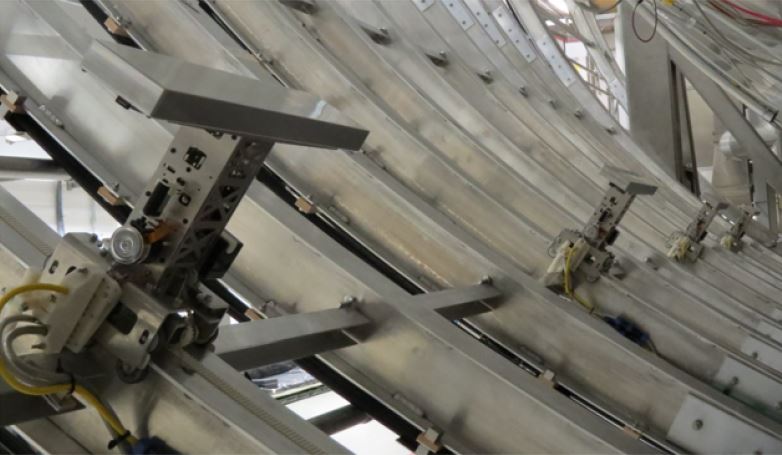}
    \includegraphics[width=0.8\textwidth]{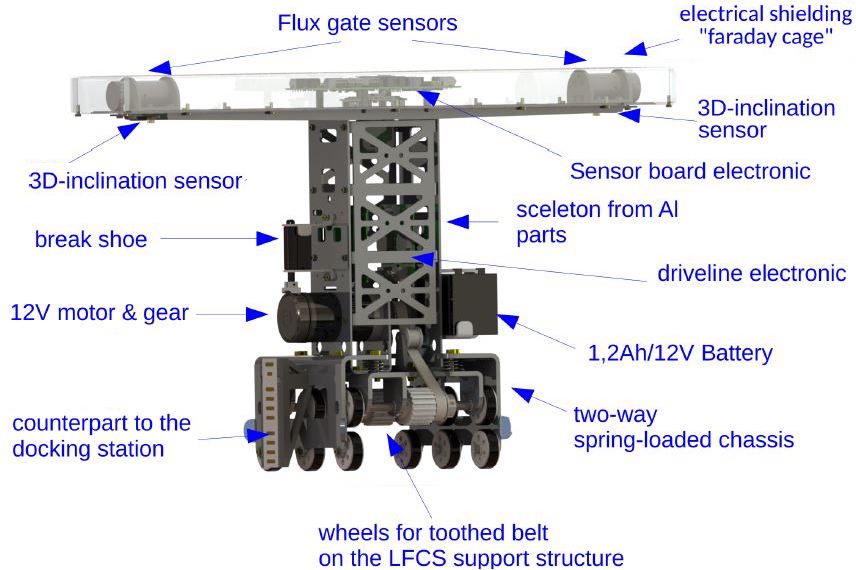}
    \caption{\textbf{Top:} The four mobile sensor units of the radial magnetic monitoring system at \gls{lfcs} support rings 3, 6, 9 and 12 in their rest positions at the docking stations. \textbf{Bottom:} Mechanical drawing of the mobile sensor unit as installed. The \SI{2.9}{\kilo\gram} unit is \SI{296}{\milli\meter} tall and \SI{532}{\milli\meter} wide. Both figures taken from \cite{Letnev2018}.}
    \label{fig:mobs_rmms_figure}
\end{figure}

Magnetometers mounted on robots, which move automatically around the \gls{ms}, form the mobile monitoring system. While this system doesn't allow for continuous magnetic monitoring, it does allow for measurement of the magnetic field in the environment of the spectrometer over large area \cite{PhDErhard2016}. The mobile magnetic monitoring system currently includes eight robots, with plans to expand to 12 in the near future. Four mobile sensor units \cite{Osipowicz2012} move on the inner side of four \gls{lfcs} support rings (rings 3, 6, 9, and 12; see \figref{fig:mobs_rmms_figure} top) and form the \gls{rmms}. The remaining four units measure the magnetic field on two vertical planes at the east and west side of the spectrometer building and form the \gls{vmms}. \cite{Letnev2018}

Each mobile units of the \gls{rmms} measures the magnetic field at 144 predefined sampling positions, close to the outer surface of the cylindrical part of the spectrometer. 
This configuration was chosen due to its axial symmetry with respect to the analyzing plane in the center of the \gls{ms} vessel. 
Each of the four \gls{lfcs} rings houses a docking station, which defines the start and end point of the mobile sensor unit for one sampling run. Furthermore, the station provides controlled charging of the installed batteries and transfers the measured data to the master module. The communication and transmission between mobile sensor units and their master module, as well as automatic charging, is realized via a modular controller system \footnote{National Instruments, {CompactRIO Platform}, \url{https://www.ni.com/en-us/shop/compactrio.html}} \cite{Letnev2018}.

The drive principle of the ``T''-shaped mobile sensor unit (\figref{fig:mobs_rmms_figure} bottom) is based on toothed-gear wheels and a toothed belt attached to the inner side of the \gls{lfcs} support ring. This design allows the unit to smoothly navigate mechanical discontinuities. 
The aluminum frame of the mobile sensor units forms a Faraday cup, which ensures the electrical safety of the unit when the \gls{ms} vessel is at high voltage. 
The two ``wings'' of the sensor unit each contain one flux gate sensor \footnote{Stefan Mayer Instruments, FL3-1000, \url{https://stefan-mayer.com/images/datasheets/Data-sheet_FL1-100.pdf}}, which has a relative systematic uncertainty of \SI{<0.5}{\percent} \cite{Letnev2018}. 
In addition, a three-dimensional inclinometer sensor system based on the 3-axis linear accelerometer\footnote{NXP Semiconductors (formerly Freescale Semiconductors), type {Fxls8471q} 3-axis linear accelerometer, \url{https://www.nxp.com/part/FXLS8471Q}} is installed in each wing. 
The inclinometer systems enable an exact transformation of the local \gls{rmms} coordinate system into the global \gls{katrin} system, despite deformations and misalignments of the \gls{lfcs} rings. 
A sophisticated control algorithm ensures the reproducibility of measurement positions with an accuracy of better than \SI{1}{\milli\meter} \cite{Letnev2018}.

\begin{figure}[!ht]
    \centering
    \includegraphics[width=0.64\textwidth]{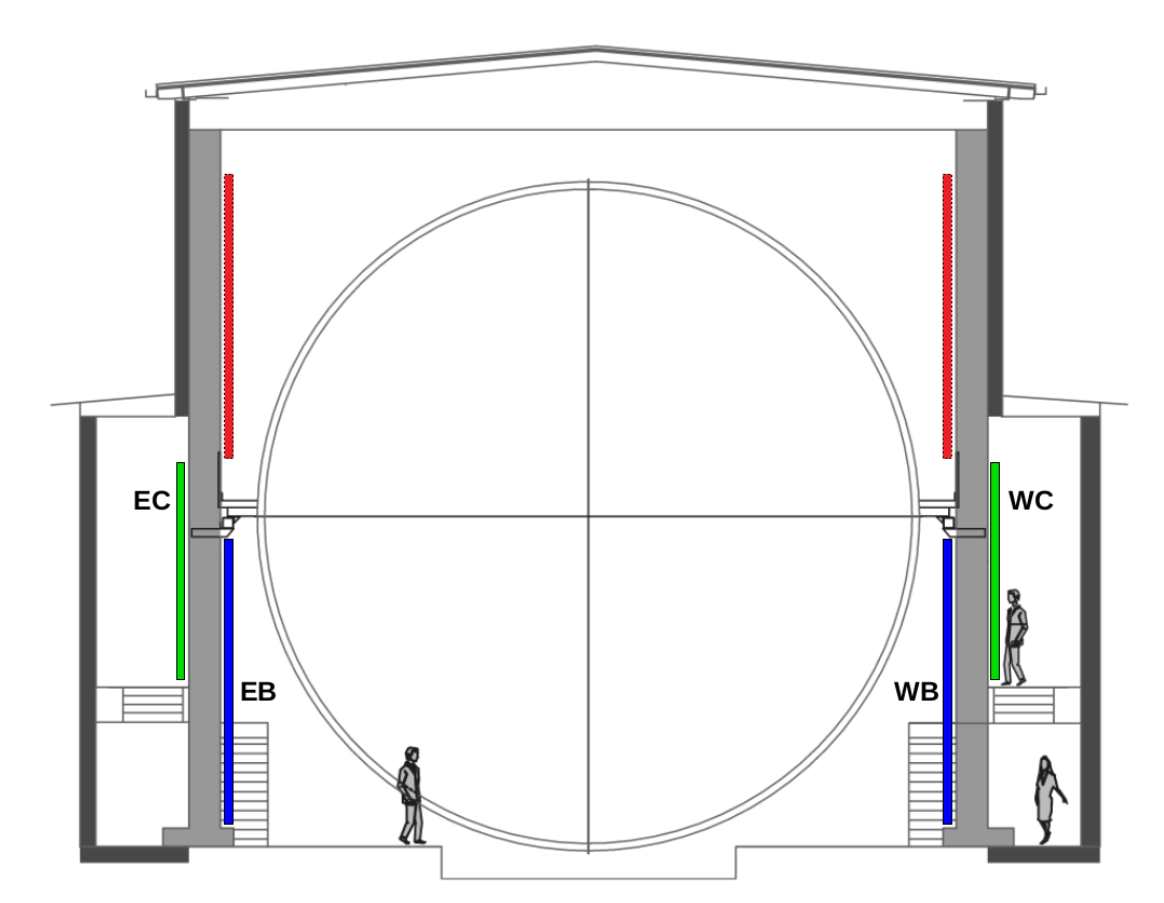}
    \includegraphics[width=0.34\textwidth]{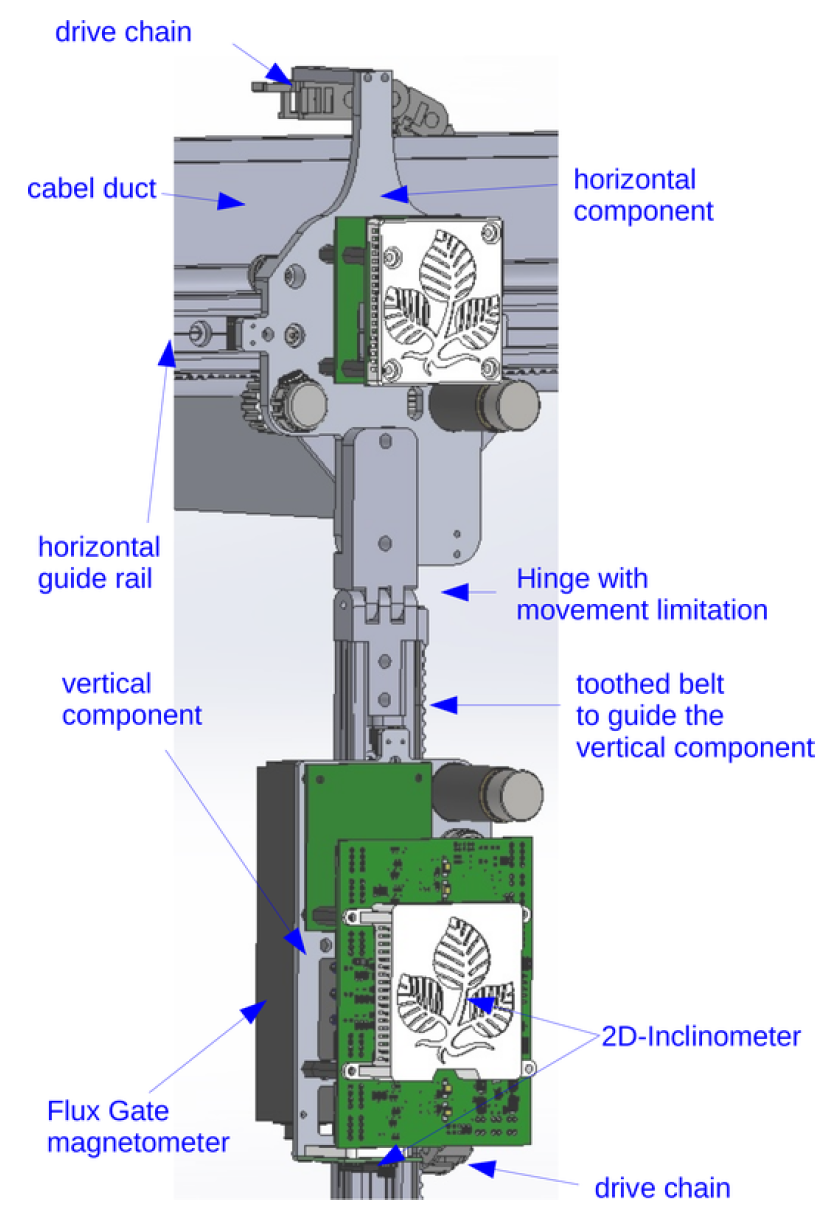}
    \caption{\textbf{Left:} Schematic representation of the vertical magnetic monitoring system position within the spectrometer hall. Individual layers are color-coded. View from the north. Individual systems are marked with EC (east center), EB (east bottom), WC (west center) and WB (west bottom). The construction of the upper system (in red) is currently in the concept phase and will be finished in the near future. \textbf{Right:} Schematic structure of a mobile sensor unit of the vertical magnetic monitoring system in CAD view. Both figures are taken from \cite{Letnev2018}.}
    \label{fig:VMMS}
\end{figure}

The \gls{vmms} covering vertical planes parallel to the walls of the spectrometer hall was developed to measure remanent and induced magnetization effects of the walls, which have a direct influence on the magnetic field in the analyzing plane \cite{Letnev2018,PhDErhard2016}. Mechanically, the vertical magnetic monitoring system is inspired by the technology of the radial system and is based on a movable construction of linear rails attached to the hall pillars. 
The measuring accuracy and positioning precision of the vertical system are the same as those of the radial system, and are all met. 
The aim is to completely cover the wall surface in the area near the cylindrical part of the \gls{ms} vessel at three height levels and to measure the magnetic field with a mesh size of \SI{20x20}{\centi\meter}. 
Four \glspl{vmms} at two height levels have been installed and commissioned. 
The left plot of \figref{fig:VMMS} shows the position of the individual systems. 

The right plot of \figref{fig:VMMS} shows the schematic of a mobile sensor unit in the \gls{vmms}. The system includes two movable components (horizontal and vertical); a movement-limiting hinge to prevent inadmissibly strong pendulum motion during sensor unit deployment; the drive chains which connect the subsystems; and the cable duct. The movement sequence is as follows: the vertical component starts at the lowest vertical position and moves upwards to the next sampling position \SI{20}{\centi\meter} away. After the measurement has been completed, it moves to the next sampling position. As soon as the end of the vertical linear rail has been reached, the horizontal component is moved to the next position and the vertical system returns to its starting point. Once the overall target position is reached, the entire procedure is repeated. In this way, a grid of magnetic field sampling points is produced until the end of the horizontal linear rail is reached \cite{Letnev2018}.

%% file: CalibrationAndMonitoringSystemMonitorSpectrometer.tex
\subsection{Monitoring High Voltage System Stability}
\label{SubSection:MonitorSpectrometer}
\glsreset{mos} 

A precise knowledge of the \gls{mace} filter retarding potential is central to obtaining an accurate integral \betaspec{}.
Previous works have found that an unaccounted-for variation of the retarding high voltage of \SI{3}{\ppm} at \SI{-18.6}{\kilo\volt} could lead to a shift of the observed neutrino mass squared by about \SI{-0.007}{\square\electronvolt}\cite{Kaspar2004,PhDThuemmler2007,PhDSlezak2015,PhDKraus2016}. Thus, the high-voltage measurement must be stable at the \si{\ppm} level to prevent a significant bias of the \gls{katrin} result. 

This requirement is met by having a second, parallel spectrometer setup \cite{Erhard2014}, the \gls{mos}, for monitoring; this is in addition to the precision high voltage divider which directly measures \cite{Rest_2019} the voltage applied to the \gls{ms} (\secref{subsec:HV-Distribution-Monitoring}).  The \gls{mos} is another \gls{mace} filter which is galvanically connected to the high voltage system of the \gls{ms}. Galvanic coupling of the \gls{ms} and \gls{mos} refers to the fact that their vessel voltages are equal and are powered by the same power supply. This ensures that data-taking is synchronized, facilitating real-time monitoring.
An ultra-stable implanted \rb{}/\kr{} source \cite{Zboril2013,PhDSlezak2015} emits monoenergetic conversion electrons, whose energy is continuously assessed at the \gls{mos}. Variation of the observed electron energy would point to an instability of the \gls{katrin} high-voltage measurement.

\subsubsection{Monitor Spectrometer Setup} 

The \gls{mos} setup is based on the hardware setup of the former Mainz neutrino mass experiment \cite{Kraus2005}.

\begin{figure}[!ht]
	\centering
	\includegraphics[width=0.7\textwidth]{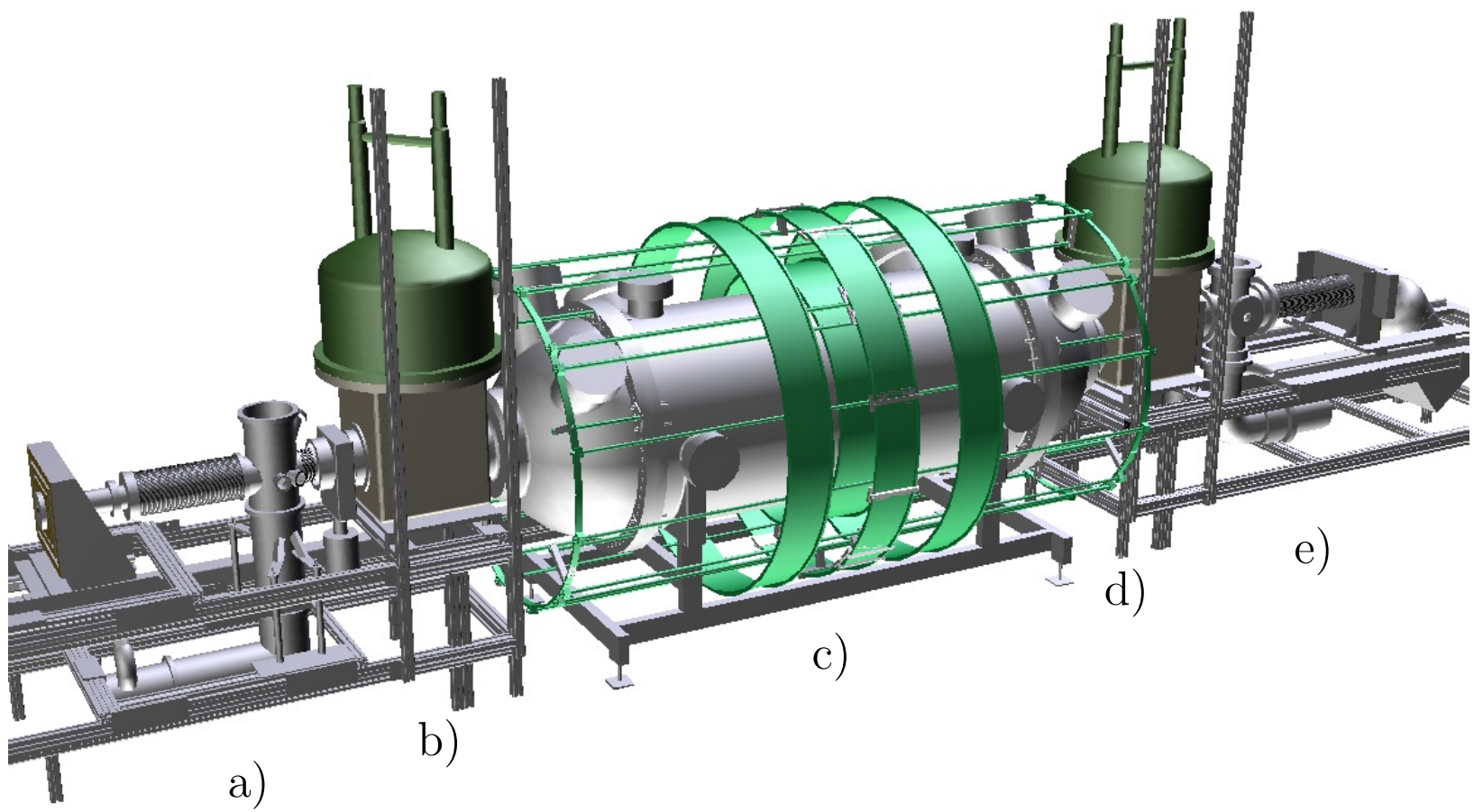}
	\caption{CAD drawing of the monitor spectrometer. Shown are the a) source chamber, b) source-side superconducting magnet, c) vessel with \gls{lfcs} and \gls{emcs}, d) detector-side superconducting magnet, and e) detector chamber \cite{Erhard2014}. }
	\label{fig:mos-cad}
\end{figure}

The \gls{mos} is a stainless steel vessel with a length of \SI{\sim 3}{\metre} and a diameter of \SI{\sim 1}{\metre}, which is maintained under ultra-high vacuum at a level of \SI{9e-11}{\milli\bar}. The retarding electric field of the \gls{mace} filter is created by a set of cylindrical and conical stainless steel solid electrodes, as well as stainless steel wire electrodes, all of which are at the same high voltage. Additionally, there is a ground electrode, which is kept at zero potential, together with the vessel hull.

The guiding magnetic field is produced by two superconducting solenoids at the spectrometer entrance and exit, and is further shaped by four normal conducting air coils in the center. There are also horizontal and vertical Earth field compensation coils around the vessel. The magnetic field is set up to achieve an energy resolution similar to that of the \gls{ms}, i.e. \SI{\sim 1}{\electronvolt} at \SI{18.6}{\kilo\electronvolt}.

The \rb{}/\kr{} source is electrically biased using a dedicated power supply by about \SI{-750}{\volt} to match the \kthirtytwo{} conversion electron energy with the endpoint of the tritium \betaspec{}. The power supply\footnote{FUG MCP 14-1250, \url{www.fug-elektronik.de}} is read out directly by a commercial 6-1/2-digit voltmeter (\figref{fig:HV-distribution}). The electron detector is a circular silicon \pin{} diode with a sensitive area of \SI{1.5}{\square\centi\metre}. Both the detector and the first pre-amplifier stage are located inside the vacuum chamber on a copper-beryllium rod, which is attached to a copper cold finger immersed in liquid nitrogen. This configuration allows the detector chamber to achieve a stable temperature of about \SI{-45}{\celsius}. For additional details, see \cite{Erhard2014}.

\subsubsection{Implanted \rb{}/\kr{} Source}  
\label{SubSubSection:ImplantedKrSourceMoS}

The electron source is a substrate made out of \gls{hopg} with a width of \SI{12}{\milli\meter}, which is ion-implanted with the generator \rb{} (half-life of \SI{\sim 86}{\day} \cite{Venos2018}).
The \rb{} production is carried out at the cyclotron of the Centre of Accelerators and Nuclear Analytical Methods (CANAM) at the Nuclear Physics Institute \v{R}e\v{z} \cite{Sentkerestiova2018}, while the implantation is done with the Bonn Isotope Separator at the Helmholtz Institute for Radiation and Nuclear Physics, Germany \cite{PhDArenz2017}. The typical implantation energy is around \SI{8}{\kilo\electronvolt}. The daughter \kr{} (half-life of \SI{\sim 1.8}{\hour} \cite{Venos2018}), which is generated by electron capture decay of \rb{}, emits conversion electrons in place due to its high retention (\SI{>90}{\percent}) in the substrate.

Due to the favorable half-lives of these two isotopes, transient equilibrium is established within a day of source production.
The long decay time and low maintenance requirements allow for continuous measurement at the \gls{mos}. A \rb{} activity of a few \si{\mega\becquerel} yields an electron rate in the no-energy-loss electron peak of the \kthirtytwo{} line on the order of \SI{10}{\kilo cps}. The actual electron rate value can be adjusted by moving the source within the magnetic field, thereby changing the \gls{mace} filter acceptance angle according to \eqnref{eq:thetaMax}.
Dedicated analysis methods were developed to describe the electron line shape of the implanted \rb{}/\kr{} source, which is influenced by solid state effects. \cite{Slezak2013,PhDSlezak2015}

\subsubsection{Electron Energy Stability} 

Extensive studies using the \gls{mos} were carried out to assess the quality of the implanted \rb{}/\kr{} sources. It was shown that the required electron energy stability can be reproduced if the \rb{} peak concentration in the substrate is not too high. Acceptable electron stability values were found to be on the order of \SI{1e20}{ion\per\centi\meter\cubed}. \cite{PhDSlezak2015}

\begin{SCfigure} 
	\centering
	\caption{Long-term stability of the \kthirtytwo{} electron energy measured at the \gls{mos}.
The labels on each of the 4 subplots denote different sources in the following format: \emph{substrate} -- \emph{implantation energy in \SI{}{\kilo\electronvolt}} -- \emph{serial number}. \\
The larger drift for HOPG-8-2 of about \SI{0.8}{\ppm\per month} is clearly visible, in contrast to the negligible drifts for the other three sources. The \rb{} peak concentration for the first sample was around \SI{9.3e20}{ion\per\centi\meter\cubed}, which is large compared to \SIlist[list-units=single]{0.2e20; 1.7e20; 1.1e20}{ion\per\centi\meter\cubed}, respectively, for the other sources. The reason for such a high concentration was the use of a new retarding setup at the ion implanter, leading to significant focusing of the ion beam. }
	\label{fig:MonSpec_K-32_stability}
	\includegraphics[width=0.5\columnwidth]{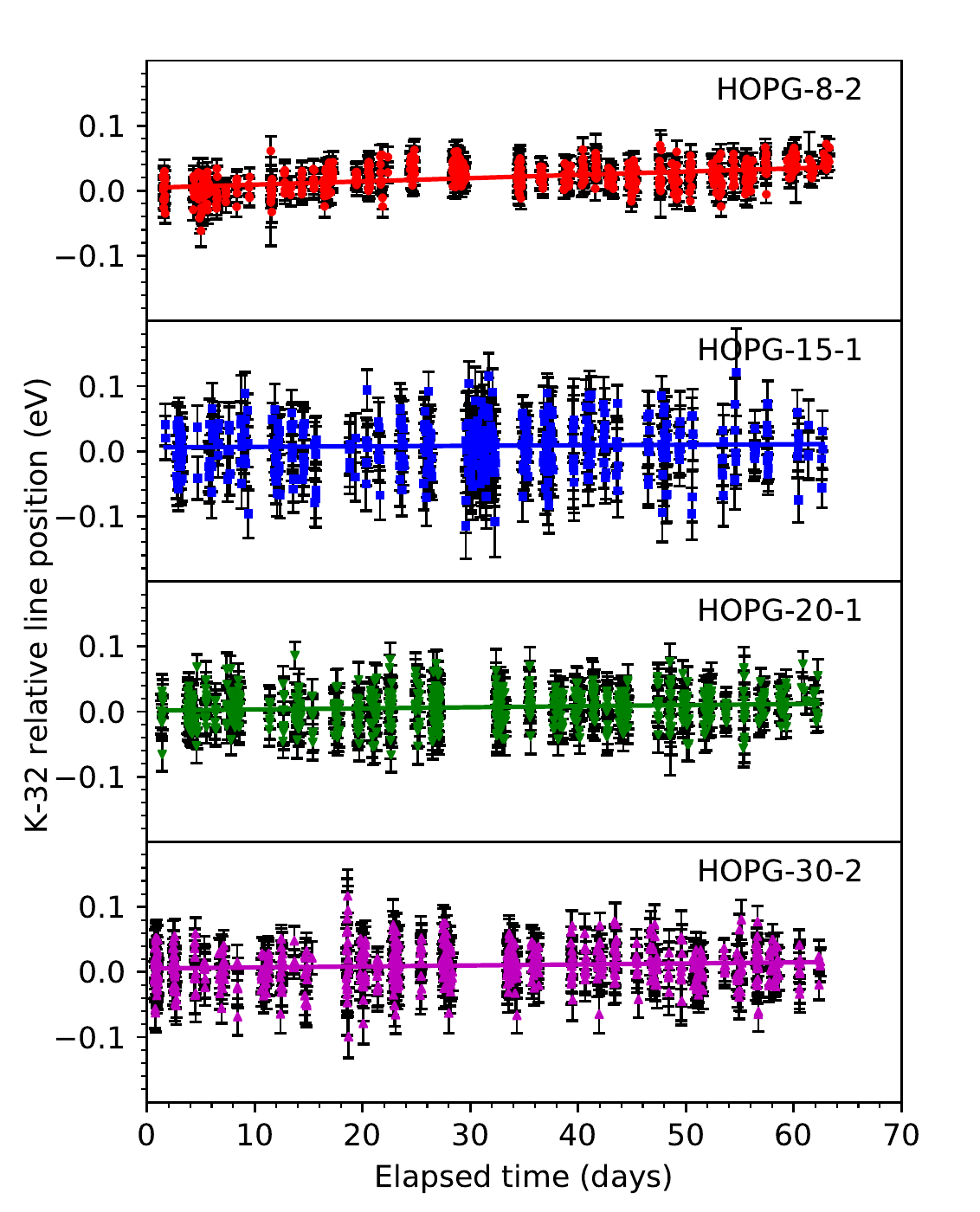}
\end{SCfigure}

The peak concentration is estimated using the source activity, radio-graphical image of the source, and a simulation of the implantation profile.
The results of a long-term stability measurement of the \kthirtytwo{} electron energy of four different sources during the same time period are shown in \figref{fig:MonSpec_K-32_stability}. The plot demonstrates reproducibility of the long-term electron energy stability in different sources, and the negative effect of high \rb{} peak concentration.

The \kthirtytwo{} line position in the spectra can be determined within a few minutes with an uncertainty of \SI{40}{\milli\electronvolt}.
The electron line position of the solid \kr{} sources is few \SI{}{\electronvolt} higher than that obtained from gaseous \kr{}. This is a known effect due to the different electron binding energies in the atom, whether bound in a solid or in the free atom. Nevertheless, the \rb{}/\kr{} \gls{hopg} source can be used as a long-term standard by measuring the energy difference of the \lthreethirtytwo{} and \lthreeninefour{} conversion electrons.

%% file: CalibrationAndMonitoringSystemElectronSources.tex
\subsection{Electron Sources}
\label{sec:electron_sources}

A number of electron sources are currently available: an electron gun (\secref{sec:egun}), a condensed krypton source (\secref{Subsubsection:CKrS}), and a gaseous krypton source (\secref{Subsubsection:GKrS}). There exists a fourth electron source, an implanted krypton source, which is part of the monitoring system at the \gls{mos}; a full description is given in \secref{SubSubSection:ImplantedKrSourceMoS}.


\subsubsection{High Resolution Angular-Selective Electron Gun}
\label{sec:egun}
\glsreset{egun} 


The \gls{egun} emits a pulsed beam of electrons to be used as probes in the \gls{katrin} setup. The design is based on the principle of electron emission from a metallic photocathode, which produces a point-like beam with electrons of narrow energy and angular distribution.  A general description of the device can be found in  \cite{Behrens2017}.
The \gls{egun} allows studying electromagnetic characteristics along the main beam line and has been used in the commissioning of various beam line sections. It is also suited to investigate and monitor source characteristics, such as electron scattering and the column density stability.

\paragraph{Principle and Requirements} 

The quantities of interest in the \gls{egun} setup are the electron rate $R_\mathrm{g}$, the energy $E$ and the electron pitch angle $\theta$.

The \gls{egun} is required to have a small energy spread of \SI{<0.5}{\electronvolt}, and a well-defined $\theta$ which covers the range of \ang{0} to \ang{90} at the pinch magnet, which is the location of highest magnetic field along the \gls{katrin} beam line. The electron rate must be stable, with \SI{<0.1}{\percent} fluctuation over several hours.
Because the energy spread is largely influenced by the photocathode work function, it is important to perform measurements of this characteristic parameter on a regular basis. This is possible by a wavelength-dependent measurement of the electron yield; see e.g. \cite{PhDBehrens2016,PhDSack2020}.

Many studies effectively measure the kinetic energy of the electrons in the source or in the analyzing plane of the \gls{mace} filter.
As explained below, this energy depends on the voltage difference between photocathode and the \gls{ms} electrodes. A precision high voltage source is therefore required to minimize fluctuations. The two systems are typically coupled, with the photocathode being supplied with a voltage offset on the order of \SI{\pm 100}{V} relative to the inner spectrometer potential.

Two such \gls{egun}s have been used at the \gls{katrin} experiment. The first was set up temporarily during the spectrometer commissioning campaign, which focused on the transmission characteristics of the \gls{ms} \cite{PhDZacher2015,PhDBehrens2016,PhDErhard2016,PhDBarrett2017}.
Additional measurements were carried out at the \gls{mos} to test the device and perform a preliminary characterization \cite{Behrens2017}.
This source was intended only for intermediate use at the spectrometer section, and it has subsequently been dismounted with the completion of the full \gls{katrin} beamline.

A second \gls{egun} is installed in the \gls{rs}, using a design similar to the first one. This \gls{rs} \gls{egun} is fully integrated with the main beam line and used for further commissioning measurements and continuous monitoring of source parameters such the column density \cite{PhDBabutzka2014,PhDSack2020,PhDSchimpf2021}.
The first commissioning phase of this setup was carried out in 2018, which included general functionality tests and characterization, investigations of electron scattering with $\mathrm{D}_2$ gas in the source section, and monitoring of the column density. After minor improvements the \gls{rs} \gls{egun} was used for electron scattering studies on $\mathrm{T}_2$ gas in the source and for characterization of the electromagnetic fields in the \gls{ms}, which will be discussed in a separate publication.

The following sections will summarize both electron source setups and highlight some important results from earlier measurement campaigns.

\paragraph{Setup} 

In both setups, the \gls{egun} design is based on the emission of electrons from a metallic photocathode illuminated by a UV light. The light is fed through an optical fiber onto the back of the thin photocathode surface, where photoemission takes place.
By using light with a photon energy $E_\gamma = h f = hc / \lambda$ that is slightly larger than the work function $\Phi$ of the photocathode, electrons are emitted with a small and narrowly distributed energy in the \SI{1}{eV} range.
The photocathode is located at an emission electrode, operated at a high electric potential of typically $U_\mathrm{start} = \SI{-18.6}{\kilo\volt}$.
The potential difference to the grounded \gls{katrin} beam line accelerates the electrons to the corresponding kinetic energy of $E \approx \SI{18.6}{\kilo\electronvolt}$. This electron acceleration is adiabatic, and the electrons follow a cyclotron motion around a magnetic field line that transports them downstream towards the detector.

\begin{figure}[h]
 \centering
 \includegraphics[width=.8\textwidth]{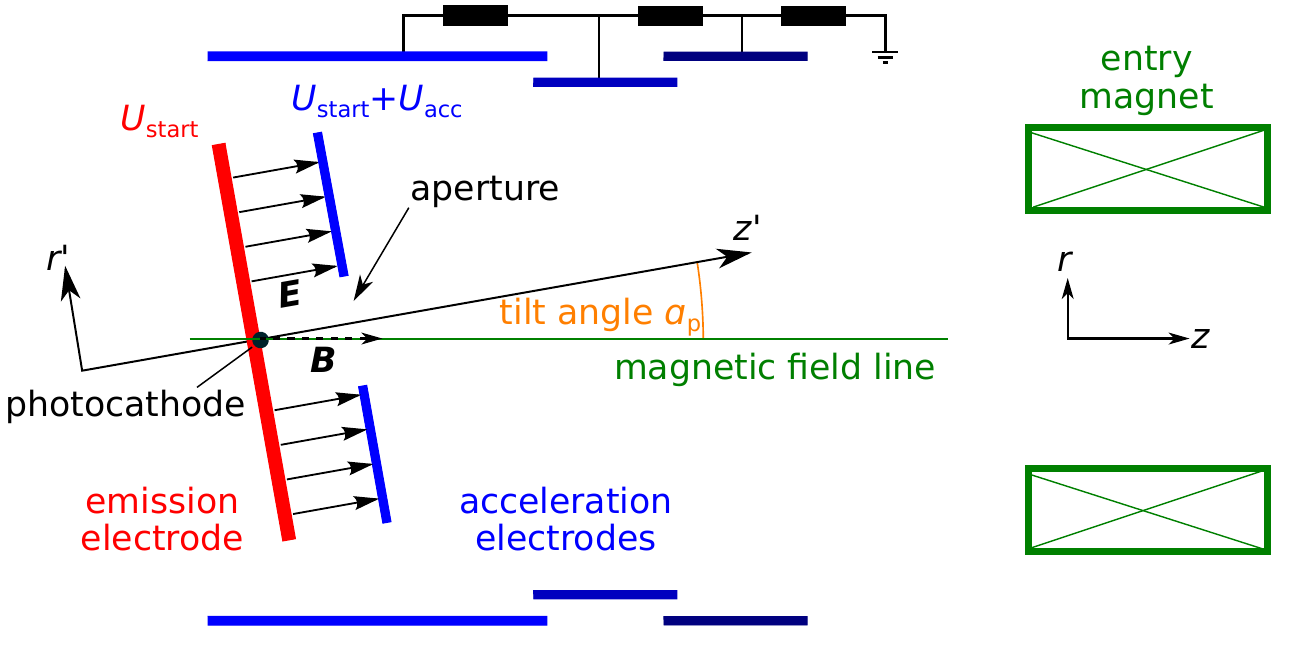}
 \caption{Electron source design, based on photo-emission of electrons with electrostatic acceleration. Shown is the \gls{rs} \gls{egun}; the similar setup used at the \gls{ms} is described in \cite{Behrens2017}.
     }
 \label{fig:egun_setup}
\end{figure}

A key feature of the \gls{egun} is \emph{angular selectivity}: the creation of a defined pitch angle $\theta = \sphericalangle(\vec{p},\vec{B})$ between electron momentum $\vec{p}$ and magnetic field $\vec{B}$.
The electrons follow an approximately $\cos\theta$ angular distribution upon emission with very low velocity. Therefore the Lorentz force is small and they can be accelerated at this early stage in a direction also non-parallel to the magnetic field. The acceleration collimates the beam into a narrow angular distribution~\cite{Behrens2017,Valerius2011}.
This non-adiabatic collimation is achieved by a electrostatic acceleration field in the order of \SI{500}{\kilo\volt\per\meter} at the photocathode surface, which is created by an acceleration electrode at a potential difference $U_\mathrm{acc} \approx \SI{+5}{\kilo\volt}$ relative to the photocathode.
The photocathode can be tilted by a few degrees against the magnetic field at the emission spot, in order to select a certain range of electron pitch angles $\theta = $\SIrange{0}{90}{\degree} at the pinch magnet.
The overall design is illustrated in \figref{fig:egun_setup} for the \gls{rs} \gls{egun}. For details, see \cite{PhDBabutzka2014,PhDSack2020}.

\paragraph{Principle and Implementation} 
\label{sec:electron_sources:setup}

To use the electron source for characterization of the \gls{ms} and other components, it is often required to move the electron beam off-center so that radial and azimuthal inhomogeneities can be investigated. This requires selection of a specific magnetic field line to define the electron trajectory.
For the \gls{ms} \gls{egun}, this was achieved by a manipulator which moves the entire source across the magnetic flux tube. By contrast, the \gls{rs} \gls{egun} uses a system of magnetic dipole coils to steer the electron beam in the vertical or horizontal direction.
Both methods allow to illuminate each of the \gls{fpd} pixels individually with the narrow electron beam.

The UV light can be produced by two methods: via a pulsed $\mathrm{Nd\!:\!YVO}_4$ laser, or a continuous light source that emits a variable range of wavelengths.
In the first method, a pulsed laser produces monochromatic UV light ($\lambda = \SI{266+-1}{\nano\meter}$) at pulse frequencies of \SIrange{20}{120}{\kilo\hertz}.
As the light intensity is high, electron rates of \SI{10}{\kilo cps} and higher are easily achieved. Depending on the detector configuration, pile-up effects can distort the observed rate. We correct such effects by a dedicated \gls{daq} filter stage and corresponding simulations; this is not further discussed here. 
By measuring the time it takes for an electron emitted via UV laser pulse to its detection, we can carry out \gls{tof} measurements.
The \gls{egun} \gls{tof} mode allows for differential electron spectroscopy with a \gls{mace} filter \cite{bonn1999ToF}, either for testing novel approaches for determining the neutrino mass \cite{Steinbrink2018,fulst2020ToF,Steinbrink_2013}, for energy loss measurements \cite{PhDSack2020}, or for calibration and commissioning measurements of the \gls{ms} and \gls{wgts} \cite{PhDBarrett2017}. Some applications of this mode are discussed in the following sections.

The second illumination method produces UV light of variable wavelengths, either by an array of UV-LEDs (at the \gls{ms}) or with a \gls{ldls} (at the \gls{rs}).
Both setups use a monochromator in the optical beam line to select a narrow interval of wavelengths, with typical FWHM \SIrange{5}{10}{\nano\meter}. The resultant light in this case is continuous and therefore not usable for \gls{tof} measurements.
However, the variable wavelength of typically \SIrange{250}{300}{\nano\meter} allows to minimize the energy spread of the emitted electrons by adjusting the wavelength such that $h f \rightarrow \Phi$. The \gls{ldls} is also used in some measurements where a higher rate stability is required.
Furthermore, the variable wavelength enables us to determine the photocathode work function via an \emph{in situ} measurement of the electron yield in dependence of wavelength. Such a measurement provides an important fix point that defines the absolute energy scale of the emitted electrons. In turn, this enables us to investigate the work function stability of the \gls{ms} \cite{PhDBehrens2016,PhDSack2020}. 

The light intensity is continuously monitored by a photodiode at the $<1\%$ level to correct for resulting rate fluctuations in the measurements. At the \gls{rs} setup, an active correction is implemented that can further stabilize the achieved electron rate to better than $0.1\%$ \cite{PhDSchimpf2021}. 

\paragraph{Commissioning Results} 

\begin{figure}[!ht]
    \centering
    \includegraphics[width=.85\textwidth]{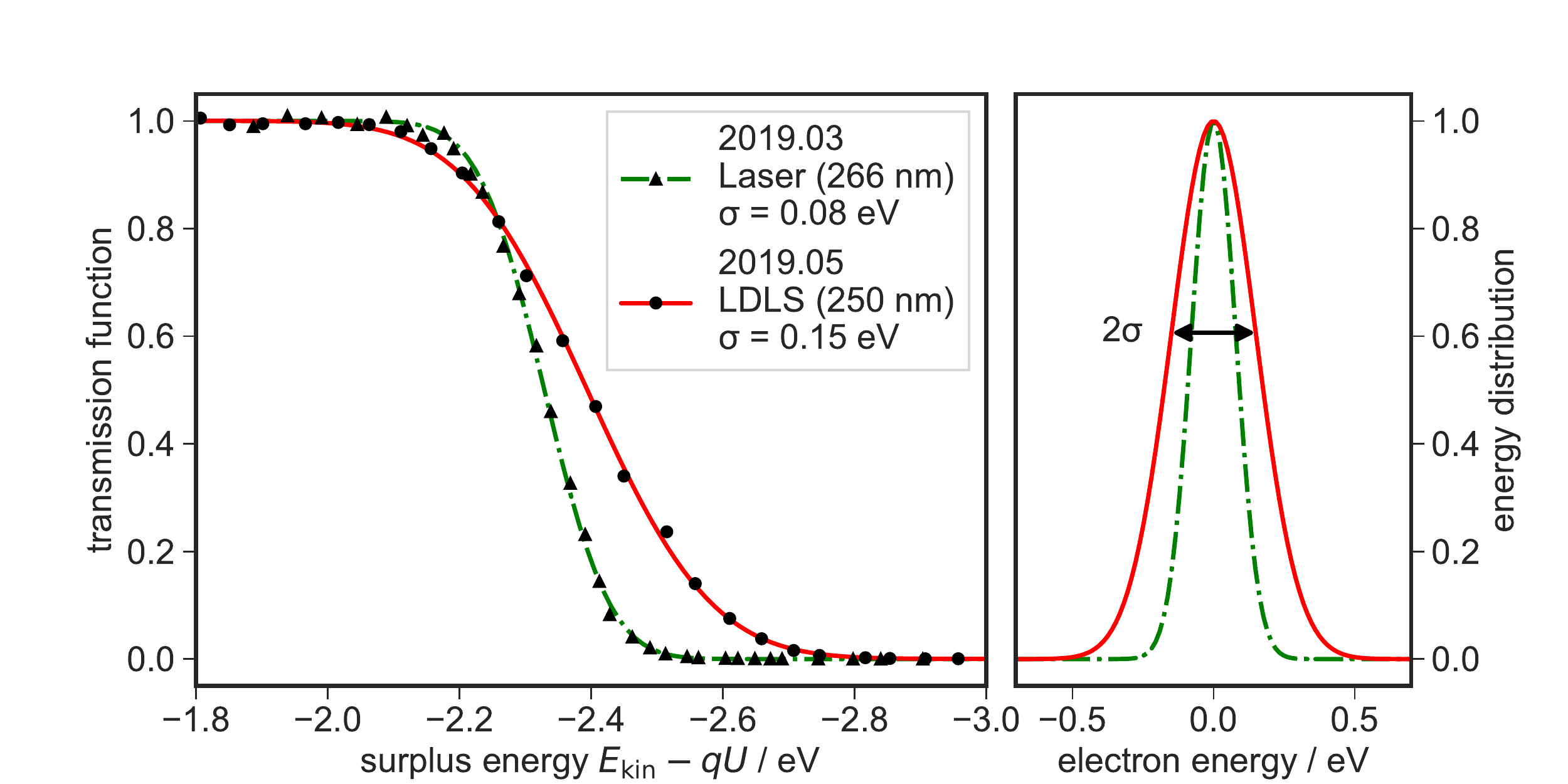}
    \caption{E-gun transmission function with different  light sources: normalized measured electron rate (black data points) and the individual best-fit results (colored lines) are shown; the corresponding energy distribution is shown separately on the right.
         }
    \label{fig:egun-linewidth}
\end{figure}

The main function of the \gls{egun} is to characterize the electron transmission through the spectrometer section. This is done by varying the surplus energy $E - qU$ relative to the retarding voltage $U$ applied to the \gls{ms} electrodes. \figref{fig:egun-linewidth} shows transmission function measurements with different light sources.
The transmission function describes the probability for an electron with energy $E$ to be transmitted, that is to overcome the retarding potential $qU$ of the spectrometer. Electrons that are produced by the \gls{egun} with slightly smaller energy require additional surplus energy to reach the detector. Because the retarding potential in the spectrometer is not completely homogeneous, electrons following the magnetic field lines will experience a slightly lower energy threshold depending on their trajectory. In \figref{fig:egun-linewidth} this is seen as a surplus energy offset of about \SI{-2.2}{eV}, corresponding to an effective retarding potential of \SI{18598.8}{eV} in the \gls{ms} center.

The fit model in this case is a simple error function, which describes the effective energy distribution of the electrons in the \gls{ap}. The energy distribution is shown on the right. For the measurement taken with the \gls{ldls}, the energy is normal distributed with a width of \SI{0.15}{eV}.
Another measurement was taken with the laser at a higher wavelength of \SI{266}{nm}, which reduces the energy spread to \SI{0.08}{eV} because the photon energy is closer to the effective photocathode work function.

In both cases, the achieved energy spread is much better than the design requirement. The setup with multiple variable light sources allows to optimize the electron emission in terms of count rate, rate stability, and energy distribution.

The potential inhomogeneity in the \gls{ms} \gls{ap} is one of the key parameters of the \gls{katrin} setup that can be studied with the \gls{egun}.
Due to the physical design of the \gls{ms}, radial and azimuthal inhomogeneities in the electromagnetic fields cannot be completely avoided.
The \emph{potential depression} $\Delta U_r(r,\phi)$ is defined as the difference between the effective retarding potential in the \gls{ap}, $U_\mathrm{ana}(r,\phi)$, and the high voltage applied to the spectrometer electrodes, $U_\mathrm{IE}$. It is typically about \SIrange{1}{3}{\volt} and depends on the spectrometer configuration.
Electrostatic simulations using the Kassiopeia software \cite{Furse2017} can determine the effective retarding potential for an arbitrary electron trajectory, but need to be validated against dedicated measurements.

\begin{figure}[!ht]
    \centering
    \includegraphics[width=.85\textwidth]{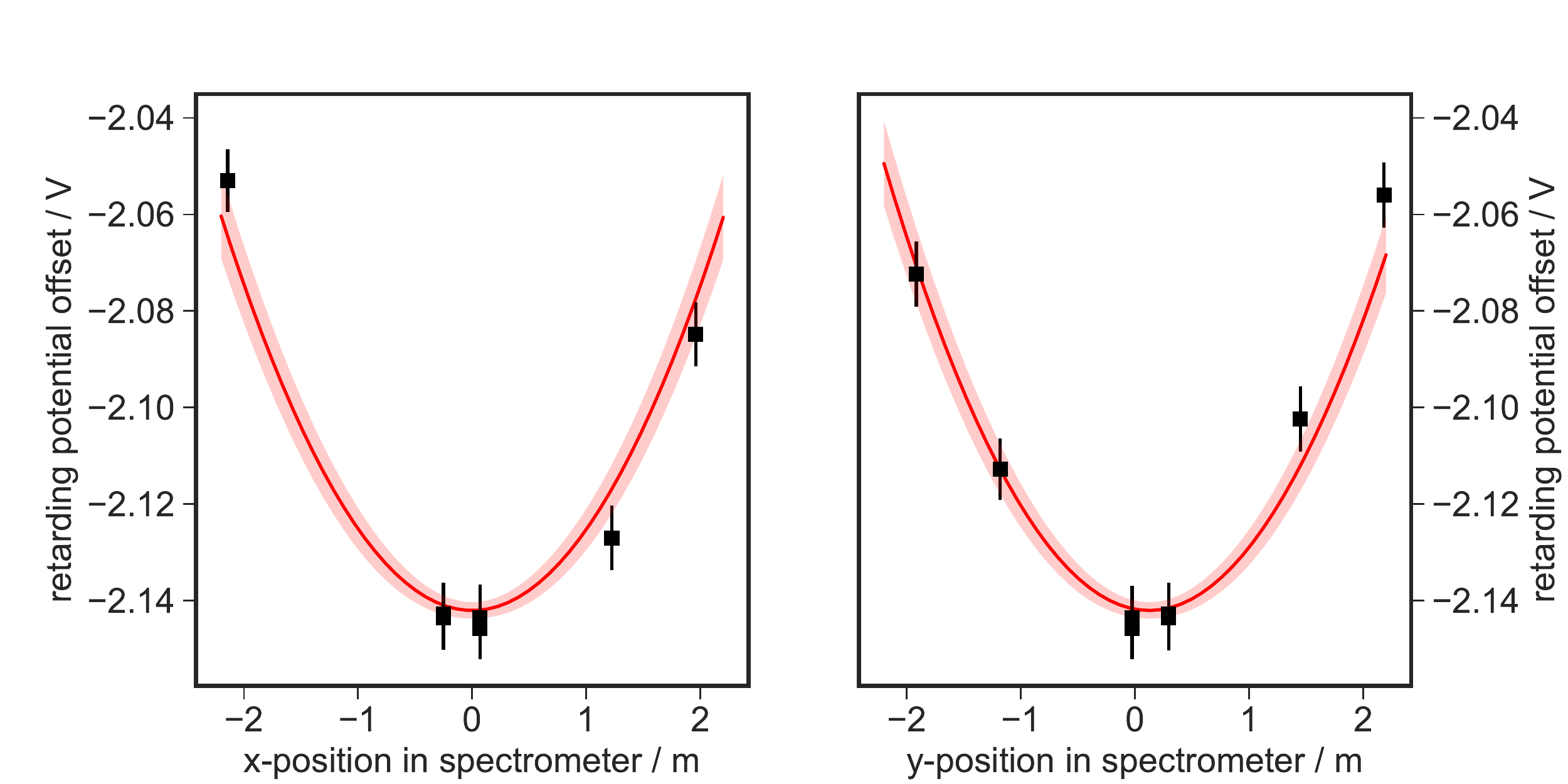}
    \caption{Potential mapping of the analyzing plane with the \gls{egun} in direction of the horizontal and vertical axes, and comparison to simulations. Shown is the determined potential inhomogeneity $\Delta U_r$ (black data points), and simulation results for the same position in the analyzing plane (colored lines with $3\sigma$ uncertainty bands). The potential inhomogeneity follows a parabolic distribution.
         }
    \label{fig:egun-ana-potential}
\end{figure}

A measurement of the retarding potential with the \gls{egun} during the first neutrino mass campaign is shown in \figref{fig:egun-ana-potential}.
The potential in the \gls{ap} is determined by measuring the transmission function for $\theta=$ \SI{0}{\degree} on different detector pixels. Each measurement is fit with a model that describes the energy and angular distribution of the electrons separately \cite{Behrens2017} and includes the characteristics of the \gls{mace} filter. Another free parameter describes the observed energy shift between the transmission functions; for details see \cite{PhDBlock2021}.
Because the electron energy $E$ is stable over time and narrowly distributed, the shift directly corresponds to the potential depression $\Delta U_r(r,\phi)$ over different detector pixels.

The measured radial inhomogeneity in horizontal and vertical direction is reproduced accurately by simulations in the given magnetic field setting, which reaches $B_\mathrm{min} = \SI{0.63}{\milli\tesla}$ in the center of the \gls{ap}. The potential calculations are based on a detailed 3D model of the \gls{ms} which includes deformations of the vessel hull.
A global potential offset of \SI{-0.181}{\volt} was added to account for work function differences between the spectrometer electrodes and the \gls{egun} photocathode, that cannot be included in the simulations.
The maximum potential depression in the \gls{ap} center then amounts to $\Delta U_{r=0} = \SI{-2.14}{\volt}$, or an absolute retarding potential of $U = \SI{-18597.86}{\volt}$ at an applied voltage of $U_\mathrm{IE} = \SI{-18600}{V}$, which is in agreement with our expectations. 
Unavoidable misalignments between beam line components lead to small deviations between the simulation and measurements that are visible in \figref{fig:egun-ana-potential}; these effects are currently being investigated.

The \gls{tof} mode that is available with a pulsed light source allows a differential measurement of the electron energy, because the \gls{tof} is determined on an event-by-event basis \cite{bonn1999ToF}.
In this case, the \gls{tof} is affected by electromagnetic fields not only in the spectrometer \gls{ap}, but also in a large region of several meters around the center of the \gls{ms}. This behavior allows us to verify the integrity of the \gls{ms} inner electrode system, and to validate the simulation model by comparing the \gls{tof} measurement results to particle-tracking simulations.

\begin{figure}[!ht]
	\centering
	\includegraphics[width=.8\textwidth]{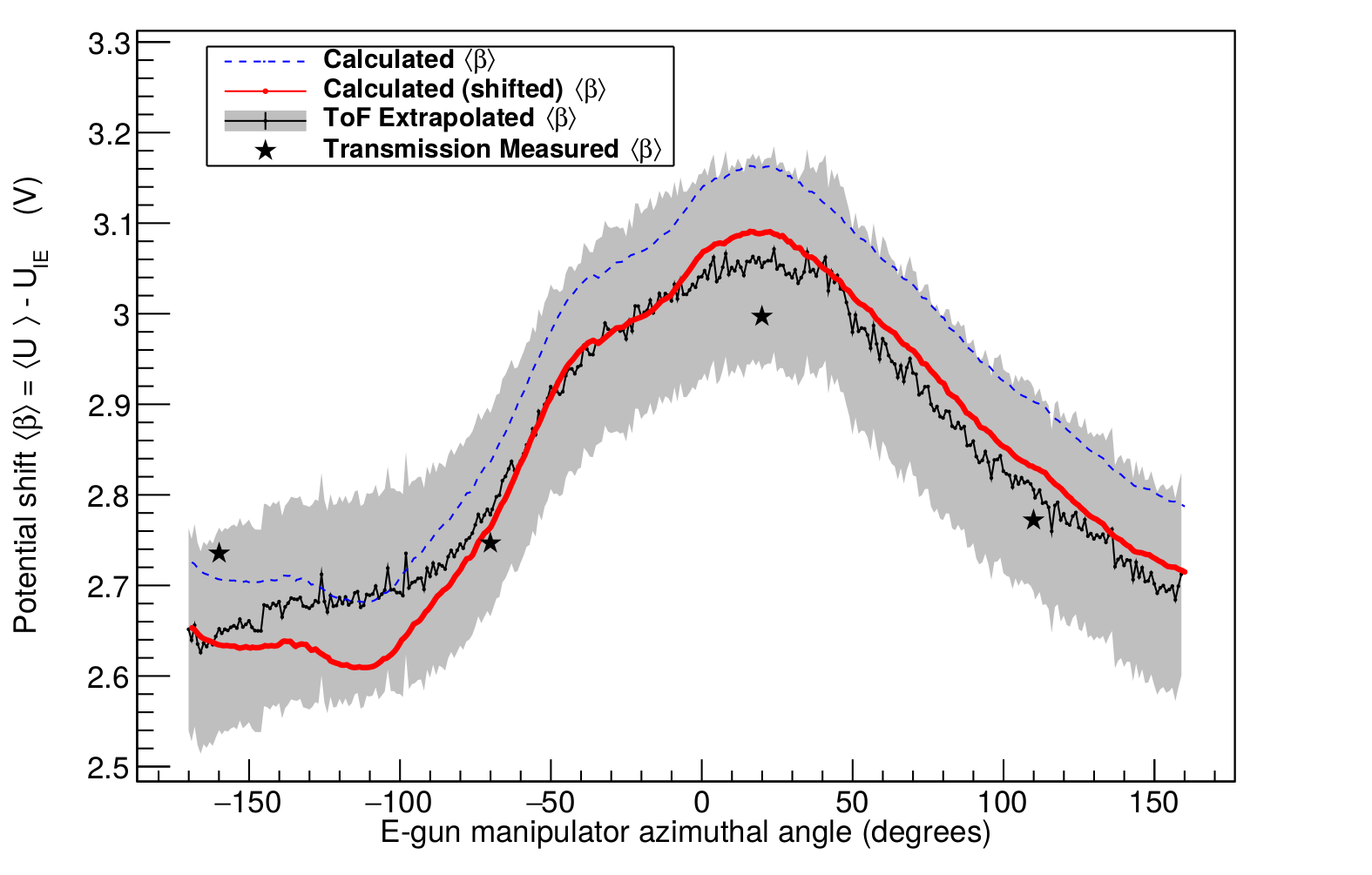}
	\caption{Variation of the spectrometer retarding potential, extrapolated from an \gls{egun} \gls{tof} measurement (black line with gray uncertainty band).
        Four additional transmission function measurements are shown as well (black stars).
        The measurements are compared to simulations (dashed blue line), to which a global potential offset was applied (red line).
        From \cite{PhDBarrett2017}.
        }
 	\label{fig:egun-tof-potential}
\end{figure}

\figref{fig:egun-tof-potential} shows the result of one such \gls{tof} measurement at different azimuthal positions of the \gls{egun} at $B_\mathrm{min} = \SI{0.38}{\milli\tesla}$. It was performed during the spectrometer commissioning campaign.
The observed electron \gls{tof} at each measurement point depends on the electron surplus energy and changes with the electric potential in the spectrometer, so that the potential shift $\beta = \langle U \rangle - U_\mathrm{IE}$ can be determined.

The scan was performed in \ang{1} steps with electron source positioned on a circle with fixed radius. The electron trajectories reach a maximum radius $r_\mathrm{max} \approx \SI{3.7}{\meter}$ in the \gls{ap}; a high radius was chosen in order to increase the sensitivity to local field variations caused by the inner electrode system. 
At each step the \gls{tof} was measured for several electron energies in the range \SIrange{-1}{10}{\electronvolt} relative to the spectrometer voltage (here reduced to $U_\mathrm{IE} = \SI{-800}{\volt}$). 
In addition, four transmission function measurements were performed at \ang{90} intervals, which allow to relate the measured \gls{tof} to the retarding potential in the \gls{ap}, which is shown in the figure.

The measurement results are in good agreement with the simulations. This validates the simulation model to a relative error of \SI{0.03}{\percent} \cite{PhDBarrett2017}. Analogous to the measurement above, a global potential offset of \SI{-0.07}{V} was applied.
A clear azimuthal dependency is observed, which is largely caused by the gravitational deformation of the vessel hull. No small-scale deviations are seen that could point to broken wires or unknown short-circuits in the inner electrode system.

In addition to gaining insight into the integrity of the \gls{ms} inner electrode system, \gls{tof} mode measurements can be used to study electron scattering in the \gls{wgts}. Results from these studies will be discussed in a separate publication.

\subsubsection{Condensed \kr{} Source}
\label{Subsubsection:CKrS}

The \acrfull{ckrs} is part of the calibration and monitoring infrastructure of the \gls{katrin} experiment. It is located at the \gls{cps} upstream from the \gls{ps}, and provides isotropic quasi-monoenergetic electrons from a thin film of \kr{}.

\paragraph{Principle and Requirements} 

The \kr{} which provides the conversion electrons is emanated from a rubidium generator and travels through a capillary towards a \SI{2x2}{\centi\metre} \gls{hopg} substrate kept at \SI{25}{\kelvin}, where it condenses in a sub-monolayer film. Because the substrate area of \SI{4}{\square\centi\metre} (further limited by a $\SI{2}{\centi\metre}$-diameter aperture) covers only a fraction of the entire flux tube, the \gls{ckrs} illuminates only about a single \gls{fpd} pixel.
This small illumination area can be advantageous: it enables a high-precision measurement of the arrival point of electrons emitted from a quasi-pointlike, monoenergetic, isotropically-emitting source.
For a full coverage of the detector, a 2D movement system is integrated into the \gls{ckrs} setup, allowing for control over the positioning of the substrate anywhere within the flux tube. It also enables complete retraction of the source from the beamline.

Using the \gls{ckrs} to calibrate the spectrometers requires a stringent energy stability of tens of \SI{}{\milli\electronvolt} over hours or even days. As the kinetic energy of the emitted conversion electrons is influenced by surface effects on the \gls{hopg} substrate, it is crucial to control and monitor the conditions of the condensed film and to minimize contaminations by residual gas molecules. To reach the UHV pressures in the \SI{e-10}{\milli\bar} range, the complete system undergoes a bake-out prior to operation and is pumped by two \glspl{tmp} and a getter pump. For reproducible starting conditions, the substrate is cleaned by heating and laser ablation before a new film is prepared.

Precise \textit{in-situ} monitoring of the film thickness and its properties is possible by means of a laser ellipsometry system included in the setup.

\paragraph{Setup} 

\begin{figure}[!ht]
	\centering
	\includegraphics[width=.75\textwidth]{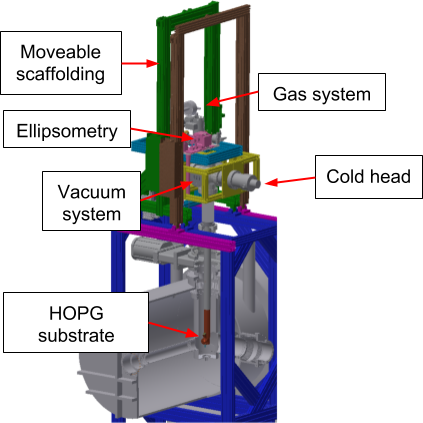}
	\caption{An overview of the \gls{ckrs} setup inside the \gls{cps}. For orientation: downstream is to the left, upstream is to the right. From \cite{PhDDyba2019}}
	\label{fig:ckrs}
\end{figure}

The \gls{ckrs}, installed on top of the \gls{cps}, can be driven into the beamline. Several subsystems, including the cold head, ellipsometry system, rubidium generator, gas and vacuum system, are placed on a movable carriage, which is electrically isolated from the rest of the scaffolding. \figref{fig:ckrs} shows a 3D model of the \gls{ckrs} inserted into the beamline.

\paragraph{Commissioning Results}

\begin{figure}[!ht]
	\centering
	\includegraphics[width=.9\textwidth]{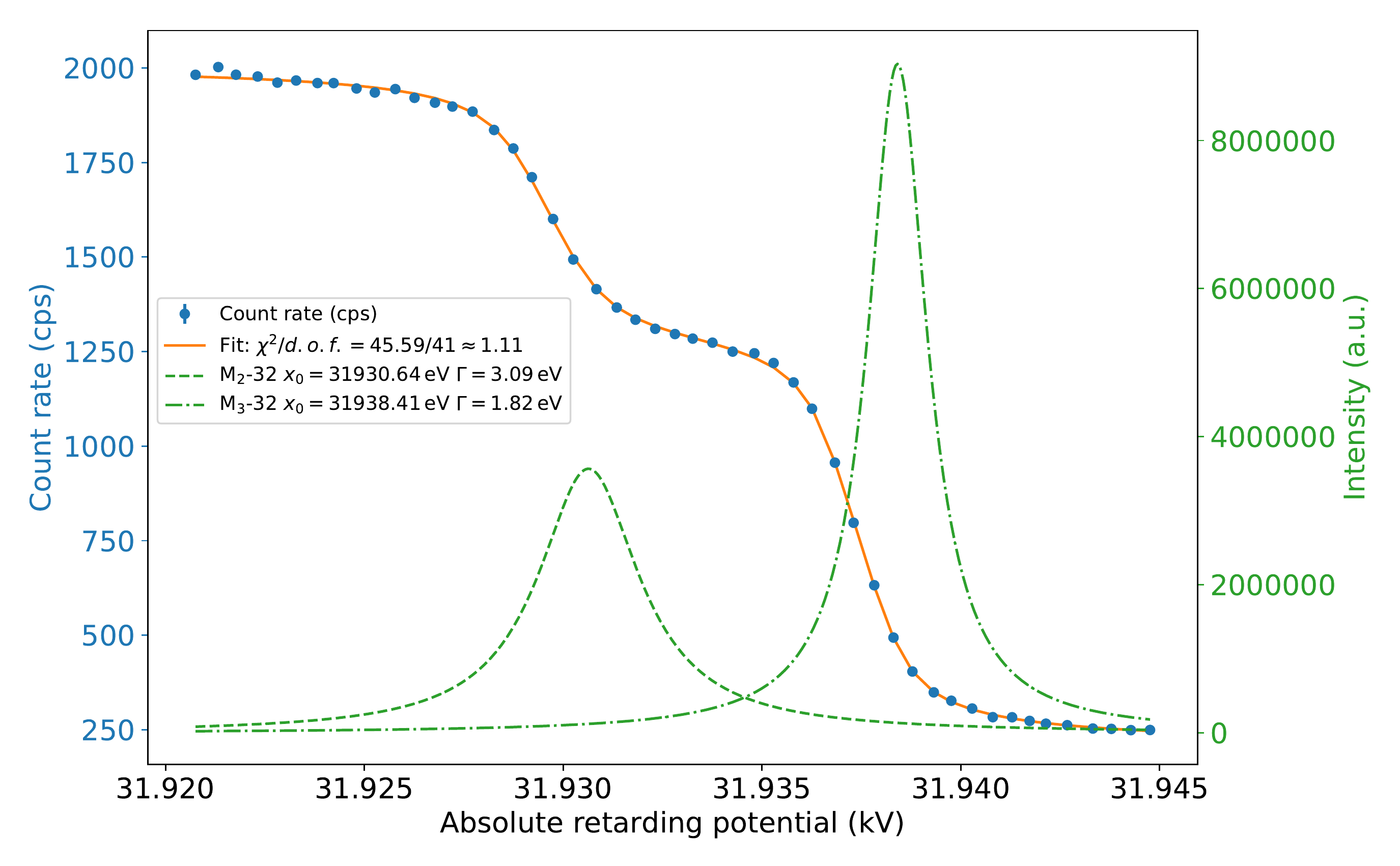}
	\caption{Fit of a single pixel measurement of the $\text{M}_2-32$ and $\text{M}_3-32$ conversion lines with the \gls{ckrs}. An additional Gaussian broadening with $\sigma=\SI{0.130}{eV}$ is included in the Lorentzian peak shapes to account for surface effects. }
	\label{fig:ckrs_3}
\end{figure}

The \gls{ckrs} was installed at the \gls{katrin} experiment in the first half of 2017, and was used during the \kr{} commissioning measurements in July 2017~\cite{Arenz2018}.
During a one-week campaign, the setup worked as intended and all available conversion electron lines were measured. \figref{fig:ckrs_3} shows an example scan for the \mthirtytwos{} doublet conversion electron line. The resulting spectrum is a convolution of the natural Lorentzian line shape with the transmission function of the \gls{ms} and the data demonstrates the good energy resolution of the spectrometer.

However, due to short preparation time, the vacuum conditions were not as good as desired.
Condensation of residual gas contaminated the film, resulting in a drift of the line position by about \SI{300}{\milli\electronvolt} in \SI{100}{\hour} (for the M2 and M3 lines). This drift seems to have subsided for later times, and the behavior can be well described by an image charge model, taking into account 3 to 4 layers of gas.

In order to improve over the first measurements, an extended bake-out of the system with more sophisticated hardware has been carried out before the four-week measurement campaign in July and August 2018. Of those four weeks, two were dedicated to improve the energy stability via pre-plating the substrate with \SI{3}{\nano\metre} of stable krypton before condensing the radioactive \kr{} to diminish the image charge influence.

\begin{figure}[!ht]
	\centering
	\includegraphics[width=.9\textwidth]{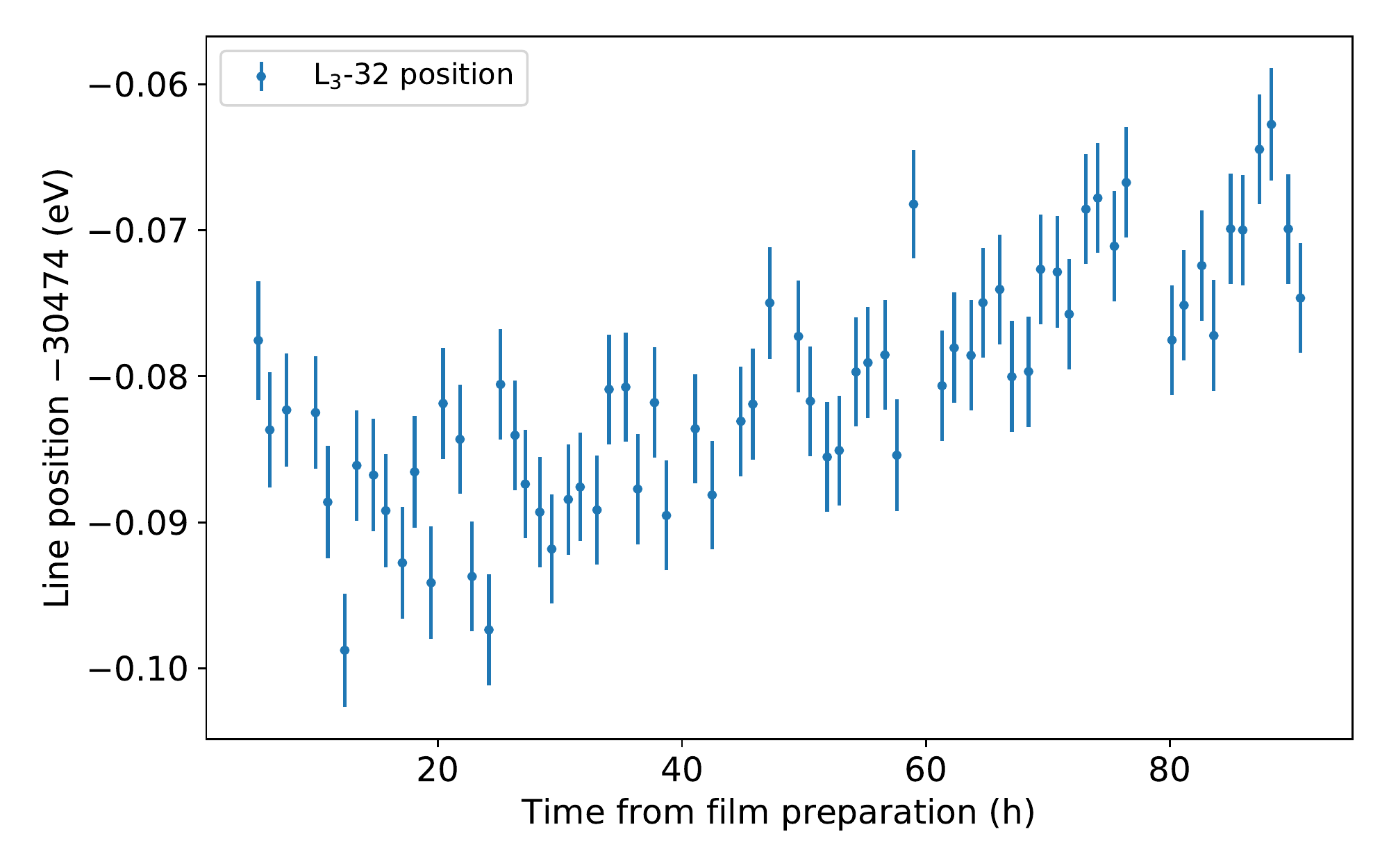}
	\caption{Fitted position of the \lthreethirtytwo{} conversion electron line over time, after an initial time span of \SI{2}{h} to allow for equilibration. The energy is stable to within \SI{30}{\milli\electronvolt} over a \SI{100}{\hour} interval.}
	\label{fig:ckrs_4}
\end{figure}

These measures improved the observed line stability of the \lthreethirtytwo{} conversion electrons to \SI{30}{\milli\electronvolt} in \SI{100}{\hour} \cite{PhDFulst2020,PhDMariiaFedkevych2019}, as shown in \figref{fig:ckrs_4}.

With the stability requirement satisfied, a scan of the analyzing plane electromagnetic fields was performed. This involves scanning the \lthreethirtytwo{} line with the \gls{ckrs} in different positions within the magnetic flux tube. Analysis of these measurements is  ongoing but should provide precise values for the magnetic fields and electric potentials in the analyzing plane, which can be used to cross-check values obtained by simulations.

The \gls{ckrs} can also be used for an absolute calibration of the HV system via measuring the difference between the \lthreeninefour{} line and the \lthreethirtytwo{} line~\cite{Arenz2018b}.


\subsubsection{Gaseous \kr{} Source}
\label{Subsubsection:GKrS}

The \acrfull{gkrs} has historically been the preferred source for energy calibrations and test measurements. It was originally applied at the Los Alamos tritium experiment~\cite{Wil87, Wil88, Sta89, Bow90, Rob91}, and later at the Livermore experiment~\cite{Sto88, Sto95}. The Troitsk group employed a \gls{gkrs} for systematic studies of the plasma effect in their \gls{wgts}~\cite{Bel08}. Recently, another such \gls{gkrs} was successful in the detection of cyclotron radiation from relativistic electrons, which is the basic technique in a future tritium-based neutrino mass experiment, Project-8 \cite{For15}.

\paragraph{Principle and Requirements}

The motivation for using the \gls{gkrs} in the \gls{katrin} experiment is straightforward: \kr{} monoenergetic electrons can be used to characterize some of the effects that influence the tritium \betael{} spectral shape. While there is much overlap with the \gls{ckrs}, there are some effects which can be explored uniquely with the \gls{gkrs}. These include the distortion of the \betaspec{} from energy losses of the \betaels{} in the tritium gas, and the stability and spatial inhomogeneity of the \gls{wgts} plasma potential induced by tritium.

The \gls{gkrs} also has several properties which make it a favorable candidate for the aforementioned tasks.
By injecting the \kr{} gas into the \gls{wgts}, the conversion electrons are subject to similar conditions as the tritium \betaels{}.
The \kr{} gas can also comingle with tritium or with \d2{} test gas in the \gls{wgts}.
There is much detailed atomic and nuclear data on the \kr{} and its radiations~\cite{Venos2018}.
Its half-life T$_{1/2} = \SI{1.8620+-0.0019}{\hour}$~\cite{Sentkerestiova2018} is short enough to avoid long-term contamination of the beamline, thus is unlikely to increase the total background rate.
Additionally, the half-life of the parent isotope \rb{} (T$_{1/2} = \SI{86.2+-0.1}{\day}$) is long enough to allow hours-long \gls{gkrs} measurements with a practically stable activity.

\begin{figure}[!ht]
	\centering
	\includegraphics[width=0.6\textwidth]{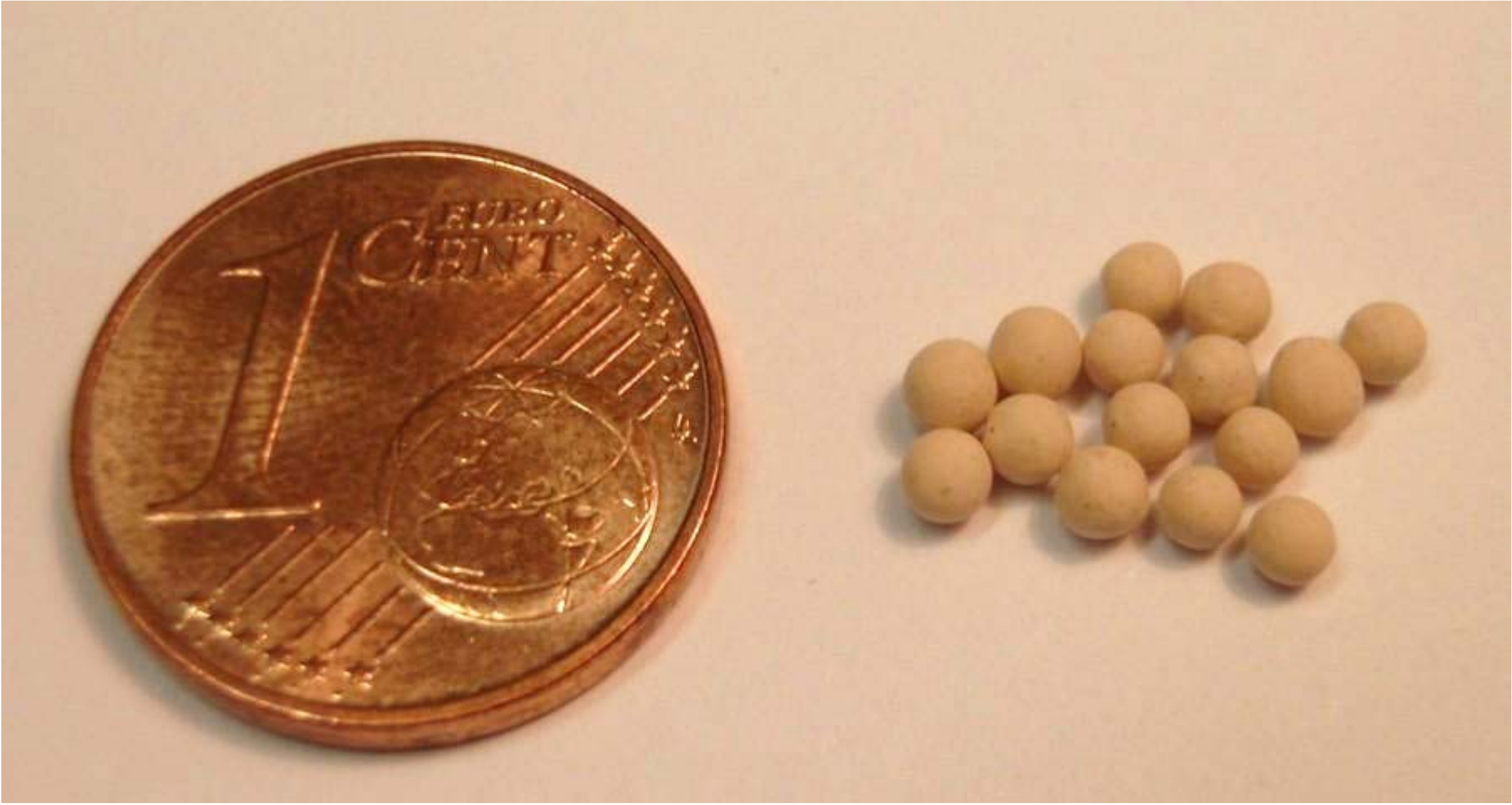}
	\caption{Photo of the \rb{}/\kr{} zeolite source. The zeolite is of type 5A, with spherical beads on average \SI{2}{\milli\meter} in diameter.}
	\label{fig:ZeoliteSource}
\end{figure}

The main component of the \gls{gkrs}, the \rb{}/\kr{} source, was developed at NPI in \v{R}e\v{z}, where \SI{1}{\giga\becquerel} of \rb{} had been deposited inside 15-30 zeolite (molecular sieve) beads (\figref{fig:ZeoliteSource}).

A large amount of the \kr{} produced in the decay of the parent isotope emanates into the vacuum at the room temperature, while the \rb{} is firmly bound in the zeolite beads. For the \gls{katrin} experiment, the \rb{} is produced at the cyclotron of the Centre of Accelerators and Nuclear Analytical Methods (CANAM) at the Nuclear Physics Institute \v{R}e\v{z}~\cite{Venos2014}.

\paragraph{Source Setup} 
\label{gkrssetup}

\begin{figure}[!ht]
	\centering
	\includegraphics[width=1.0\textwidth]{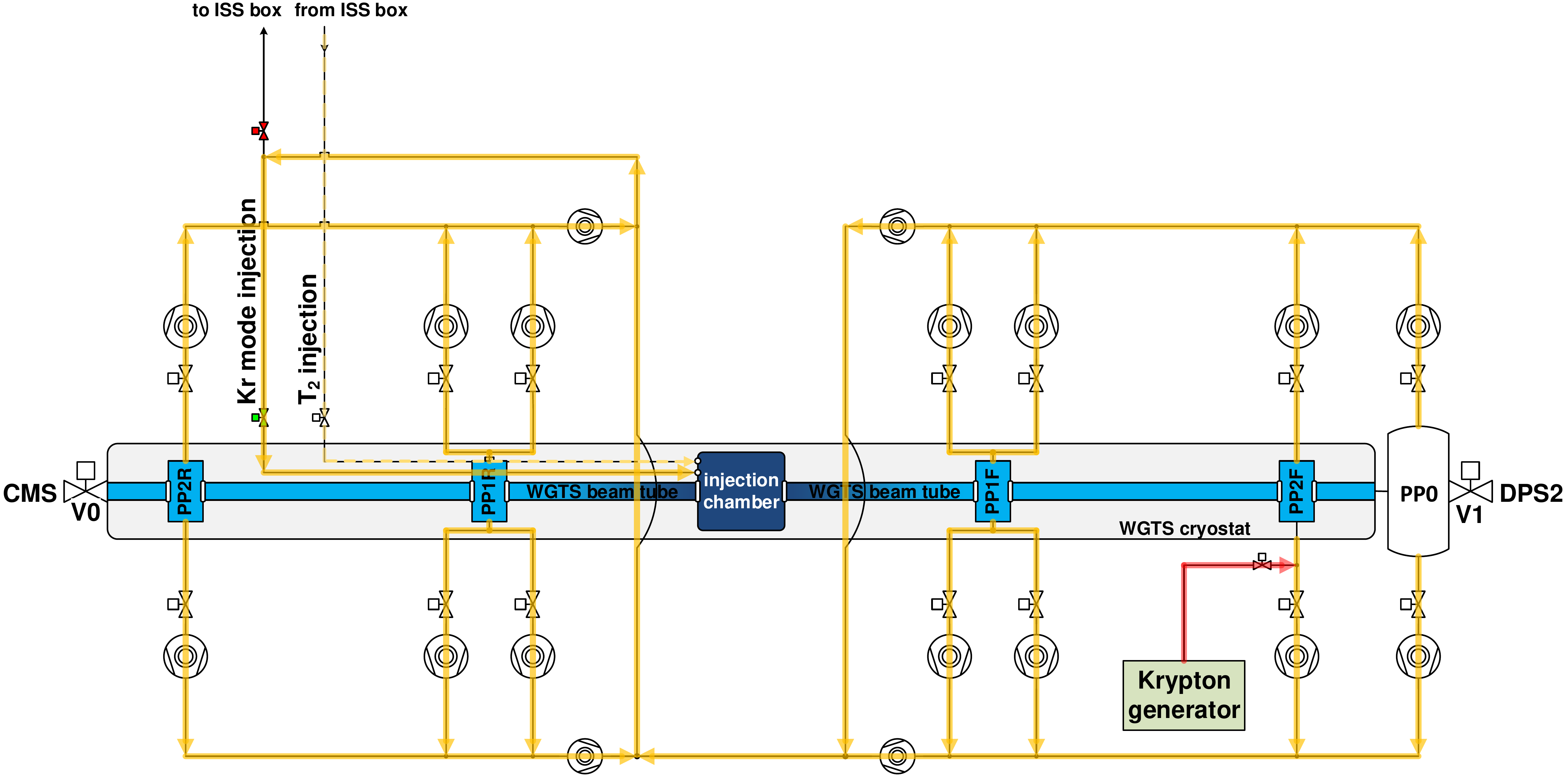}
	\caption{The scheme of \kr{} injection into the \gls{wgts}: \kr{} generator connection to \gls{pp2f} (red), krypton circulation tubing (yellow), pumps and loop valves (green). The abbreviations CMS, V0, V1, ISS, RS, PP and DPS2 stand for the Control and Monitoring System, beamline Valves 0 and 1, Isotope Separation System, Rear Section, PumpPort and Differential Pumping Section 2, respectively. }
	\label{fig:KrInjection}
\end{figure}

The \gls{gkrs} consists of the \kr{} generator, \gls{wgts} tube, connection tubing and chambers in between. The output of the generator is attached to the outlet of the second \gls{wgts} \gls{pp2f} (\figref{fig:KrInjection}).

\begin{figure}[!ht]
	\centering
	\includegraphics[width=0.7\textwidth]{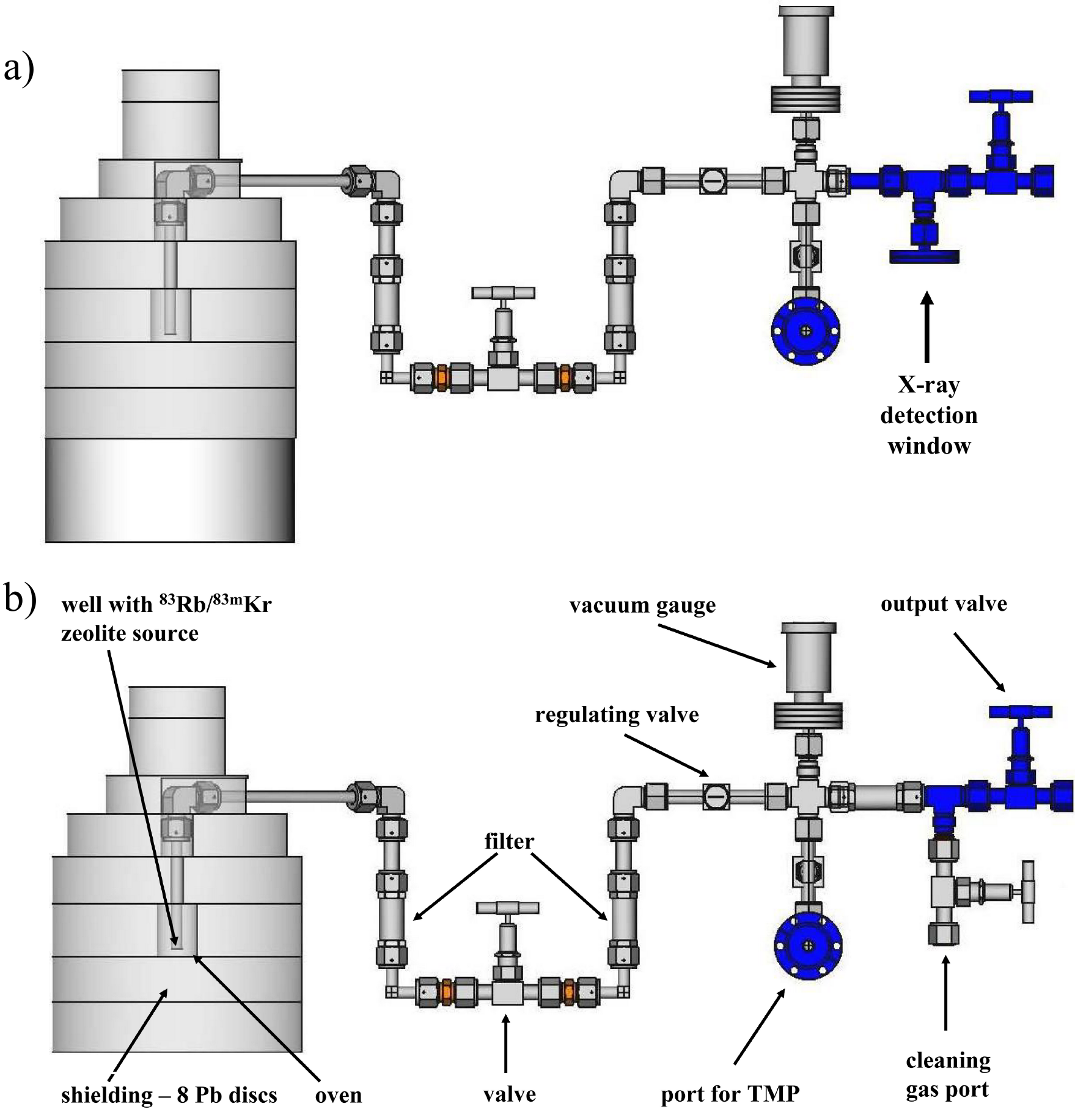}
	\caption{Schematics of the \kr{} generator. a) During commissioning an acrylic window for X-ray detection was installed.  b) Final setup as installed at the \gls{katrin} beamline, the X-ray detection window is replaced by the cleaning gas port.
	}
	\label{fig:Gen}
\end{figure}

\begin{figure}[!ht]
	\centering
	\includegraphics[width=0.8\textwidth]{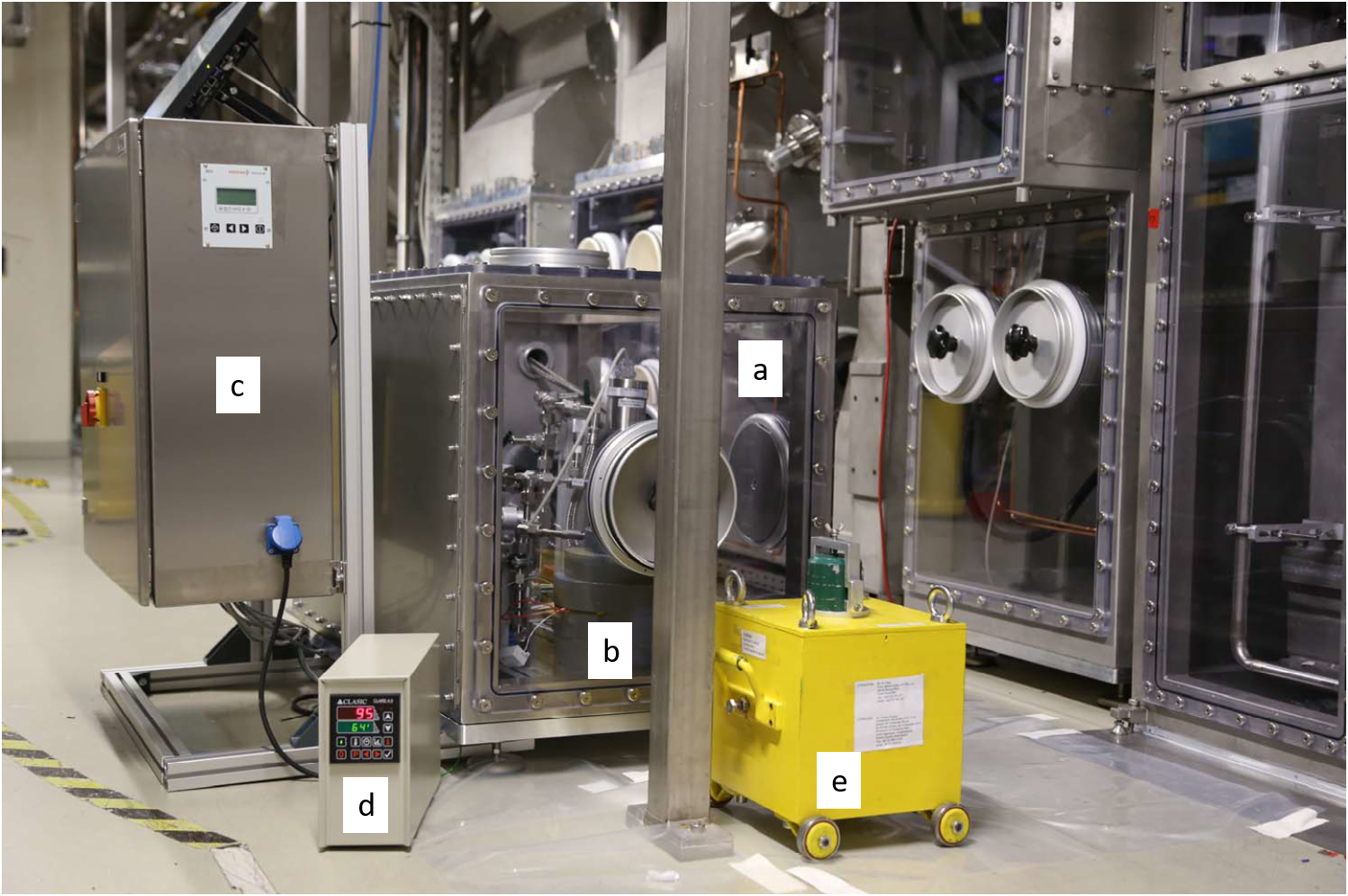}
	\caption{The \kr{} generator and transport container of the \rb{}/\kr{} zeolite source. Shown here are: (a) secondary containment with the generator (b) installed inside, (c) power distribution cabinet, (d) oven controller, and (e) transport container. }
	\label{fig:GenPhoto2}
\end{figure}

\begin{figure}[!ht]
	\centering
	\includegraphics[width=0.8\textwidth]{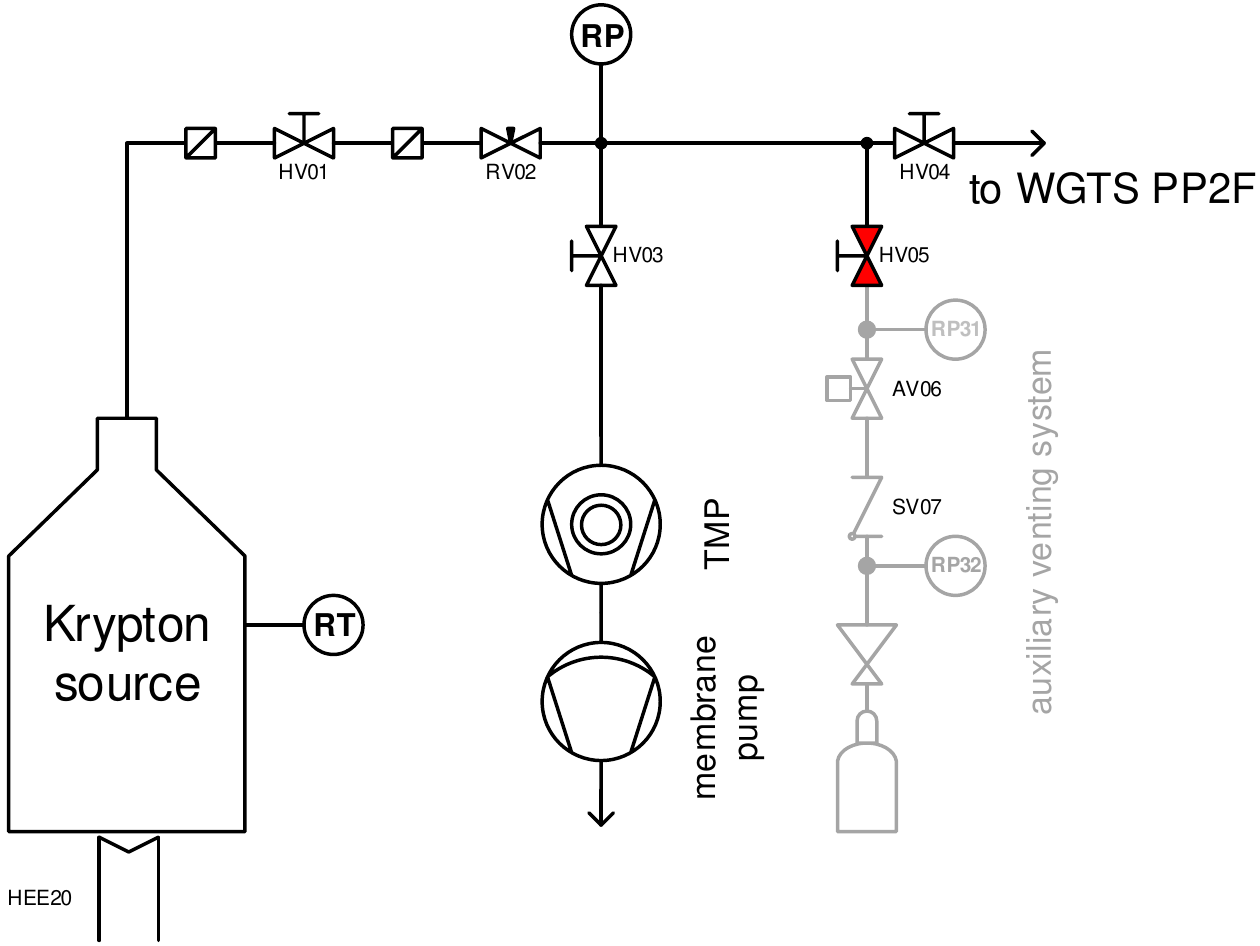}
	\caption{The schematics of the \kr{} generator control computer, which monitors the status of the vacuum valves and the set of \gls{tmp} and membrane pumps. }
	\label{fig:GenFotoScheme2}
\end{figure}

The \kr{} gas from the generator is pumped with the \gls{tmp} at \gls{pp2f} into the \gls{wgts} loops and further through a \SI{4}{\milli\meter}-diameter capillary to the tritium injection chamber at the middle of the \gls{wgts} tube.
In the krypton calibration mode, the capillary and the \gls{wgts} tube are kept at \SI{100}{\kelvin}. Operating at these higher temperatures is necessary, as a substantial loss of \kr{} would arise at lower temperatures due to freeze-out on the walls.
The \kr{} generator scheme is shown in \figref{fig:Gen}. The generator setup uses stainless-steel \SI{12.7}{\milli\meter} VCR components \footnote{Swagelok, VCR fittings, \url{https://www.swagelok.com/en/product}}.
The \rb{}/\kr{} zeolite source is situated at the bottom of the well, embedded in a cylindrical oven which is also nested in lead shielding.
The oven is used to bake the zeolite beads (\SI{3}{\hour} at \SI{200}{\celsius}) before activating the generator.
This allows for the removal of residual gases absorbed by the beads during storage in air.
A secondary benefit of baking is to avoid the deterioration of the \gls{wgts} vacuum after the generator is connected.
Two sintered filters with \SI{0.5}{\micro\meter} pores prevent aerosol and small zeolite abrasions, which might contain \rb{}, from diffusing into the \gls{wgts}.
A regulating valve controls the amount of \kr{} gas that flows into the \gls{wgts}.
For generator maintenance, a port for cleaning the gas is also installed.
The generator itself is shown in \figref{fig:GenPhoto2}; its flow diagram in \figref{fig:GenFotoScheme2}.

As displayed in \figref{fig:GenPhoto2}, the generator is inside the secondary containment, which prevents the release of tritium into the environment.

\paragraph{Commissioning Results} 

The \kr{} generator itself was commissioned at \gls{npi} \v{R}e\v{z}. For this, a T-piece equipped with a  \SI{1.4}{\milli\meter}-thick acrylic window was connected to the generator tubing. The amount of \kr{} within the generator was determined from the X-ray intensity with a Si(Li) detector (see \figref{fig:Gen}). The zeolite \rb{}/\kr{} source had an initial \rb{} activity of \SI{1.3}{\giga\becquerel}.

The commissioning of the \gls{gkrs} was accomplished in two steps. In the first step, the \kr{} generator was attached and the count rates at a fixed \gls{ms} retarding voltage of \SI{-31800}{\volt} were monitored. In the second step, the \kthirtytwo{} electron conversion line was scanned and its energy spectrum was reconstructed. Additional details on conducting this operation safely can be found in \cite {Arenz2018}.


The first commissioning step was successfully passed without a complete \gls{katrin} beamline. The \kr{} from the generator, which was temporarily equipped with a beryllium window for monitoring the activity using an \gls{sdd}, diffused into the \gls{wgts} via \gls{pp2f}.

\begin{figure}[!ht]
	\centering
	\includegraphics[width=0.8\textwidth]{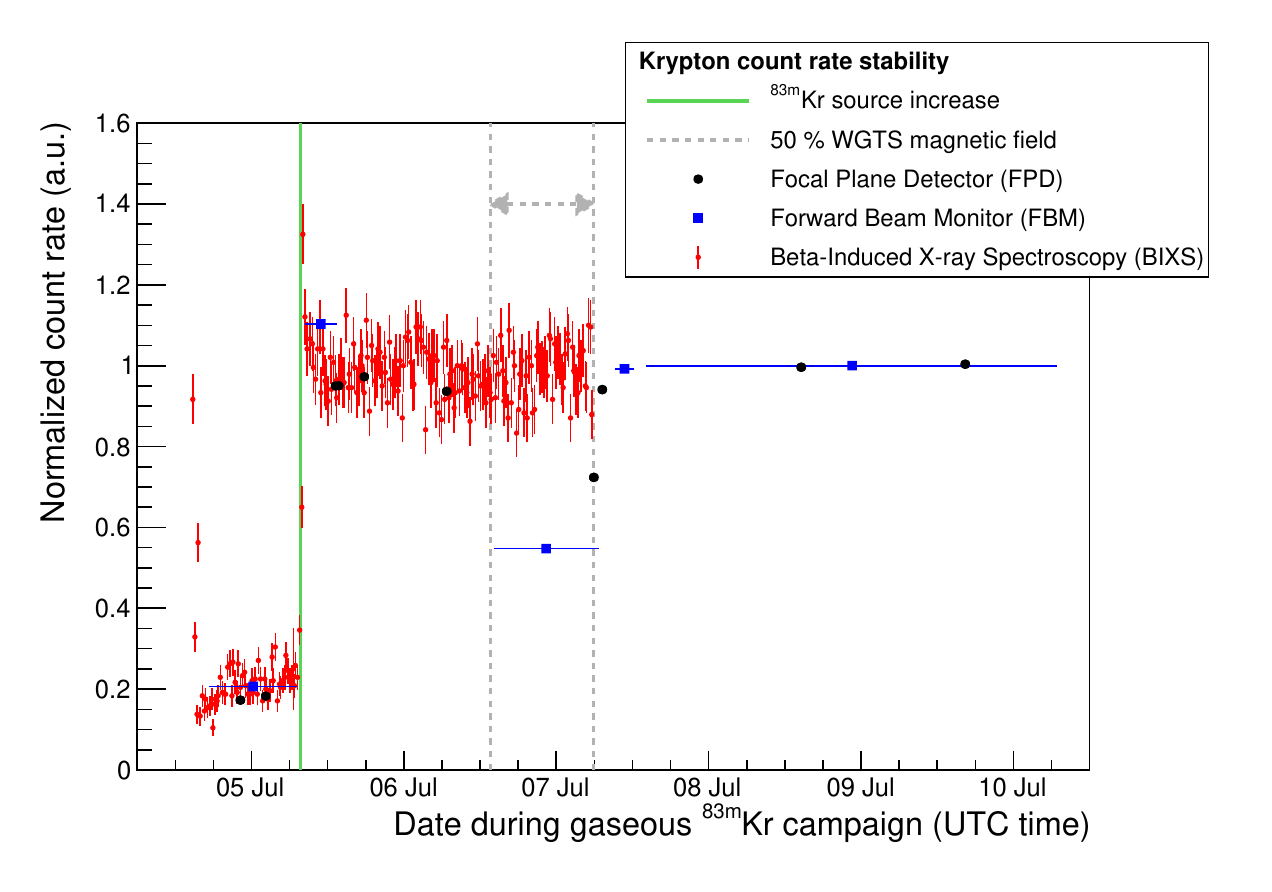}
	\caption{Time dependence of the normalized count rates from \gls{fpd}, \gls{fbm} and \gls{bixs} detectors. The data from all three detectors are normalized to those in their respective stable regions. The fivefold increase in count rate on all detectors is visible after the \kr{} flow was increased. The vertical lines indicate the dates when there were hardware changes. The statistical uncertainties in the \gls{fpd} and \gls{fbm} data are too small to be visible.}
	\label{fig:KrRateStab2017}
\end{figure}

Four detectors were operational during commissioning: three along the beamline (\gls{bixs}, \gls{fbm} and \gls{fpd}), and the \gls{sdd} detector at the generator.
After the generator outlet valve was opened, the measured count rates at each detector stabilized within \SI{\sim 30}{minutes}.
The normalized count rates from three \gls{katrin} detectors are shown in \figref{fig:KrRateStab2017}.

As expected, the observed count rate in the \gls{fpd} and \gls{fbm} decreased when the magnetic field of the \gls{wgts} was lowered. The \gls{bixs} system is not sensitive to changes in magnetic field, so its observed count rate did not change.
At an \gls{ms} retarding voltage of \SI{-31800}{\volt}, the conversion electrons from the \SI{32}{\kilo \electronvolt} decay's M and N shells reached the \gls{fpd}, at a count rate of \SI{\sim 23}{\kilo cps}. 
Taking into account the probability of the parent \rb{} decaying to a particular \kr{} isomeric state, the emanation factor, and several other correction factors, the final calculated yield of \kr{} is \SI{0.16}{\percent}.

The \gls{fpd} background rate at this setting was measured before and after the \gls{gkrs} application, and was found to be \SI{532+-12}{\milli cps} and \SI{565+-12}{\milli cps}, respectively. The increase in background rate is not so significant, demonstrating that the long-lived \rb{} is effectively trapped in the zeolite beads or in the generator filters, and therefore does not contaminate the \gls{katrin} experiment.


The second step of the \gls{gkrs} commissioning proceeded with the standard setup as described in \secref{gkrssetup}, except that the activity of \rb{} in the source was \SI{0.89}{\giga\becquerel}.

After the generator outlet valve was opened, the \gls{wgts} pressure increased from \SI{1.14e-7}{\milli\bar} to \SI{1.24e-7}{\milli\bar} for about \SI{30}{\second}.
This demonstrates that baking was sufficient to remove residual gases in the zeolite beads.
The electron count rate measured at the \gls{fpd} stabilized within \SI{\sim 2}{\hour}, and the M and N conversion electron rate reached \SI{40}{\kilo cps}. Using a similar procedure as in the first commissioning step, the \kr{} activity visible to the \gls{fpd} reached \SI{2.1}{\mega\becquerel}, with a yield of \SI{0.4}{\percent}.

\begin{figure}[!ht]
	\centering
	\includegraphics[width=0.8\textwidth]{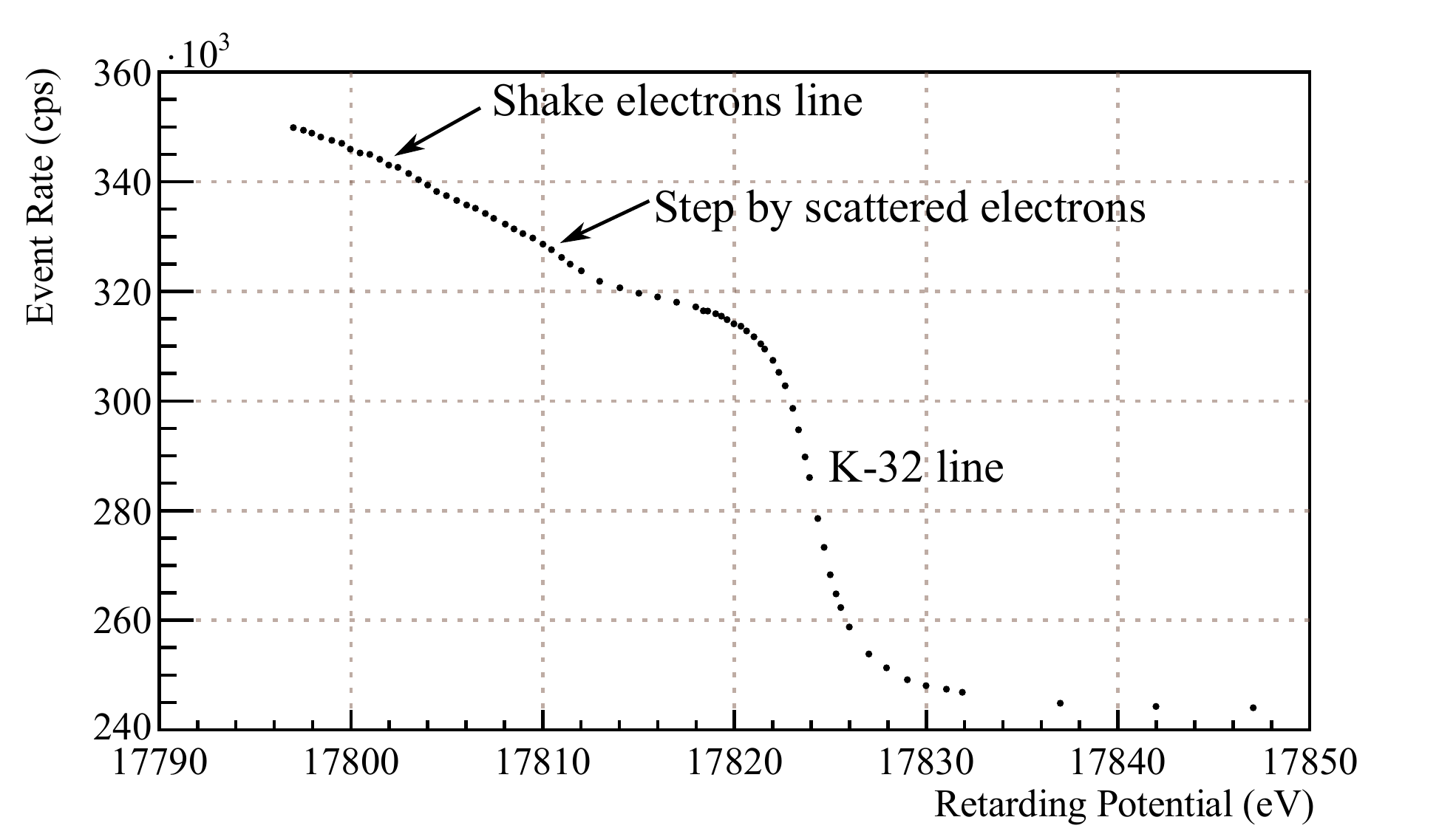}
	\caption{An example of the integral spectrum of \kthirtytwo{} conversion electrons measured in the broad energy range \SIrange{-26}{18}{\electronvolt}, relative to the line position at \SI{17824.2}{\electronvolt}. Besides the main K line, the shake line and the step due to the inelastically scattered electrons are visible. }
	\label{fig:K32Scan}
\end{figure}

The analysis of a typical spectrum of \kthirtytwo{} electrons taken within \SI{15}{\minute} (shown in \figref{fig:K32Scan}) gave a nearly negligible statistical uncertainty in the line position (\SI{4}{\milli\electronvolt}), which indicates an adequate \gls{gkrs} intensity.
Because the measurement was made with an admixture of \d2{} gas, a characteristic step due to electron energy loss is visible in the spectrum.
The \gls{fpd} background rate was measured before and after \gls{gkrs} application: \SI{345+-5}{\milli cps} and \SI{348.4+-2.3}{\milli cps}, respectively, indicating that the contamination of the \gls{katrin} setup by \rb{} is unlikely.

After stopping both the generator pumping and zeolite bead baking, a fast growth of \kr{} was observed, followed by a decrease with the \rb{} half-life in just \SI{18}{\hour}. This confirmed the expected half-life behavior.
In terms of emanation, it was found that \SI{\sim 81}{\percent} of \kr{} produced in \rb{} decay emanates into the vacuum if the zeolite source was baked at \SI{200}{\celsius} and pumped for \SI{3}{\hour}.

Further details on the development and the properties of the \rb{}/\kr{} zeolite source and \kr{} generator can be found in \cite {Venos2014} and \cite{Sentkerestiova2018}.

\subsubsection{CERMAX lamp} 

The CERMAX lamp as described in \secref{sssec:rw_uv} was installed at the \gls{katrin} experiment in early 2019. First tests show that the setup is working well under the given conditions. When the lamp is operated with \SI{1000}{\watt} of electrical power, the photo-current at the \gls{rw} is on the order of \SI{1.1}{\micro\ampere}, which corresponds to \SI{6.9e12}{e^-\per\second}. This is about two orders of magnitude more than the number of \betaels{} produced in the \gls{wgts}, or one order of magnitude more than the total electron rate in the \gls{wgts} (\betael{} + secondaries). In order to check that the required homogeneity of  \SI{+-10}{\percent} over the whole \gls{rw} can be achieved with the final setup, a measurement of the photoelectron distribution is necessary. However, due to the high rate, the spatial distribution cannot be measured with the \gls{fpd} easily.

In contrast to the \gls{fpd}, the \gls{fbm} was designed to monitor the \gls{wgts} source stability and can handle high rates better. It has an energy threshold of about \SI{5}{\kilo\electronvolt}, but low-energy photoelectrons can only be accelerated up to \SI{500}{\electronvolt} by applying a corresponding voltage to the \gls{rw}. While this makes the \gls{fbm} unsuitable for detecting individual electrons at low energies, it can still be used for homogeneity measurements because the high rate of electrons affects a change in the DC-offset of the \pin diodes.

\begin{figure}[!ht]
	\centering
	\includegraphics[width=0.8\textwidth]{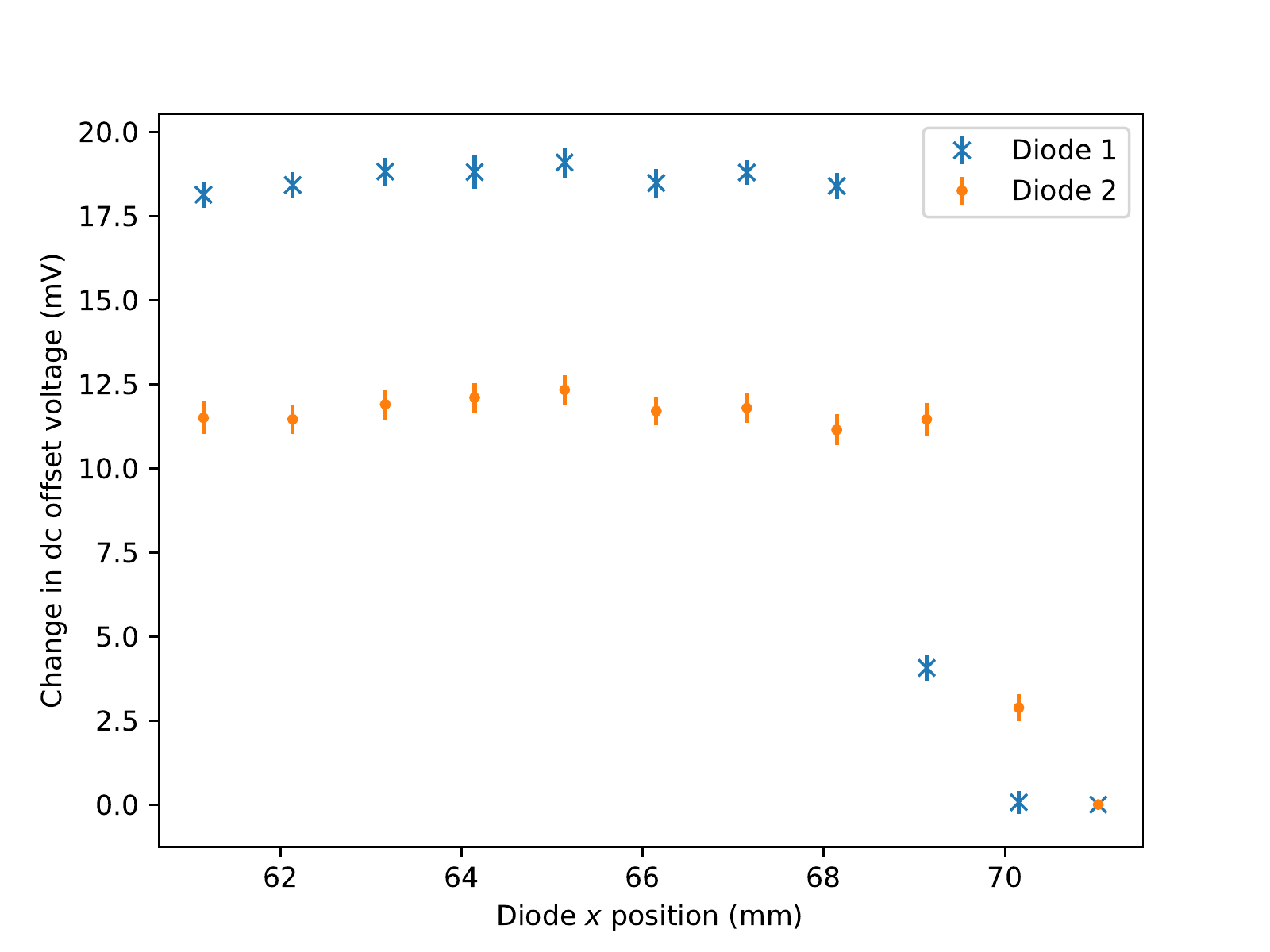}
	\caption{ An \gls{fbm} scan near the edge of the \gls{katrin} flux tube with photoelectrons from the UV-illuminated \gls{rw}. Only the change in the DC offset voltage is plotted relative to the value outside of the flux tube. }
	\label{fig:fbm-cermax-scan}
\end{figure}

A first proof-of-principle test has been performed and the result can be seen in \figref{fig:fbm-cermax-scan}. A change of offset voltage on the order of \SIrange{10}{20}{\milli\volt} is observed, depending on the diode, when the \gls{fbm} is moved into the flux tube.

This method can be used to make a full 2D scan of the flux tube, thereby investigating the performance of the UV illumination. Results from these studies are expected in the near future.

%% file: DataManagement.tex
\section{Data Management and Control Systems}
\label{sec:data_handling}
\label{Subsection:SCAndRunControl}

\subsection{Overview}
\label{Subsection:SCOverview}

\begin{figure}[!ht]
    \includegraphics[width=\textwidth]{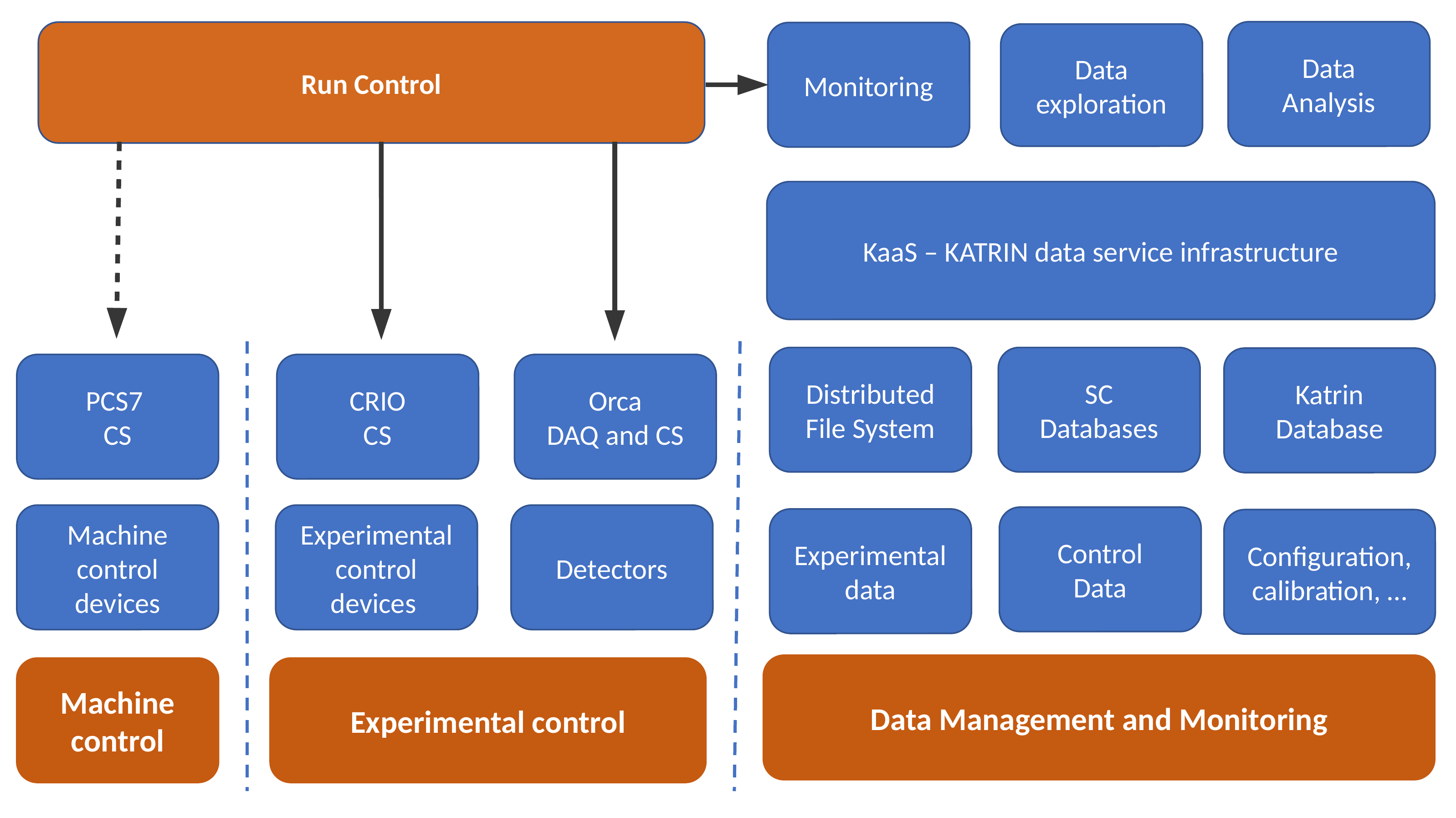}
    \caption{Overview of user interfaces (top); communication services (middle) and field-level devices and data storage (bottom) in \emph{machine control}, \emph{experimental control} and \emph{data analysis system}.  {\bf fig:data-overview}}
    \label{fig:data-overview}
\end{figure}

The complex \gls{katrin} experimental system is integrated in a modular and hierarchical data and control infrastructure. The requirements of this infrastructure are threefold. The system has to ensure the safe operation for personnel and the experimental system; it needs to be flexible in enabling smart control of all physically relevant parameters, and it must record in detail the status of the entire setup for data analysis and early data quality checks.
In order to realize these requirements, the data handling and control was split into three systems called \emph{machine control}, \emph{experimental control} and \emph{data analysis}. The main elements of all three systems are shown in Figure \ref{fig:data-overview}. In all three systems, user interfaces are connected to field-level devices and data storage via dedicated communication layers.

The \emph{machine control} system is a closed system that ensures the highest safety requirements; for example, all tritium components need to fulfill strict tritium safety regulations. But there are also other critical components, such as the high voltage system or the heating and cooling system, whose failure can cause severe harm to operators or damage to the experimental system. All these components are integrated into the machine control system, or they can be deactivated by safety interlocks.

The \emph{experimental control} system has been designed to give scientists a comprehensive overview of the experimental status, realize programmable control of any experimental control parameter, and read data from the central and auxiliary detector systems.
An essential feature of the experimental control system is the uniform communication and data handling layer that extends to all integrated systems. Connected components at the user side are status displays and monitoring screens, data archival tools, and the run control system. Detector systems and monitoring and control devices are integrated on the field-level side.
Where safety protocols need to be enforced, an interlock controlled by the machine control system is realized. Data from all devices, detectors, and the machine control system are stored in the central data storage.

The \emph{data analysis} system is a framework tailored to the needs to process experimental data automatically and to perform rapid data evaluations. The analysis system is independent of the two control systems: All recorded data is replicated from the data storage located in the experimental control system.
The automatic data analysis tasks process control system data and combine it with detector data. All this data is available for further analysis by a dedicated data infrastructure consisting of several databases and web services. Access to the experimental data is provided via web front-ends used for manual monitoring, and a set of software libraries that allow analysis tools to process run and slow-control data.
Data storage and computing resources of the analysis system are embedded in a cloud platform called \gls{kaas} to ensure reliability and scalability of the analysis services.

To ensure a reliable connection between the control systems and the integrated field-level devices, a redundant local network ring is constructed around the experimental site. In total 24 optical fibers are available to realize 12 independent network segments. This network infrastructure is used for machine control and experimental control.

\begin{figure}[!ht]
    \includegraphics[width=\textwidth]{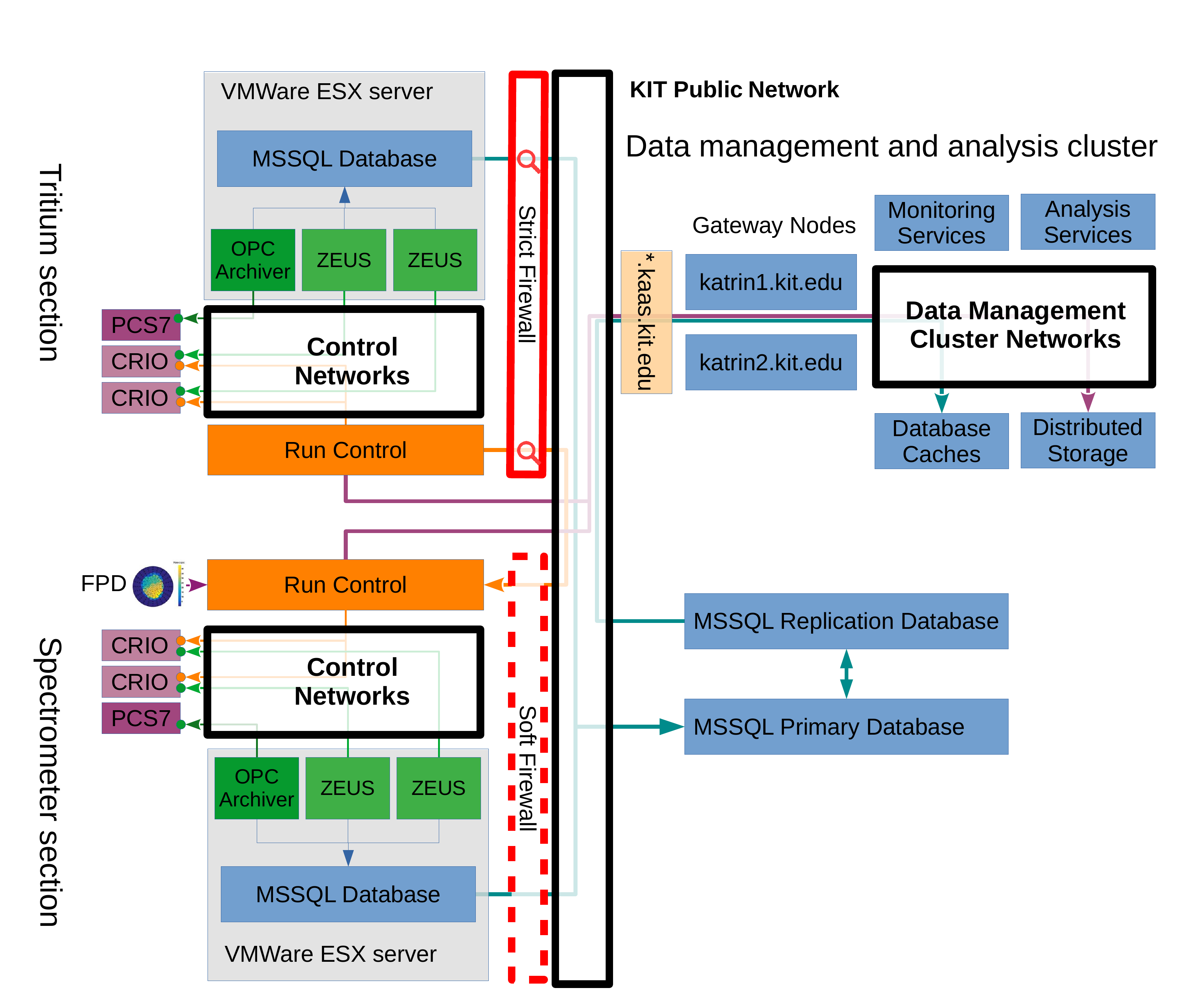}
    \caption{Simplified scheme of \gls{katrin} control networks and data flow. \textbf{fig:katrin-network}}
    \label{fig:katrin-network}
\end{figure}

\figref{fig:katrin-network} illustrates the \gls{katrin} control networks and the data flow in the archival system. A significant part of the infrastructure runs in dedicated virtual machines on a single powerful server equipped with VMWare ESX virtualization stack. The virtual machines include the primary slow control database, \gls{zeus}, and \gls{opc} archiving appliance. The strict separation of the \gls{katrin} experiment in two segments inside (see sections \ref{sec:source_system}, \ref{sec:transport_system}, \ref{SubSection:GasCompositionMonitoring}, \ref{Subsection:ActivityMonitoring},  \ref{SubSection:IonBlocking}, \ref{sec:electron_sources}) and outside (see sections \ref{sec:spectrometer_system}, \ref{sec:detector_system}, \ref{SubSection:SpectrometerMagneticFieldMonitoringSystem}, \ref{SubSection:MonitorSpectrometer}) of \gls{tlk} required the doubling of the control infrastructure. These are two identical subsystems, where one monitors components in the tritium part and the other is responsible for non-tritium devices. Both systems archive the data and provide access to control variables for the \gls{katrin} run control system. The control access is restricted to the non-tritium sections of the \gls{katrin} experiment. The archived data is available for monitoring purposes within the local \gls{katrin} experiment network but is also immediately transferred to the main \gls{katrin} data servers in the data center using replication mechanisms integrated into Microsoft SQL Server.

Section \ref{Subsection:machine_control} describes the machine control system.
Section \ref{Subsection:experimental_control} presents the experimental control system, including monitoring, run control, archival, and the data and communication infrastructure.
Section \ref{Subsection:data_analysis} introduces the concepts and realization of the analysis system, the \gls{kaas} cloud infrastructure, data organization, automatic data processing, and the analysis tools.

\subsection{Machine control system}
\label{Subsection:machine_control}

The machine control system can be operated by on-site experts only and controls critical hardware such as beamline valves and cryogenic components. The system consists of two parts: source and transport section (tritium components), and spectrometer and detector section (non-tritium components).

For both components, a custom Siemens \gls{pcs7} library has been developed. It meets the stringent regulations of nuclear facility operations in Germany. In order to reduce the number of software environments and to have the same look-and-feel for the operators, the same library was also used for the non-tritium part of \gls{katrin}.
The machine control system integrates all tritium source components, and the spectrometer vacuum and heating/cooling system. These components are typically operated in a static mode and do not require scripting. The operation of these components is limited to access by dedicated \gls{pcs7} stations in the \gls{katrin} control room. Remote operation is not allowed.

The interface of the machine control system is simplified to guarantee the safe operation of the devices in machine control. Still, a few interfaces are necessary for the normal operation of \gls{katrin}. This exception allows for data archival and control by scripting of certain devices. These devices need to be synchronized with the experimental control system, and at the same time must be supervised because of safety requirements.

An example of safety-critical systems that require experimental control is the high voltage (HV) system (\secref{sec:SDS_HV}). The experimental control system (\secref{Subsection:experimental_control}) is used to define the setpoints of individual voltages, usually by predefined scripts in the run control system.
However, an emergency shutdown of the HV system components could be triggered by the machine control system. For instance, the HV is turned off in case of a sudden pressure rise in the vacuum system. The control system therefore continuously evaluates the global failure status and deactivates HV operation and the experimental control of the HV system if necessary.

All sensor data are recorded in an internal \gls{pcs7} database and are archived by the \gls{katrin} data management system via an \gls{opc} server.
This information from the machine control system is made available to the experimental control system for monitoring and data analysis. An interface is integrated into the data service layer of the experimental control system.
In total, \num{2752} sensors in \num{108} groups are integrated in the machine control system. The typical sampling rates are in the range of \SIrange{1}{10}{\second}.

\subsection{Experimental control system}
\label{Subsection:experimental_control}

The experimental control system is responsible for defining the experimental parameters and for acquiring all relevant information for data analysis. The operation of the \gls{katrin} experiment is carried out by automated control scripts and, if necessary, can be manually adjusted by control room operators.
The system includes devices for the monitoring and control of \gls{katrin} instrumentation and the data archival system (slow control system). It also includes the run control system, which manages the detector readout and controls certain experimental parameters that need to be adjusted dynamically.

The main experimental devices are calibration sources, the monitor spectrometer, high voltage system, \gls{fbm}, the air-coil system, and the detector system. Most of the slow control systems are implemented in LabView using \gls{crio} hardware from National Instruments. In total, \num{54} devices with a total of \num{2994} sensors are connected to the system. Typical sampling rates are in the range of \SIrange{1}{10}{\second}. 

The monitoring and control systems are integrated into the data management infrastructure. The status of all affiliated devices from both the experimental and machine control sections is continuously recorded in the slow control database. The data logging is performed using a legacy \gls{zeus}~\cite{Amsbaugh2015} control system for \gls{crio} devices.
The parameters of \gls{pcs7} subsystems are queried using \gls{opc} protocol by a special archiving appliance. In both cases, Microsoft SQL Server instances are used to store the data (see \figref{fig:data-sc-overview}). \gls{crio} devices with adjustable parameters are also integrated with the run control system using a standard network connection and a custom TCP-based protocol. No experimental control is foreseen for \gls{pcs7} devices. The parameters of the system can only be changed from within the restricted local network. The values can be changed either manually by control room operators using the implemented \gls{hmi} interfaces, or through the run control system with automated scripts.  

The run control is implemented using the \gls{orca} \gls{daq} software \cite{Howe2004}. Scripting and graphical interfaces enable the control of the detector system and the configuration of experimental parameters.
All current and historical data stored in the \gls{katrin} data management system is available to \gls{orca} through dedicated \gls{rest} services. To adjust slow control parameters, a special \gls{orca} object implements a TCP server that supplies the configured parameter set values to the \gls{crio} systems upon request.
In addition, dedicated objects in \gls{orca} communicate directly with some devices integrated into the experimental control system, such as the high voltage system and the detector front-end electronics. Parameters of these objects are controlled in run scripts, e.g. the \gls{ms} voltage to perform a neutrino mass scan.

The run files acquired during the experiment are first stored locally and are then transferred to the data management cluster with \texttt{rsync}\footnote{\url{https://rsync.samba.org}}. This step is repeated in regular intervals of \SIrange{1}{15}{\minute}.

\begin{figure}
    \includegraphics[width=\textwidth]{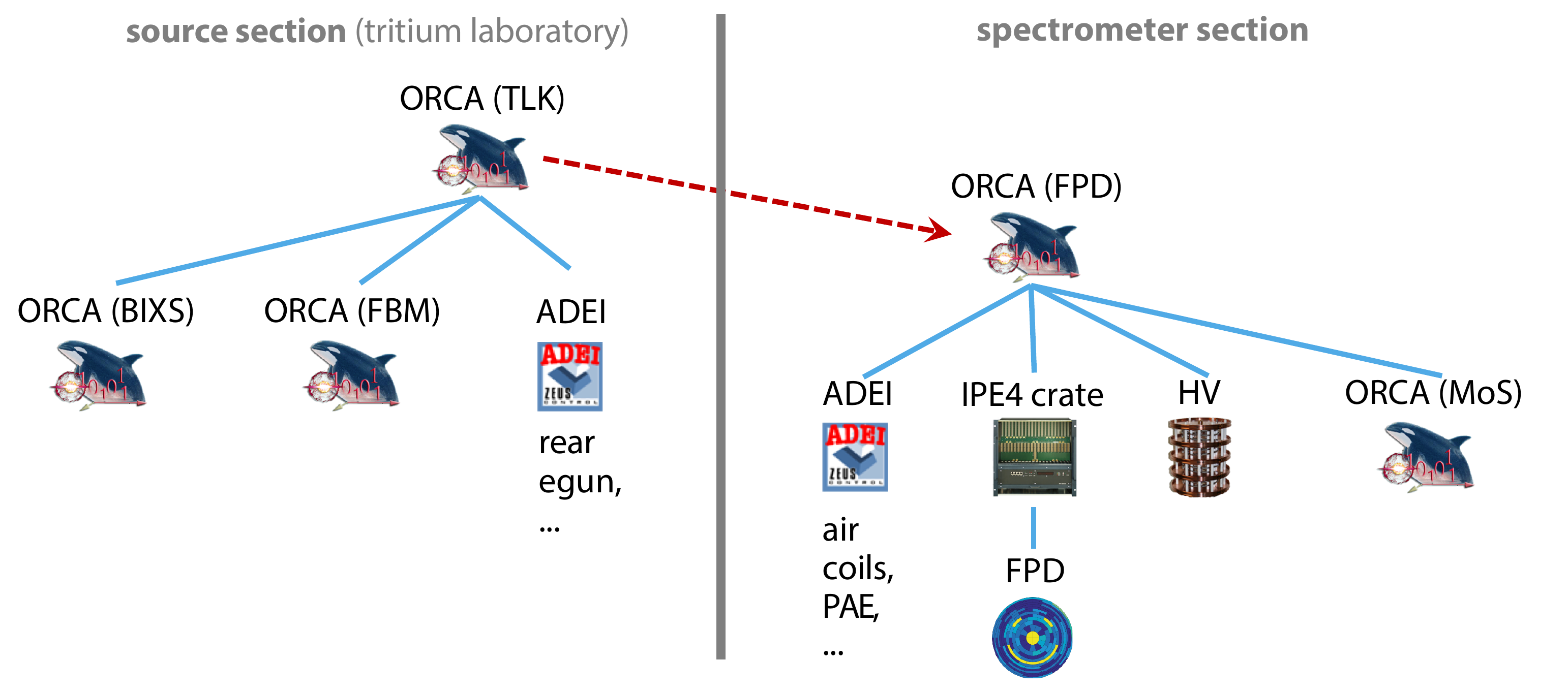}
    \caption{Hierarchical setup of the run control system. It integrates devices from the tritium and the non-tritium parts of \gls{katrin}. \textbf{fig:ORCA-Run-Control}}
    \label{fig:ORCA-Run-Control}
\end{figure}

In order to access all relevant systems of the \gls{katrin} experiment, a hierarchical run control setup has been realized as shown in \figref{fig:ORCA-Run-Control}. The \gls{orca} DAQ system at the \gls{fpd} takes detector data and enacts the voltages at the main spectrometer via a direct connection to the high voltage system.
The \gls{fpd} can be remotely controlled from a source section \gls{orca} system in the \gls{tlk} for dedicated measurements. The \gls{tlk} system also controls other detector and control systems inside the \gls{tlk}. The \gls{fpd} system controls the \gls{mos} system during neutrino mass runs to coordinate data acquisition.

\subsection{Data management, monitoring, and analysis system}
\label{Subsection:data_analysis}

\begin{figure}[!ht]
     \centering
    \includegraphics[width=.9\textwidth]{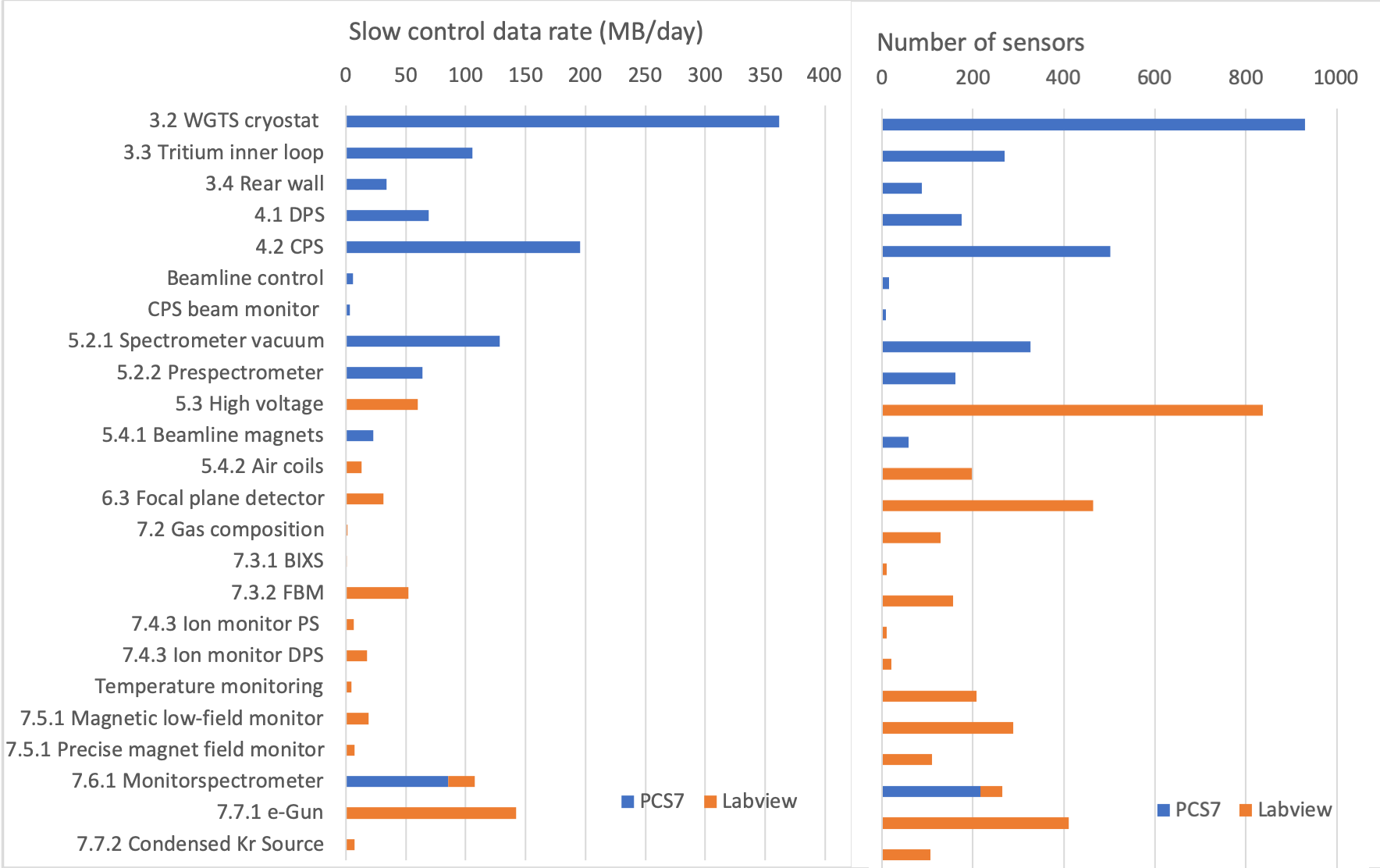}
    \caption{Overview of the partitioning of machine and experimental control system. For each component described in the previous sections, the number of slow control sensors and the typical data rate are given. {\bf fig:data-sc-overview}}
    \label{fig:data-sc-overview}
\end{figure}

The design of the \gls{katrin} data management system uses an on-premise cloud architecture aimed to simplify maintenance and ensure resilience against both hardware and software failures. The system is characterized by a high intensity of data flow and a relatively large archive of historical data.
The three main data sources are the machine control system, experimental control system, and the \gls{fpd}. About \num{6000} sensors are continuously monitoring the operating parameters of all components of the \gls{katrin} infrastructure.
Both machine and experimental control systems contribute about equally to the total data rates and the number of connected sensors, as shown in \figref{fig:data-sc-overview}. The cumulative bandwidth from sensor data is roughly \SI{6}{TB} per year.
The run data recorded at the \gls{fpd} and \gls{mos} detectors are stored in a distributed file system. The data is taken in the raw \gls{orca} data format and then converted to ROOT \cite{Brun1997} files by automatic processing scripts. For other detector systems, the data is processed in a similar manner. The combined average data rate is about \SI{10}{TB} per year and expected to grow while the experiment is running. By the end of 2020 the experiment had accumulated a total data volume of about \SI{50}{TB}.

\begin{figure}[!ht]
    \centering
    \includegraphics[width=\textwidth]{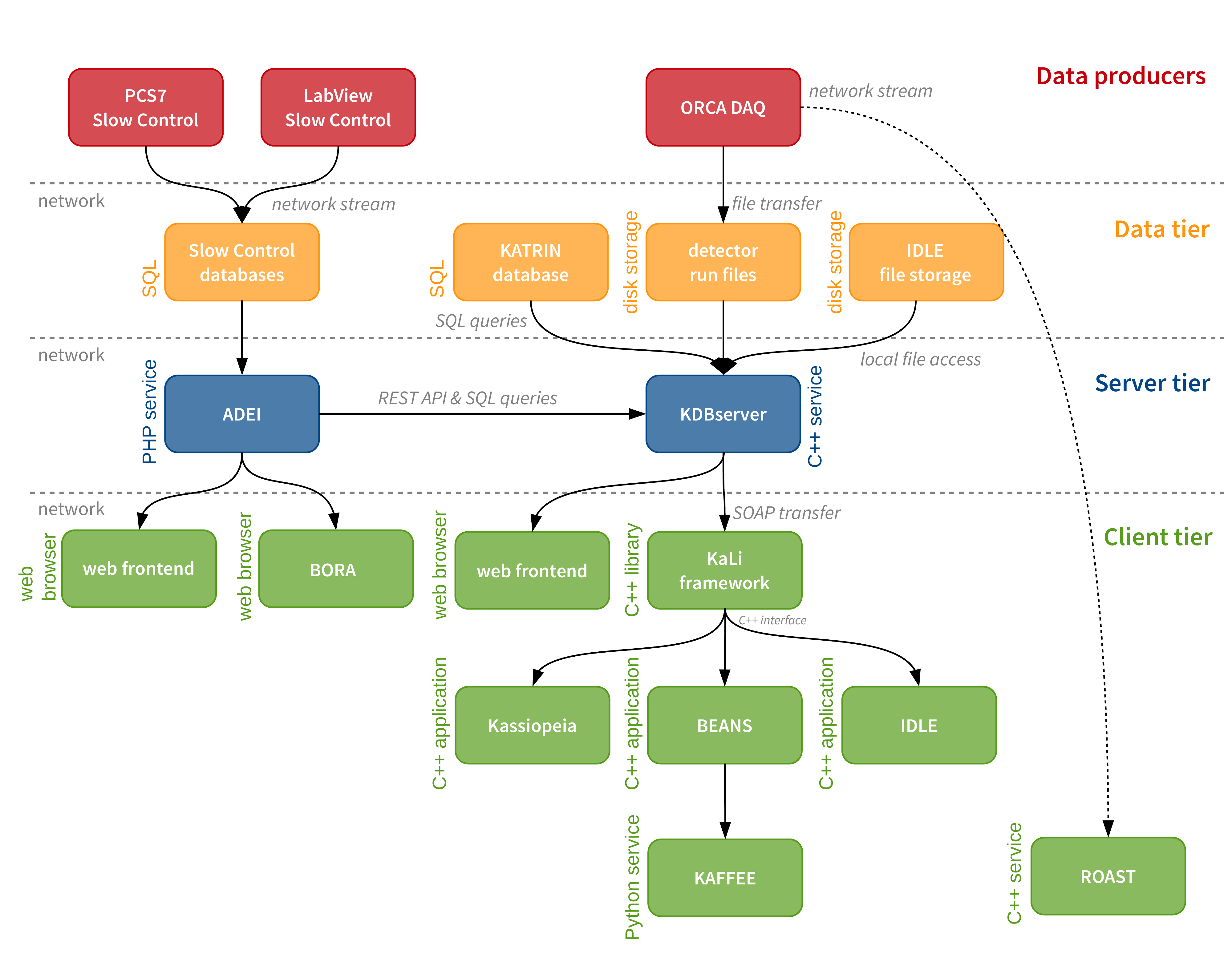}
    \caption{Data handling components, data layers, format and protocols. \textbf{fig:data-logic-tiers}}
    \label{fig:data-logic-tiers}
\end{figure}

The architecture of the analysis platform is show on \figref{fig:data-logic-tiers} and is described in detail in the following subsections.

\subsubsection{KaaS cloud platform}

The \gls{kaas} platform implements the data management and analysis cluster (\figref{fig:katrin-network}), providing both database functionalities and applications for data access. Some of these applications are further detailed in the following sections.

\gls{kaas} is operated on top a 3-node cluster and is based on the RedHat OpenShift\footnote{\url{https://www.openshift.com/}}~3.7 orchestration technology. 
The file storage is realized using the RedHat GlusterFS distributed file system, where 3-way data replication is used to ensure data safety. The slow control data is stored in Microsoft MSSQL servers with 2-way replication to increase both data redundancy and performance.
The software running in the OpenShift cloud consists of several distinct subsystems and includes both offline data processing tasks and online web services to handle user requests. The offline tasks are generally used for automatic data processing, quality control, and data compression. For instance, all incoming data from the slow control systems are validated against configured critical thresholds.

To provide users with quick overviews of the system behavior over extended periods, the slow-control data is continuously aggregated to generate minutely, hourly, and daily time-series. These generated statistics are cached in the replicated MySQL database and are used by the \gls{adei} platform \cite{Chilingaryan2010} for real-time visualization of data which may span multiple terabytes in the raw form. 
\gls{adei} allows users to explore, manage, and visualize large archives of time series. It also offers a \gls{rest} \gls{api} to provide structured access to slow control data for all other system components. The \gls{bora} framework is used to build status monitoring displays for \gls{katrin} operators. Currently, more than 20 status displays periodically query data from \gls{adei} to display different aspects of the experiment.
The monitoring systems generate a significant load on the system, producing several hundred queries per second with an average data rate of roughly \SI{20}{MB/s} from the caching database. The \gls{kdb} server integrates access to slow control and detector data for data analysis and provides a web interface to view the integrated data sets. Due to its role in the analysis of neutrino-mass runs, \gls{kdb} experiences uneven, large load spikes after major data-taking campaigns.

The OpenShift infrastructure allows us to reliably run a large collection of heterogeneous components developed by the international collaboration over more than a decade. These applications often depend on conflicting versions of software components, including deprecated and sometimes unstable versions.
The container technology allows us not only to isolate the environment of each component but also to limit the impact that failures in one component might have on others and on overall system stability. The OpenShift system further provides high availability and scalability. Failed components are restarted automatically, and running components can be easily migrated in case of failing nodes in the cluster.

\subsubsection{ADEI data management platform}

The \gls{katrin} control systems integrate custom and commercial components from various vendors. Many different data formats, underlying storage engines, and workflows are used to store the data. The \gls{adei} platform integrates all available data sources and makes them available to users in a uniform, comprehensible, and easy-to-use fashion~\cite{Chilingaryan2010}.
It includes a layer of abstraction to ensure a stable \gls{api} over the course of the experiment and the development of its analysis, regardless of changes to the data format, the number of data streams, or the underlying databases. \gls{adei} is also fully integrated with the OpenShift cloud platform to provide a highly available service to users and to auto-scale under increased load.

\begin{figure}[htb]
    \includegraphics[width=.9\textwidth]{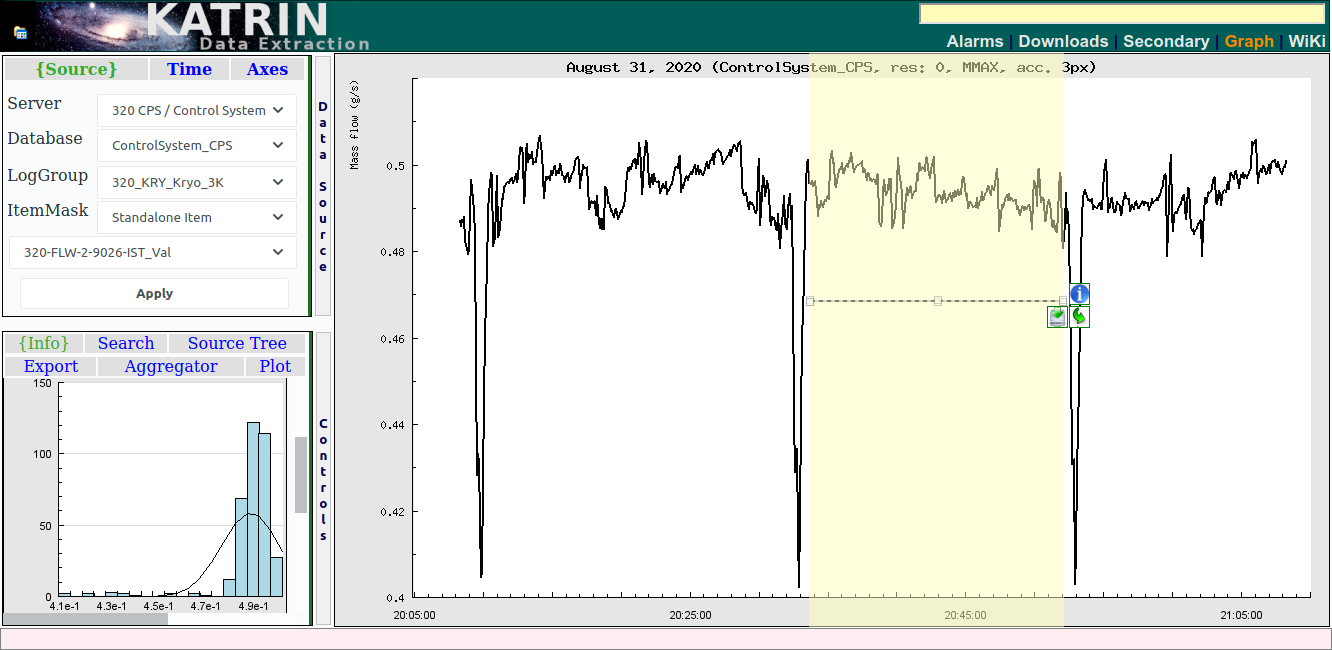}
    \caption{\gls{katrin} data web portal for near-time access to recorded slow-control data. The user can select a time interval and value range graphically in the display plot area (yellow box). \textbf{fig:data-adei}}
    \label{fig:data-adei}
\end{figure}

The most prominent implementation of the \gls{adei} web services is the interactive \gls{katrin} data portal, which provides graphical access to data from all slow control systems. \figref{fig:data-adei} shows a screenshot of this portal and its hierarchical list of the available data sources. The user can select a time interval and value range graphically in the display plot area.
To improve performance, we use intelligent aggregation techniques to distill a few thousand data points from the millions that may span the time interval. As a result, the complete plot-generation time usually does not exceed \SI{500}{\milli\second}. Every generated view of the data portal can be retrieved with a unique, persistent reference that can be included in documentation or electronic communication.

\subsubsection{BORA status monitoring framework}

To simplify the overall architecture, the monitoring subsystem is designed as part of the data management infrastructure and executed on the \gls{kaas} platform. Data from a variety of \gls{katrin} components is queried using \gls{adei}  interfaces. The introduced latency of typically \SIrange{10}{20}{\second} from this approach is within the requirements for our monitoring systems.

\begin{figure}[htb]
    \includegraphics[width=.9\textwidth]{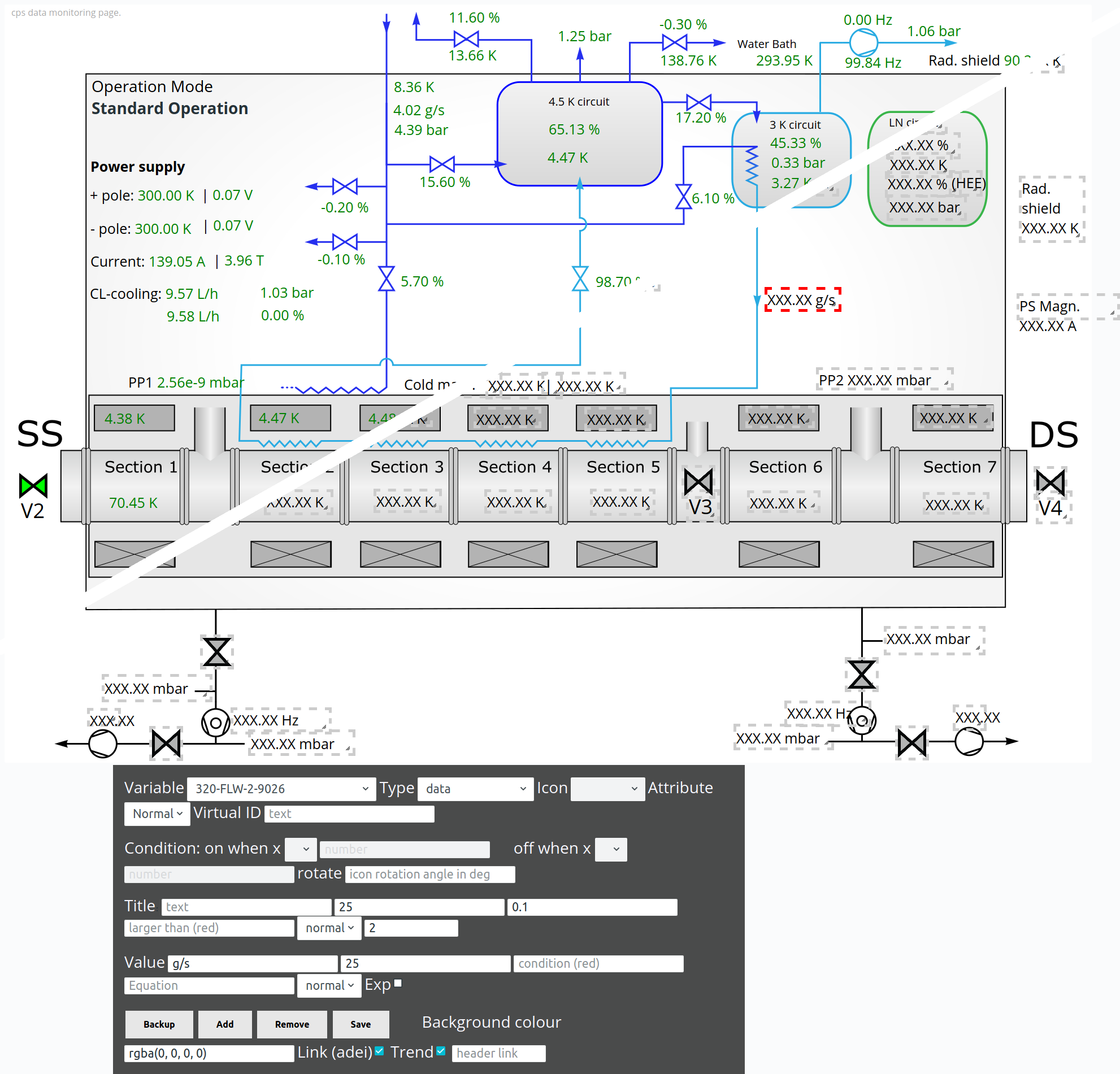}
    \caption{\gls{bora} status display for \gls{adei}; an operation example for the \gls{cps}. Composite picture of data display (upper left) and design mode (bottom right). \textbf{fig:data-bora-status}}
    \label{fig:data-bora-status}
\end{figure}

Due to the complexity of the \gls{katrin} setup, individual status displays are required for each component. Different modes of operation often require adjusted displays for the same subsystem. The \gls{bora} framework is provided to simplify the design and configuration of status displays. It offers graphical design tools to associate sensors in \gls{adei} with a configured visual representation, which can be viewed by users in their web browser.
For each display, several hundreds of sensors can be selected from the sensor inventory and placed as needed on the process image. \figref{fig:data-bora-status} show a \gls{cps} status page in the designer view and in operation. Once configured, the status is updated every \SI{10}{\second} and is available to the control room operators, but also to off-site collaborators via internet access.

\subsubsection{Processing of run data}

Measurement runs are taken with the \gls{orca} software at the \gls{fpd}, the \gls{mos} and the \gls{fbm} sites\footnote{In 2019, the FBM switched from \gls{orca} to a different data acquisition system.}. The data is first stored locally on the \gls{daq} machine and then transferred to the \gls{kaas} cluster for automated processing and data distribution.
Here a processing chain is implemented where the run files are converted into the ROOT~\cite{Antcheva2009} data format and registered in a dedicated database~\cite{PhDKleesiek2014} to make these runs available for further analysis.

The run processing works as follows:
\begin{enumerate}
    \item The run files are transferred to the cluster via the \texttt{rsync} software and stored on a distributed filesystem. This step is repeated in regular intervals to synchronize with the individual remote detector systems.
    
    \item Each run file is then converted from the raw \gls{orca} file format into a ROOT file, which stores all data fields in a tree structure where they can be easily accessed. Most runs are taken in so-called energy mode, where one entry corresponds to a single electron event; other data structures are used in some cases (see \tabref{table:DAQModes}). During the conversion process, the events are also sorted by acquisition time and the data is checked for consistency. This step improves the performance of user-space analysis tools, which typically iterate over detector events.
    
    \item \gls{orca} assigns sub-run numbers to specific time intervals of programmable length within a run. Sub-runs are typically used to represent a retarding voltage step during the \gls{katrin} $\upbeta$-scans. The ROOT files are indexed on the sub-run level to improve performance when accessing data during analysis. Indices are added for the first and last events of a given type in each sub-run, allowing fast access to the corresponding sub-set of events in an entire run.
    
    \item The processed runs are added to a central \gls{sql} run database that is part of the \gls{kdb} server. Each run and sub-run is added to a data table that contains the run number, start and end time, and other parameters. The corresponding run files (in \gls{orca} and ROOT formats) are transferred to a central location on the distributed filesystem, where they can be accessed by users through the \gls{kdb} interface.
    
    \item A web service provides access to the run database and file storage. Analysis tools can look up and access any processed run. When an analysis is carried out, the corresponding ROOT file is downloaded to the local system where it can be accessed by the user. Checksum and caching mechanisms ensure data consistency and reduce unnecessary server load.
\end{enumerate}

\subsubsection{Databases for run and sensor data}

In addition to the run data, there is also data from various sensor and monitoring systems which is taken continuously when the experiment is running. Examples are temperature and pressure readings, electric currents and voltages, or measurements from magnetic field sensors. This slow-control data is stored into \gls{sql} database tables and can be accessed through \gls{adei} interfaces.

To make this data available to analysis tools, an intermediate database layer is added in \gls{kdb} to map individually named sensors to the corresponding table of the slow-control database~\cite{PhDKleesiek2014}. This is necessary because the data paths in \gls{adei} can change with the experimental setup, while the analysis tools should not be aware of such low-level modifications.

Sensors and components are therefore referenced by a static label, the so-called \gls{katrin} number, which is constructed with several groups and generally follows the format \texttt{NNN-XXX-M-nnnn-ZZZZ}. The groups are: \texttt{NNN} a 3-digit number referencing a sub-system (e.g. 200 for the \gls{wgts}); \texttt{XXX} a 3-character string for the sensor type (e.g. RTP for a temperature sensor); \texttt{M} a number for safety information (e.g. 3 for parts of the main beam line); \texttt{nnnn} a 4-digit number referencing an individual sensor system; and \texttt{ZZZZ} a 4-digit number referencing an individual channel of the sensor. For example, the reading of the central temperature sensor at the \gls{wgts} beam tube has \gls{katrin} number \texttt{200-RTP-3-5112-0001}.

The sensor database is implemented with \gls{sql} and part of \gls{kdb}, along with the run database explained earlier. The advantage of this system is that it provides a consistent mapping, which can be adjusted in case the structure of the underlying slow-control database changes, for example, when new sensor channels are added after hardware upgrades or when sensors are regrouped. This applies both to LabView and \gls{pcs7} sensors.
Furthermore, the database allows the definition of calibration functions for each mapped sensor channel, and analysis tools can choose between raw or calibrated data depending on their specific use case. Additional database tables exist for additional parameters of a sensor, such as geometrical position or alignment (useful for the definition of the simulation model) or definitions for automated data quality filtering of measurements.

The organization of this data in a central database provides a consistent access method for the analysis tools and allows the sensor mappings, calibration curves, and other content to be updated as necessary. To provide a means of consistency, all database entries are defined with an \emph{entry timestamp} (when they were added to the database) and a user-defined \emph{validity start timestamp}. A given entry is valid when the current time is equal to or later than the validity start timestamp. If multiple entries exist for this criterion, the entry time is used to select the most recent entry. Thus, only one valid entry can exist in the database at any given time.
The database does not allow to change existing entries, and in order to update the database, a new entry must be created with the same or newer validity start timestamp. By this mechanism, it is ensured that existing entries and their validity time intervals are not modified accidentally, and all database changes can be traced back over time.

Access to the database is provided through the \gls{kali} software \cite{PhDKleesiek2014}.
It is written in C++ and provides an object-oriented class structure that resembles the database contents; for example, there are classes that correspond to runs, sub-runs, sensor data, and so on.
When the data is accessed by analysis tools, it is first serialized at the web service layer into an \gls{xml} data stream via the \gls{soap}~\cite{Engelen2008} framework. The stream is then transferred to the client over an HTTP connection, and de-serialized on the client side into C++ class objects that are then further processed by the analysis software. The implementation using \gls{soap} allows for consistent data handling on the server and user side by the same software framework.
In addition, a web interface for managing database contents is implemented using the Wt library\footnote{\url{https://www.webtoolkit.eu/wt}}. This interface can be used to list and update database contents such as sensor mappings or calibration functions. Additional user-space applications exist to quickly list information from the database, for example, to view sensor data over the period of a given run.

\subsubsection{User-side analysis tools}
\label{Subsection:analysis_tools}

The user-side analysis is divided into two layers. The first layer analyzes event-level and time-series level data and digests them into tables, histograms, graphs, and/or reduced data sets in \gls{json} (or equivalent formats) to be further processed by the second analysis layer. Typical second-layer analyses include fitting and Python-based scripting on data graphs and histograms.
In between the analysis layers lies the \gls{idle} software, which facilitates storing, managing, and distributing user analysis outputs. Some simulation outputs are also stored in \gls{idle}. The \gls{idle} storage and database system is implemented as part of \gls{kdb}.

A core tool in the first-layer analysis is the \gls{beans} C++ library that reads \gls{orca} event data (from an ORCA-ROOT file) and slow control time-series data from \gls{kali} and produces analysis results in a ROOT file and/or in \gls{json} (or equivalent format) document.
Analysis logic in \gls{beans} is described as a linear chain of unitary analysis actions, following concepts similar to the functional programming paradigm\footnote{Functional programming is a declarative programming paradigm where programs are constructed by expressions that each return a value, rather than by a sequence of imperative statements that change the internal state of the program.} as opposed to imperative programming, which is common in nuclear and particle physics data analysis with C++ \cite{hughes:matters-cj}. This allows users to quickly construct analysis programs out of existing elements, and at the same time, it serves as an intuitive framework on which analysts can collaborate and exchange their developments.

User analysis logic in \gls{beans} can be constructed without C++ coding. Instead, a configuration file in \gls{json} (or equivalent format) is used to implement an identical analysis chain. A web-based GUI tool (VisualBEANS) to build \gls{beans} configuration files is provided to quickly construct analysis scripts without coding and compiling. As a dedicated platform software, \gls{roast} connects \gls{beans} scripts and real-time data broadcasting from \gls{orca} systems.

By a combination of these tools, various analysis scripts could be developed quickly on-site during normal operation. They are extensively used for near-time and real-time analysis and monitoring of \gls{katrin} commissioning runs.

Routine user analysis tasks are automated by the \gls{kaffee} software, originally developed for \gls{fpd} commissioning. \gls{kaffee} automatically identifies the types and contents of the data files, and automatically runs analysis programs according to the configuration. For example, it generates reduced data files (so-called ``Run Summary Documents'') for tritium runs, applies quality filtering according to a list of measurement types, and performs automated analysis for detector calibration runs.
The output files of the analyses are then scanned by \gls{kaffee} and indexed in a document-oriented database that serves as an analysis catalog. A web interface is provided to browse the catalog contents, along with quick interactive analysis functions such as histogram fitting in the web browser.

The Run Summary Documents, which are \gls{json} (or equivalent format) documents primarily used as input to tritium spectrum fitting, contain highly digested (sub-)run data such as the detector counts with efficiency corrections, the retarding voltage, the source temperature, tritium pressure, gas throughput, isotopologue composition provided by \gls{lara}, and so on.
These are typically produced by \gls{kaffee} automation with \gls{beans} scripts, and stored in \gls{idle}. Some simulation outputs and calibration analysis results used for tritium spectrum fitting, such as field maps of the analyzing plane or the column density determined from routine calibrations, are also stored in \gls{idle} in the same format.
\gls{idle} implements file version control with validity periods and file access permission levels. Data blinding to tritium runs makes use of these control features. The files stored in \gls{idle} are distributed through \gls{kali}, taking advantage of \gls{kali}'s cache mechanism.

\clearpage

%% file: Summary.tex
\section{Summary}
\label{sec:summary}

This paper outlines the technical challenges overcome and the milestones achieved in getting the experimental setup ready to take data.
All systems, including monitoring and calibration subsystems, necessary for \gls{katrin} neutrino mass measurements are complete and have passed (or surpassed) their design requirements.

Since work on this paper began, the \gls{katrin} experiment has published results from a number of successful measurement campaigns~\cite{Arenz2018,Aker2020_firstOperationTritium,gkrs2019} and established a new, improved upper limit on the neutrino mass~\cite{Aker2019-PRL}.

And while the possibility for future upgrades is left open, this paper represents a snapshot of the state of the \gls{katrin} experiment, as of the first neutrino mass campaign: ready for operation, ready for data.